\documentclass[rmp,aps,nofootinbib,twocolumn]{revtex4}

\usepackage{graphics}
\usepackage{epsfig}

\usepackage{amssymb,bm}

\usepackage{color}

\newif\ifrandbem
\randbemtrue
\ifrandbem
\newcommand{\randbem}[1]{\marginpar{\small\raggedright \bf #1}}
\else
\newcommand{\randbem}[1]{}
\fi

\newcommand\alf{\alpha}
\newcommand\eps{\epsilon}

\newcommand\Gam{\Gamma}
\newcommand\lam{\lambda}
\newcommand\Lam{\Lambda}
\newcommand\om{\omega}
\newcommand\sg{\sigma}
\newcommand\Sg{\Sigma}

\newcommand\bG{{\bf G}}
\newcommand\bGam{{\bf\Gam}}

\newcommand\bk{{\bf k}}

\newcommand\bp{{\bf p}}
\newcommand\bq{{\bf q}}
\newcommand\bQ{{\bf Q}}
\newcommand\bR{{\bf R}}
\newcommand\bS{{\bf S}}

\newcommand\bSg{{\bf\Sigma}}
\newcommand\cA{{\cal A}}
\newcommand\cB{{\cal B}}
\newcommand\cD{{\cal D}}
\newcommand\cG{{\cal G}}
\newcommand\cH{{\cal H}}

\newcommand\cS{{\cal S}}
\newcommand\cV{{\cal V}}
\newcommand\cW{{\cal W}}
\newcommand\cZ{{\cal Z}}

\newcommand\psib{\bar\psi}
\newcommand\Psib{\bar\Psi}
\newcommand\etab{\bar\eta}
\newcommand\chib{\bar\chi}
\newcommand\Xib{\bar\Xi}

\newcommand\bra{\langle}
\newcommand\ket{\rangle}
\newcommand\up{\uparrow}
\newcommand\down{\downarrow}
\newcommand\sgn{\mathrm{sgn}}

\newcommand\dps{\displaystyle}
\newcommand{\sfrac}[2]{{\textstyle\frac{#1}{#2}}}

\begin{document}

\title{Functional renormalization group approach to correlated fermion systems}

\author{Walter Metzner}
\affiliation{Max-Planck-Institute for Solid State Research,
 Heisenbergstra{\ss}e 1, D-70569 Stuttgart, Germany}
\author{Manfred Salmhofer}
\affiliation{Institut f\"ur Theoretische Physik, Universit\"at Heidelberg,
 Philosophenweg 19, D-69120 Heidelberg, Germany}
\author{Carsten Honerkamp}
\affiliation{Institut f\"ur Theoretische Festk\"orperphysik and 
 JARA-Fundamentals of Future Information Technology, \\
 RWTH Aachen University, D-52056 Aachen, Germany}
\author{Volker Meden}
\affiliation{Institut f\"ur Theorie der Statistischen Physik and 
 JARA-Fundamentals of Future Information Technology, \\
 RWTH Aachen University, D-52056 Aachen, Germany}
\author{Kurt Sch\"onhammer}
\affiliation{Institut f\"ur Theoretische Physik, Universit\"at G\"ottingen,
 Friedrich-Hund-Platz 1, D-37077 G\"ottingen, Germany}

\begin{abstract}
Numerous correlated electron systems exhibit a strongly 
scale-dependent behavior. 
Upon lowering the energy scale, collective phenomena, bound
states, and new effective degrees of freedom emerge. 
Typical examples include
(i) competing magnetic, charge, and pairing instabilities 
in two-dimensional electron systems, 
(ii) the interplay of electronic excitations and order parameter
fluctuations near thermal and quantum phase transitions in metals, 
(iii) correlation effects such as Luttinger liquid behavior 
and the Kondo effect showing up in linear and non-equilibrium 
transport through quantum wires and quantum dots.
The functional renormalization group is a flexible and unbiased 
tool for dealing with such scale-dependent behavior.
Its starting point is an exact functional flow equation, which 
yields the gradual evolution from a microscopic model action to the 
final effective action as a function of a continuously decreasing 
energy scale.
Expanding in powers of the fields one obtains an exact hierarchy
of flow equations for vertex functions.
Truncations of this hierarchy have led to powerful new approximation 
schemes.
This review is a comprehensive introduction to the functional
renormalization group method for interacting Fermi systems.
We present a self-contained derivation of the exact flow equations
and describe frequently used truncation schemes.
Reviewing selected applications we then show how approximations
based on the functional renormalization group can be fruitfully
used to improve our understanding of correlated fermion systems.
\end{abstract}   

\vspace*{3mm}

\maketitle

\tableofcontents


\section{INTRODUCTION}
\label{sec:I}

\subsection{Motivation}

The Coulomb interaction between electrons in solids leads to a 
virtually unlimited variety of phenomena, such as magnetic correlations 
and magnetic order, high-temperature superconductivity, metal-insulator 
transitions, phase separation and stripes, and the formation of exotic 
quantum liquid phases. 
The latter include Luttinger liquids, quantum critical points, and 
fractional quantum Hall states.

\begin{figure}[ht]
\centerline{\includegraphics[width = 6cm]{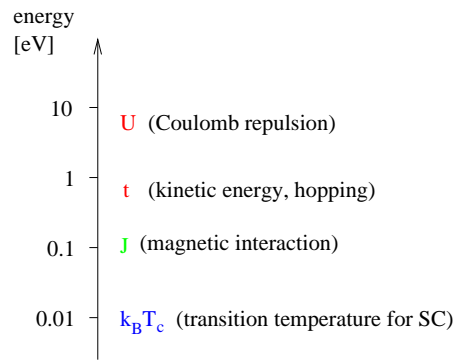}}
\caption{(Color online)
 Important energy scales in high-temperature superconductors 
 of the cuprate family. Magnetic interactions and superconductivity are
 generated from the kinetic energy (hopping) and the Coulomb 
 repulsion.}
\label{htscscales}
\end{figure}

Interacting electron systems usually exhibit very distinct behavior on
different energy scales. Composite objects and collective phenomena emerge
at scales far below the bare energy scales of the microscopic Hamiltonian.
For example, in cuprate high-temperature superconductors one bridges three 
orders of magnitude from the highest scale, the bare Coulomb interaction, 
via the intermediate scale of short-range magnetic correlations, down to 
the lowest scale of $d$-wave superconductivity and other ordering phenomena 
(see Fig.~\ref{htscscales}).
This diversity of scales is a major obstacle to a straightforward
numerical solution of microscopic models, since the most
interesting phenomena emerge only at low temperatures and in systems with
a large size. It is also hard to deal with by conventional many-body
methods, if one tries to treat all scales at once and within the same
approximation, for example by summing a subclass of Feynman diagrams.
Perturbative approaches which do not separate different scales are plagued 
by infrared divergences, and are therefore often inapplicable even at
weak coupling, especially in low dimensions.

It is thus natural to treat degrees of freedom with different energy 
scales successively, descending step by step from higher to lower scales. 
This is the main idea behind the {\em renormalization group} (RG).

\subsection{RG for interacting Fermi systems}

Renormalization group methods have a long tradition in the theory
of interacting Fermi systems.
Already in the 1970s, various versions of the RG have been used to
deal with infrared singularities arising in one-dimensional Fermi 
systems \cite{Solyom79}.
Naturally, the RG was also applied to (mostly bosonic) effective
field theories describing critical phenomena at continuous
classical or quantum phase transitions in interacting Fermi systems
\cite{Fradkin91,Sachdev99}.

Renormalization group approaches dealing with interacting fermions
in arbitrary dimensions $d$ have been developed much later.
Due to the extended (not point-like) geometry of the Fermi surface 
singularity in dimensions $d>1$, the renormalization group flow 
cannot be reduced to a finite number of running couplings.
However, the main reason for the delayed development of a 
comprehensive RG approach for interacting Fermi systems in higher
dimensions was probably not this difficulty, but rather a lack of
motivation.
The few infrared singularities appearing in three-dimensional
Fermi systems could usually be handled by simple resummations of 
perturbation theory \cite{Abrikosov63,Nozieres64}.
Triggered by the issue of non-Fermi liquid behavior in two-dimensional
systems, and the related discussion on the validity of perturbation
theory, systematic RG approaches to interacting Fermi systems in
arbitrary dimensions have been developed by various groups in the
early 1990s.

Aiming at a mathematical control of interacting Fermi systems,
\textcite{FT1,FT2}, and independently \textcite{Benfatto90a,
Benfatto90b}, have formulated a rigorous fermionic version of 
Wilson's momentum-shell RG \cite{Wilson74}.
Important rigorous results have indeed been obtained in 
one-dimensional \cite{Benfatto94} and two-dimensional 
\cite{FMRT92,FST1,FKT03,FKT04,Disertori00,BGM06} systems.
The essential message from these results is that no hitherto
unknown instabilities or non-perturbative effects occur in Fermi
systems with sufficiently weak short-range interactions, at least
in the absence of special features such as Van Hove singularities 
at the Fermi level.

The Wilsonian RG for interacting Fermi systems was popularized among 
(non-mathematical) physicists by \textcite{Shankar91,Shankar94} 
and \textcite{Polchinski93}, who presented some of the main ideas 
in a pedagogical style.
In particular, they provided an intuitive RG perspective of Fermi
liquid theory.
Subtleties associated with the singularities of the interaction 
vertex for forward scattering were clarified a bit later
\cite{Chitov95,Metzner98}.
A Hamiltonian-based RG interpretation of Fermi liquid theory was
presented by \textcite{Hewson94}, who discussed not only translation 
invariant systems but also models for magnetic impurities in metals.

As an alternative to the Wilsonian RG one may also use flow 
equations for Hamiltonians based on infinitesimal unitary 
transformations, which make the Hamiltonian successively more 
diagonal \cite{Wegner94}. 
This approach has been used successfully for quantum impurity 
models and other systems \cite{Kehrein06}.
A weak coupling truncation of the flow equations has been 
applied to identify instabilities of the two-dimensional Hubbard
model \cite{Grote02}.

There is much current interest in RG methods for correlated fermions 
in non-equilibrium. The perturbative RG \cite{Rosch01,Mitra06}, 
Wilson's numerical RG \cite{Anders05}, as well as Wegner's flow 
equation approach \cite{Kehrein06} were extended to non-equilibrium, 
and real-time RG methods were developed
\cite{Schoeller00a, Schoeller00b,Schoeller09a}.

\subsection{Functional renormalization group}

The Wilsonian RG is not only useful for a deeper and partially 
even rigorous understanding of interacting fermion systems.
A specific version of Wilson's RG known as {\em exact} or
{\em functional} RG turned out to provide a valuable framework 
for computational purposes.
Approximations derived from exact functional flow equations 
have played an increasingly important role in the last decade.
These developments are the central topic of this review.

Exact flow equations describe the evolution of a generating
functional for all many-particle Green or vertex functions
as a function of a flow parameter $\Lam$, usually an infrared 
cutoff.
They can be derived relatively easily from a functional integral 
representation of the generating functional.
Exact flow equations have been known since the early years of 
the RG, starting with the work of \textcite{Wegner73}.
\textcite{Polchinski84} employed an exact flow equation to 
formulate a relatively simple proof of renormalizability of 
the $\Phi^4$-theory in four dimensions.
Renormalizability proofs can be further simplified by using a
Wick-ordered variant of Polchinski's equation \cite{Wieczerkowski88}.

For computational purposes the exact flow equation for the 
effective action, first derived in the context of bosonic 
field theories by \textcite{Wetterich93} turned out to be most 
convenient.
The effective action $\Gam^{\Lam}[\phi]$ is the generating 
functional for one-particle irreducible vertex functions.
The latter are obtained by taking derivatives with respect to
the source field $\phi$.
The flow parameter $\Lam$ describes a regularization of the 
underlying bare action, which regularizes infrared divergencies 
in perturbation theory.
The regularization is removed at the end of the flow, say for
$\Lam \to 0$.
The initial regulator (for $\Lam = \Lam_0$) can be chosen such that
$\Gam^{\Lam_0}[\phi]$ is given by the bare action.
The flow of $\Gam^{\Lam}[\phi]$ then provides a smooth interpolation
between the bare action of the system and the final effective action 
$\Gam[\phi]$, from which any desired information can be extracted.
This flow is determined by an exact functional differential equation
\cite{Wetterich93,Morris94,Ellwanger94}.
Expanding in the fields one obtains a hierarchy of flow equations
for the one-particle irreducible vertex functions.
The advantage of that hierarchy compared to others, obtained, for 
example, from Polchinski's equation, is that self-energy feedback
is included automatically and no one-particle reducible terms 
appear.

The expression {\em functional} RG stems from the feature that
the exact flow equations describe the flow of a functional or
(equivalently) of a hierarchy of functions.
An important difference compared to Wilson's original formulation
is that a complete set of source fields is kept in the flowing 
generating functionals, not only those corresponding to scales 
below $\Lam$.
Hence, the full information on the properties of the system 
remains accessible, not only the low energy or long wavelength 
behavior.

Exact flow equations can be solved exactly only in special cases, 
where the underlying model can also be solved exactly (and more 
easily) by other means.\footnote{An instructive example is provided
by the exact solution of the Tomonaga-Luttinger model via functional 
RG flow equations \cite{Schuetz05}.}
However, the functional RG is a valuable source for devising powerful 
new approximation schemes, which can be obtained by truncating the 
hierarchy and/or by a simplified parametrization of the Green or 
vertex functions.
These approximations have several distinctive advantages: i) they have 
a renormalization group structure built in, that is, scales are handled
successively and infrared singularities are thus treated properly; 
ii) they can be applied directly to microscopic models, not only to 
effective field theories which capture only some asymptotic behavior; 
iii) they are physically transparent, for example one can see 
directly how and why new correlations form upon lowering the scale; 
iv) one can use different approximations at different scales.
Small steps from a scale $\Lam$ to a slightly smaller scale $\Lam'$
are much easier to control than an integration over all degrees of
freedom in one shot, and one can take advantage of the flexibility
provided by the choice of a suitable flow parameter.

Approximations derived from exact flow equations have been applied
in many areas of quantum field theory and statistical physics
\cite{Berges02}.
In the context of interacting Fermi systems, functional RG methods
were first used for an unbiased stability analysis of the 
two-dimensional Hubbard model 
\cite{Zanchi98,Zanchi00,Halboth00a,Honerkamp01d}.
Since then, approximations derived within the functional RG
framework have been applied to numerous interacting fermion
systems.

\subsection{Scope of the review}

This review provides a thorough introduction to the functional
RG in the context of interacting Fermi systems.
It should serve as a manual and reference for many-body theorists
who wish to apply approximations based on the functional RG to 
their own problem of interest.
We will first describe the functional RG framework and derive in 
particular the exact flow equations, which are the starting point 
for approximations.
We will discuss general aspects related to the flow equations
such as the choice of cutoffs, power counting and truncations.
Links to the use of flow equations in the mathematical literature 
will be pointed out along the way.
We will then review some of the most interesting applications of 
truncated functional RG equations (see table of content).
Our aim is not to deliver an exhaustive overview of all applications,
but rather to show via selected applications how the functional RG
method can be fruitfully used.

The functional RG was recently extended to Fermi systems out of
equilibrium \cite{Jakobs03,Gezzi07,Jakobs07a,Jakobs10a,Jakobs10c,
Karrasch10a,Karrasch10c}.
In the derivation of the flow equations in Sec.~II we restrict 
ourselves to the equilibrium formalism. 
Functional RG flow equations for non-equilibrium Keldysh Green and 
vertex functions can be derived in close analogy 
\cite{Jakobs03,Gezzi07,Jakobs10b,Karrasch10b}.
The necessary extensions are briefly mentioned when discussing the 
application of this method to finite bias steady state transport 
through correlated quantum wires and quantum dots in Sec.~VI.

A number of reviews with a focus on the functional RG are already 
available.
Mathematically rigorous developments until the end of the last
millenium were summarized in a book by \textcite{Salmhofer99},
a large portion of which is dedicated to interacting Fermi 
systems.
Examples of approximations derived from the exact flow equation 
for the effective action with many applications in quantum field
theory and statistical physics were presented in the review 
article by \textcite{Berges02}.
A detailed introduction to the functional RG in a textbook style
supplemented by selected applications (including Fermi systems)
can be found in the recent book by \textcite{Kopietz10}.


\section{FUNCTIONAL FLOW EQUATIONS}
\label{sec:II}

In this section we present the general functional RG framework.
The reader should be familiar with the functional integral formalism
for quantum many-body systems, as described in classic textbooks 
such as \textcite{Negele87}.
After introducing the generating functionals for Green and vertex
functions in Sec.~II.A, we derive the exact functional flow 
equations in Sec.~II.B. 
The flow equation (\ref{floweqGam}) for the effective action
$\Gam^{\Lam}$ is the central equation of this review.
Expanding in the fields we derive the hierarchy of flow equations
for vertex functions in Sec.~II.C, which is the starting point
for approximations.
Possible choices of flow parameters are reviewed in Sec.~II.D.
The general structure of the RG hierarchy and power counting
are discussed in Sec.~II.E, with various references 
to the closely related mathematical literature.
Sec.~II.F is dedicated to flow equations for observables and 
correlation functions. 
Coupled flow equations for fermions and bosons, which are useful
for studies of spontaneous symmetry breaking and quantum 
criticality, are derived in Sec.~II.G.


\subsection{Generating functionals}
\label{sec:functionals}

We consider a system of interacting fermions which can be 
described by Grassmann fields $\psi$, $\bar\psi$, and an action 
of the form
\begin{equation} \label{bareaction}
 \cS[\psi,\psib] = 
 - (\psib, G_0^{-1} \psi) + V[\psi,\psib] \; ,
\end{equation}
%
%
where $V[\psi,\psib]$ is an arbitrary many-body interaction,
and $G_0$ is the propagator of the non-interacting system.
The bracket $(.,.)$ is a shorthand notation for the sum
$\sum_x \psib(x) \, (G_0^{-1} \psi)(x)$, where
$(G_0^{-1} \psi)(x) = \sum_{x'} G_0^{-1}(x,x') \, \psi(x')$.
The Grassmann field index $x$ collects the quantum numbers 
of a suitable single-particle basis set and imaginary time
or frequency.
In case of continuous variables, the sum over $x$ includes
the appropriate integrals.
Prefactors such as temperature or volume factors depend on
the representation (e.g.\ real or momentum space) and are
therefore not written in this general part.
A two-particle interaction has the general form
\begin{equation} \label{bareinteraction}
 V[\psi,\psib] = 
 \frac{1}{4} \sum_{x_1,x_2 \atop x'_1,x'_2}  
 V(x'_1,x'_2;x_1,x_2) \, 
 \psib(x'_1) \psib(x'_2) \psi(x_2) \psi(x_1) \; .
\end{equation}

In particular,
for spin-$\frac{1}{2}$ fermions with a single-particle basis 
labeled by momentum $\bk$ and spin orientation $\sg$, one has 
$x = (k_0,\bk,\sg)$, where $k_0$ is the fermionic Matsubara 
frequency.
If the bare part of the action is translation and spin-rotation
invariant, the bare propagator has the diagonal and 
spin-independent form $G_0(x,x') = 
\delta_{k_0k'_0} \delta_{\bk\bk'} \delta_{\sg\sg'} G_0(k_0,\bk)$ 
with
\begin{equation} \label{g0k}
 G_0(k_0,\bk) = \frac{1}{ik_0 - \xi_{\bk}} \; ,
\end{equation}
where $\xi_{\bk} = \eps_{\bk} - \mu$ is the single-particle 
energy relative to the chemical potential. 

Connected Green functions can be obtained from the generating
functional \cite{Negele87}
\begin{equation} \label{calG}
 \cG[\eta,\etab] = 
 - \ln \int \cD\psi \cD\psib \, 
 e^{-\cS[\psi,\psib]} \, e^{(\etab,\psi) + (\psib,\eta)} \; ,
\end{equation}
where
$\int \cD\psi \cD\psib \dots = 
 \int \prod_x d\psi(x) d\psib(x) \dots \;$.
Completing squares yields the identity
\begin{equation} \label{gauss}
 \int \cD\psi \cD\psib \, e^{(\psib,G_0^{-1} \psi)}
 e^{(\etab,\psi) + (\psib,\eta)} = 
 \cZ_0 \; e^{(-\etab,G_0 \eta)} \; ,
\end{equation}
where $\cZ_0 =
 \int \cD\psi \cD\psib \, e^{(\psib, \, G_0^{-1} \psi)}$
is the partition function of the non-interacting system.
Hence $\cG[\eta,\etab] = -\ln\cZ_0 + (\etab, G_0 \eta)$ in 
the non-interacting case $V[\psi,\psib] = 0$.
For vanishing source fields, $\cG[0,0] = - \ln\cZ$, where
\begin{equation} \label{Z}
 \cZ =
 \int \cD\psi \cD\psib \, e^{-\cS[\psi,\psib]} 
\end{equation}
is the partition function of the interacting system.
The connected $m$-particle Green functions are given by
\begin{eqnarray} \label{G2m}
 \hskip -5mm
 && G^{(2m)}(x_1,\dots,x_m;x'_1,\dots,x'_m) = 
 \nonumber \\
 && - \bra \psi(x_1) \dots \psi(x_m) 
             \psib(x'_m) \dots \psib(x'_1) \ket_c =
 \nonumber \\
 && (-1)^m \! \left.
 \frac{\partial^{2m} \cG [\eta,\etab]}{\partial\etab(x_1) \dots 
 \partial\etab(x_m)
 \partial\eta(x'_m) \dots 
 \partial\eta(x'_1)}
 \right|_{\eta,\etab = 0} , \hskip 5mm
\end{eqnarray}
where $\bra \dots \ket_c$ is the connected average of the 
product of Grassmann variables between the brackets.
The one-particle Green function $G^{(2)}$ is the propagator of
the interacting system, which we will usually denote without the
superscript by $G$.
Expanding $\cG[\eta,\etab]$ in the fields yields a formal power 
series with the connected Green functions as coefficients,
\begin{widetext}
\begin{equation} \label{Gexpansion}
 \cG[\eta,\etab] = - \ln\cZ + (\etab, G \eta) +
 \frac{1}{(2!)^2} \sum_{x_1,x_2,x'_1,x'_2} 
 G^{(4)}(x_1,x_2;x'_1,x'_2) \,
 \etab(x_1) \etab(x_2) \eta(x'_2) \eta(x'_1) + \dots
 \; .
\end{equation}
\end{widetext}

Renormalization group equations are most conveniently formulated 
for the Legendre transform of $\cG[\eta,\etab]$, the socalled 
{\em effective action}
\begin{equation} \label{Gampsi}
 \Gam[\psi,\psib] = 
 (\etab,\psi) + (\psib,\eta) + \cG[\eta,\etab] \; ,
\end{equation}
%
%
with $\psi = - \partial\cG/\partial\etab$ and
$\psib = \partial\cG/\partial\eta \,$, which generates 
one-particle irreducible vertex functions \cite{Negele87}
\begin{eqnarray} \label{Gam2m}
 && \Gam^{(2m)}(x'_1,\dots,x'_m;x_1,\dots,x_m) =
 \nonumber \\[2mm]
 && \left.
 \frac{\partial^{2m} \Gam[\psi,\psib]}
 {\partial\psib(x'_1) \dots \partial\psib(x'_m)
 \partial\psi(x_m) \dots \partial\psi(x_1)}
 \right|_{\psi,\psib = 0} . \hskip 5mm
\end{eqnarray}
%
%
In the non-interacting case one obtains
$\Gam[\psi,\psib] = - \ln\cZ_0 - (\psib, G_0^{-1} \psi)$.
The Legendre correspondence between the functionals $\cG$ and $\Gam$
yields relations between the connected Green functions $G^{(2m)}$ 
and the vertex functions $\Gam^{(2m)}$.
In particular, 
\begin{equation} \label{Dyson}
 \Gam^{(2)} = G^{-1} = G_0^{-1} - \Sg \; ,
\end{equation}
where $\Sg$ is the self-energy.
The connected two-particle Green function is related to the 
two-particle vertex by
\begin{eqnarray} \label{G4Ga4}
 && G^{(4)}(x_1,x_2;x'_1,x'_2) = 
 \sum_{y_1,y_2,y'_1,y'_2} \! G(x_1,y'_1) \, G(x_2,y'_2) \,
 \nonumber \\
 && \times \, \Gam^{(4)}(y'_1,y'_2;y_1,y_2) \, G(y_1,x'_1) \, G(y_2,x'_2)
 \; ,
\end{eqnarray}
while the three-particle Green function
$G^{(6)} = G^3 \Gam^{(6)} G^3 + G^3 \Gam^{(4)} G \Gam^{(4)} G^3$
involves $\Gam^{(4)}$ and $\Gam^{(6)}$.
More generally, the connected $m$-particle Green functions are
obtained by adding all possible trees that can be formed with
vertex functions of equal or lower order and $G$-lines
\cite{Negele87}.

The effective action obeys the reciprocity relations
\begin{equation} \label{reciproc1}
 \frac{\partial\Gam}{\partial\psi} = - \etab \; , \quad
 \frac{\partial\Gam}{\partial\psib} = \eta \; .
\end{equation}
The second functional derivatives of $\cG$ and $\Gam$ with respect 
to the fields are also reciprocal \cite{Negele87}.
We define the matrices of second derivatives at finite fields
\begin{widetext}
\begin{equation} \label{bG2}
 \bG^{(2)}[\eta,\etab] = - \left( \begin{array}{cc} 
 \frac{\partial^2\cG}{\partial\etab(x)\partial\eta(x')} &
 - \frac{\partial^2\cG}{\partial\etab(x)\partial\etab(x')} \\[3mm]
 - \frac{\partial^2\cG}{\partial\eta(x)\partial\eta(x')} &
 \frac{\partial^2\cG}{\partial\eta(x)\partial\etab(x')}
 \end{array} \right)
 = - \left( \begin{array}{cc} 
 \bra \psi(x) \psib(x') \ket & \bra \psi(x) \psi(x') \ket \\
 \bra \psib(x) \psib(x') \ket & \bra \psib(x) \psi(x') \ket
 \end{array} \right) \, ,
\end{equation}
and
\begin{equation} \label{bGam2}
 \bGam^{(2)}[\psi,\psib] = \left( \begin{array}{cc} 
 \frac{\partial^2\Gam}{\partial\psib(x')\partial\psi(x)} &
 \frac{\partial^2\Gam}{\partial\psib(x')\partial\psib(x)} \\[3mm]
 \frac{\partial^2\Gam}{\partial\psi(x')\partial\psi(x)} &
 \frac{\partial^2\Gam}{\partial\psi(x')\partial\psib(x)}
 \end{array} \right) =
 \left( \begin{array}{cc}
 \bar\partial\partial\Gam[\psi,\psib](x',x) &
 \bar\partial\bar\partial\Gam[\psi,\psib](x',x) \\
 \partial\partial\Gam[\psi,\psib](x',x) &
 \partial\bar\partial\Gam[\psi,\psib](x',x) 
 \end{array} \right) \; ,
\end{equation}
\end{widetext}
where the matrix elements in the second matrix of the last equation
are just a more conventient notation for those in the first matrix.
The reciprocity relation for the second derivatives reads
\begin{equation} \label{reciproc2}
 \bGam^{(2)}[\psi,\psib] = 
 \left( \bG^{(2)}[\eta,\etab] \right)^{-1} \; .
\end{equation}
Note that anomalous components are involved as long as the 
source fields are finite. 
Only at $\eta=\etab=0$ and $\psi=\psib=0$, and in the absence
of $U(1)$ charge symmetry breaking one has the simple relation
$\Gam^{(2)} = \big( G^{(2)} \big)^{-1}$.

Another useful generating functional is the 
{\em effective interaction} \cite{Salmhofer99}
\begin{equation} \label{effpot}
 \cV[\chi,\chib] = 
 - \ln \left\{ \frac{1}{\cZ_0} 
 \int \cD\psi \cD\psib \, e^{(\psib,G_0^{-1}\psi)}
 e^{-V[\psi+\chi,\psib+\chib]} \right\} \; .
\end{equation}
A simple substitution of variables yields the relation
\begin{equation} \label{VG}
 \cV[\chi,\chib] = 
 \cG[\eta,\etab] + \ln\cZ_0 - (\etab,G_0 \eta) \; ,
\end{equation}
where $\chi = G_0 \eta$ and $\chib = G_0^t \etab$.
Here $G_0^t$ is the transposed bare propagator, that is,
$G_0^t(x,x') = G_0(x',x)$.
Hence, functional derivatives of $\cV[\chi,\chib]$ with
respect to $\chi$ and $\chib$ generate connected Green 
functions with bare propagators amputated from external 
legs in the corresponding Feynman diagrams. 
The term $\ln\cZ_0 - (\etab,G_0 \eta)$ cancels the 
non-interacting part of $\cG[\etab,\eta]$ such that 
$\cV[\chi,\chib] = 0$ for $V[\psi,\psib] = 0$.
The effective interaction $\cV$ can also be expressed via 
functional derivatives, instead of a functional integral:
\begin{eqnarray} \label{expV}
 e^{-\cV[\chi,\chib]} &=&
 \frac{1}{\cZ_0}
 \int \cD\psi \cD\psib \, e^{(\psib,G_0^{-1}\psi)} \, 
 e^{-V[\psi+\chi,\psib+\chib]} \nonumber \\
 &=& \frac{1}{\cZ_0} \, 
 e^{-V[\partial_{\etab},\partial_{\eta}]} 
 \int \cD\psi \cD\psib \, e^{(\psib,G_0^{-1}\psi)}
 \nonumber \\
 && \times \left.
 e^{(\etab,\psi+\chi) + (\eta,\psib+\chib)} \right|_{\eta,\etab=0} 
 \nonumber \\
 &=& e^{-V[\partial_{\etab},\partial_{\eta}]} 
 \left. e^{(\etab,G_0 \eta)} e^{(\etab,\chi) + (\eta,\chib)} 
 \right|_{\eta,\etab=0} 
 \nonumber \\
 &=& e^{-V[\partial_{\etab},\partial_{\eta}]} 
 \left. e^{(\partial_{\chi},G_0 \partial_{\chib})} 
 e^{(\etab,\chi) + (\eta,\chib)} 
 \right|_{\eta,\etab=0} 
 \nonumber \\
 &=& e^{\Delta_{G_0}} \, e^{-V[\chi,\chib]} \; ,
\end{eqnarray}
with the functional Laplacian
\begin{equation} \label{DeltaG0}
 \Delta_{G_0} = 
 \big( \partial_{\chi}, G_0 \partial_{\chib} \big) =
 \sum_{x,x'} \frac{\partial}{\partial\chi(x)} \, G_0(x,x') \,
 \frac{\partial}{\partial\chib(x')} \; .
\end{equation}

It is sometimes convenient (see Sec.~II.G) to combine the 
fields $\psi$ and $\psib$ in a Nambu-type field
\begin{equation} \label{nambuPsi}
 \Psi(x) = \left( \begin{array}{c} \psi(x) \\
 \psib(x) \end{array} \right) \; ,
\end{equation}
and similarly for the source fields $\eta$ and $\etab$,
\begin{equation} \label{nambuH}
 H(x) = \left( \begin{array}{c} {\phantom -} \eta(x) \\ 
 - \etab(x) \end{array} \right) \; .
\end{equation}
The minus sign in the definition of $H$ makes sure that the
source term $(\etab,\psi) + (\psib,\eta)$ appearing in the 
definition of $\cG$, and also in the Legendre transform relating 
$\cG$ and $\Gam$, can be written concisely as $(\bar H, \Psi)$.
In Nambu notation, the matrices of second derivatives of 
$\cG$ and $\Gam$ have the compact form
\begin{equation} \label{bG2H}
 \bG^{(2)}[H] = 
 - \frac{\partial^2 \cG}{\partial\bar H(x) \partial H(x')}
\end{equation}
and
\begin{equation} \label{bGam2Psi}
 \bGam^{(2)}[\Psi] = 
 \frac{\partial^2 \Gam}{\partial\Psib(x') \partial\Psi(x)}
 \; ,
\end{equation}
respectively.


\subsection{Exact fermionic flow equations}
\label{sec:floweqs}

\begin{figure}[ht]
\centerline{\includegraphics[width = 4.5cm]{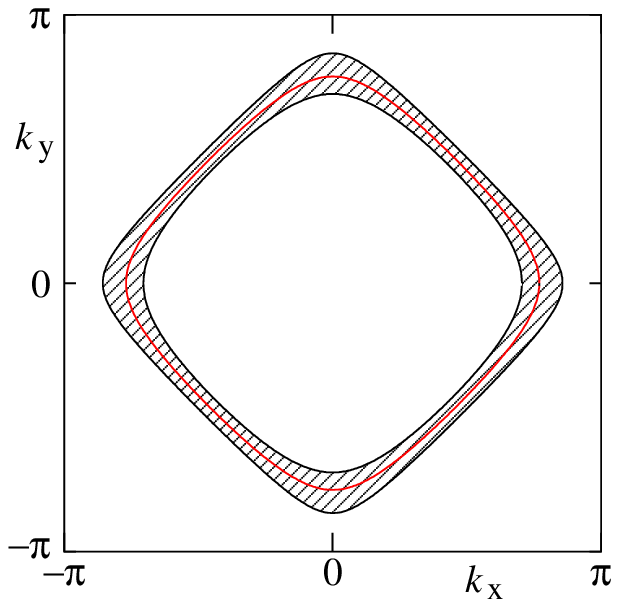}}
\caption{(Color online)
 Momentum space region around the Fermi surface excluded
 by a sharp momentum cutoff for fermions with a tight-binding
 dispersion on a two-dimensional square lattice (lattice 
 constant = one).}
\label{fig:cutoff}
\end{figure}

In this section we derive exact flow equations describing
the evolution of the generating functionals defined above, as 
a function of a flow parameter $\Lam$ which parametrizes a
modification of the bare propagator $G_0$. 
Usually $\Lam$ is an infrared cutoff or another scale
dependence.
For example, in a translation invariant system one may impose 
a momentum cutoff, modifying $G_0$ to
\begin{equation} \label{G0Lamk}
 G_0^{\Lam}(k_0,\bk) = 
 \frac{\theta^{\Lam}(\bk)}{ik_0 - \xi_{\bk}} \; ,
\end{equation}
where $\theta^{\Lam}(\bk)$ is a function that vanishes for
$|\xi_{\bk}| \ll \Lam$ and tends to one for $|\xi_{\bk}| \gg 
\Lam$. In this way the infrared singularity of the propagator 
at $k_0 = 0$ and $\xi_{\bk} = 0$ (corresponding to the 
non-interacting Fermi surface in $\bk$-space) is cut off at
the scale $\Lam$.
A simple choice for $\theta^{\Lam}(\bk)$, which was often
used in numerical solutions of truncated flow equations, is
\begin{equation} \label{thetaLamk}
 \theta^{\Lam}(\bk) = \Theta(|\xi_{\bk}| - \Lam) \; ,
\end{equation}
where $\Theta$ is the step function. With this choice
momenta close to the Fermi surface are strictly excluded,
as illustrated in Fig.~\ref{fig:cutoff} for a two-dimensional 
lattice fermion system.
Alternatively, one may also use a smooth cutoff function.
In the absence of translation invariance it is more convenient
to use a frequency cutoff instead of a momentum cutoff.
The cutoff excludes ''soft modes'' below the scale $\Lam$ from 
the functional integral.
Instead of a cutoff one may also choose other flow parameters
such as temperature. The various possibilities will be discussed
more extensively in Sec.~\ref{sec:flowparameters}.
For the derivation of the flow equations it does not matter how
$G_0^{\Lam}$ depends on $\Lam$.

The bare action constructed with $G_0^{\Lam}$ (instead of $G_0$) 
will be denoted by $\cS^{\Lam}[\psi,\psib]$, and the generating 
functionals introduced in Sec.~\ref{sec:functionals} by 
$\cG^{\Lam}[\eta,\etab]$, $\cV^{\Lam}[\chi,\chib]$, and 
$\Gam^{\Lam}[\psi,\psib]$, respectively. 
The original functionals $\cG$, $\cV$ and $\Gam$ are recovered 
in the limit $\Lam \to 0$.

In the presence of a cutoff, Eq.~(\ref{expV}) becomes
\begin{equation} \label{expVLam}
 e^{-\cV^{\Lam}} = e^{\Delta_{G_0^{\Lam}}} \, e^{-V} \; .
\end{equation}
At the highest energy scale $\Lam_0$ one has $G_0^{\Lam_0} = 0$,
and thus $\cV^{\Lam_0} = V$.
Hence, $\cV^{\Lam}$ interpolates smoothly between the bare 
interaction $V$ and the generating functional $\cV$.
Introducing the soft mode propagator 
\begin{equation} \label{barG0Lam}
 \bar G_0^{\Lam} = G_0 - G_0^{\Lam} \; ,
\end{equation}
which has support on scales {\em below} $\Lam$, we can write
\begin{equation} \label{duality} 
 e^{-\cV} = e^{\Delta_{G_0}} e^{-V} =
 e^{\Delta_{\bar G_0^{\Lam}} + \Delta_{G_0^{\Lam}}} \, 
 e^{-V} =
 e^{\Delta_{\bar G_0^{\Lam}}} \, e^{-\cV^{\Lam}} \; .
\end{equation}
$\cV^{\Lam}$ obviously plays a dual role: It is the generating 
functional for (amputated) Green functions of a system with a
cutoff $\Lam$, and at the same time the interaction for the 
remaining low energy degrees of freedom 
\cite{Morris94,Salmhofer99}.

The effective interaction satisfies the following exact 
renormalization group equation \cite{Salmhofer99,Brydges88}
\begin{equation}\label{floweqV}
 \frac{d}{d\Lam} \cV^{\Lam}[\chi,\chib] =
  - \left( \frac{\partial\cV^{\Lam}}{\partial\chi} \, ,
    \dot{G}_0^{\Lam} \,
    \frac{\partial\cV^{\Lam}}{\partial\chib} \right)
  - {\rm tr} \left( \dot{G}_0^{\Lam} \,
        \frac{\partial^2 \cV^{\Lam}}
        {\partial\chib \partial\chi} \right) \; ,
\end{equation}
where $\dot{G}_0^{\Lam} = \frac{d}{d\Lam} G_0^{\Lam}$ and
tr denotes the trace ${\rm tr} A = \sum_x A(x,x)$.
Its derivation is simple:
\begin{eqnarray} \label{floweqVderiv} 
 \frac{d}{d\Lam} {\cV^{\Lam}} &=& 
 - e^{\cV^{\Lam}} \frac{d}{d\Lam} e^{-\cV^{\Lam}} \nonumber \\
 &=& - e^{\cV^{\Lam}} \frac{d}{d\Lam} 
 \big( e^{\Delta_{G_0^{\Lam}}} e^{-V} \big) =
 - e^{\cV^{\Lam}} \Delta_{\dot{G}_0^{\Lam}} \, e^{-\cV^{\Lam}} 
 \nonumber \\
 &=& \mbox{right-hand side of Eq.~(\ref{floweqV})} \; .
 \nonumber
\end{eqnarray}
In the second step we have used Eq.~(\ref{expVLam}).
With the initial condition 
\begin{equation} \label{VLam0}
 \cV^{\Lam_0}[\chi,\chib] = V[\chi,\chib] \; ,
\end{equation}
the RG equation determines the flow of $\cV^{\Lam}$ uniquely 
for all $\Lam < \Lam_0$.
The initial value $\Lam_0$ must be chosen such that 
$G_0^{\Lam_0}$ vanishes.
For a sharp momentum cutoff, $\Lam_0$ can be chosen as the 
maximal value of $|\xi_{\bk}|$; for a frequency cutoff
$\Lam_0 = \infty$.

An expansion of the functional $\cV^{\Lam}[\chi,\chib]$ in the 
renormalization group equation (\ref{floweqV}) in powers of $\chi$
and $\chib$ leads to the fermionic analog of Polchinski's 
\cite{Polchinski84} flow equations for amputated connected 
$m$-particle Green functions $V^{(2m) \Lam}$.

From the flow equation for $\cV^{\Lam}$, Eq.~(\ref{floweqV}), and 
the relation (\ref{VG}) applied to $\cV^{\Lam}$ and $\cG^{\Lam}$,
one obtains an exact flow equation for $\cG^{\Lam}$:
\begin{equation} \label{floweqG}
 \frac{d}{d\Lam} \cG^{\Lam}[\eta,\etab] =
 \left( \frac{\partial \cG^{\Lam}}{\partial\eta} ,
        \dot{Q}_0^{\Lam}  
        \frac{\partial \cG^{\Lam}}{\partial\etab} \right) +
 {\rm tr} \left(\dot{Q}_0^{\Lam} \,
 \frac{\partial^2 \cG^{\Lam}}
      {\partial\etab \partial\eta} \right) \; ,
\end{equation}
where $Q_0^{\Lam} = (G_0^{\Lam})^{-1}$ and the dot denotes a
$\Lam$-derivative.
This flow equation can also be derived more directly, by 
applying a $\Lam$-derivative to the functional integral
representation of $\cG^{\Lam}$.

The flow equations for $G^{(2m)\Lam}$ and $V^{(2m)\Lam}$ generate,
among others, also one-particle reducible terms, which require 
some special care. 
In this respect the flow equations for one-particle irreducible 
vertex functions $\Gam^{(2m)\Lam}$, obtained from the 
scale-dependent effective action,
\begin{equation} \label{effactionLam}
 \Gam^{\Lam}[\psi,\psib] = 
 (\etab^{\Lam},\psi) + (\psib,\eta^{\Lam}) +
 \cG^{\Lam}[\eta^{\Lam},\etab^{\Lam}] \; ,
\end{equation}
are easier to handle. 
Note that $\eta^{\Lam}$ and $\etab^{\Lam}$ are $\Lam$-dependent 
functions of $\psi$ and $\psib$, as they are determined by the
$\Lam$-dependent equations 
$\psi = - \partial\cG^{\Lam}/\partial\etab$ and 
$\psib = \partial\cG^{\Lam}/\partial\eta$.
Since the $\Lam$-dependence does not change the structure of the
action as a function of the fields, all standard relations
between the connected Green functions $G^{(2m)}$ and the vertex
functions $\Gam^{(2m)}$ carry over to the ones for $G^{(2m)\Lam}$
and $\Gam^{(2m)\Lam}$.

The $\Lam$-derivative of $\Gam^{\Lam}$ can be written as
$\frac{d}{d\Lam} \Gam^{\Lam}[\psi,\psib] =
 (\frac{d}{d\Lam} \etab^{\Lam},\psi)
 + (\psib,\frac{d}{d\Lam} \eta^{\Lam})
 + \frac{d}{d\Lam} \, \cG^{\Lam}[\eta^{\Lam},\etab^{\Lam}]$,
where the derivative in front of $\cG^{\Lam}$ acts also 
on the $\Lam$-dependence of $\eta^{\Lam}$ and $\etab^{\Lam}$.
Due to the relations
$\partial\cG^{\Lam}/\partial\eta = \psib$ and
$\partial\cG^{\Lam}/\partial\etab = - \psi$, most terms
cancel and one obtains
\begin{equation} \label{dLamdG}
 \frac{d}{d\Lam} \Gam^{\Lam}[\psi,\psib] =
 \frac{d}{d\Lam} \left.
 \cG^{\Lam}[\eta^{\Lam},\etab^{\Lam}]
 \right|_{\eta^{\Lam},\etab^{\Lam} \; \mbox{fixed}} \; .
\end{equation}
Inserting the flow equation (\ref{floweqG}) for $\cG^{\Lam}$ and
using the reciprocity relations (\ref{reciproc1}) and 
(\ref{reciproc2}), one obtains the exact functional flow 
equation for the effective action
\begin{equation} \label{floweqGam}
 \frac{d}{d\Lam} \Gam^{\Lam}[\psi,\psib] =
 - \big(\psib, \dot Q_0^{\Lam} \psi \big)
 - \frac{1}{2} {\rm tr} \left[ \dot \bQ_0^{\Lam} \left(
   \bGam^{(2)\Lam}[\psi,\psib] \right)^{-1} \right] \; .
\end{equation}
Here $\bGam^{(2)\Lam}[\psi,\psib]$ is the matrix of second
functional derivatives defined in Eq.~(\ref{bGam2}), and
\begin{equation} \label{bQ0Lam}
 \bQ_0^{\Lam} = \left( \begin{array}{cc}
 Q_0^{\Lam} & 0 \\ 0 & - Q_0^{\Lam t} 
 \end{array} \right) =
 \mbox{diag} (Q_0^{\Lam},- Q_0^{\Lam t}) \; ,
\end{equation}
where $Q_0^{\Lam t}(x,x') = Q_0^{\Lam}(x',x)$.

Alternative definitions of the effective action $\Gam^{\Lam}$, 
differing by interaction-independent terms, have also been used. 
One variant is to normalize the functional integral defining
$\cG^{\Lam}$ at $V=0$, dividing by $\cZ_0^{\Lam}$. 
This yields an additional contribution $\ln\cZ_0^{\Lam}$ to 
$\cG^{\Lam}$ and to its Legendre transform $\Gam^{\Lam}$.
In the flow equation for $\Gam^{\Lam}$ this leads to an
additional term ${\rm tr}(\dot Q_0^{\Lam} G_0^{\Lam})$, 
which is field independent and therefore does not couple
to the other contributions \cite{Salmhofer01}.
Another variant is \cite{Ellwanger94,Berges02}
\begin{equation} \label{GamRLam}
 \Gam_R^{\Lam}[\psi,\psib] = 
 \Gam^{\Lam}[\psi,\psib] + (\psib,R^{\Lam}\psi) \; ,
\end{equation}
where $R^{\Lam} = Q_0^{\Lam} - Q_0$. The additional 
quadratic term cancels the first (trivial) term in the
flow equation (\ref{floweqGam}) for $\Gam^{\Lam}$, and one
obtains the equivalent flow equation
\begin{equation} \label{floweqGamR}
 \frac{d}{d\Lam} \Gam_R^{\Lam}[\psi,\psib] =
 - \frac{1}{2} {\rm tr} \left[ \dot \bR^{\Lam} \left(
   \bGam_R^{(2)\Lam}[\psi,\psib]
   + \bR^{\Lam} \right)^{-1} \right] \; ,
\end{equation}
where 
$\bR^{\Lam} = 
 \mbox{diag} \left( R^{\Lam},- R^{\Lam t} \right)$.
The functional $\Gam_R^{\Lam}$ and its analog for bosonic 
fields is known as {\em effective average action} in the 
literature \cite{Berges02}.
Both $\Gam_R^{\Lam}$ and $\Gam^{\Lam}$ tend to the same
effective action $\Gam$ in the limit $\Lam \to 0$, where
$R^{\Lam}$ vanishes. 
At the initial scale $\Lam_0$, one has
$\Gam_R^{\Lam_0}[\psi,\psib] = \cS[\psi,\psib]$,
while 
\begin{eqnarray} \label{GamLam0}
 \Gam^{\Lam_0}[\psi,\psib] 
&=&
 - (\psib, Q_0^{\Lam_0} \psi) + V[\psi,\psib]  
 =
 \cS^{\Lam_0}[\psi,\psib] 
 \nonumber\\
 &=&
 \cS[\psi,\psib] - (\psib,R^{\Lam_0}\psi) \; .
\end{eqnarray}
Hence, $\Gam_R^{\Lam}$ has the attractive feature that it
interpolates smoothly between the (unregularized) bare action 
$\cS$ and the final effective action $\Gam$, while $\Gam^{\Lam}$
interpolates between the {\em regularized} bare action $\cS^{\Lam_0}$
and $\Gam$. 
On the other hand, the functional $\Gam^{\Lam}$ has the 
advantage that its second functional derivative directly yields 
the inverse propagator $(G^{\Lam})^{-1}$ without the need to 
add $R^{\Lam}$.

In Appendix \ref{sec:wick} we present yet another version of 
exact flow equations, based on a {\em Wick ordered} effective 
interaction. That version also contains one-particle 
reducible contributions, but it has the distinct advantage that 
the vertices are connected by propagators with an energy scale
at or below $\Lam$. This facilitates a systematic power counting
\cite{Salmhofer99}, and also a numerical evaluation of flow 
equations, since the integration regions shrink upon lowering 
$\Lam$.

It is instructive to compare the functional RG flow equations 
with the traditional Wilsonian momentum shell RG \cite{Wilson74}, 
which was applied to Fermi systems by \textcite{Shankar91,Shankar94} 
and \textcite{Polchinski93}.
In the commonly used version of Wilson's RG, the flow of the 
effective action is computed only for soft fields, that is, for 
fields with energy or momentum variables {\em below} the scale $\Lam$, 
while in the functional RG the effective action with unrestricted 
source fields is computed.
This allows for a direct calculation of correlations functions
with arbitrary external variables such as momenta or Matsubara
frequencies.
Furthermore, in the traditional implementations of Wilson's RG
the integration of degrees of freedom is combined with a rescaling
of momenta and fields, which is chosen such that the momentum 
space and certain terms in the quadratic part of the action remain
invariant during the flow. 
This facilitates the classification of interactions as relevant, 
marginal or irrelevant, and helps to identify fixed points of
the flow.
The functional RG flow equations derived above do not involve
any rescaling.
Rescaling momentum space in a shell around the Fermi surface
requires a non-linear transformation in dimensions $d > 1$, 
which spoils the simple linear form of momentum conservation 
\cite{Shankar94,Metzner98,Kopietz01}, and is therefore of 
questionable value.
Power-counting can be done also without rescaling, as shown in 
Sec.~II.E.
Rescaling of the fields can be implemented easily by a simple
substitution of variables \cite{Shankar94,Kopietz01}. 
However, in many applications of the functional RG, quantitative 
results including power-laws with anomalous scaling dimensions 
are obtained simply by direct calculation of the (unscaled) 
physical quantities.


\subsection{Expansion in the fields}
\label{sec:expansion}

\subsubsection{Hierarchy of flow equations}
\label{sec:hierarchy}

\begin{figure} [ht]
\centerline{\includegraphics[width = 7cm]{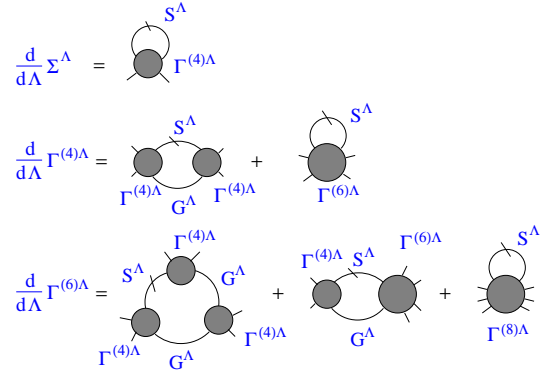}}
\caption{(Color online)
 Diagrammatic representation of the flow equations for
 the self-energy $\Sg^{\Lam}$, the two-particle vertex
 $\Gam^{(4)\Lam}$, and the three-particle vertex $\Gam^{(6)\Lam}$
 in the one-particle irreducible version of the functional RG. 
 Lines with a dash correspond to the single scale propagator 
 $S^{\Lam}$, the other lines to the full propagator $G^{\Lam}$.}
\label{fig:floweq1pi+}
\end{figure}

The functional flow equation for the effective action can be 
expanded in powers of the fields. 
To this end we expand the effective action as
\begin{equation} \label{GamA}
 \Gam^{\Lam}[\psi,\psib] = 
 \sum_{m=0}^{\infty} \cA^{(2m)\Lam}[\psi,\psib] \; ,
\end{equation}
where $\cA^{(2m)\Lam}[\psi,\psib]$ is homogeneous of degree $2m$
in the fields,
\begin{widetext}
\begin{equation} \label{A2m}
 \cA^{(2m)\Lam}[\psi,\psib] = \frac{(-1)^m}{(m!)^2} 
 \sum_{x_1,\ldots,x_m \atop x'_1,\ldots,x'_m}
 \Gam^{(2m)\Lam}(x'_1,\dots,x'_m;x_1,\dots,x_m) \,
 \psib(x'_1) \dots \psib(x'_m) \psi(x_m) \dots \psi(x_1) \; ,
\end{equation}
\end{widetext}
for $m \geq 1$. 
The field-independent constant $\cA^{(0)\Lam}$ yields the grand 
canonical potential: 
\begin{equation} \label{grandpot}
 \cA^{(0)\Lam} = T^{-1} \Omega^{\Lam} \; .
\end{equation}
Here we have restored the explicit temperature factor, since it 
is independent of the representation of the fields.
To expand the inverse of $\bGam^{(2)\Lam}$ on the right hand side
of the flow equation, we isolate the field-independent part of
$\bGam^{(2)\Lam}$ as
\begin{equation} \label{bGam2dec}
 \bGam^{(2)\Lam}[\psi,\psib] = 
 (\bG^{\Lam})^{-1} - \tilde\bSg^{\Lam}[\psi,\psib] \; ,
\end{equation}
where
\begin{equation} \label{bGLam}
 \bG^{\Lam} =
 \left(
 \left. \bGam^{(2)\Lam}[\psi,\psib] \right|_{\psi,\psib = 0} 
 \right)^{-1} =
 \mbox{diag} (G^{\Lam},- G^{\Lam t}) \; 
\end{equation}
is the full propagator, and (cf. Eq.~(\ref{bGam2}))
\begin{equation} \label{bGam2tilde}
 \tilde\bSg^{\Lam}[\psi,\psib] = 
 - \left( \begin{array}{cc}
 \bar\partial\partial\Gam^{\Lam}[\psi,\psib] &
 \bar\partial\bar\partial\Gam^{\Lam}[\psi,\psib] \\
 \partial\partial\Gam^{\Lam}[\psi,\psib] &
 \partial\bar\partial\Gam^{\Lam}[\psi,\psib] 
 \end{array} \right) + 
 \big( \bG^{\Lam} \big)^{-1} \; .
\end{equation}
Note that $\tilde\bSg^{\Lam}[\psi,\psib]$ contains all 
contributions to $\bGam^{(2)\Lam}[\psi,\psib]$ which are 
at least quadratic in the fields. We can now expand
$\left( \bGam^{(2)\Lam} \right)^{-1} =
 \big(1 - \bG^{\Lam} \tilde\bSg^{\Lam} \big)^{-1} 
 \, \bG^{\Lam}$
as a geometric series. Inserted in (\ref{floweqGam}), 
this yields
\begin{eqnarray} \label{floweqGamexp}
 \frac{d}{d\Lam} \Gam^{\Lam}[\psi,\psib] =
 - {\rm tr} \left( \dot Q_0^{\Lam} G^{\Lam} \right) -
 \big(\psib, \dot Q_0^{\Lam} \psi \big) + \hskip 1.5cm
 \nonumber \\
 \frac{1}{2} {\rm tr} \left[ \bS^{\Lam} 
 \left( \tilde\bSg^{\Lam}[\psi,\psib] +
 \tilde\bSg^{\Lam}[\psi,\psib] \, \bG^{\Lam} 
 \tilde\bSg^{\Lam}[\psi,\psib] + \dots \right) \right] \; ,
 \nonumber \\
\end{eqnarray}
where
\begin{equation} \label{bSLam}
 \bS^{\Lam} = 
 \mbox{diag} (S^{\Lam},-S^{\Lam t}) =
 - \bG^{\Lam} \dot{\bQ}_0^{\Lam} \bG^{\Lam} \; .
\end{equation}
Using the Dyson equation 
$(G^{\Lam})^{-1} = Q_0^{\Lam} - \Sg^{\Lam}$, the socalled 
{\em single-scale propagator} $S^{\Lam}$ can also be written as 
$\Lam$-derivative of the propagator at fixed self-energy,
\begin{equation} \label{SLam}
 S^{\Lam} =
 \frac{d}{d\Lam} \left. G^{\Lam} 
 \right|_{\Sg^{\Lam} \; \mbox{fixed}} \; .
\end{equation}
The expansion of the flow equation in powers of $\psi$, $\psib$
is now straightforward and leads to a hierarchy of flow equations 
for $\Sg^{\Lam}$, the two-particle vertex $\Gam^{(4)\Lam}$, 
and the higher-order vertices $\Gam^{(6)\Lam}$, $\Gam^{(8)\Lam}$,
etc.
The first three equations in this hierarchy are shown 
diagrammatically in Fig.~\ref{fig:floweq1pi+}.
Note that only one-particle irreducible one-loop diagrams
contribute, and internal lines are dressed by self-energy
corrections.
The hierarchy does not close at any finite order, since the
flow of each vertex $\Gam^{(2m)\Lam}$ receives a contribution 
from a tadpole diagram involving $\Gam^{(2m+2)\Lam}$, and 
$m$-particle vertices with arbitrary $m$ are generated by the 
flow, irrespective of their presence in the bare action.

\begin{figure}[ht]
\centerline{\includegraphics[width = 4cm]{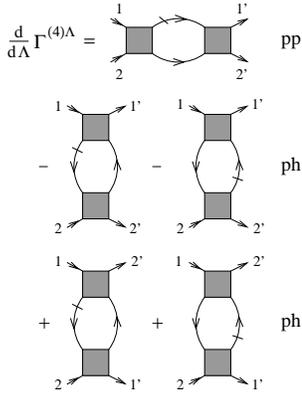}}
\caption{Contributions to the flow of the two-particle vertex
 with particle-particle and particle-hole channels written
 explicitly, without the contribution from $\Gam^{(6)\Lam}$.}
\label{fig:floweqgamma4.eps}
\end{figure}

Let us derive explicitly the first two flow equations from the
hierarchy.
Comparing coefficients of quadratic contributions (proportional
to $\psib\psi$) to the exact flow equation yields
\begin{equation} \label{floweqA2}
 \frac{d}{d\Lam} \cA^{(2)\Lam} = 
 - (\psib,\dot Q_0^{\Lam} \psi)
 - {\rm tr} \left( 
 S^{\Lam} \bar\partial \partial \cA^{(4)\Lam} \right) \; .
\end{equation}
Inserting Eq.~(\ref{A2m}), and using 
$\Gam^{(2)\Lam} = Q_0^{\Lam} - \Sg^{\Lam}$, one obtains the
flow equation for the self-energy,
\begin{equation} \label{floweqSigma}
 \frac{d}{d\Lam} \Sg^{\Lam}(x',x) =
 \sum_{y,y'} S^{\Lam}(y,y') \, \Gam^{(4)\Lam}(x',y';x,y)
 \; .
\end{equation}
Comparing coefficients of quartic contributions (proportional
to $(\psib\psi)^2$) yields
\begin{widetext}
\begin{eqnarray} \label{floweqA4}
 \frac{d}{d\Lam} \cA^{(4)\Lam} &=& 
 \frac{1}{2} {\rm tr} \left(
 S^{\Lam} \bar\partial\partial\cA^{(4)\Lam} G^{\Lam}
 \bar\partial\partial\cA^{(4)\Lam} +
 S^{\Lam t} \partial\bar\partial\cA^{(4)\Lam} G^{\Lam t}
 \partial\bar\partial\cA^{(4)\Lam} \right)
 \nonumber \\
 &-& \frac{1}{2} {\rm tr} \left(
 S^{\Lam} \bar\partial\bar\partial\cA^{(4)\Lam} G^{\Lam t}
 \partial\partial\cA^{(4)\Lam} +
 S^{\Lam t} \partial\partial\cA^{(4)\Lam} G^{\Lam}
 \bar\partial\bar\partial\cA^{(4)\Lam} \right)
 - {\rm tr} \left(S^{\Lam} 
 \bar\partial\partial\cA^{(6)\Lam} \right) \, .
 \hskip 8mm
\end{eqnarray}
Inserting Eq.~(\ref{A2m}), one obtains the flow equation
for the two-particle vertex,
\begin{eqnarray}\label{floweqGamma4}
 \frac{d}{d\Lam} \Gam^{(4)\Lam}(x'_1,x'_2;x_1,x_2) &=&
 \sum_{y_1,y'_1} \sum_{y_2,y'_2} 
 G^{\Lam}(y_1,y'_1) \, S^{\Lam}(y_2,y'_2) \nonumber \\
 &\times& \Big\{ 
 \Gam^{(4)\Lam}(x'_1,x'_2;y_1,y_2) 
 \Gam^{(4)\Lam}(y'_1,y'_2;x_1,x_2)  \nonumber \\
 &-& \Big[ \Gam^{(4)\Lam}(x'_1,y'_2;x_1,y_1) 
 \Gam^{(4)\Lam}(y'_1,x'_2;y_2,x_2) 
 + (y_1 \leftrightarrow y_2, y'_1 \leftrightarrow y'_2) \Big]
 \nonumber \\
 &+& \Big[ \Gam^{(4)\Lam}(x'_2,y'_2;x_1,y_1) 
 \Gam^{(4)\Lam}(y'_1,x'_1;y_2,x_2) 
 + (y_1 \leftrightarrow y_2, y'_1 \leftrightarrow y'_2) \Big]
 \Big\} \nonumber \\
 &-& \sum_{y,y'} S^{\Lam}(y,y') \,
 \Gam^{(6)\Lam}(x'_1,x'_2,y';x_1,x_2,y) \; .
\end{eqnarray}
\end{widetext}
Note that there are several distinct contributions involving 
two two-particle vertices, corresponding to the familiar 
particle-particle, direct particle-hole, and crossed 
particle-hole channel, respectively, as shown diagrammatically 
in Fig.~\ref{fig:floweqgamma4.eps}. 
Similarly, one can obtain the flow equation for $\Gamma^{(6)}$
and all higher vertices.

Since $\Gam[\psi,\psib]$ at $\psi = \psib = 0$ is essentially
(up to a factor $T$) the grand canonical potential $\Omega$,
the flow equation (\ref{floweqGam}), evaluated at vanishing
fields, yields also a flow equation for the grand canonical
potential:
\begin{equation} \label{floweqOmega}
 \frac{d}{d\Lam} \Omega^{\Lam} = 
 - T \, {\rm tr} \left( \dot Q_0^{\Lam} G^{\Lam} \right) \; .
\end{equation}

The flow equation (\ref{floweqGam}) and the ensuing equations
for the vertex functions can be easily generalized to cases 
with $U(1)$-symmetry breaking by allowing for off-diagonal 
elements in the matrices $\bQ_0^{\Lam}$, $\bG^{\Lam}$ and
$\bS^{\Lam}$.

\subsubsection{Truncations}
\label{sec:truncations}

The exact hierarchy of flow equations for the vertex functions
can be solved only for systems which can also be solved more 
directly, that is, without using flow equations.
Usually truncations are unavoidable.
A natural truncation is to neglect the flow of all vertices 
$\Gam^{(2m)\Lam}$ beyond a certain order $m_0$.
We call this the {\em level-$m_0$ truncation}.
The structure of the resulting equations and general properties
of their solution will be discussed in Sec.~II.E.
Note that the level-$m_0$ truncation contains all perturbative
contributions to order $m_0$ in the bare two-particle 
interaction.

In practice, in applications to physically interesting systems, 
vertices $\Gam^{(2m)\Lam}$ with $m > 3$ have so far 
been neglected, and the contributions from $\Gam^{(6)\Lam}$
to the flow of $\Gam^{(4)\Lam}$ are usually restricted to
self-energy corrections (see below) or discarded completely.
In particular, the analysis of competing instabilities 
(see Sec.~III) is based entirely on a level-2 truncation given 
by the flow equation (\ref{floweqGamma4}) for the two-particle 
vertex, with $\Gam^{(6)\Lam}$ replaced by zero, where in 
addition the self-energy feedback is neglected.
This seemingly simple approximation captures the complex
interplay of fluctuations in the particle-particle and
particle-hole channel, which leads to interesting effects
such as the generation of $d$-wave superconductivity from
antiferromagnetic fluctuations.
In the quantum transport phenomena reviewed in Sec.~VI, the
self-energy as given by the flow equation (\ref{floweqSigma})
plays a crucial role. Some of the phenomena described there
are already obtained by a level-1 approximation where the 
flowing two-particle vertex in Eq.~(\ref{floweqSigma}) is 
approximated by the bare one. 
That truncation might look like a Hartree-Fock approximation,
but it is in fact very different, and it works well in cases
where Hartree-Fock fails completely.

The truncated flow equations are still rather complicated.
They involve the flow of functions, not just a limited number
of running couplings. For example, the effective two-particle
interaction in a translation invariant system is a function
of three independent momentum and energy variables.
Hence, a simplified parametrization of effective interactions
is necessary even for a numerical solution.
A useful strategy is to neglect dependences which become
irrelevant in the low-energy limit, that is, whose contributions 
to the flow scale to zero. 

Contributions to the effective action are called {\em ``relevant''}, 
{\em ``marginal''}, and {\em ``irrelevant''}, if their importance 
increases, stays fixed, or decreases, respectively, upon lowering
the scale $\Lam$.
This classification can be obtained from power counting.
To this end, one traditionally considers a renormalization group
transformation where one rescales momenta and fields after the 
integration over fields in a momentum shell of width $d\Lam$ such 
that a certain quadratic part of the action (the Gaussian fixed 
point) remains invariant \cite{Wilson74}.
From the behavior of the other terms of the action under this
transformation one can assess directly whether they increase,
remain invariant, or decrease compared to the quadratic part.

For Fermi systems in dimensions $d > 1$ the conventional RG
transformation is not applicable, since the reduction of 
momentum space by the mode elimination cannot be compensated
by a linear rescaling of momenta \cite{Shankar91,Shankar94}.
However, one can perform the power counting more directly by 
estimating the scale dependences of Feynman diagrams on the 
right hand side of the flow equations.
As described in Sec.~II.E.3, this can be done rigorously and 
to all orders.
At the crudest level the power counting is independent of
dimensionality and corresponds to what one would get from
the above-mentioned RG transformation applied to one-dimensional 
systems \cite{Shankar94}, that is:
(i) the self-energy has a relevant piece describing a Fermi
surface shift, while linear dependences on frequency and
momentum perpendicular to the Fermi surface are marginal;
(ii) a regular two-particle interaction is marginal; its 
dependences on frequencies and momenta perpendicular to the 
Fermi surface are irrelevant, such that one can parametrize
it by its static value on the Fermi surface;
(iii) regular $m$-particle interactions with $m \geq 3$ are
irrelevant.
This basic classification does not depend on dimensionality
because the bare propagator $G_0$, Eq.~(\ref{g0k}), is singular
on a $(d-1)$-dimensional surface, such that the codimension
of the singularity in the $(d+1)$ dimensional space spanned
by momentum and frequency is always two.

One should, however, not jump to the conclusion that the
$m \geq 3$ terms can simply be discarded from the RG hierarchy
in general. 
This is because effective interactions with $m \geq 3$ may diverge
for small $\Lam$ even in case of finite two-particle interactions.
For example, the first contribution to the flow of $\Gam^{(6)\Lam}$ 
in Fig.~3 generates a three-particle interaction of order 
$\Lam^{-1}$ if the external momenta add up to zero at each vertex.
When inserted into the equation for $\Gam^{(4)\Lam}$ in Fig.~3, this 
may give rise to a marginal term of third order in $\Gam^{(4)\Lam}$.
For $d=1$, this term is indeed marginal.
However, if $d \geq 2$ and the Fermi surface is curved, this and
other contributions are suppressed below the basic power counting 
estimate due to geometrically reduced integration volumes 
\cite{FT1,Shankar94}.
This improved power counting is described in App.~B.3.
It can also be used to give a precise, scale-dependent meaning to 
nesting of the Fermi surface. 

\begin{figure}[ht]
\centerline{\includegraphics[width = 8cm]{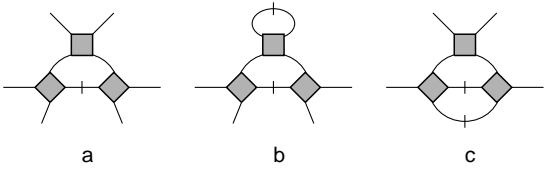}}
\caption{(a) Third order graph contributing to $\Gamma^{(6)\Lam}$.
(b) Tadpole contraction. 
(c) Contraction to form a graph with overlapping loops.}
\label{fig:3rd}
\end{figure}

A less obvious effect is that this improvement becomes uniform, 
that is, independent of the external momenta, in graphs with 
overlapping loops \cite{FST1, Salmhofer98a}, so that their contribution 
gets further suppressed (see also App.~B.3). It is this effect which 
implies that the derivative of the self-energy is not marginal, but irrelevant
for curved Fermi surfaces in dimension $d \ge 2$. 
Moreover, it justifies truncated flows beyond the weak-coupling regime,
as follows. Consider again the first contribution to the flow of 
$\Gam^{(6)\Lam}$, shown also in Fig.~\ref{fig:3rd} (a). 
When this term is inserted in the equation for $\Gam^{(4)\Lam}$, 
the two lines can be joined in two ways, shown in Fig.\ \ref{fig:3rd} (b) and (c). 
The graph in (b) gets no extra small factor, but 
the graph in (c) has overlapping loops and for positively curved Fermi surfaces
in $d=2$, its contribution gets suppressed by an additional small factor 
$\sim \frac{\Lambda}{\Lambda_I} \log \frac{\Lambda_I}{\Lambda}$ 
at scales below a scale $\Lambda_I$, which depends only on the geometry of 
the constant energy surfaces of the initial dispersion $\eps_{\bk}$ \cite{FST2}.
This suppression holds uniformly for all values of the external momenta. 
For $d=3$, a similar bound holds, without the logarithm. 
Similar (in general, weaker) estimates are shown in \cite{FST1} for general non-nested 
regular Fermi surfaces in $d \ge 2$ and for Fermi surfaces with Van Hove singularities 
in \cite{FS1,FS2}.
The contribution from graph (c) remains small compared to the second-order term if
$|\Gam^{(4)\Lam}|\, \frac{\Lambda}{\Lambda_I} \log\frac{\Lambda_I}{\Lambda}$
is small.
Note that this condition does {\em not} require $|\Gam^{(4)\Lam}| $
itself to be small: curvature effects justify dropping these terms 
beyond the weak-coupling regime, provided that the above condition is satisfied.
This will be used in Sec.~\ref{sec:III}. 
The detailed argument and a discussion of the consequences for the 
functional RG flow, are given in Sections 1 and 5 of \cite{Salmhofer01}.

In the theory of interacting Fermi systems, one is not only interested 
in low-energy fixed points and scaling, but also in the behavior at 
intermediate scales, and formally irrelevant terms may play an 
important role. 
There are cases where one would like to know the full temperature,
momentum, or frequency dependence of physical quantities, because a 
low-energy expansion contains insufficient information.
One of the advantages of the functional RG framework is that such
dependences can be computed directly.

In many situations, a comparison to standard resummations of the
perturbation expansion is desirable, and it is also an interesting 
question to what extent such resummations can be reproduced by
truncations of the functional RG flow equations. 
A very important observation regarding this was made by 
\textcite{Katanin04a}, who showed that a partial inclusion of the 
six-point vertex in the flow allows to recover approximations 
of the type Hartree-plus-ladder summations (in cases where these
approximations are a good starting point). 
This also allows to continue fermionic flows into symmetry-broken 
phases (see Sec.~\ref{sec:IV}). 
If we drop the eight-point vertex from the equation for 
$\Gamma^{(6)\Lam}$, it is determined by a Feynman graph containing 
three four-point vertices, depicted in Fig.~\ref{fig:3rd}a. 
When backsubstituted into the equation for the four-point vertex, 
two external legs get contracted in all possible ways.
We have just discussed that the contribution from the graph in (c)
is suppressed by improved power counting.
When two legs of a single-four-point vertex are contracted 
to form a tadpole (see Fig.~\ref{fig:3rd}b), the value of the 
thus obtained subgraph is $\dot \Sigma^\Lambda$, by the flow equation for 
the self-energy. Thus a factor $G^\Lambda \dot \Sigma^\Lambda G^\Lambda$ 
appears in the integral for the value of the graph. 
By Dyson's equation, 
\begin{equation}
 \dot G^\Lambda =
 G^\Lambda \dot \Sigma^\Lambda G^\Lambda +
 S^\Lambda ,
\end{equation}
so this can be combined with the second-order contribution to 
replace $S^\Lambda$ by $\dot G^\Lambda$.
If all other effects of the six-point function, corresponding to 
graphs of the type shown in Fig.~\ref{fig:3rd}c are dropped, the 
equation for the four-point function gets changed to one where 
the product 
$G^\Lambda(k) S^\Lambda(k') + S^\Lambda(k) G^\Lambda(k')$
is replaced with 
$G^\Lambda(k) \dot G^\Lambda(k') + \dot G^\Lambda(k) G^\Lambda(k')
 = \frac{d}{d\Lambda} \left(G^\Lambda(k) G^\Lambda (k') \right)$.
If one now restricts further to a single channel in the four-point 
equation, it becomes explicitly solvable by a ladder summation in 
that channel. 
Backsubstitution in the equation for the self-energy gives the 
corresponding Hartree-type term. 
This is explained in \cite{Katanin04a}, and, also in its extension 
to flows with symmetry breaking, in \cite{Salmhofer04}.


\subsection{Flow parameters}
\label{sec:flowparameters}

In the derivation of the exact functional flow equation, the scale 
dependence of the bare propagator $G_0^{\Lam}$ was not specified. 
The derivation holds for any choice of $G_0^{\Lam}$, provided all 
functions involved are indeed differentiable with respect to $\Lam$, 
and provided that the resulting flow equation is well defined.
These conditions are not trivial; in fact, badly chosen flow
parameters may lead to divergences on the right hand side of the
flow equation.
On the other hand, one can exploit the flexibility provided by the
choice of the $\Lam$-dependence to ones own advantage.
Besides regularity issues, the scale dependence of $G_0^{\Lam}$ 
is only constrained by the initial condition
\begin{equation} \label{G0initial}
 G_0^{\Lam_0} = 0 \; ,
\end{equation}
and the final condition
\begin{equation} \label{G0final}
 G_0^{\Lam \to 0} = G_0 \; .
\end{equation}
The functional $\Gam = \Gam^{\Lam \to 0}$ reached at the end of 
the exact flow is independent of the choice of $G_0^{\Lam}$.
However, in most practical calculations, where approximations are
unavoidable, a judicious choice of $G_0^{\Lam}$ is mandatory.
Important aspects related to the choice of $G_0^{\Lam}$ are:
regularization of infrared singularities, minimization of truncation 
errors, respecting symmetries, technical convenience.
In the following we will review the most frequently used cutoff
schemes along with their merits and drawbacks.


\subsubsection{Momentum and frequency cutoffs}
\label{sec:cutoffs}

For the sake of a concise discussion, let us focus on translation
and spin-rotation invariant one-band systems, such that the bare
propagator $G_0$ can be written as a simple function of frequency 
and momentum as in Eq.~(\ref{g0k}). 
The scale dependence can then be introduced by multiplying
$G_0$ with a suitable cutoff function $\theta^{\Lam}$,
\begin{equation} \label{multcutoff}
 G_0^{\Lam}(k_0,\bk) = 
 \theta^{\Lam}(k_0,\bk) \, G_0(k_0,\bk) \; ,
\end{equation}
with $\theta^{\Lam_0} = 0$ and $\theta^{\Lam \to 0} = 1$.
To regularize the infrared divergence of $G_0$ at zero frequency
and for momenta on the Fermi surface ($\xi_{\bk} = 0$), the cutoff
function $\theta^{\Lam}(k_0,\bk)$ has to vanish sufficiently 
quickly for $k_0 \to 0$, $\xi_{\bk} \to 0$ at fixed $\Lam > 0$.
The most frequently used cutoff functions are either pure momentum 
cutoffs of the form $\theta^{\Lam}(\bk) = \vartheta(|\xi_{\bk}|/\Lam)$ 
or frequency cutoffs $\theta^{\Lam}(k_0) = \vartheta(|k_0|/\Lam)$,
where $\vartheta(x)$ is a function that vanishes for $x \ll 1$ and
tends to one for $x \gg 1$. 
Mixed momentum and frequency cutoffs of the form
$\theta^{\Lam}(k_0,\bk) = \vartheta[(k_0^2 + \xi_{\bk}^2)/\Lam^2]$
are preferred in the mathematical literature, as they facilitate
power counting and rigorous estimates.

A technical advantage of momentum cutoffs compared to frequency
cutoffs is that Matsubara sums on the right hand side of the 
flow equations can often be performed analytically.
Furthermore, a momentum cutoff does not spoil the analytic structure
of propagators and vertex functions in the complex frequency plane.
However, there are also serious drawbacks, which are specific to Fermi
systems.
Once self-energy effects are taken into account, the Fermi surface is
usually deformed in the course of the flow, such that the momentum
cutoff has to be continuously adapted to the new Fermi surface, which
complicates the flow equations considerably.
Second, particle-hole excitations with a small momentum transfer $\bq$ 
are suppressed by the momentum cutoff for
$|\xi_{\bk+\bq} - \xi_{\bk}| < 2\Lam$. 
As a consequence, the limit of vanishing momentum transfer $\bq \to 0$
in interaction vertices and response functions does not commute with
the limit $\Lam \to 0$ \cite{Metzner98}. 
In other words, forward scattering interactions and the response to 
homogeneous fields can be obtained only by taking the limit $\bq \to 0$ 
at the end of the flow, at $\Lam = 0$ \cite{Honerkamp01c}.
This is a serious drawback in stability analyses (see Sec.~\ref{sec:III}), 
where one compares the increase of the effective interaction in 
different momentum channels (including forward scattering), or different
susceptibilities, upon lowering $\Lam$ until a divergence occurs in at
least one channel at a finite scale $\Lam_c > 0$.

A frequency cutoff has the advantage that it does not interfere with
Fermi surface shifts, and that particle-hole processes with small
momentum transfers are captured smoothly by the flow \cite{Husemann09a}. 
It can also be used in systems without translation invariance
\cite{Andergassen04}, where a momentum cutoff is less useful since 
the propagator is not diagonal in momentum space.
However, a frequency cutoff affects the analytic properties of 
propagators and vertex functions in the complex frequency plane. 
Depending on the sort of truncation used, this may pose a serious
problem if one likes to continue results to real frequency.
For a frequency cutoff the initial cutoff is $\Lam_0 = \infty$.
Since the contributions to the self-energy flow are of order $\Lam^{-1}$
at large $\Lam$, one has to retain the convergence factor $e^{ik_0 0^+}$ 
on the right hand side of the flow equation \cite{Andergassen04},
analogously to the convergence factor in the perturbation expansion
of the self-energy \cite{Negele87}; 
for a rigorous justification, see \textcite{Pedra08}. 

For a bare propagator $G_0(k_0,\bk) = (ik_0 - \xi_{\bk})^{-1}$ and a
multiplicative cutoff as in Eq.~(\ref{multcutoff}), the full
propagator has the form
\begin{equation} \label{GLam}
 G^{\Lam}(k_0,\bk) = 
 \frac{\theta^{\Lam}(k_0,\bk)}{ik_0 - \xi_{\bk} -
 \theta^{\Lam}(k_0,\bk) \Sg^{\Lam}(k_0,\bk)} \; ,
\end{equation}
and the single-scale propagator 
$S^{\Lam} = - G^{\Lam} \dot Q_0^{\Lam} G^{\Lam}$, 
see Eq.~(\ref{bSLam}), reads
\begin{equation} \label{SLam2}
 S^{\Lam}(k_0,\bk) = 
 \frac{(ik_0 - \xi_{\bk}) \partial_{\Lam} \theta^{\Lam}(k_0,\bk)}
 {\left[ ik_0 - \xi_{\bk} -  
  \theta^{\Lam}(k_0,\bk) \Sg^{\Lam}(k_0,\bk) \right]^2} \; .
\end{equation}
For a sharp cutoff function such as
$\theta^{\Lam}(k_0) = \Theta(|k_0| - \Lam)$, the single-scale
propagator seems ill-defined, since 
$\partial_{\Lam} \Theta(|k_0| - \Lam) = - \delta(|k_0| - \Lam)$,
so that a delta peak in the numerator of Eq.~(\ref{SLam2})
coincides with a discontinuity (due to the step function) in the 
denominator.
However, this ambiguity can be easily removed by viewing the 
step function $\Theta(x)$ as a limit of increasingly sharp 
regularized step functions $\Theta_{\eps}(x)$, where the 
discontinuity is smeared over a width $\eps$ \cite{Morris94}.
With $\delta_{\eps}(x) = \partial_x \Theta_{\eps}(x)$, a simple
substitution of variables yields
\begin{equation} \label{morrislemma}
 \delta_{\eps}(x) \, f(x,\Theta_{\eps}(x)) 
 \stackrel{\eps \to 0}{\longrightarrow}
 \delta(x) \int_0^1 du f(0,u) \; ,
\end{equation}
for any continuous function $f$. Note that the right hand side
is unique, that is, it does not depend on the shape of the 
smeared step function $\Theta_{\eps}(x)$ for $\eps > 0$.
For a sharp frequency cutoff 
$\theta^{\Lam}(k_0) = \Theta(|k_0| - \Lam)$, for example,
the single-scale propagator thus simplifies to
\begin{equation} \label{SLamsharp}
 S^{\Lam}(k_0,\bk) = 
 - \frac{\delta(|k_0| - \Lam)}
 {ik_0 - \xi_{\bk} - \Sigma^{\Lam}(k_0,\bk)} \; ,
\end{equation}
as long as it does not appear in products where other factors
are also discontinuous at $|k_0| = \Lam$.
Otherwise, for example in products of the form 
$S^{\Lam}(k_0,\bk) [G^{\Lam}(k_0,\bk)]^m$, one has to apply
Eq.~(\ref{morrislemma}) to the entire product.

A sharp cutoff has the obvious technical advantage that the 
integration over the cutoff variable ($k_0$ or $\xi_{\bk}$)
can be carried out analytically, thanks to the delta-function
in the numerator of $S^{\Lam}$. 
On the other hand, a sharp cutoff generates discontinuities in 
the momentum or frequency dependences of the vertex functions, 
corresponding to a pronounced non-locality of the effective 
action \cite{Morris94}, which is often not amenable to a 
simple parametrization.
At finite temperature the flow equations are ill-defined for
a sharp frequency cutoff, since the Matsubara frequencies are
discrete: $k_0 = (2n+1)\pi T$ with integer $n$. 
Continuous cutoff functions at $T>0$ are conveniently chosen 
such that the $\Lam$-derivative is non-zero only in a frequency 
range of width $2\pi T$, since then only two frequencies 
contribute to the Matsubara sum on the right hand side of the 
flow equation \cite{Enss05b}.

There are useful cutoff schemes which are formulated more 
naturally by {\em adding}\/ a regulator function $R^{\Lam}$ to 
the inverse propagator (instead of multiplying):
\begin{equation}
 Q_0^{\Lam}(k_0,\bk) = 
 \big[ G_0^{\Lam}(k_0,\bk) \big]^{-1} =
 Q_0(k_0,\bk) + R^{\Lam}(k_0,\bk) \; ,
\end{equation}
with $R^{\Lam_0} = \infty$ and $R^{\Lam \to 0} = 0$.
In particular, regulator functions of the form \cite{Litim01}
\begin{equation} \label{litimcutoff} 
 R^{\Lam}(\bk) = 
 - Z^{\Lam} [\sgn(\xi_{\bk})\Lam - \xi_{\bk}] 
 \Theta(\Lam - |\xi_{\bk}|) \; ,
\end{equation}
or its frequency dependent analogue,
$R^{\Lam}(k_0) = 
 i Z^{\Lam} [\sgn(k_0) \Lam - k_0] \Theta(\Lam - |k_0|)$,
have some distinct advantages. 
The prefactor $Z^{\Lam}$ is initially one and is then
determined by a momentum (or frequency) derivative of the 
flowing self-energy $\Sg^{\Lam}(k_0,\bk)$.
The Litim cutoff satisfies a criterion of ''optimal''
regularization of the infrared singularity of the propagator
\cite{Litim01}. 
For simple truncations it also leads to a very convenient
form of the integrands, facilitating the integrations.

It is easy to choose the cutoff function in a way that does 
not affect the global symmetries of the system, such as global 
charge conservation or global spin rotatation invariance.
However, {\em local}\/ conservation laws are typically spoiled.
The corresponding Ward identities are modified by cutoff
dependent additional terms, which vanish only in the limit 
$\Lam \to 0$ \cite{Enss05a}. 
It is very hard to devise truncations which satisfy the 
modified Ward identities at each scale $\Lam$, and hence 
truncated flows often violate Ward identities also in the 
limit $\Lam \to 0$ \cite{Katanin04a}.
In these cases it is better to compute only independent 
quantities from the flow, and determine the remaining 
quantities, which are fixed by local conservation laws,
via the Ward identity.


\subsubsection{Temperature and interaction flows}
\label{sssec:tflow}

For fermion systems the infrared singularity of the bare 
propagator can also be regularized by temperature, instead 
of a cutoff, since the fermionic Matsubara frequencies stay
away from zero at a distance $\pi T$.
A flow equation with temperature as a flow parameter can be 
obtained from the general flow equation derived in 
Sec.~\ref{sec:floweqs}, if one manages to shift all temperature 
dependences of the bare action to the quadratic part.
This is indeed possible by a simple rescaling of the fields
\cite{Honerkamp01c}.
Let us consider a translation invariant system of 
spin-$\frac{1}{2}$ fermions for definiteness, where the 
fields depend on a momentum $\bk$, a spin index $\sg$, and
a Matsubara frequency $\om_n = (2n+1)\pi T$.
Rescaling the fields as
$\psi'_{\sg}(n,\bk) = T^{3/4} \psi_{\sg}(\om_n,\bk)$ and
$\psib'_{\sg}(n,\bk) = T^{3/4} \psib_{\sg}(\om_n,\bk)$
removes all explicit $T$-factors from the (quartic) interaction
in the bare action.
The temperature dependence is thereby shifted entirely to the
quadratic part of the action, given by the inverse bare 
propagator for the rescaled fields,
\begin{equation} \label{QprimeT}
 Q_0^T(n,\bk) = \frac{T^{1/2}}{i\om_n - \xi_{\bk}} \; .
\end{equation}
The effective action $\Gam^T[\psi',\psib']$ for the rescaled 
fields obeys the exact flow equation Eq.~(\ref{floweqGam}), 
with temperature as the flow parameter. 
The unscaled vertex functions $\Gam^{(2m)}$ are recovered 
from the vertex functions $\Gam^{(2m)T}$ by multiplying with 
$T^{3m/2}$.
The temperature flow has several advantageous features. 
First, it generates directly a temperature scan of the 
computed quantities. In cutoff schemes one has to run a full
flow for each temperature separately. 
Second, the temperature flow includes particle-hole 
excitations with small momentum transfers uniformly at each
scale.
Third, local symmetries and the corresponding Ward identities
are respected at each step at least for the exact flow, 
which makes the still difficult issue of Ward identities in 
truncated flows at least more transparent.

A particularly simple choice of a flow parameter is provided
by a uniform factor $\lam$ scaling the bare propagator
\cite{Honerkamp04},
\begin{equation} \label{G0lam}
 G_0^{\lam} = \lam G_0 \; ,
\end{equation}
with $\lam_0 = 0$, and $\lam \to 1$ at the end of the flow.
By a simple rescaling of the fields one can see that this is
equivalent to multiplying the bare quartic interaction with
a factor $\lam^2$, which means that the interaction is scaled
up continuously from $0$ to its full strength in the course 
of the flow. Hence the name ''interaction flow'' for this 
scheme.
In the absence of self-energy feedback the interaction flow
has the technical advantage that the propagator has the same
form at each scale, such that certain loop integrals need 
to be done only once.
However, the global scaling of the propagator does not 
regularize the infrared singularities, such that one easily
runs into infrared divergences.
Nevertheless, for suitable problems and simple truncations
the interaction flow has been shown to yield results similar
to flows with a cutoff, and with less computational effort
\cite{Honerkamp04}.


\subsection{General properties of the RG equations}
\label{sec:powercount}

\newcommand{\Higher}{{\cal H}}
\newcommand{\Same}{{\cal K}}
\newcommand{\Lower}{{\cal L}}
\newcommand{\Four}[1]{V^{#1}}
\newcommand{\bP}{{\bf P}}
\newcommand{\bsg}{{\underline{\sigma}}}
\newcommand{\km}{{\underline{k}}}
\newcommand{\aex}{\kappa}
\newcommand{\snorm}[1]{s_{#1}}
\newcommand{\Sa}{a}
\newcommand{\Sb}{b}
\newcommand{\Sc}{c}
\newcommand{\vphi}{\varphi}
\newcommand{\db}{{\mkern2mu\mathchar'26\mkern-2mu\mkern-9mud}}
\newcommand{\mshell}[1]{{\cal F}_{#1}}
\newcommand{\vFmin}{v_{\rm F,min}}
\newcommand{\twokF}{{\cal F}^{(2)}}
\newcommand{\Gausscurvature}{\kappa_{\rm Gauss}}
\newcommand{\ereno}{e_1}
\newcommand{\ebare}{e_0}

In this section we discuss the general structure of the RG hierarchy 
of equations and provide power counting bounds for its solution.
These bounds are simple, but mathematically exact, and they 
provide a strict sense to the notion of relevant and irrelevant terms. 
We shall also briefly discuss improved power counting bounds, 
which provide sharper estimates for bulk Fermi systems in $d \ge 2$ 
and exhibit the role of Fermi surface geometry.

The generating functionals were introduced to obtain the
Green functions and vertex functions of the model by differentiation,  
cf.\ (\ref{G2m}) and (\ref{Gam2m}). 
In the framework of the RG as an iterated convolution, 
they acquire an independent importance.
Indeed, in many situations in bosonic field theory, an expansion 
in the fields is avoided in favor of a gradient expansion
\cite{Berges02} or other types of parametrization (see also Sec.~V), 
and the flow may lead to a non-analytic function of the fields.
Functions of Grassmann variables are, however, defined only 
by power series expansions in these variables, so in this case 
the meaning of the RG is strictly that of the infinite hierarchy. 
This is only a seeming disadvantage because by  the anticommutation
properties of Grassmann variables, the fully regularized 
functionals (as they appear in the flow equations) have convergent
expansions in the fields 
\cite{FMRS,GKGN,Lesniewski87,Abdesselam98,FKT98,Salmhofer00,FKT02}. 
In contrast, the expansion in the fields of bosonic functionals 
is almost always divergent, even in the regularized theory. 
Convergent expansions then take the form of cluster expansions
that distinguish between regions of small and large fields 
(see, for example, \textcite{BKFT10} and references therein).

\subsubsection{Inductive structure of the RG hierarchy}

The functional $\tilde\bSg^{\Lam}[\psi,\psib]$ 
appearing in (\ref{floweqGamexp}) has an expansion 
similar to (\ref{GamA}), namely 
$\tilde\bSg^{\Lam}[\psi,\psib] (x',x) =
\sum_{m\ge 1} \tilde\bSg^{(2m)\Lam}[\psi,\psib] (x',x)$,
where $\tilde\bSg^{(2m)\Lam}$ is homogeneous of degree $2m$ 
in the fields, hence has a representation with coefficient 
functions $\tilde\bSg^{(2m)\Lam}$ similar to (\ref{A2m}).
By definition (\ref{bGam2tilde}) of $\tilde\bSg$, the 
$\tilde\bSg^{(2m)\Lam}$ are determined by 
$\Gamma^{(2m+2)\Lam}$, for example
\begin{widetext}
\begin{eqnarray}
 \left( \tilde\bSg^{(2m)\Lam} (x',x) \right)_{11} \, 
 (x'_1, \dots, x'_m;x_1, \ldots, x_m) =
 - \Gam^{(2m+2)\Lam} (x',x'_1,\dots, x'_m; x_1,,\dots,x_m,x)
 \, .
\end{eqnarray}
Here the indices refer to the matrix structure of 
(\ref{bGam2tilde}). The other matrix elements are
given by similar expressions. 

We use this to expand (\ref{floweqGamexp}) in homogeneous parts 
in $\psi$ and $\psib$ and compare coefficients. This gives
\begin{eqnarray}
\sfrac{d}{d \Lam} \cA^{(2m)\Lam} [\psi,\psib] 
&=&
\sfrac12 \; {\rm tr} 
\left( 
\bS^\Lambda \tilde\bSg^{(2m)\Lambda} [\psi,\psib]
\right)
+
\sfrac12 {\rm tr}
\left(
\bS^\Lambda \tilde\bSg^{(2)\Lambda} [\psi,\psib]
\bG^\Lambda \tilde\bSg^{(2m-2)\Lambda} [\psi,\psib]
\right)
\nonumber\\
&+&
\sfrac12
\sum_{p \ge 2} 
\sum_{m_0, \ldots, m_p \ge 1 \atop m_0+ \ldots + m_p =m}
{\rm tr}
\left(
\bS^\Lambda \tilde\bSg^{(2m_0)\Lambda} [\psi,\psib]
\prod\limits_{q=1}^p
\bG^\Lambda \tilde\bSg^{(2m_q)\Lambda} [\psi,\psib]
\right) \, .
\label{mqexp}
\end{eqnarray}
\end{widetext}
In the sum over $p$, each of $m_0, \ldots , m_p$ is at least one 
because $\tilde\bSg$ only contains field-dependent terms, and
$m_0 + \ldots + m_p = m$ must hold since $\cA^{(2m)\Lambda}$
is homogeneous of degree $2m$ in the fields $[\psi,\psib]$. 
These two conditions imply that $p \le m$ and that 
$m_q \le m-p$ for all $0 \le q \le p$, 
so that for every given $m$, the sum only runs over finitely many terms. 
Since the coefficient in $\cA^{(2m)\Lambda}$ is $\Gamma^{(2m)\Lambda}$ 
and $\tilde\bSg^{(2m)\Lambda} \sim \Gamma^{(2m+2)\Lambda}$,
comparing coefficients of powers of $\psi$ and $\psib$ in (\ref{mqexp}) 
gives a hierarchy of differential equations for the $\Gamma^{(2m)\Lambda}$, 
labelled by $m$. 
We rewrite (\ref{mqexp}) as
\begin{eqnarray}\label{HSL}
\sfrac{d}{d \Lambda}
\Gamma^{(2m)\Lambda}
&=&
\Higher^\Lambda \Gamma^{(2m+2)\Lambda}
+
\Same^\Lambda \left(\Gamma^{(4)\Lambda}\right)
\;\Gamma^{(2m)\Lambda}
\nonumber\\
&+&
\sum_{p=2}^m
\Lower^\Lambda_p \left(\Gamma^{(<2m)\Lambda}\right) \, .
\end{eqnarray}
The three summands on the right hand side are obtained from 
the three terms in (\ref{mqexp}), 
and it is understood that both sides are functions of $2m$
variables $x_1, \ldots, x_{2m}$. The action of the operator $\Higher^\Lambda$ 
on $\Gamma^{(2m+2)\Lambda}$  is linear, 
as is that of $\Same^\Lambda(\Gamma^{(4)\Lambda})$ 
on $\Gamma^{(2m)\Lambda}$, while $\Lower^\Lambda_p$ is nonlinear in the 
lower-$m$ vertex functions
$\Gamma^{(<2m)\Lambda}= \Gamma^{(4)\Lambda}, \ldots, \Gamma^{(2m-2)\Lambda}$.
Specifically, the action of $\Higher^\Lambda$
is given by a tadpole-type contraction and summation, the action of 
$\Same^\Lambda(\Gamma^{(4)\Lambda})$ 
is given by the evaluation of a one-loop diagram 
formed from $\Gamma^{(2m)\Lambda}$ and the four-point function 
$\Gamma^{(4)\Lambda}$, and $\Lower^\Lambda_p$ is given by a sum 
over one-loop diagrams involving $p+1$ vertex functions, 
each of which has $m_q < m$.
Thus $\Higher^\Lambda$, $\Same^\Lambda(\Gamma^{(4)\Lambda})$
and $\Lower^\Lambda_p$ also
depend on $\Lambda$ and on the self-energy $\Sigma^\Lambda$
via the propagators $S^\Lambda$ and $G^\Lambda$.
The $\Higher^\Lambda$-term couples the higher vertex function $\Gamma^{(2m+2)\Lambda}$
into the equation for $\Gamma^{(2m)\Lambda}$.
Thus the hierarchy does not close among finitely many $m$, 
and therefore truncations need to be employed to obtain solutions.

\subsubsection{Truncated hierarchies and their iterative solution}

If for some $m_0 \ge 1$,  the initial vertex functions 
$\Gamma^{(2m)\Lambda_0}$ vanish for all $m > m_0+1$,
one may employ the approximation of setting 
$\Gamma^{(2m)\Lambda} = \Gamma^{(2m)\Lambda_0}$
for all $m \ge m_0 +1$. That is, all vertices with $m > m_0 +1$
remain zero, and the $(m_0+1)$-particle vertex is kept fixed
at its initial value. 
This {\em level-$m_0$ truncation} reduces the infinite hierarchy 
to a system of finitely many differential equations for 
$(\Gamma^{(2m)\Lambda})_{m \le m_0}$. 
The vertex $\Gamma^{(2m_0+2)\Lambda_0}$ enters in the equation for $\Gamma^{(2m_0)\Lambda}$. 
Specifically, in the level-$1$ truncation, the two-particle vertex $\Gamma^{(4)\Lambda}$ is fixed to its bare value 
$\Gamma^{(4)\Lambda_0}$, and the self-energy is
the solution of (\ref{floweqSigma}).
The level-$2$ truncation is given by (\ref{floweqGamma4}), 
with $\Gamma^{(6)\Lambda}$ fixed to its initial value
$\Gamma^{(6)\Lambda_0}$ (which may vanish), 
together with (\ref{floweqSigma}).
The term $\Same^\Lambda (\Gamma^{(4)\Lambda}) \Gamma^{(4)\Lambda}$
is quadratic in $\Gamma^{(4)\Lambda}$.

In the level-$m_0$ truncation of the hierarchy, with $m_0 > 2$, 
and {\em at given $\Sigma^\Lambda$ and 
$(\Gamma^{(2m')\Lambda})_{m'<m_0}$},
Eq.\ (\ref{HSL}) for $\Gamma^{(2m_0)\Lambda}$
becomes a linear inhomogeneous differential equation 
for $\Gamma^{(2m_0)\Lambda}$, which can be solved by 
an operator version of the standard method of variation of the constant:
when all sums and integrals corresponding to the traces in 
(\ref{mqexp}) are written out, it takes the form of a linear 
integro-differential equation which is, viewed more abstractly, 
a linear ordinary differential equation in a suitable space of 
functions, to which standard techniques apply. 
Together with the initial condition $\Gamma^{(2m)\Lambda_0}$,
this determines $\Gamma^{(2m)\Lambda}$ uniquely in terms 
of $(\Gamma^{(2m')\Lambda})_{m'<m_0}$. 
Backsubstitution of this solution into the $\Higher^\Lambda$-term for the 
equation for $\Gamma^{(2m_0-2)\Lambda}$ then yields an equation
for $\Gamma^{(2m_0-2)\Lambda}$, which can be solved, 
in terms of the not yet determined lower vertex functions 
$(\Gamma^{(2m')\Lambda})_{m'<m_0-1}$.
Proceeding downwards in $m$ in this way, one can formally solve
the truncated hierarchy, with the final equation determining 
$\Sigma^\Lambda$. We write ``formally'' here because after
at most two steps of this iteration, the differential equations become
nonlinear, so that existence of the solution is typically known only for short flow 
times, and because the question of blowup of solutions is rather nontrivial. 
Indeed, we shall see below that blowup generically occurs in RG
equations if relevant terms have not been taken into account.
This phenomenon is related to the infrared divergences of 
unrenormalized perturbation theory. 
The major advantage of the RG method
is that the growing terms can be identified and studied long before they get singular,
and then removed by taking into account appropriately chosen relevant parts
in the flowing action. 

Increasing $m_0$ to improve the accuracy is then a natural 
strategy for approximation of the true solution; however, 
explicit and numerical calculations can be done only for
small $m_0$, because the number of variables increases rapidly with $m$. 
Nevertheless, one can get useful information in the form of bounds
for the maximal possible value of the vertex 
functions  (or other norms that measure their size). 
This is done in the following section. 

\subsubsection{Running coupling expansion and power counting}
\label{ssec:runcoup}

We turn to the standard situation of a model with two-body interactions,
where the initial interaction of the fermion
system is quartic, i.e.\ $\Gamma^{(2m)\Lambda_0} = 0 $ for all $m \ge 3$.
We also assume that this interaction is short-range so that its Fourier
transform is bounded  (e.g.\ an unscreened Coulomb interaction is long-range).
In a perturbative expansion in powers of the initial four-point interaction 
$\Four{\Lambda_0} = \Gamma^{(4)\Lambda_0}$, the vertices are given by
sums over irreducible graphs. 
An irreducible Feynman graph formed with $r$ four-legged vertices
can have at most $2r$ external legs, so that in order $r$ in that expansion,
all vertex functions with $m > r$ vanish. 

As we shall now explain, one can solve the RG hierarchy in terms of a similar
expansion in the {\em scale-dependent} four-point function 
$\Four{\Lambda}=\Gamma^{(4)\Lambda}$, again by integrating 
the RG hierarchy downwards in scale, but keeping the $2m$-point functions
for $m > 2$ only to a fixed order in $\Four{\Lambda}$. The equation for
$\Four{\Lambda}$ itself then becomes an integro-differential equation with 
a power $r$ nonlinearity on the right hand side (the equation for 
$\Sigma^\Lambda$ remains unchanged). 
This leads in a natural way to power counting estimates
for the higher $2m$-point functions in terms of the maximal value of
the four-point vertex that occurs in the flow. 

We denote the $O\left( (\Four{\Lambda})^r \right)$ contribution to 
$\Gamma^{(2m)\Lambda}$ by $\Gamma_r^{(2m)\Lambda}$.
Its scale derivative equals
\begin{eqnarray} \label{mprimeflow}
 \sfrac{d}{d\Lambda} \Gamma_r^{(2m)\Lambda}
 &=& \Higher^{\Lambda} \Gamma_r^{(2m+2)\Lambda} +
 \Same^{\Lambda} (\Four{\Lambda})\, \Gamma_{r-1}^{(2m)\Lambda}
\nonumber \\
 &+&
 \sum_{p \ge 2}
 \sum {}'\;
 \Lower^\Lambda_p 
 \left(
 \Gamma_{r_0}^{(2m_0)\Lambda},
 \ldots,
 \Gamma_{r_p}^{(2m_p)\Lambda}
 \right) . \hskip 5mm
\end{eqnarray}
The primed sum runs over all sequences $(m_0, \ldots m_p)$ 
and all sequences $(r_0, \ldots, r_p)$ with
$m_q \ge 1$ and $r_q \ge 1$ for all $1 \le q \le p$, 
$\; m_0 + \ldots + m_p = m+ p$, and $r_0 + \ldots + r_p =r$.
The solution of the RG hierarchy for an initial quartic interaction has
the property that $\Gamma_r^{(2m)\Lambda}=0$ for all $m > r$. 
Therefore, for $m=r$, the $\Higher^\Lambda$ term drops out of (\ref{HSL}),
and all remaining terms contain only $\Four{\Lambda}$ or terms of order
at most $r-1$ in $\Four{\Lambda}$. Thus, given these lower-order
$\Gamma$'s, $\Gamma_r^{(2r)\Lambda}$ can be obtained by integration. 
Then the right hand side of the equation for $m=r-1$ is determined, so
$\Gamma_r^{(2r-2)\Lambda}$ can be determined, and so on. 
Successive backsubstitution then leads to a system of equations where
$\frac{d}{d\Lambda} \Gamma_r^{(2m)\Lambda}$ gets contributions
from a sum of graphs with $r$ vertices of type $\Four{\Lambda'}$, where
$\Lambda_0 \ge \Lambda' \ge \Lambda$, and propagators $G^{\Lambda''}$
and $S^{\Lambda'''}$, and all the intermediate scales $\Lambda'$ etc.\ are 
integrated.
Thus the equation becomes nonlocal in the flow parameter $\Lambda$
but the right hand side is known once $\Four{\Lambda}$ and $\Sigma^\Lambda$
are known. $\Four{\Lambda}$ is given by a degree $r$ nonlinear equation,
with a similar graphical background as discussed above, and $\Sigma^\Lambda$
by the standard self-energy equation (\ref{floweqSigma}).
While more restricted than the level-$m$ truncation,
the running coupling scheme also captures effects
that cannot be seen in any fixed order of bare perturbation theory,
such as screening or asymptotic freedom of certain coupling functions.

We use this inductive structure to derive basic power counting bounds 
for the vertex functions in terms of the flowing four-point function, 
for a $d$-dimensional bulk fermion system ($d \ge 1$). 
For simplicity, we focus on spin-$\frac12$ fermions with translation-invariant
action, so that we can use $x_i = (k_i,\sg_i) = (k_{0,i},\bk_i,\sg_i)$, 
as discussed at the beginning of Section \ref{sec:functionals}. 
We also assume that the symmetries of the action remain unbroken.
These specific assumptions are for presentation only; power counting can
be done without them.
By translation invariance, 
$\Gamma_r^{(2m)\Lambda} \left((k_1,\sg_1), \ldots (k_{2m},\sg_{2m})\right)
 = \delta \left( \sum_i k_i \right)\, 
 \hat \Gamma_r^{(2m)\Lambda}(\km,\bsg)$,
where the delta function forces conservation of the spatial momentum
$\bk$ (up to reciprocal lattice vectors) and conservation of the frequency
variable $k_0$, and we have introduced the abbreviations
$\bsg = (\sg_1,\ldots \sg_{2m})$ and $\km = (k_1, \ldots, k_{2m-1} )$.
For $\Lambda=\Lambda_0$, 
the function $\hat \Gamma_r^{(2m)\Lambda}$ is smooth and bounded
because the initial interaction is short-range,
and this stays so during the flow above critical scales. 

Consider the maximal size of the vertex functions, 
$\Vert \Gamma_r^{(2m)\Lambda} \Vert =
 \sup\limits_{\km, \bsg} 
 | \hat \Gamma_r^{(2m)\Lambda}(\km,\bsg) |$. 
Then, for $m \ge 3$,
\begin{equation} \label{poco}
 \Vert \Gamma_r^{(2m)\Lambda} \Vert \le \gamma_r^{(2m)}\; 
 {\snorm{\Lambda}}^{r-m+1}
 {f_\Lambda}^r \; 
 \Lambda^{2-m} \; ,
\end{equation}
where $\gamma_r^{(2m)}$ is independent of $\Lambda$ and $\beta$, 
\begin{equation}
 f_\Lambda = \sup\limits_{\Lambda \le \ell \le \Lambda_0}  
 \Vert\Four{\ell}\Vert
\end{equation}
is the maximal value of the four-point coupling on all scales 
between $\Lambda$ and $\Lambda_0$, and 
\begin{equation}
 \snorm{\Lambda}  =  \max\limits_\alpha \sum\limits_{\alpha'} \;
 \int \db k \; |\hat S^\Lambda_{\alpha,\alpha'} (k) | \; ,
\end{equation}
with 
$\int \db k \dots = T \sum_{k_0} \int \frac{d^dk}{(2\pi)^d} \dots$
(for the general power counting, we do not need to assume that the 
propagator is diagonal in the spin indices $\alpha,\alpha'$, so $\hat S^\Lambda$
also carries these indices). 
The dependence of $\snorm{\Lambda}$ on $\Lambda$ is determined
by the shape of the Fermi surface. 
As shown in Appendix \ref{appssec:propbounds}, $\snorm{\Lambda}$ 
is of order one for regular Fermi surfaces. If the Fermi surface 
contains Van Hove points, $\snorm{\Lambda}$ grows logarithmically
in $\Lambda$ for $\Lambda \to 0$. 

At first sight, one may worry about the factor $\Lambda^{2-m}$, 
which diverges for $m \ge 3$ as $\Lambda \to 0$. For the maximum value 
of the vertex functions, it is indeed true -- and easily verified in
examples -- that there are always particular values of the external 
momenta where these vertex functions become very large in $\beta = 1/T$
(and diverge at zero temperature).
However, this happens only on a ``small'' set of momenta. 
For a general $m$-point function, it is very involved to determine this
set, but this is not necessary for power counting. 
One can use the $L^1$ norm instead, i.e.\ consider 
$\Vert \Gamma_r^{(2m)\Lambda} \Vert_1 =
 \sum_{\bsg} \int \db k_1 \ldots \db k_m \; 
 |\Gamma_r^{(2m)\Lambda} (\km,\bsg)|$.
Using generalizations of (\ref{poco}), 
one can then show that if $f_\Lambda$ remains finite 
\begin{equation}
 \Vert
 \Gamma_r^{(2m)\Lambda}
 \Vert_1
 \le
 c_r^{(2m)} \; f_\Lambda^r
\end{equation}
with constants $c_r^{(2m)}$ that are independent of $\Lambda$, $\beta$ 
and the system size $L$ (see \textcite{Salmhofer99}, Section 4.4.3). 
This implies that, even in the limit $\beta \to \infty$, 
the $2m$-point vertices can become singular only on a set of 
zero Lebesgue measure in momentum space. In general, this set can be
rather complicated, but, loosely speaking, it will have codimension at
least one. 

It is one of the appealing features of the flow-equation RG that exact
statements like (\ref{poco}) can be proven in a few lines,
see Appendix \ref{app:powercounting}.
The argument given there also implies an at-most logarithmic growth of
the coefficients in the equation for $f_\Lambda$ itself. 
The self-energy then comes out of order $f_\Lambda$, 
provided that renormalization is done correctly, see
Section \ref{sssec:selfenergy}.
At small $f_\Lambda$, the size of the vertices $\Gamma^{(2m)\Lambda}$
with $m \ge 3$ is thus determined by $f_\Lambda$. 
The terms with $m \ge 3$ are the RG-irrelevant ones, $m=2$ is marginal, 
and $m=1$ is relevant. 
This classification is explained in detail in
Appendix \ref{app:powercounting}.
In a Taylor expansion of the vertex functions in the Matsubara 
frequency around zero, and in momentum around the Fermi surface, 
additional small factors arise, which cancel the small denominators 
of the propagators; at the same time, the vertex function is replaced 
by a differentiated one.
Hence, the flow obtained by projecting the frequencies to zero and the 
momenta to the Fermi surface gives the dominant contribution for small 
$\Lambda$.
This is expected from a simple counting of bare scaling dimensions,
and can be established more rigorously by power-counting arguments
similar to those used above and in Appendix \ref{app:powercounting}.

In the case of a curved Fermi surface in $d \ge 2$, $f_\Lambda$ indeed 
stays small in a weakly interacting system above a BCS-like temperature 
(see, e.g., \textcite{Salmhofer98b}), indicating the absence of 
symmetry-breaking.
At zero temperature, $f_\Lambda$ grows as $\Lambda$ decreases, and 
the four-point function has singularities at points corresponding to 
nesting vectors of the Fermi surface; 
for details, see Appendix \ref{ssec:imppoco}.
This growth of $f_\Lambda$ with decreasing scale $\Lambda$ is 
usually called the ``flow to strong coupling'' in RG studies, 
and is described in more detail in Section \ref{sec:III}.
A singularity of the two-particle vertex in momentum space
means that the interaction becomes long-range in position space.
This is associated with the formation of critical fluctuations,
and in case of spontaneous breaking of a continuous symmetry, 
with the appearance of Goldstone bosons (see Section \ref{sec:IV}).

\subsubsection{Self-energy and Fermi surface shift}
\label{sssec:selfenergy}

The self-energy is important for all effective one-particle properties 
of the system, and it can cause drastic effects, as compared to the 
non-interacting fermions. 
Accordingly, in the RG flow, the self-energy is a relevant term. 
In absence of symmetry breaking, it modifies the inverse 
propagator to $ik_0 - \xi_{\bk} - \Sigma (k_0,\bk)$. 
The long-distance behaviour of the fermionic propagator is determined
by the behaviour of this function around its zero set. A Taylor expansion
around $k_0=0$ gives
\begin{eqnarray}
 ik_0 - \xi_{\bk} - \Sigma (k_0,\bk) =
 \frac{ik_0- e_{\bk}}{Z_{\bk}} + \rho(k_0,\bk)
\end{eqnarray}
with 
$Z_{\bk}^{-1} = 1 + i (\partial_0 \Sigma) (0,\bk)$,
$Z_{\bk}^{-1} e_{\bk} = \xi_{\bk} + \Sigma (0,\bk)$
and a Taylor remainder $\rho$. If $\rho$ vanishes
faster than linearly in $k_0$ as $k_0 \to 0$, we thus obtain 
an effective description in terms of quasiparticles with
dispersion relation $e_\bk$, hence ``interacting'' Fermi surface
$\left\{\bk: \xi_{\bk} + \Sigma(0,\bk) = 0 \right\}$, Fermi velocity
$\nabla e_{\bk}$, and quasiparticle weight $Z_{\bk}$.
If $Z_{\bk}$ is bounded and nonvanishing for all $\bk$, 
the long-distance decay of the fermion propagator in position space
is the same as for the free theory. 

The question whether $\Sigma$ is smooth enough for the above to hold 
is nontrivial. In the one-dimensional Luttinger model, 
$\partial_0 \Sigma(0,k_F)$ diverges, and the small-$k_0$ behaviour of
the self-energy, $\Sigma (k_0,k_F) \sim |k_0|^\alpha$ with $\alpha < 1$
depending on the interaction strength, implies that the self-energy effects
dominate at small $k_0$ and the decay in position space becomes more rapid,
so that the occupation number density $n(\bk)$ becomes a continuous function of
$\bk$ even at zero temperature \cite{Giamarchi04}. 
An even more drastic change is spontaneous symmetry breaking, where
the propagator cannot be written any more in the simple form given above
(see Section \ref{sec:IV}).

In the RG flow, $\Sigma$ is replaced by the $\Lambda$-dependent 
self-energy $\Sigma^\Lambda = (G_0^\Lambda)^{-1} - (G^\Lambda)^{-1}$.
An important phenomenon in this
context is the shift in the Fermi surface entailed by $\Sigma^\Lambda$. 
In terms of power counting, this shift is the most relevant term.
Cutting off the propagator around the free Fermi surface then fails to
regularize the propagator, 
which leads to spurious singularities in the RG flow.
A convenient method to avoid this is to introduce a counterterm. 
In the context of the bare perturbation expansion, the counterterm 
method was already described in \cite{Nozieres64}.
The main idea is to anticipate the form that the propagator takes 
at the end of the flow and to rearrange the flow such that this form, 
not the bare one is used as the starting point for the RG analysis, 
hence the Fermi surface is fixed to that of the interacting system in 
the flow. 
The difference between the two dispersion functions appears as a (finite)
counterterm. To obtain a one-to-one relation between the model given by
the Hamiltonian and the one with fixed interacting Fermi surface, 
one has to solve a self-consistency equation. 
\textcite{FT1} used the counterterm method for the radius shift of
a circular Fermi surface in a RG flow. 
\textcite{FST1,FST2,FST3,FST4} generalized this to the case of 
non-circular curved Fermi surfaces, solved the self-consistency equation, 
and showed that $Z_{\bk}$ remains finite for $d \ge 2$ to all orders. 
The corresponding fixed-point problem for the Fermi surface was also
considered by \textcite{Ledowski03}.
The role of Van Hove singularities was analyzed by \textcite{FS1,FS2}.
\textcite{FT2} also used the counterterm method to
derive the equation for the superconducting gap from an RG flow; 
for further work in that direction, see also Section \ref{sec:IV}.
In alternative to counterterms, one can try to avoid a momentum space 
cutoff altogether \cite{Honerkamp01b,Honerkamp01c,Husemann09a}, or to 
use an adaptive scheme (see the appendix in \textcite{Honerkamp01d} and 
\textcite{BGM06,Salmhofer07}).


\subsection{Flow equations for observables and correlation functions}
\label{sec:responsefcts}

All observables of the fermionic system are given by polynomials 
in the fields, so they can be calculated from the connected Green 
functions $G^{(2m)\Lambda}$, hence by the above-mentioned tree 
relations also from the irreducible vertex functions 
$\Gamma^{(2m)\Lambda}$.
It is nevertheless convenient, and due to the limitations of 
approximations often mandatory, to calculate the flow of 
observables and their correlation (or response) functions 
by separate flow equations, which we derive and discuss now. 

For simplicity we restrict the presentation to the particularly 
important class of observables that are correlations of fermionic 
bilinears.
Charge-invariant bilinears are of the form
\begin{equation}
\cB(x)
=
\sum_{y,y'} \psib (y) \, B(x; y,y') \psi (y') .
\end{equation}
Charge-non-invariant bilinears are of the form 
\begin{equation}
\cB(x)
=
\sum_{y,y'} 
\left(
\psi (y) \, B(x; y,y') \psi (y')
+
\psib (y) \, \tilde B(x; y,y') \psib (y')
\right) .
\end{equation}
The functions $B$ and $\tilde B$  determine the spatial and spin 
structure of these bilinears. 
For translation-invariant systems, we can choose a momentum 
representation where $x = (k_0,\bk)$ and $y = (p_0,\bp,\sigma)$,
as explained above Eq.~(\ref{g0k}).
With the notations $p = (p_0,\bp)$ and 
$\int \db p = T \sum_{p_0} \int d\bp$, 
a charge invariant bilinear is of the form 
\begin{equation}
 \cB(k) =
 \int \db p\;
 \psib_\sigma (p)\, 
 b_{\sigma,\sigma'} (p,k) \, \psi_{\sigma'} (p+k) \, .
\end{equation}
The frequency $k_0$ is an integer multiple of $2\pi T$.
The case $k = 0$ and $b = \sigma_i$, where $\sigma_i$ denotes
the $i^{\rm th}$ Pauli matrix, corresponds to a uniform spin
density. The case $k_0 =0$, $\bk = (\pi,\pi, \ldots, \pi)$ 
and $b = \sigma_i$ corresponds to a staggered spin density. 
Similarly, the same choices of $k$ but with $b_{\sigma,\sigma'}
= \delta_{\sigma,\sigma'}$ correspond to charge densities. 
Other choices of $\bk$ can be used to test tendencies towards 
noncommensurate magnetic or charge ordering. 
Cooper pair fields correspond to the non-charge invariant 
combinations
\begin{eqnarray}
\cB(k) 
&=&
\int \db p\;
\big[\, 
\psib_\sigma (p)\, 
\Delta_{\sigma,\sigma'} (p,k) \, \psib_{\sigma'} (-p+k)
\nonumber\\
&+&
\psi_\sigma (p)\, 
\overline{\Delta}_{\sigma',\sigma} (-p+k,k) \, \psi_{\sigma'} (-p+k)\, 
\big]\; .
\end{eqnarray}
Again, the simplest choice is uniform singlet pairing, 
where $k=0$ and 
$\Delta_{\sigma,\sigma'} (p,k) = \Delta (\bp) \varepsilon_{\sigma,\sigma'}$. In this case 
$\Delta (\bp)$ is the gap function. 
Triplet pairing, extended Cooper pairs, and spatially nonuniform gaps, are described
by suitable generalizations. 

A convenient way of generating correlation functions of the 
bilinears $\cB$ is to couple the $\cB (x)$ to external source fields 
$J (x)$, i.e.\ to add a term $\left(J,\cB\right) = \sum_x J(x) \cB(x)$
to the action. The external field $J$ is not an integration variable, 
so it can be regarded as a (functional) parameter on which 
$\cG$ depends. Writing $\cG=\cG(J,\eta,\etab)$, we then have
\begin{equation}
 \langle \cB(x) \, \cB(y) \rangle
 -
 \langle \cB(x) \rangle \, \langle \cB(y) \rangle
 =
 - \frac{\partial^2 \cG (J,\eta,\etab)}{\partial J(x) \partial J(y)} \,
 \Big\vert_{J=0\atop\eta,\etab=0} \; .
\end{equation}
In presence of $J$, the effective action 
$\Gamma = \Gamma (J, \psi,\psib)$, 
as well as all other quantities appearing in the fermionic 
Legendre transform (\ref{Gampsi}), depend on $J$ as well. 
Since relations such as 
$\frac{\partial \Gamma}{\partial \psib} (J, \psi,\psib)=\eta (J, \psi,\psib)$
remain valid for any $J$, straightforward differentiation yields
\begin{equation}\label{gaGaJ}
 \frac{\partial^2 \cG (J, \eta,\etab) }{\partial J(x) \partial J(y)} \,
 \Big\vert_{J=0 \atop \eta,\etab=0} 
 =
 \frac{\partial^2 \Gamma (J, \psi,\psib) }{\partial J(x) \partial J(y)} \,
 \Big\vert_{J=0 \atop \psi,\psib=0} \; .
\end{equation}
Graphically, this relation is intuitive in that the bilinears 
always couple to the (effective) vertices by two lines, 
and the fermionic vertices are all even, so that the graphs that
contribute are automatically irreducible. 

\begin{figure}[ht]
\centerline{\includegraphics[width = 5cm]{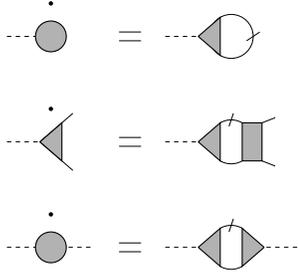}}
\caption{The truncation of the hierarchy for the response function
that corresponds to keeping only the irreducible two-particle 
vertex in the fermionic hierarchy.}
\label{fig:response2}
\end{figure}

Again, because $J$ plays the role of a parameter, 
the flow equation (\ref{floweqGam}) is unchanged. 
Flow equations for the response functions are then 
obtained simply by expanding $\Gamma$ in the 
fields $J$ and comparing coefficients. This again leads
to a hierarchy of equations for the vertex functions
$\Gamma^{(2m,n)}$ that have $2m$ fermionic and 
$n$ external bosonic lines. The $J$-independent term
corresponds to the standard fermionic hierarchy for the 
$\Gamma^{(2m,0)} = \Gamma^{(2m)}$, which therefore remains 
unchanged. 
When counting powers in the fermionic fields, 
each $J$ corresponds to a bilinear, so that the truncation 
$\Gamma^{(2m)} = 0$ for $m \ge m_0$ for the fermionic 
vertices corresponds to a truncation $\Gamma^{(2m,n)} = 0$
for $m+n \ge m_0$. 
The flow equations remaining after a truncation for $m_0 = 3$ 
are shown diagrammatically in Figure \ref{fig:response2}.

Clearly, the two-point correlation
$\langle \cB(x)\, \cB(y)\rangle 
 - \langle \cB(x) \rangle\, \langle \cB(y)\rangle$
of any fermionic bilinear $\cB$ only involves the fermionic 
two- and four-point functions, hence could simply be calculated 
from the knowledge of $\Gamma^{(2)}$ and $\Gamma^{(4)}$
by the reciprocity relation and (\ref{G4Ga4}). 
Eq.\ (\ref{gaGaJ}) shows that the route via external fields 
in the one-particle irreducible equations is strictly equivalent 
to this if the hierarchy is treated exactly.
When making truncations to the hierarchy and other approximations, 
the two are no longer the same.
Anomalous scaling dimensions of fermionic bilinears (or other
composite objects) are captured easily by separate flow equations
for these quantities, while they are hard to obtain from 
$\Gam^{(2)}$ and $\Gam^{(4)}$, if the latter are computed from
a truncated flow equation.
An instructive example is given by the calculation of the 
density profile near a static impurity in a Luttinger liquid 
in \textcite{Andergassen04}. 


\subsection{Flow equations for coupled boson-fermion systems}
\label{sec:bosefermi}

The focus of this review is on fermion systems. 
However, even if the bare action involves only fermionic fields,
bosonic degrees of freedom are frequently generated as fermion
composites and order parameter fields. For example, Cooper pairs
and the order parameter in a superconductor are bosons.
Technically, bosonic fields are introduced in an originally purely
fermionic theory by a Hubbard-Stratonovich decoupling of an 
interaction between fermions \cite{Popov87}.
Often the fermionic fields are subsequently intregrated out, such
that an effective action involving only bosons remains.
Otherwise one has to deal with a coupled theory of fermions and
bosons. 
In this section we generalize the flow equations derived in 
Sec.~\ref{sec:floweqs} to interacting boson-fermion systems.
Flow equations for coupled boson-fermions systems have been
derived by various groups, with slight differences in the
notation \cite{Berges02,Kopietz10}.

We first introduce some notation for bosons, and write down the
bosonic analogues of some of the most important equations from 
Sec.~\ref{sec:floweqs}.
Bosonic particles are described by complex fields $\phi$.
It is convenient to combine $\phi$ and its complex conjugate 
$\phi^*$ in a bosonic Nambu field
\begin{equation} \label{nambuPhi}
 \Phi(x) = \left( \begin{array}{l} \phi(x) \\
 \phi^*(x) \end{array} \right) \; .
\end{equation}
The generating functional for connected Green functions can be
written as \cite{Negele87}
\begin{equation} \label{calGb}
 \cG[H] = 
 - \ln \int \cD\Phi \, e^{-\cS[\Phi]} \, e^{(H^*,\Phi)}
 \; ,
\end{equation}
where $\cS[\Phi]$ is the bare action, and 
\begin{equation} \label{nambuHb}
 H(x) = \left( \begin{array}{l} h(x) \\
 h^*(x) \end{array} \right) \; .
\end{equation}
the source field.
Connected Green functions are obtained as functional derivatives
\begin{eqnarray} \label{G2mb}
 && G^{(2m)}(x_1,\dots,x_m;x'_1,\dots,x'_m) =
 \nonumber \\[1mm]
 && \bra \phi(x_1) \dots \phi(x_m) 
             \phi^*(x'_m) \dots \phi^*(x'_1) \ket_c =
 \nonumber \\[1mm]
 && - \left.
 \frac{\partial^{2m} \cG[H]}
 {\partial h^*(x_1) \dots \partial h^*(x_m)
 \partial h(x'_m) \dots \partial h(x'_1)}
 \right|_{H = 0} . \hskip 5mm
\end{eqnarray}
The effective action is defined as Legendre transform
\begin{equation} \label{GamPhi}
 \Gam[\Phi] = (H^*,\Phi) + \cG[H] \; ,
\end{equation}
with $\Phi = - \partial\cG/\partial H^*$. 
Functional derivatives of $\Gam[\Phi]$ yield the bosonic 
$m$-particle vertex functions 
\begin{eqnarray} \label{Gam2mb}
 && \Gam^{(2m)}(x_1,\dots,x_m;x'_1,\dots,x'_m) =
 \nonumber \\[1mm]
 && \left.
 \frac{\partial^{2m} \Gam[\Phi]}
 {\partial\phi^*(x_1) \dots \partial\phi^*(x_m)
 \partial\phi(x'_m) \dots \partial\phi(x'_1)}
 \right|_{\Phi = 0} . \hskip 5mm
\end{eqnarray}
The matrices of second derivatives at finite fields
\begin{eqnarray} \label{bG2b}
 \bG^{(2)}[H] &=& 
 - \frac{\partial^2 \cG}{\partial H^*(x) \partial H(x')}
 \nonumber \\[1mm]
 &=& \left( \begin{array}{cc} 
 \bra \phi(x) \phi^*(x') \ket & \bra \phi(x) \phi(x') \ket \\
 \bra \phi^*(x) \phi^*(x') \ket & \bra \phi^*(x) \phi(x') \ket
 \end{array} \right) 
\end{eqnarray}
and
\begin{equation} \label{bGam2b}
 \bGam^{(2)}[\Phi] = 
 \frac{\partial^2 \Gam}{\partial\Phi^*(x) \partial\Phi(x')}
\end{equation}
obey the reciprocity relation
$\bGam^{(2)}[\Phi] = (\bG^{(2)}[H])^{-1}$.

Endowing the bare propagator $G_0$ with a cutoff or another 
scale dependence, one can derive exact flow equations for the 
generating functionals in complete analogy to the fermionic
case. 
In particular, the flow equation for the effective action
$\Gam^{\Lam}[\Phi]$ has the form
\begin{equation} \label{floweqGamb}
 \frac{d}{d\Lam} \Gam^{\Lam}[\Phi] = 
 \frac{1}{2} (\Phi^*,\dot\bQ_0^{\Lam} \Phi) +
 \frac{1}{2} {\rm tr} \left[ \dot\bQ_0^{\Lam}
 \left( \bGam^{(2)\Lam}[\Phi] \right)^{-1} \right] \; ,
\end{equation}
where
$\bQ_0^{\Lam} = \mbox{diag} (Q_0^{\Lam}, Q_0^{\Lam t})$
with $Q_0^{\Lam} = (G_0^{\Lam})^{-1}$.
Note that the first term on the right hand side can also be 
written as $(\phi^*,\dot Q_0^{\Lam} \phi)$.
The above flow equation is equivalent to the frequently
used flow equation for the effective average action
\cite{Berges02}
\begin{equation} \label{GamRLamb}
 \Gam_R^{\Lam}[\Phi] = \Gam^{\Lam}[\Phi] -
 \frac{1}{2} (\Phi^*,\bR^{\Lam} \Phi) \; ,
\end{equation}
with $R^{\Lam} = Q_0^{\Lam} - Q_0$, which reads
\cite{Wetterich93}
\begin{equation} \label{floweqGamRb}
 \frac{d}{d\Lam} \Gam_R^{\Lam}[\Phi] =
 \frac{1}{2} {\rm tr} \left[ \dot \bR^{\Lam} \left(
   \bGam_R^{(2)\Lam}[\Phi]
   + \bR^{\Lam} \right)^{-1} \right] \; .
\end{equation}
Order parameters are often associated with real (not complex)
bosonic fields. In that (simpler) case the above equations 
are still valid if one replaces the complex Nambu fields
$\Phi$ and $H$ by the real fields $\phi$ and $h$.

A generalization to coupled fermion-boson systems is now
straightforward. 
Bosonic and fermionic fields are conventiently collected in 
a ''super-field''
\begin{equation} \label{Xi}
 \Xi = \left( \begin{array}{c}
 \Phi \\ \Psi \end{array} \right) \; ,
\end{equation}
where $\Phi$ and $\Psi$ are the bosonic and fermionic Nambu
fields defined above (see Sec.~II.A). 
The conjugate super-field is given by
\begin{equation} \label{Xib}
 \Xib = \left( \begin{array}{l}
 \Phi^* \\ \Psib \end{array} \right) \; .
\end{equation}
The generating functional for connected Green functions
involving both bosons and fermions reads
\begin{equation} \label{calGbf}
 \cG[H_b,H_f] = 
 - \ln \int \cD\Phi \cD\Psi \, 
 e^{-\cS[\Phi,\Psi]} \, e^{(H_b^*,\Phi) + (\bar H_f,\Psi)}
 \; ,
\end{equation}
where $\cS[\Phi,\Psi]$ is the bare action, and $H_b$ and $H_f$
are the Nambu source fields for bosons and fermions, respectively.
Functional derivatives with respect to the source fields generate
connected Green functions with an arbitrary number of bosonic and
fermionic fields, the only general constraint being that the number 
of fermion fields is always even. 

The effective action $\Gam[\Phi,\Psi]$ is given by the 
Legendre transform
\begin{equation} \label{GamPhiPsi}
 \Gam[\Phi,\Psi] = 
 (H_b^*,\Phi) + (\bar H_f,\Psi) + \cG(H_b,H_f) \; ,
\end{equation}
where $\Phi = - \partial\cG/\partial H_b^*$ and 
$\Psi = - \partial\cG/\partial \bar H_f$.
The source fields may also be collected in a super-field
\begin{equation} \label{calH}
 \cH = \left( \begin{array}{c}
 H_b \\ H_f \end{array} \right) \; .
\end{equation}
The Legendre transform can then be written more
concisely as $\Gam[\Xi] = (\bar\cH,\Xi) + \cG(\cH)$.

The matrix of second functional derivatives of $\cG$ at finite 
fields
\begin{eqnarray} \label{bG2bf}
 \bG^{(2)}[\cH] &=& 
 - \frac{\partial^2 \cG}{\partial\bar\cH(x) \partial\cH(x')}
 \nonumber \\[1mm]
 &=& \left( \begin{array}{cc} 
 \bra \Phi(x) \Phi^*(x') \ket & - \bra \Phi(x) \Psib(x') \ket \\
 \bra \Psi(x) \Phi^*(x') \ket & - \bra \Psi(x) \Psib(x') \ket
 \end{array} \right)
\end{eqnarray}
involves also mixed boson-fermion propagators, which vanish
only for $\cH = 0$.
The matrix of second derivatives of the effective action
\begin{eqnarray} \label{bGam2bf}
 \bGam^{(2)}[\Xi] &=& 
 \frac{\partial^2 \Gam}{\partial\Xib(x)\Xi(x')} 
 \nonumber \\[1mm] 
 &=& \left( \begin{array}{cc}
 \dps \frac{\partial^2\Gam}{\partial\Phi^*(x)\partial\Phi(x')} &
 \dps \frac{\partial^2\Gam}{\partial\Phi^*(x)\partial\Psi(x')} \\[3mm]
 \dps \frac{\partial^2\Gam}{\partial\Psib(x)\partial\Phi(x')} &
 \dps \frac{\partial^2\Gam}{\partial\Psib(x)\partial\Psi(x')}
 \end{array} \right) \; . \hskip 5mm
\end{eqnarray}
is related to $\bG^{(2)}[\cH]$ by the reciprocity relation
$\bGam^{(2)}[\Xi] = (\bG^{(2)}[\cH])^{-1}$.

A flow of the generating functionals is generated by modifing the 
bare propagators for bosons and fermions, $G_{b0}$ and $G_{f0}$,
such that they depend on some scale parameter $\Lam$. 
We denote the scale dependent bare propagators by $G_{b0}^{\Lam}$ 
and $G_{f0}^{\Lam}$, and their inverse by $Q_{b0}^{\Lam}$ 
and $Q_{f0}^{\Lam}$.
The generalization of the exact flow equations for the effective 
action in purely bosonic or fermionic systems to coupled boson-fermion 
systems reads
\begin{eqnarray} \label{floweqGambf}
 \frac{d}{d\Lam} \Gam^{\Lam}[\Phi,\Psi] &=& 
 \frac{1}{2} (\Phi^*,\dot\bQ_{b0}^{\Lam} \Phi) -
 \frac{1}{2} (\Psib,\dot\bQ_{f0}^{\Lam} \Psi) 
 \nonumber \\
 &+& \frac{1}{2} {\rm Str} \left[ \dot\bQ_0^{\Lam}
 \left( \bGam^{(2)\Lam}[\Phi,\Psi] \right)^{-1} \right] \; ,
\end{eqnarray}
where
\begin{equation} \label{bQ0Lambf}
 \bQ_0^{\Lam} = \left( \begin{array}{cc}
 \bQ_{b0}^{\Lam} & 0 \\ 0 & \bQ_{f0}^{\Lam}
 \end{array} \right) \; .
\end{equation}
The supertrace Str incorporates a minus sign in the fermionic sector.
The flow equation (\ref{floweqGambf}) is equivalent to the flow
equation \cite{Berges02}
\begin{equation} \label{floweqGamRbf}
 \frac{d}{d\Lam} \Gam_R^{\Lam}[\Phi,\Psi] =
 \frac{1}{2} {\rm Str} \left[ \dot \bR^{\Lam} \left(
   \bGam_R^{(2)\Lam}[\Phi,\Psi]
   + \bR^{\Lam} \right)^{-1} \right] \, ,
\end{equation}
with $\bR^{\Lam} = \bQ_0^{\Lam} - \bQ_0$, 
for the effective average action 
\begin{equation}
 \Gam_R^{\Lam}[\Phi,\Psi] = \Gam^{\Lam}[\Phi,\Psi] -
 \frac{1}{2}(\Phi^*,\bR_b^{\Lam} \Phi) +
 \frac{1}{2}(\Psib,\bR_f^{\Lam} \Psi) \, .
\end{equation}

\begin{figure}[ht]
\centerline{\includegraphics[width = 8cm]{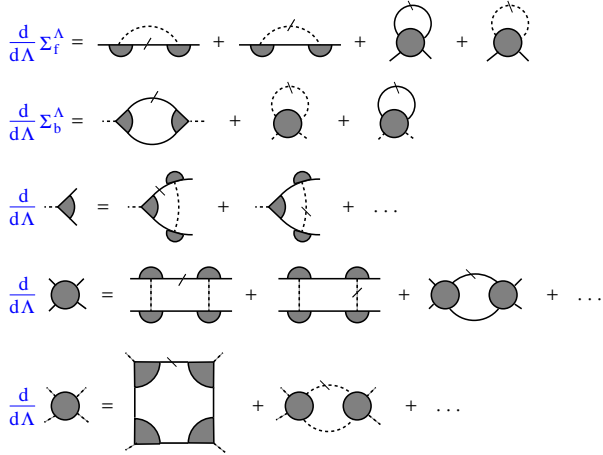}}
\caption{Diagrammatic representation of the flow equations for
 the (fermionic and bosonic) self-energies and some of the 
 interaction vertices in a coupled boson-fermion theory.
 Solid lines denote fermionic, dashed lines bosonic propagators.
 Propagators with a dash are single-scale propagators.}
\label{fig:floweq1pi_bf}
\end{figure}

The expansion of the exact functional flow equation (\ref{floweqGambf})
proceeds in complete analogy to the purely fermionic case.
Inserting
\begin{equation} \label{bGam2decbf}
 \bGam^{(2)\Lam}[\Phi,\Psi] = 
 (\bG^{\Lam})^{-1} - \tilde\bSg^{\Lam}[\Phi,\Psi] \; ,
\end{equation}
with
\begin{equation} \label{bGLambf}
 \bG^{\Lam} =
 \left(
 \left. \bGam^{(2)\Lam}[\Phi,\Psi] \right|_{\Phi = \Psi = 0} 
 \right)^{-1} =
 \mbox{diag} (\bG_b^{\Lam},\bG_f^{\Lam}) \; ,
\end{equation}
into the functional flow equation (\ref{floweqGambf}), one obtains
\begin{widetext}
\begin{eqnarray} \label{floweqGambfexp}
 \frac{d}{d\Lam} \Gam^{\Lam}[\Phi,\Psi] &=&
 \frac{1}{2} {\rm Str} \left( \dot \bQ_0^{\Lam} \bG^{\Lam} 
 \right) +
 \frac{1}{2} (\Phi^*,\dot\bQ_{b0}^{\Lam} \Phi) -
 \frac{1}{2} (\Psib,\dot\bQ_{f0}^{\Lam} \Psi) 
 \nonumber \\ &-&
 \frac{1}{2} {\rm Str} \left[ \bS^{\Lam} 
 \left( \tilde\bSg^{\Lam}[\Phi,\Psi] +
 \tilde\bSg^{\Lam}[\Phi,\Psi] \, \bG^{\Lam} 
 \tilde\bSg^{\Lam}[\Phi,\Psi] + \dots \right) \right] \; ,
\end{eqnarray}
\end{widetext}
with the single-scale propagator
\begin{equation} \label{SLambf}
 \bS^{\Lam} = - \bG^{\Lam} \dot\bQ_0^{\Lam} \bG^{\Lam} =
 \mbox{diag}(\bS_b^{\Lam},\bS_f^{\Lam}) \; .
\end{equation}
The expansion in powers of the fields is now straightforward 
and leads to a hierarchy of flow equations for all vertex 
functions. The first few terms are shown diagrammatically in
Fig.~\ref{fig:floweq1pi_bf}.

The flow equations derived above are also valid in case of 
$U(1)$-symmetry breaking, if one allows for off-diagonal elements
in the matrices $\bQ_{b0}^{\Lam}$, $\bQ_{f0}^{\Lam}$, 
$\bG_b^{\Lam}$, $\bG_f^{\Lam}$ etc. 
\cite{Berges02,Schuetz06}.

Coupled flow equations for fermions and bosonic Hubbard-Stratonovich
fields are particularly convenient to treat fluctuations associated 
with spontaneous symmetry breaking (see Sec.~IV) and quantum 
criticality (see Sec.~V), but they may also be used to study 
Luttinger liquids and other symmetric states in interacting Fermi
systems \cite{Schuetz05,Bartosch09a,Ledowski07a,Ledowski07b}.


\section{COMPETING INSTABILITIES}
\label{sec:III}

\newcommand{\Last}{\Lambda_*}
\newcommand{\Tst}{T_*}
\renewcommand{\bQ}{{\bf Q}}
\newcommand{\pk}{\kappa}

In this section we describe how one can apply the level-$2$ truncation 
of the fermionic RG, mainly without self-energy corrections, 
to two-dimensional fermion systems, 
and study the interplay of ordering tendencies. 
In resummations of perturbation theory, 
their manifestation are singularities in the four-point function
and in certain susceptibilities. 
In the RG, the precursor to a singularity
is the growth of some parts of the vertex function 
(often termed ``flow to strong coupling''). 
Since singularities in the vertex function change the power counting drastically, 
this truncated flow then has to be stopped before a singularity happens, 
at a scale $\Last > 0$, 
where one can read off the dominant interactions and infer a tentative phase diagram
(in this, susceptibilities are used to compare the strength of different 
ordering tendencies and to determine $\Last$).
As discussed in Sec.~\ref{sec:truncations}, curvature effects of the Fermi 
surface imply that the truncations discussed here can be used also when the 
interaction is no longer small, provided that the power counting
improvement factor times the interaction strength remains small.
To obtain a true phase diagram, however, one needs to integrate over all 
degrees of freedom, also those with scales below $\Last$. 
This has been achieved in some cases (see Section \ref{sec:IV}), but much remains
to be done.

This first step of monitoring the flow to strong coupling above $\Last$, 
as described in this section, is important for the following reasons.
(1) It allows to determine the effective interaction just above transition scales
from the given microscopic model
without any additional a priori assumptions about the nature of symmetry-breaking,
and thereby provides an initial condition for the integration at scales below $\Last$.
(2) It exhibits how the interplay of the scale-dependent scattering processes 
on different parts of the Fermi surface gradually builds up the effective 
interaction.
(3) It has by now become a versatile tool for analyzing models with 
an elaborate microscopic structure, such as multiple bands.

\subsection{Hubbard model and $N$-patch RG schemes}
\label{sec:2dHubbard}

The Hubbard model and its extensions have become standard in correlated fermion systems:
on the square lattice as a candidate model for  high-temperature superconducting cuprates 
\cite{Anderson97,Fulde91}, in a multiband-generalization, for the newly discovered iron 
superconductors \cite{Miyake10}, on triangular lattices for organic crystals 
\cite{kino,mckenzie}, on the honeycomb lattice for graphene  \cite{herbut,lopez-sancho}.
The Hamiltonian for the simplest one-band Hubbard model reads
\begin{equation}\label{hubbard2dsq}
H = - \sum_{i,j,s} t_{i-j} c_{i,s}^\dagger c_{j,s} + 
U \sum_i n_{i,\uparrow} n_{i, \downarrow } 
\end{equation}
where $t_{i-j}=t_{j-i}$ is the hopping amplitude between sites $i$ and $j$ and 
$U$ is the Hubbard on-site repulsion. 
We consider here mainly the case with only nearest-neighbor hopping 
$t$ and next-to-nearest neighbor hopping $t'$ on a square lattice. 
Additional hopping terms can be added if a more detailed description of the band 
structure is required, and other interaction terms may be added. 
The chemical potential $\mu$ and $t$ and $t'$ determine the band structure 
$\xi_{\bk} = -2t ( \cos k_x + \cos k_y ) - 4t' \cos k_x \cos k_y - \mu$, 
and hence the shape of the Fermi surface.  

Resummations of perturbation theory in $U$ suggest singularities in different 
channels, arising from Fermi surface nesting and Van Hove singularities 
\cite{Schulz87}, hence competing effects, which are best treated by RG methods.
After two-patch studies, which provided a very crude approximation to 
the momentum dependence of the four-point vertex
\cite{Guinea,Furukawa98,Lederer,Dzyaloshinskii87,Schulz87}, 
more careful analyses with momentum-dependent vertices were done 
using the Polchinski \cite{Zanchi97,Zanchi98,Zanchi00}, the Wick ordered  
\cite{Halboth00a,Halboth00b}, and the one-particle irreducible flow equations  
\cite{Honerkamp01d}, all with a momentum space regulator.
To include ferromagnetism, the temperature flow was introduced by 
\textcite{Honerkamp01b,Honerkamp01c} and \textcite{Honerkamp01a}, and further 
developed by \textcite{Katanin03}. 
The results of these studies at Van Hove filling were confirmed using a refined 
parametrization of the wavevector dependence \cite{Husemann09a}.
The decoupling of the various ordering tendencies in the limit of small $U$ very 
close to the instability and the influence of non-local interactions were discussed 
by \textcite{Binz02,Binz03}. 

\begin{figure}
\centerline{\includegraphics[width = 0.4\columnwidth]{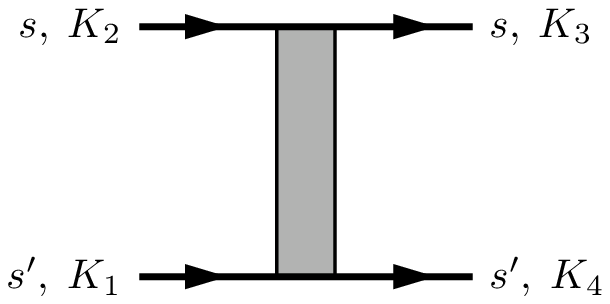}
\includegraphics[width=0.55\columnwidth]{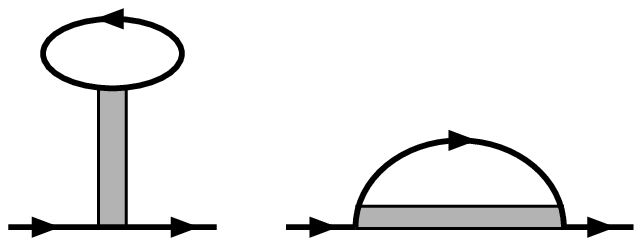}}
\centerline{\includegraphics[width = 0.8\columnwidth]{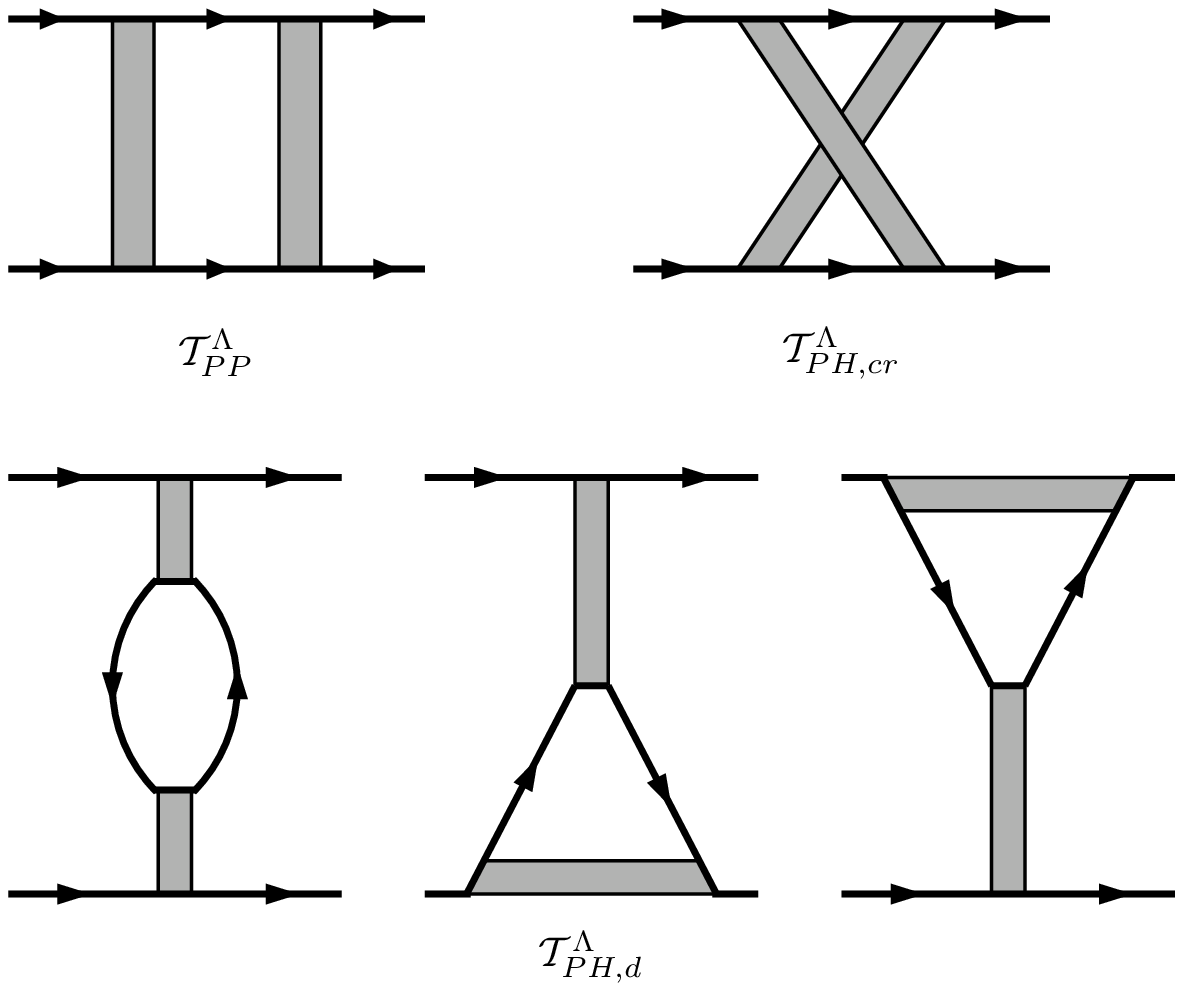}}
\caption{Top row: The coupling function $V^\Lambda(K_1,K_2,K_3)$ with the 
spin convention, and the diagrams entering in the flow equation for 
the self-energy (middle and right diagram). Middle and bottom row: 
The diagrams for the flow of the coupling function. 
The internal lines are either full propagators $G^\Lambda$ or single-scale 
propagators $S^\Lambda$.}
\label{1loopRGDE}
\end{figure}

In the general RG setup of Section \ref{sec:II}, the fermion fields now carry 
a spin index $s$ and a multiindex $K$ consisting of Matsubara frequencies $\omega$, 
wavevectors $\bk$, and possibly a band index $b$.  
To avoid bias, the action is required to retain all symmetries of the initial action. 
This implies (see \textcite{Honerkamp01d,Salmhofer01}) that 
\begin{widetext}
\begin{eqnarray}
\Gamma^{(4)\Lambda}_{s_1s_2s_3s_4} (K_1,K_2;K_3,K_4) &=&
V^\Lambda(K_1,K_2;K_3,K_4) \delta_{s_1s_3}\delta_{s_2s_4} -
V^\Lambda(K_2,K_1;K_3,K_4) \delta_{s_1s_4}\delta_{s_2s_3}  
\label{fierz}
\end{eqnarray}
for a spin-rotation invariant system.
By lattice- and time-translation invariance, $K_4$ is fixed by $K_1,K_2$ and $K_3$ 
in the one-band model (in multiband models, the fourth band index $b_4$ still 
remains free). We therefore abbreviate notation to $V^\Lambda(K_1,K_2,K_3)$.
In the truncation $\Gamma^{(6)\Lambda} =0$, the flow equations for the self-energy 
and for the coupling function become
\begin{equation} 
\sfrac{d}{d\Lambda} \Sigma^\Lambda (K) 
= - \int dK' \,\left[  2 V^\Lambda (K,K',K) - V^\Lambda(K,K',K') \right] \, 
S^\Lambda (K') \, , \quad
\sfrac{d}{d\Lambda} V^\Lambda 
= {\cal T}^\Lambda_{PP} + {\cal T}^\Lambda_{PH,d} +
{\cal T}^\Lambda_{PH,cr} \label{sigmavdot}
\end{equation} 
with the particle-particle term $ {\cal T}^\Lambda_{PP}$ and the direct and crossed
particle-hole terms ${\cal T}^\Lambda_{PH,d} $ and ${\cal T}^\Lambda_{PH,cr}$:
\begin{eqnarray}
 {\cal T}^\Lam_{PP} (K_1,K_2;K_3,K_4) 
 &=& 
 \int dK \, V^{\Lam}(K_1,K_2,K) \; L^{\Lam}(K,-K+K_1+K_2) \, 
 V^{\Lam}(K,-K+K_1+K_2 ,K_3) \label{PPdia} \; ,
 \\
 {\cal T}^{\Lam}_{PH,d} (K_1,K_2;K_3,K_4) 
 &=&
 \int dK\, \biggl[- 2 V^{\Lam}( K_1,K,K_3 ) \, L^{\Lam}(K,K+K_1-K_3) \,
 V^{\Lam}(K+K_1-K_3,K_2,K) \nonumber
 \\ && \qquad +
 V^{\Lam}(K_1,K,K+K_1-K_3) \, L^{\Lam}(K,K+K_1-K_3) \, V^{\Lam}(K+K_1-K_3,K_2,K)
 \nonumber  \\ && \qquad +
 V^{\Lam}(K_1,K,K_3) \, L^{\Lam}(K,K+K_1-K_3) \, V^{\Lam}(K_2,K+K_1-K_3,K)
 \biggr]\; ,
 \label{PHddia}
 \\
 {\cal T}^{\Lam}_{PH,cr}(K_1,K_2;K_3,K_4) 
 &=&
 \int dK \, V^{\Lam}(K_1,K+K_2-K_3,K) \, L^{\Lam}(K,K+K_2-K_3) \,
 V^{\Lam}(K,K_2,K_3 ) \; .
\label{PHcrdia}
\end{eqnarray}
Here 
$L^{\Lam}(K,K') = S^\Lam(K) G^{\Lam}(K') + G^{\Lam}(K) S^{\Lam}(K')$ 
is the product of single-scale propagators $S^\Lambda$ and full propagators 
$G^\Lambda$ with momentum assignments corresponding to the diagrams in 
Fig.~\ref{1loopRGDE}. 
\end{widetext}

For the Hubbard Hamiltonian (\ref{hubbard2dsq}), the initial condition is 
$V^{\Lam_0}(K_1,K_2,K_3) = U$. 
Other interactions can be dealt with by modifying this initial condition.
The truncation $\Gam^{(6)\Lam} = 0$ is justified only for a sufficiently
small bare coupling, since a contribution to $\Gam^{(6)\Lam}$ is generated
at third order in the two-particle interaction, which leads to third order 
contributions to the flow of $V^{\Lam}$ (see Sec.~II).
In most studies the self-energy feedback into the flow of $V^{\Lam}$ was
also neglected, since it also affects the flow only at third order in
$V^{\Lam}$.

The coupling function $V^\Lambda (K_1,K_2,K_3)$ depends on three wavevectors and 
three Matsubara frequencies, so that the RG equation for a two-dimensional system 
is a differential equation in a 9-dimensional space.  
As discussed in Section~\ref{sec:powercount}, its most singular part sits at
zero Matsubara frequency. Hence one may neglect the frequency dependence.
Then $V^\Lambda$ defines an effective Hamiltonian. 
Similarly, the $\bk$-dependence is most important in the angular direction 
along the Fermi surface. 
This dependence can then be taken into account by a discretization, 
i.e. by devising patches in the Brillouin zone in which the coupling function is 
kept constant. 
\textcite{FMRT92} showed that using $N$ patches leads to a natural $N$-vector model
in two dimensions. \textcite{Zanchi98,Zanchi00} were the first to use it in studies 
of the Hubbard model. 

\begin{figure}
\begin{center}
\resizebox{0.65\columnwidth}{!}{\includegraphics{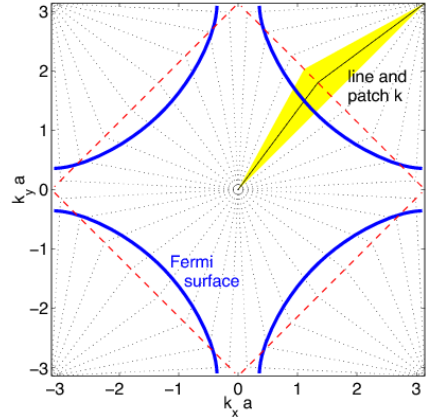} }

\end{center}
\caption{(Color online)
 $N$-patch discretization of the Brillouin zone for the one-band 
 Hubbard model on the 2D square lattice. The colored region is a 
 patch in which the coupling function is approximated as a constant.}
\label{setup}
\end{figure}

Usually one forms elongated patches that extend roughly perpendicular to the
Fermi surface but are rather narrow parallel to the Fermi surface 
(see Fig.~\ref{setup}). 
The coupling function is then computed for wavevectors $\bk_1$ to $\bk_3$ at 
the Fermi surface in the center of the patches.  
We label the patches by $\pk_i=1, \dots N$. The function $V^\Lambda$ is thus approximated
by $O(N^3)$ interpatch couplings $V^\Lambda (\pk_1,\pk_2,\pk_3)$. 
Even if $\bk_1,\bk_2$ and $\bk_3$ are on the Fermi surface, $\bk_4$ can be anywhere. 
In the calculation of the loop integrals it is however necessary to assign a patch 
number $\pk_4$ to $\bk_4$, which amounts to an approximation of projecting $\bk_4$ 
on the Fermi surface. 
Note that this projected $N$-patch discretized coupling function 
$V^\Lambda (\pk_1,\pk_2,\pk_3)$ then has fewer symmetries;
for instance $V^\Lambda (\pk_1,\pk_2,\pk_3) \not= V^\Lambda (\pk_2,\pk_1,\pk_4)$ 
in general, as in the latter object $\bk_3$ is not necessarily on the
Fermi surface. 
For sufficiently large $N$, this discretization captures the angular variation 
of the coupling function along the Fermi surface with good precision. 

The results obtained within this approximation, described in the following, 
have been found to be robust when the dependence on frequencies $\omega_i$ 
\cite{Honerkamp07,Klironomos06} and the component of $\bk_i$ transversal 
to the Fermi surface  \cite{Halboth00a,Honerkamp01a,Honerkamp04} are included. 
\textcite{Katanin09} performed a flow to third order in the scale-dependent 
four-point-vertex (see Section \ref{ssec:runcoup}), with the frequency 
dependence in the same approximation as \textcite{Honerkamp03}.

\subsection{Results for the two-dimensional Hubbard model}

Starting from the initial condition given by the Hubbard model, 
the flow is run from $\Lambda_0$ down to a characteristic scale $\Last$, 
where the largest coupling reaches some multiple $\alpha$ of the bandwidth. 
The choice of $\alpha$ varies widely in the literature; the discussion here is based on 
the comparably cautious choice $\alpha =2$ or $3$, as well as on the consistency check
that the results do not change drastically as $\alpha$ is changed. 
The characteristic scale $\Last$ corresponds to a temperature $\Tst$.
If $T$ is clearly above $\Tst$, the flow can be integrated to scale zero 
without any instabilities. $\Tst$ is only an upper bound for the temperature
where ordering can set in because of order parameter fluctuations 
at scales below $\Last$.  
In two dimensions they are so strong that long-range order that breaks 
continuous symmetries does not occur at any $T>0$, 
thus ``ordering'' is to mean either short-range order with a very large 
correlation length, or ordering in a related system with a small coupling in the third
direction, as is present in most materials. 

\subsubsection{Antiferromagnetism and Superconductivity}

The results discussed here are obtained with a slightly smeared-out step-function 
as cutoff on $\bk$ (no cutoff on the frequencies) and by dropping the self-energy. 

\begin{figure}
\centering
\resizebox{1.0\columnwidth}{!}{\includegraphics{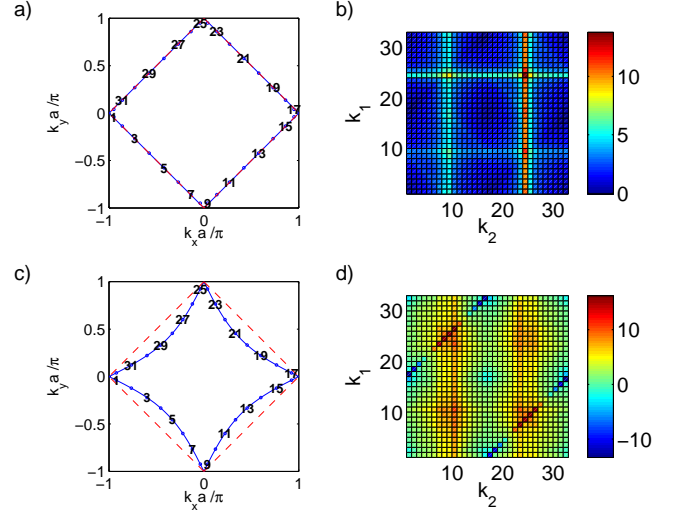} }
\caption{(Color online)
$N$-patch functional RG data obtained with the momentum-shell functional RG 
for the repulsive Hubbard model on the 2D square lattice. 
Upper plots: $\mu=0$, $t'=0$ and initial $U=2t$, 
lower plots: $\mu=1.2t$, $t'=-0.3t$, $U=3t$. 
To the left: Fermi surfaces for the two cases and the $N=32$ discretization 
points for the two incoming $\bk_1$, $\bk_2$ and the 1st outgoing wavevector $\bk_3$. 
To the right: the coupling function $V^{\Last} (\pk_1,\pk_2,\pk_3)$ with $\pk_3=1$ and 
$\pk_1$ and $\pk_2$ moving around the Fermi surface.
The colorbars on the right indicate the values of the interactions.}
\label{comp0_120}
\end{figure}

\noindent
{\em Antiferromagnetism.}
For $t'=0$ and $\mu=0$, 
the band is half-filled and the Fermi surface a perfect square. 
Every vector connecting parallel sides of the Fermi surface is a nesting 
vector, and $\nabla \xi_{\bk} = 0$ at $(\pi,0)$ and $(0,\pi)$. 
This strongly enhances particle-hole terms at wavevector $\bQ = (\pi, \pi)$. 
A random-phase approximation summation of these bubbles results in a divergent static 
spin susceptibility at $\bQ$  for any $U>0$ at sufficiently low $T$, indicating the 
formation of  an antiferromagnetic (AF) spin-density wave (SDW), in accordance with 
mean-field studies \cite{Fulde91}.
The basic RG results at low $T$ are shown for $U=2t$ in Fig.~\ref{comp0_120}.
The labelling of the $N=32$ patches along the Fermi surface can be read off 
Fig.~\ref{comp0_120} a). 
Fig.~\ref{comp0_120} b) shows $V^\Lambda$ as a function of the patch indices 
$\pk_1$ and $\pk_2$, at  $\Last  \sim 0.16t$ and with $\pk_3 = 1$ 
(i.e.\ $\bk_3$ near $(-\pi,0)$).
Strongly enhanced repulsive interactions appear as a vertical line at $\pk_2=24$
(i.e.\ for $\bk_2 - \bk_3 = \bQ$), almost $\pk_1$-independent,
and as a horizontal line at $\pk_1=24$  (corresponding to  $\bk_1 - \bk_3= \bQ$) with 
only a weak dependence on $\pk_2$, roughly half as large as the vertical feature. 
In an extrapolation where the regular profiles are narrowed down to delta functions 
with an appropriate prefactor $J$,
$V^\Lambda (\pk_1,\pk_2,\pk_3) = 
 \frac{J}{4} ( 2 \delta_{\bk_2- \bk_3,\bQ} +  \delta_{\bk_1- \bk_3,\bQ})$,
corresponding to a mean-field AF spin interaction Hamiltonian 
$J \sum_{\langle i,j \rangle } e^{i \bQ\cdot({\bf R}_i-{\bf R}_j)} {\bf S}_i \cdot {\bf S}_j$, 
with ${\bf S}_i = \frac12 c^+_i \mbox{\boldmath$\sigma$} c^{\hphantom{+}}_i$. 
The effective Hamiltonian consisting of the low-scale hopping term and this interaction
exhibits AF long-range order at sufficiently low $T$. 
An analysis of the flow of susceptibilities\cite{Halboth00a,Honerkamp01d} as described 
in Sec.~\ref{sec:responsefcts} confirms this picture. 

The extrapolation to a mean-field Hamiltonian is a drastic oversimplification, 
in which the spin fluctuations are lost, but they are retained in the $V^\Lambda$ obtained 
by the RG flow. 
As the leading instability is clearly exposed by this analysis, 
one can also resort to a bosonized description that treats the collective infrared 
physics \cite{Baier04}. 

\noindent
{\em $d$-wave Cooper pairing. }
For $t'=-0.3t$ and $\mu = -1.2t$, the Fermi surface still contains the saddle
points $(\pi,0)$ and $(0,\pi)$
but  is curved away from these points (Fig.~\ref{comp0_120}c)). 
Now Cooper pair scattering dominates, well visible in  Fig.~\ref{comp0_120}d)
on the diagonal lines $\bk_1+ \bk_2=0$ ( $|\pk_1-\pk_2|=N/2$ in terms of patch indices).
It is attractive when the incoming pair $\bk_1, -\bk_1$ is near the same saddle point 
$(\pm \pi, 0)$ as the outgoing pair $\bk_3, - \bk_3$, and repulsive when incoming and
outgoing pairs are at different saddle points. 
This is the symmetry of the formfactor $d(\bk) = d_0 (\cos k_x - \cos k_y)$ for
$d_{x^2-y^2}$-Cooper pairing. In an extrapolation as above, 
$V^\Lambda (\bk_1, \bk_2, \bk_3)$ gives rise
to the mean-field Hamiltonian
\[ H^\Lambda_{d\mathrm{SC}}  = V^{\phantom{\Lambda}}_{d\mathrm{SC}} \sum_{\bk,\bk' } \, 
\, d(\bk) d(\bk') \, c^\dagger_{\bk',\uparrow} c^\dagger_{-\bk',\downarrow} 
c_{-\bk,\downarrow} c_{\bk,\uparrow} \,  . \] 
which has a $d$-wave singlet-paired ground state. 
This $d$-wave pairing instability was found in a number of studies using different 
functional RG schemes \cite{Zanchi98,Zanchi00,Halboth00a,Halboth00b,Honerkamp01d,
Tsai01,Honerkamp01b,Honerkamp01c,Honerkamp01a}, in a rather large parameter region. 
This constitutes convincing evidence that the weakly coupled Hubbard model possesses 
a $d$-wave superconducting ground state. 

\noindent
{\em Interplay of AF and SC. } 
In Fig.~\ref{comp0_120}d), 
the sign structure of the $d$-wave term goes together, and fits perfectly with, 
enhanced repulsive interactions near $\pk_1 = 8$ and $\pk_2 = 24$, 
which are the remnants of the SDW feature in Fig.~\ref{comp0_120}b).
Their larger width is due to the Fermi surface curvature. As $\Lambda$ is decreased, 
these SDW features appear first, due to approximate nesting at high scales, and then
create an attractive component in the $d_{x^2-y ^2}$-pairing channel, which then grows 
as $\Lambda$ is lowered further, while the SDW is cut off by Fermi surface curvature, 
as discussed also in Appendix \ref{ssec:imppoco}.
When the SDW-enhancing terms are removed by hand from the right hand side of the 
RG equation, the $d$-wave terms are suppressed as well. 
Thus the $d$-wave pairing interaction is induced by AF-spin fluctuations that 
appear on higher scales. 

At fixed $U$, $t$ and $t'$, there is a sizable interval of $\mu$ for which
the Fermi surface remains close to the saddle points. Since both AF-SDW and 
$d$-wave SC are driven by repulsive scattering between $(\pi,0)$ and $(0,\pi)$,
both grow and reinforce one another. In the {\em saddle point regime},
it becomes impossible to single out one over the other in the truncation used here.
By analogy with the quasi-one-dimensional ladder systems, it has been argued
that in this regime, the Fermi surface gets truncated 
\cite{Furukawa98,Honerkamp01d,Laeuchli04}.
 
\subsubsection{Ferromagnetism vs.\ Superconductivity}

At the Van Hove filling, ferromagnetic (FM) tendencies are enhanced by the logarithmic
divergence of the density of states, and the Stoner criterion for the bare
interaction suggests an FM ordered state at arbitrarily small
$U$.  However, the Van Hove singularities also make the $O(U^2)$ Cooper pair scattering 
$\log^2$-divergent, hence put the two terms into direct competition. 

\begin{figure}
\centering
\resizebox{0.48\columnwidth}{!}{\includegraphics{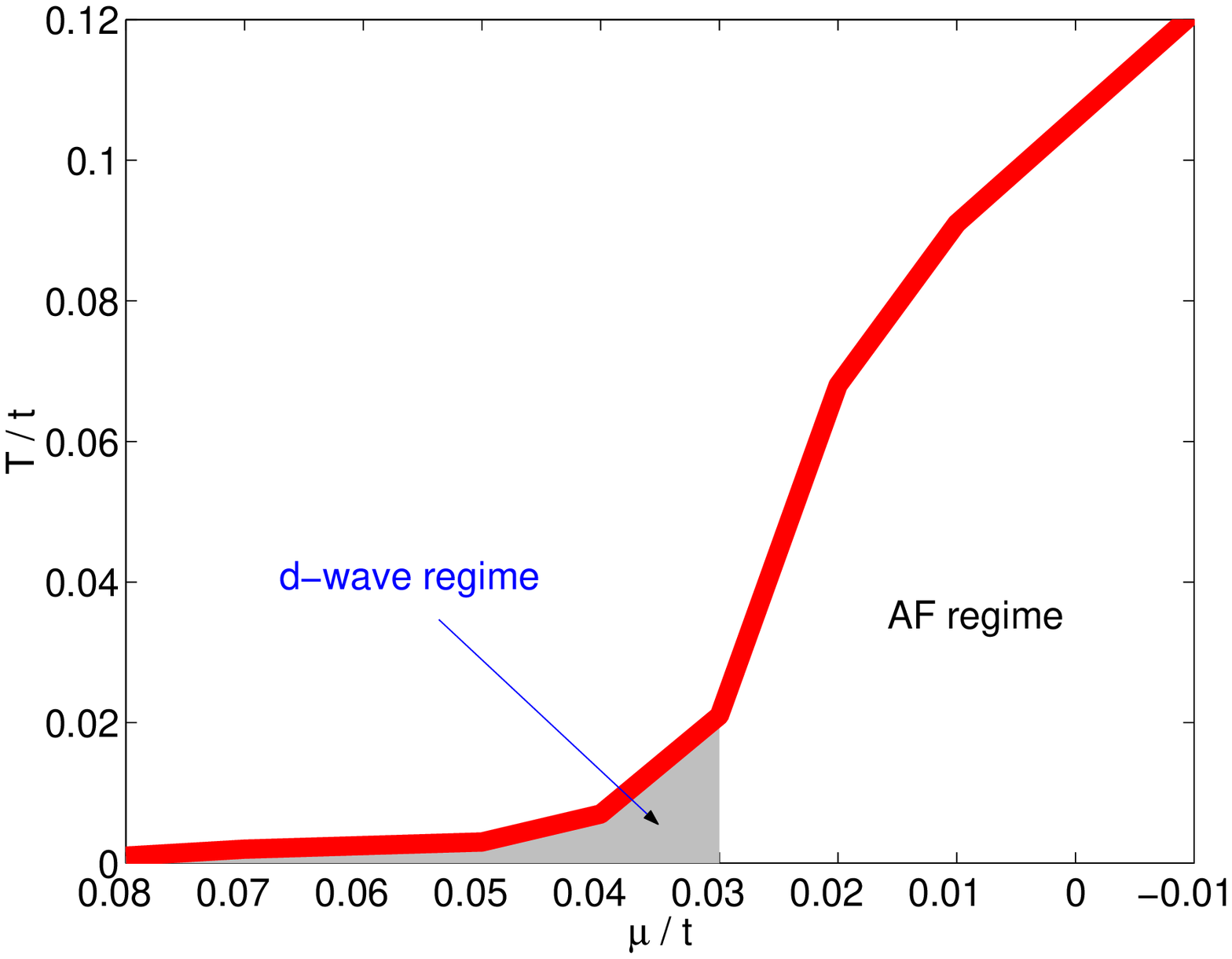} }
\hfill
\resizebox{0.47\columnwidth}{!}{\includegraphics{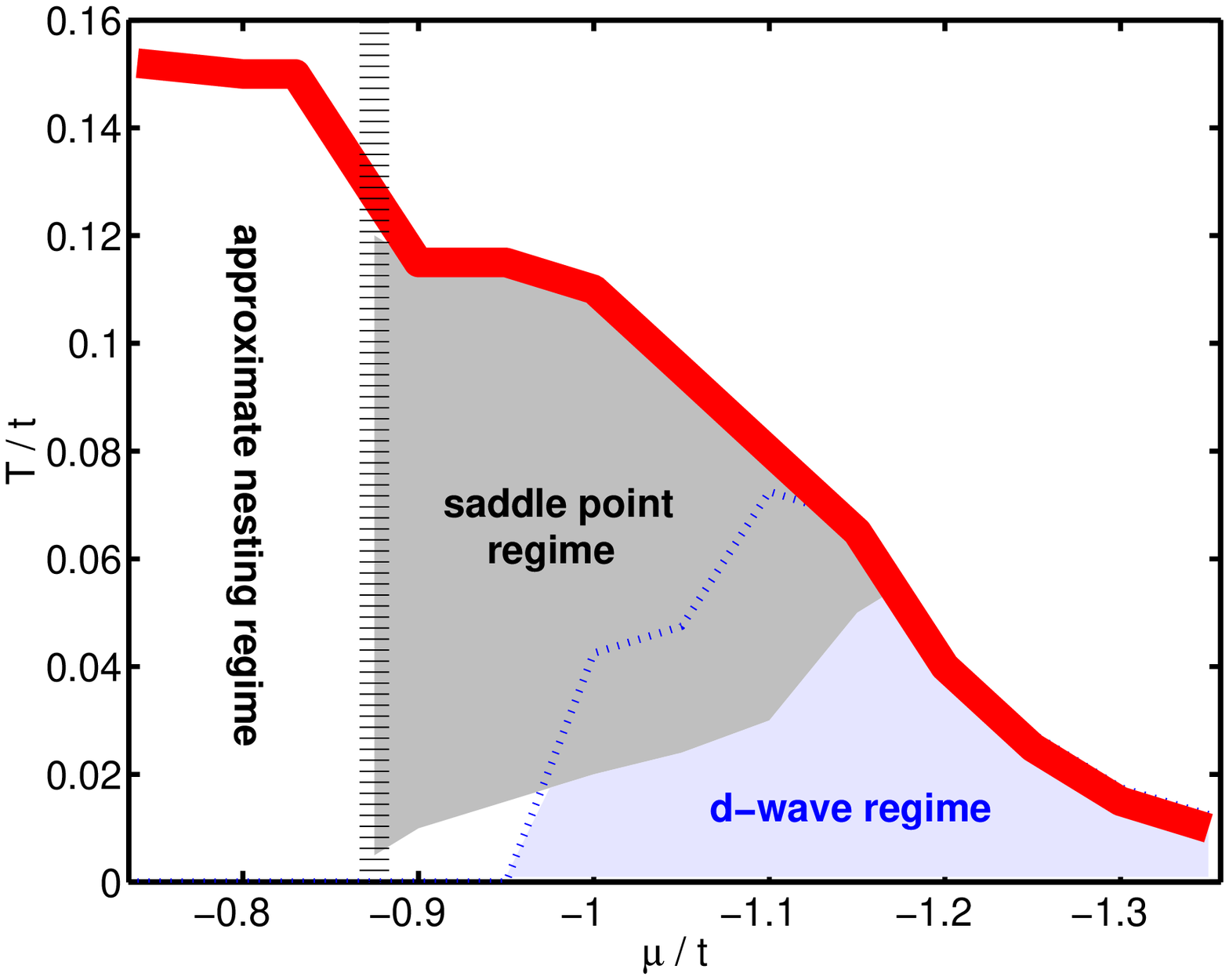} } 
\resizebox{0.48\columnwidth}{!}{\includegraphics{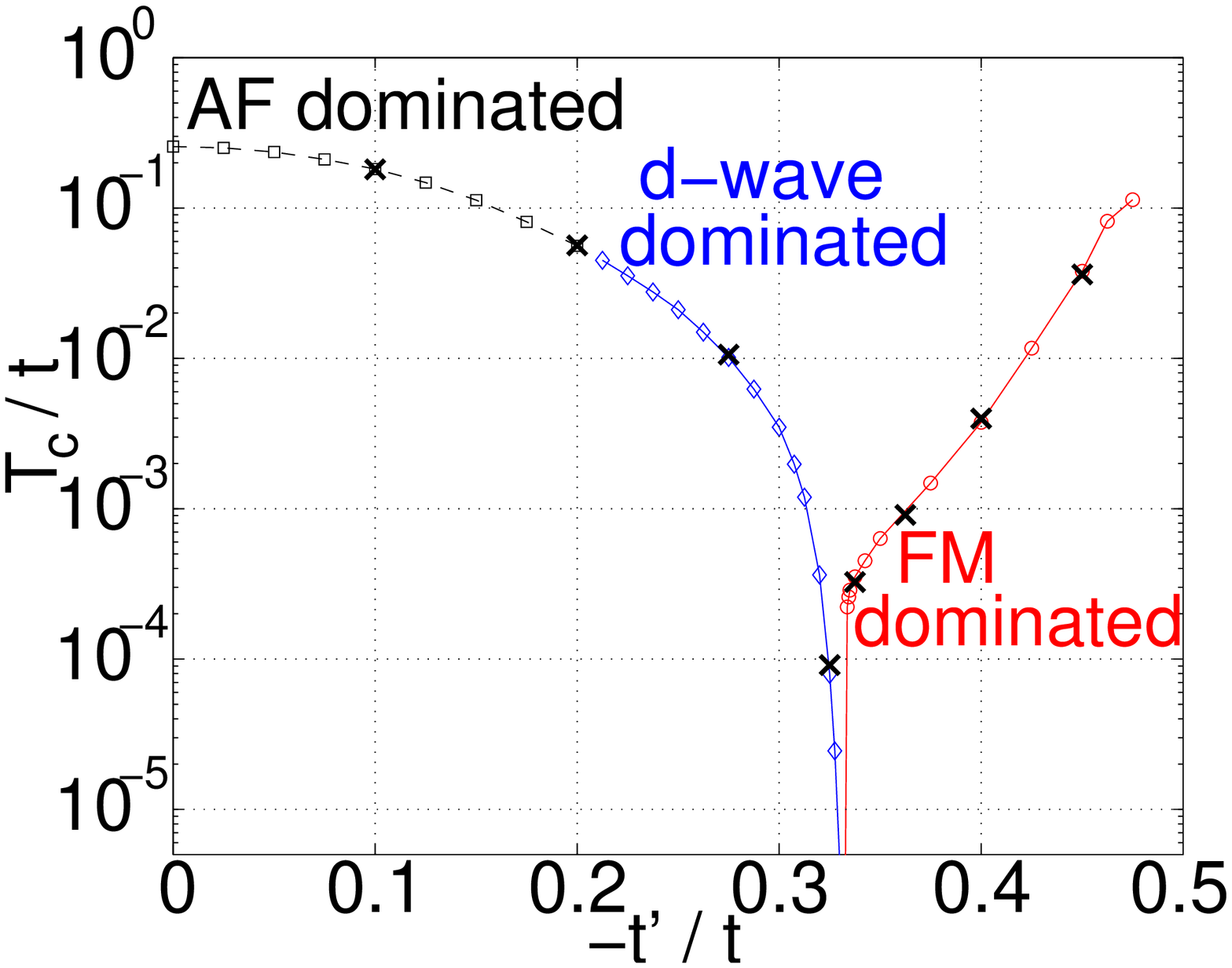} } \hfill 
\resizebox{0.48\columnwidth}{!}{\includegraphics{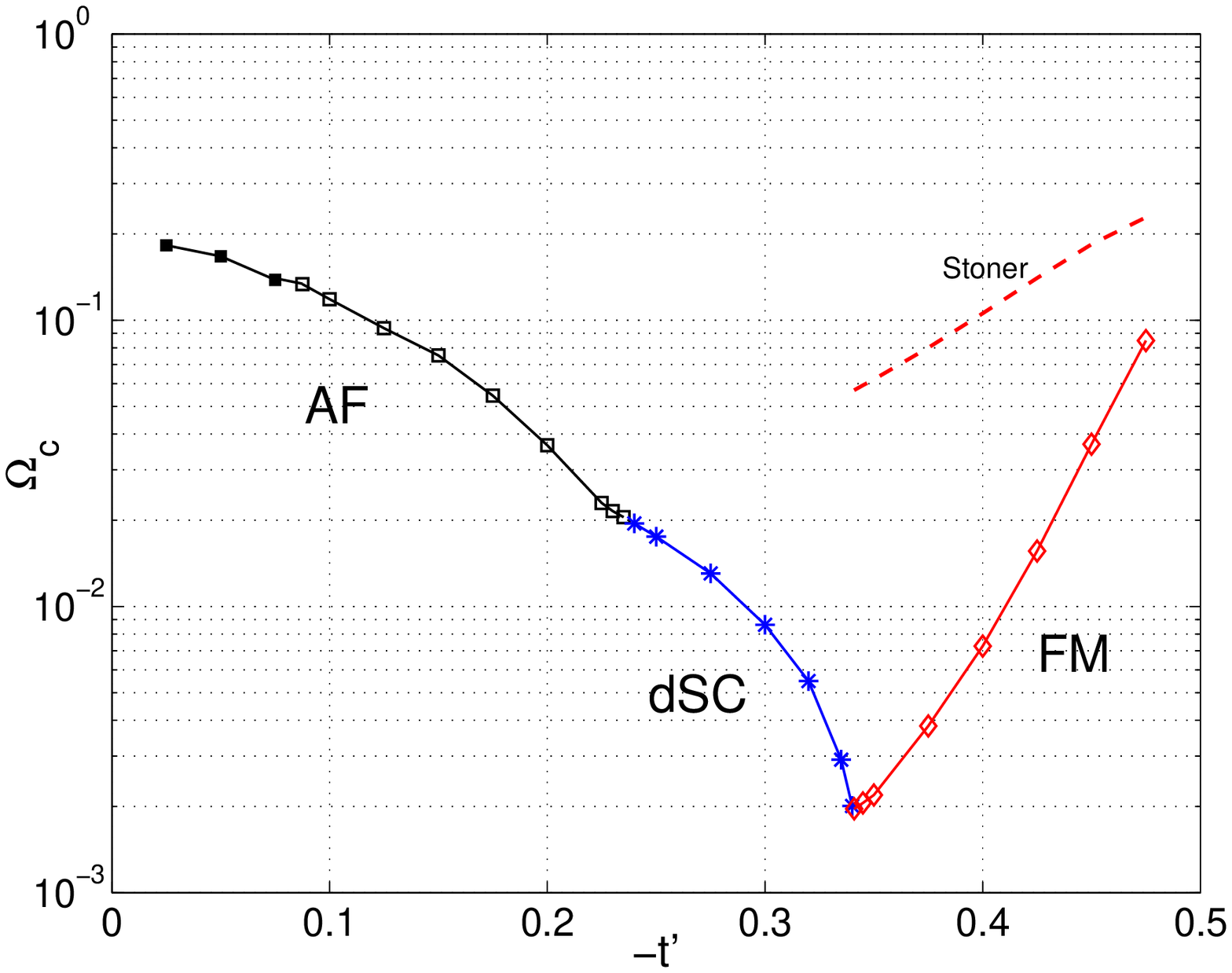} } 
\caption{(Color online)
Leading instabilities as found by $N$-patch functional RG in the $t$-$t'$-Hubbard model. 
Left upper plot: 
$\Tst$ vs.\  $\mu$ for band filling larger than unity, at $t'=-0.3t$ and $U=3t$. 
There is a high-energy-scale AF SDW instability with a weaker $d_{x^2 -y^2}$-wave 
pairing instability when the AF-SDW is cut off. Data from  \textcite{Honerkamp01a}.
Right upper plot: Data for the same $t'$ and $U$ on the 'hole-doped' side with band 
fillings smaller than one, from  \textcite{Honerkamp01d}. 
Now there is a broad crossover 'saddle point regime' between the nesting-driven 
AF-SDW instability and the $d_{x^2 -y^2}$-wave pairing regime.  
Lower left plot: $\Tst$ vs. $t'$ at the Van Hove filling where the Fermi surface 
contains the 
points $(\pi,0)$ and $(0,\pi)$. For large $t'$ one finds a ferromagnetic instability. 
Data from \textcite{Honerkamp01b,Honerkamp01c} obtained with the $T$-flow. 
Right lower plot: $\Omega_*$  vs.\ $t'$ at Van Hove filling, now obtained with the
simplified vertex parametrization of \textcite{Husemann09a} and with a soft 
frequency regulator $\Omega$.}
\label{pds}
\end{figure}

As discussed in Section \ref{sec:cutoffs},
the momentum-shell cutoff artificially suppresses FM. For this reason,
the $T$-flow (see Section \ref{sssec:tflow}) was invented  
\cite{Honerkamp01b,Honerkamp01c}, and we discuss results
obtained by $T$-flow here.
The main difference to the AF/SC scenario discussed above is 
that at zero transfer momentum, scattering processes driving
FM must have the opposite sign from those driving singlet SC, 
hence mutually suppress one another.  This simple picture is confirmed
by the RG with momentum-dependent vertices, in a study where
 $t'$ and $\mu$ are varied at fixed $U$ and $t$, such that the Fermi surface
always contains the saddle points: near to $t' = - t/3$,
$\Tst$ gets strongly suppressed, hinting at a quantum critical point between
the dSC and FM phases (lower left plot in Fig.~\ref{pds}). 
These results were later confirmed by a two-particle self-consistent approach 
\cite{Hankyevich2003} and in the so-called $\Omega$-scheme, which employs a 
soft infrared regulator on the Matsubara frequencies  \cite{Husemann09a}; 
see the lower right plot in Fig.~\ref{pds}. 
In the latter study, the $N$-patch scheme was replaced by a parametrization
of the vertex functions in terms of exchange bosons. 
The much higher value of $\Last$ in the transitional regime near $t' = - t/3$ 
is believed to be due to a form factor that was not fully resolved there. 

\subsubsection{Charge  instabilities}

The effective interaction develops a pronounced momentum dependence
also in the charge sector.
In the forward scattering channel, this amounts to the formation
of non-uniform contributions to the Landau interaction.
If strong enough, the latter can lead to a {\em Pomeranchuk instability} 
\cite{Pomeranchuk58}, that is, a symmetry-breaking deformation of the 
Fermi surface.

In particular, the antiferromagnetic peak drives the combination of 
couplings $V^\Lambda_c (\pk_1,\pk_2,\pk_3) = 
2V^\Lambda (\pk_1,\pk_2,\pk_3) - V^\Lambda( \pk_2,\pk_1,\pk_3)$ 
at certain $\bQ = \bk_3-\bk_1$.
Near to $\bQ\approx {\bf 0}$ and $\bQ \approx (\pi, \pi)$, 
\begin{equation}
V^\Lambda_c (\pk_1,\pk_2,\pk_3)  \approx 
- f_d(\bk_1) f_d(\bk_2) V_d (\bk_3-\bk_1) \, , 
\end{equation}
where $f_d(\bk)$ has the same symmetries as $d (\bk) = \cos k_x - \cos k_y$, 
but is more strongly peaked near the saddle points. 
For $\bQ=(\pi,\pi)$ the corresponding mean-field state is the 
{\em $d$-density wave state}, which breaks time-reversal invariance  
\cite{Nayak01} and gaps the single-particle states, except at nodal points
on the Brillouin zone diagonal. 
For forward scattering, $\bQ={\bf 0}$, the mean-field state only breaks 
the lattice rotational symmetry of the electronic dispersion and hence of 
the Fermi surface. 
This tendency to form a {\em nematic} state \cite{Fradkin10}
via a $d$-wave Pomeranchuk instability driven by forward scattering 
interactions was discovered using functional RG \cite{Halboth00b}. 
Although the Pomeranchuk instability is not leading in the flow 
for the Hubbard model \cite{Honerkamp02}, a nematic state can coexist 
with the superconducting state \cite{Neumayr2003,Yamase07}, 
and it may get less suppressed by fluctuations since it breaks no 
continuous symmetry. 
The $d$-wave Pomeranchuk instability has been investigated as a 
possible source of nematicity of the electronic state in relation with 
experiments on various correlated electron systems
\cite{Honerkamp05a,Yamase06,Yamase09,Metlitski10,Okamoto10}.

\subsubsection{Flows with self-energy effects}

We briefly summarize functional RG studies where the self-energy has been included.
If a frequency-independent vertex function $V^\Lambda$ is directly inserted in
the right hand side of Eq.~(\ref{floweqSigma}), 
then $\Sigma^\Lambda$ is real and independent of the frequency, 
hence only changes the dispersion. 
This was taken into account in the appendix of \textcite{Honerkamp01d}, where the 
adaptive scale decomposition method later detailed in \textcite{Salmhofer07} was used. 
To keep the density fixed, $\mu$ is adjusted as a function of $\Lambda$. 
Since the interaction grows in the flow, it is a nontrivial check 
of the validity of the truncation that the feedback from the interaction 
does not lift the low-kinetic-energy modes 
to high energies, which would drastically shift the Fermi surface and lead to 
spurious divergences.
The first study by \textcite{Honerkamp01d} showed that the Fermi surface tends
to become flatter as $\Lambda$ decreases, but that it indeed shifts very little
before the flow is stopped at $\Last$. 
Thus including the real part of the self-energy does not lead  to any essential changes
in the AF/SC scenario described above. However, correlations that only feed on the immediate
vicinity of the saddle points, like FM, are affected more strongly, and a full analysis 
of the coupled flow of self-energy and vertex directly at the saddle points remains an 
open problem, in spite of partial results \cite{FS2}.

The imaginary part and the frequency-dependence of the self-energy can be 
approximated by inserting the integrated flow of the interaction vertex in the 
self-energy equation \cite{Honerkamp01a}. 
This effectively includes two-loop frequency-dependence effects,
and captures the $T^2$-dependence of the quasiparticle scattering rate 
in a Fermi-liquid situation and
the exponent of the  vanishing quasiparticle weight in the Luttinger liquid 
up to second order in the bare couplings \cite{Honerkamp03}. 
For the 2D Hubbard model, the quasiparticle lifetime and renormalization 
factor was calculated in \textcite{Honerkamp01a,Honerkamp03}, 
exhibiting a strongly $\bk$-dependent 
quasiparticle degradation as $\Last$ is approached. This trend was also found by 
\textcite{Zanchi01} in a slightly different approximation for the self-energy, and is 
also robust in a more elaborate treatment  \cite{Katanin09}, where the six-point vertex
was included partially. 
The anisotropy of the quasiparticle 
lifetime was found to have a non-Fermi-liquid temperature dependence and to correlate 
with the strength of the generated $d$-wave pairing interaction \cite{Ossadnik08}, 
similar to what is observed experimentally in overdoped cuprates. More refined studies of 
the frequency-dependence revealed, however, that a simple parametrization in terms of a 
quasiparticle weight is insufficient  \cite{Katanin04b,Rohe05}. 
It was shown that near $\Last$, 
the small-$|\omega|$-behavior of $\Sigma^\Lambda (\omega,\bk)$ leads to a
split-up of the quasiparticle peak. 
All these findings are 
consistent with an anisotropic break-up of the Fermi surface that one would like to 
connect with the phenomenology of the high-$T_c$ cuprates \cite{Honerkamp01d,Lee06},
but a quantitative comparison is difficult due to the strongly coupled nature of the 
cuprates.

\subsection{Pnictide superconductors}

\begin{figure}
\centering
\resizebox{0.99\columnwidth}{!}{\includegraphics{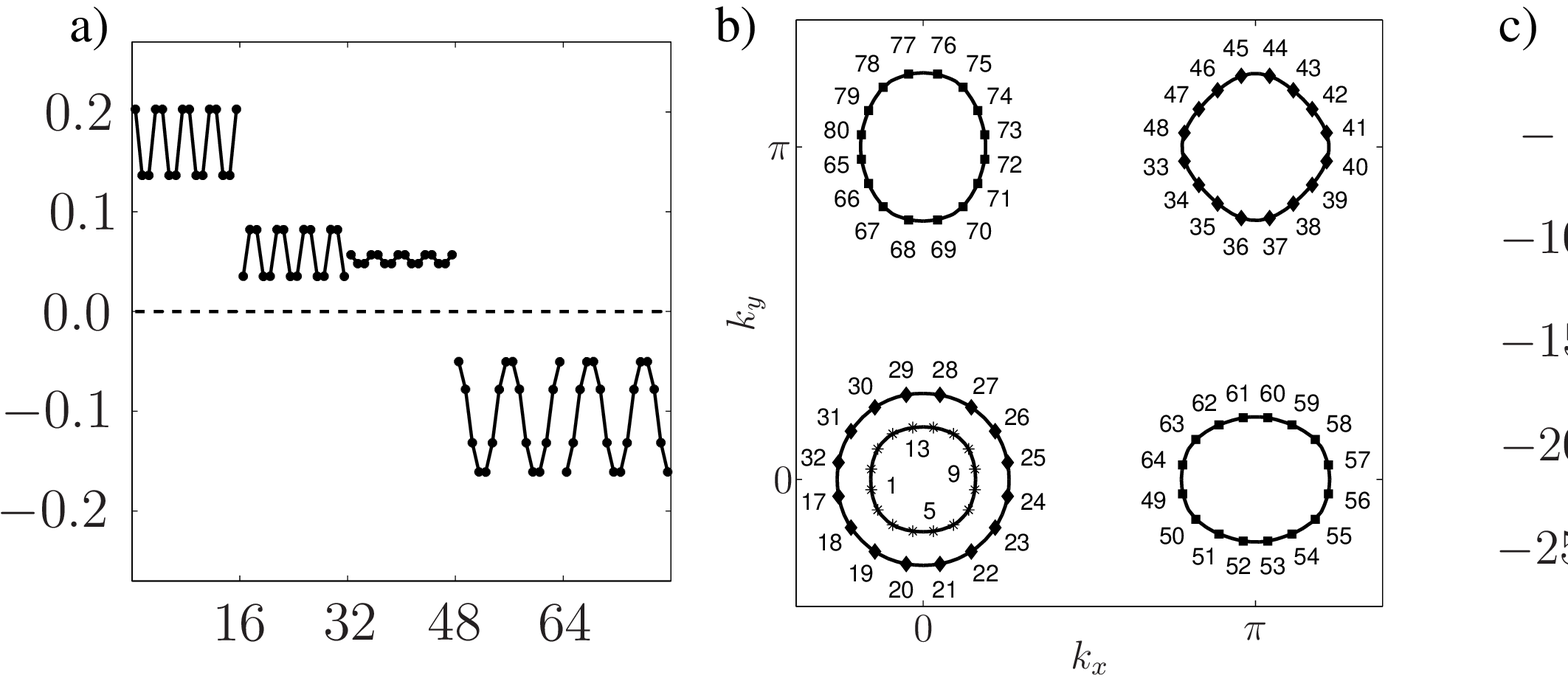} }
\caption{(Color online)
Functional renormalization group results for the iron pnictide compound LaFeAsO 
at moderate hole doping. a) Superconducting form factor as the outcome of functional RG, 
plotted versus the position on the hole pockets at $\Gamma$ and $M$ and electron pockets 
at $X$, numbered as depicted in b). 
The competing fluctuations manifest themselves in diverging ordering susceptibilities
at low RG scales as shown in c), including in particular spin density wave 
(SDW), superconductivity (SC), Pomeranchuk (PI) and charge density wave (CDW) 
instabilities \cite{Thomale10}.}
\label{feas}
\end{figure}

The functional RG has been very useful in the study of the newly discovered iron 
pnictide superconductors \cite{Norman08,Ishida09,Hirschfeld10}. 
Here the functional RG may work even 
better, as the pnictides are less strongly correlated than the high-$T_c$ cuprates. This 
can already be inferred from the experimental phase diagram, where one only finds 
metallic antiferromagnetic phases (if at all), but never Mott insulating 
antiferromagnetism. Theoretical works that try to assess the 
iron $d$-orbital onsite-interaction strengths find values that put the materials into the 
range of weak to moderate correlations \cite{Anisimov09,Miyake10}. Regarding the electronic 
structure, the pnictides are more complex than the cuprates. 
At least three of the five 
iron $d$-orbitals have non-negligible weight near the Fermi level \cite{Mazin08,Daghofer10}. 
Therefore, even if one is only interested in the vicinity of the Fermi 
surface, the multi-band character has to be kept.  
The Fermi surface (see Fig.~\ref{feas} b)) is divided into two hole pockets, 
centered around the origin of the Brillouin zone at $\bk=0$, and two 
electron pockets around $\bk=(\pi,0)$ and $\bk=(0,\pi)$ in the unfolded zone 
corresponding to the small unit cell with one iron atom (or $\bk=(\pi,\pi)$ in the folded 
zone corresponding to the large unit cell with two iron atoms). As pointed out 
early \cite{Mazin08,Kuroki08}, there is approximate nesting of electron- and hole pockets 
which enhances particle-hole susceptibilities with the wavevector connecting these 
pockets. In addition, depending on the parameters and approximations \cite{Ikeda10}, there 
can be a third hole pocket at $(\pi,\pi)$ in the unfolded zone.

The first $N$-patch studies of the pnictides were performed by \textcite{Wang09a,Wang09b,Wang10} 
for a five-band model. These authors obtained a sign-changing $s$-wave pairing 
instability driven by AF fluctuations as the dominant pairing instability. Further they 
found strongly anisotropic gaps around the electron pockets, with possibility of node 
formation.
The basic structure of the phase diagram with the sign-changing pairing gap between 
electron- and hole-pockets can be understood already from simplified few-patch RG 
approaches \cite{Chubukov08}. This would however predict isotropic gaps around these 
pockets \cite{Platt09}. To understand the gap anisotropy one has to take into account the 
multi-orbital nature of the electronic spectrum in the iron pnictides, as was done already 
in the initial studies \cite{Wang09a,Wang09b,Wang10}. In order to understand this point, 
let us start with a single-particle Hamiltonian in wavevector-Fe-$d$-orbital space
\begin{equation} H = \sum_{\bk,s,o} h(\bk)_{oo'} c^\dagger_{\bk,o,s} c_{\bk,o',s} 
\end{equation} where the matrices $h(\bk)_{oo'} $ take into account intra-  and 
inter-orbital terms for orbital index $o=o'$ or $o\not=o'$ respectively.  $s$ is the spin 
quantum number. The energy bands are obtained by a unitary transformation from orbital to 
band operators (index $b$), $c_{\bk,b,s} = \sum_{o} u_{bo}(\bk) c_{\bk,o,s}$.
The standard choice for the interaction between the electrons is to introduce 
orbital-dependent intra- and inter-orbital onsite repulsions, plus Hund's rule and pair 
hopping terms. While these local terms lead to $\bk$-independent interactions in the 
orbital basis, parametrized by a tensor $V_{o1,o2,o3,o4}$, after the transformation to 
bands one arrives at a $\bk$-dependent interaction function
\begin{widetext}
\begin{eqnarray} V_{b1,b2,b3,b4} (\bk_1,\bk_2,\bk_3,\bk_4) &= & \sum_{o1,o2,o3,o4}
V_{o1,o2,o3,o4}  
\; u_{b1,o1} (\bk_1) u_{b2,o2} (\bk_2) u^*_{b3,o3} (\bk_3)  u^*_{b4,o4} (\bk_4) \,  . 
\label{omu} \end{eqnarray} \end{widetext}
The combination of $u_{bo}$s behind the interaction tensor is sometimes called the 
'orbital make-up' \cite{Maier09,Graser09}. 
These prefactors cause a marked $\bk$-structure already in the initial interaction 
which is then renormalized during the functional RG flow. 
It turns out that this orbital make-up has an essential 
influence on the competition between different channels in the flow and is responsible 
for the gap anisotropies found in the multi-band functional RG studies by 
\textcite{Wang09a,Wang09b,Wang10} and in subsequent functional RG studies 
\cite{Thomale10,Platt10}. A typical result for the predicted pairing 
gaps is shown in Fig.~\ref{feas} a). Note that according to the functional RG 
analysis, the pairing state should be strongly doping-dependent 
\cite{Thomale09,Thomale10,Thomale11}.

Summarizing this brief section, the iron superconductors pose an interesting problem 
where the functional RG has been instrumental in 
obtaining the main ordering tendencies in good agreement with 
current experiments. For future 
research, one goal should be to make the functional RG a  useful bridge between ab-initio 
descriptions providing the effective model at intermediate energy scales and the many-body 
effects seen in the experiments at low scales. In particular it will be interesting to relate 
experimentally observed materials trends in, e.g., the gap structure or the energy scales 
of the different systems, to changes in the microscopic Hamiltonian taken from ab-initio 
descriptions. Furthermore, the functional RG studies may have to be extended to include the 
dispersion orthogonal to the iron-pnictide planes, as this would yield additional 
possibilities for nodes in the gap function \cite{Norman08,Ishida09,Hirschfeld10,Platt11}.

\label{sec:pnictides}

\subsection{Other systems}
\label{sec:othersys}

Besides the above-described two larger fields of application, 
the functional RG truncations described in this Section have also been 
employed in a number of other models in strongly correlated electron physics. Here we 
briefly list some of these activities.

In relation to possible unconventional superconductivity in organic crystals and 
layered cobaltates, Hubbard-type models on the triangular lattice have been studied 
\cite{Tsai01,Honerkamp03tria}. At large $U$, the spin exchange interaction between the 
sites of the triangular lattice is geometrically frustrated, leading to a much weaker 
appearance of antiferromagnetism and a possible non-magnetic insulating 
phase \cite{Morita02,Sahebsara08,Yoshioka09}.
At weak coupling and for nearest-neighbor hopping, 
Fermi surface nesting is absent, so that near to or at half band filling,
only low-scale  Kohn-Luttinger-like superconducting instabilities occur out of an 
innocuous Fermi liquid. 
However, there appears to be a strong dependence on details of the microscopic modelling.

To study interaction effects in graphene, the $N$-patch functional RG has 
been applied to the extended Hubbard model on the honeycomb lattice. 
In  nominally undoped graphene, the Fermi surface becomes a set of {\em Dirac points} 
where the density of states vanishes, and
no instabilities are found for sufficiently small interactions. 
If the interaction strength exceeds a certain value, 
various instabilities driven by particle-hole fluctuations between the two Dirac 
points \cite{Honerkamp08} are found. 
Interestingly, for larger second-nearest neighbor 
interactions, there is the possibility of an instability towards a quantum spin Hall 
phase \cite{Raghu08}. However, 
a spin-liquid phase for 
intermediate strength of the Hubbard onsite repulsion that was recently  found in
quantum Monte Carlo calculations \cite{Meng2010} is not reflected in the functional RG 
results on this level of approximation.
When the Fermi level is moved away from the Dirac points, the functional RG again 
detects pairing instabilities. 
In the case of dominant nearest neighbor repulsion, the leading pairing 
tendency is in the $f$-wave triplet channel \cite{Honerkamp08}.

The unbiasedness of the functional RG, and the access it gives to $\bk$- and
$\omega$-dependence of vertex functions, is also of great use in (quasi-)one-dimensional models. 
The half-filled  extended Hubbard model in one dimension has been studied in the search for 
bond-order-wave phases, which could indeed be found with a refined patching of the 
$\bk$-dependence of the interaction away from the Fermi points \cite{Tam06}.
For quasi-1D models with a small transverse hopping in a second direction the change
from a gapless Luttinger liquid in a strictly one-dimensional
situation to Fermi liquid instabilities 
toward ordering can be monitored as a function of the transverse 
hopping \cite{Honerkamp03}. 
The Fermi surface in coupled metallic chains was studied by 
\textcite{Ledowski05,Ledowski07a,Ledowski07b}. 
The possibility of triplet pairing driven by density wave 
fluctuations has been explored in such situations \cite{Nickel05,Nickel06}.
In these quasi-one-dimensional systems, including the frequency dependence of the  interaction 
vertex is numerically more feasible than in two dimensions. This has been used to study 
the interplay of phonon-mediated and direct electron-electron interactions for chains 
\cite{Tam07C}, ladders \cite{Tam07L} and systems with small transverse 
hopping \cite{Bakrim10}.

Many-fermion lattice Hamiltonians can also be realized with ultracold atoms in optical 
lattices, opening up new directions. 
For example,
mixtures of more than two hyperfine states \cite{HonerkampHofstetter04} 
and boson-mediated pairing on two-dimensional  lattices  \cite{Mathey06,Mathey07,Klironomos07} 
have been investigated using fermionic $N$-patch methods.

Another promising development is the application of the functional RG to quantum spin 
systems \cite{reuther1}. Here, an
 auxiliary-fermion representation is used for the spins in generalized Heisenberg models, 
and the functional RG can be formulated in terms of these fermions.
 As important difference to systems of itinerant electrons, in the quantum spin system 
the kinetic energy for the pseudo-fermions is zero and the interactions only depend on 
one spatial or wavevector variable. This allows one to keep the full frequency dependence 
of the self-energy and interaction vertex on the imaginary axis, in the usual truncation 
where the six-point vertex is neglected. 
The Katanin modification (\textcite{Katanin04a}, see also Section II.C.2) 
of the flow hierarchy turns out to be crucial here. 
If it is employed, the auxiliary-fermion functional RG describes the transitions 
from N\'eel order to collinear order 
through an intermediate paramagnetic phase in the $J_1$-$J_2$ spin-$1/2$ model on the 
square lattice as function of $J_1/J_2$ in good agreement with numerical 
approaches.  Furthermore, similar systems on the triangular lattice \cite{reuther2} and 
with longer-ranged couplings \cite{reuther3} were studied.  The success of a relatively simple
truncation in such an intrinsically strongly coupled system is explained by these authors
in that the diagrams summed in this flow contain the leading 
contributions in both $1/N$- and $1/S$-expansions plus particle-particle diagrams, 
hence those contributions that are believed to be most important.


\section{SPONTANEOUS SYMMETRY BREAKING}
\label{sec:IV}

In many interacting Fermi systems a symmetry of the bare action
is spontaneously broken at sufficiently low temperatures and, in
particular, in the ground state. 
In the fermionic flow equations, the common types of spontaneous
symmetry breaking such as magnetic order or superconductivity are
associated with a divergence of the effective two-particle
interaction at a finite scale $\Lam_c > 0$, in a specific 
momentum channel.
In Sec.~III we discussed several examples for such divergences.
The truncation for the effective two-particle vertex leading to
the $N$-patch scheme described and used in Sec.~III is insufficient 
to describe the symmetry-broken phase.
To continue the flow below the scale $\Lam_c$, an appropriate
order parameter has to be introduced.

There are two distinct ways of implementing spontaneous symmetry
breaking in the functional RG.
In one approach the fermionic flow is computed in presence of
a small (ideally infinitesimal) symmetry breaking term added to
the bare action, which is promoted to a finite order parameter
below the scale $\Lam_c$ \cite{Salmhofer04}.
A relatively simple truncation of the exact flow equation 
captures spontaneous symmetry breaking in mean-field models 
such as the reduced BCS model exactly, although the effective 
two-particle interactions diverge.
Another possibility is to decouple the interaction by a bosonic
order parameter field, via a Hubbard-Stratonovich transformation,
and to study the coupled flow of the fermionic and order
parameter fields \cite{Baier04}.

In case of competing instabilities a reliable calculation based
on either of the above-mentioned routes to symmetry breaking is
quite involved. 
For a rough estimate of order parameters and phase diagrams, one 
may also neglect low energy fluctuations and combine flow 
equations at high scales with a mean-field treatment at 
low scales. 
In this functional RG + mean-field approach, one stops the flow 
of the effective two-particle interaction at a scale 
$\Lam_{\rm MF} > \Lam_c$, that is, before it diverges.
The remaining low energy degrees of freedom are treated in mean-field 
approximation, with a reduced effective interaction extracted from 
the effective two-particle vertex $\Gam^{(4)\Lam_{\rm MF}}$.
In a first application of this ``poor man's'' approach to 
symmetry breaking the interplay and possible coexistence of
antiferromagnetism and $d$-wave superconductivity in the (repulsive) 
two-dimensional Hubbard model were studied \cite{Reiss07}.
As in any hybrid method, the results depend quantitatively on 
the choice of the intermediate scale $\Lam_{\rm MF}$ (except for 
mean-field models), and there is no unique criterion for this choice.

We now review the purely fermionic and the Hubbard-Stratonovich
approaches to spontaneous symmetry breaking in the functional
RG.
The methods will be explained for the case of a superconductor
as a prototype for continuous symmetry breaking, and the reader 
is referred to the literature on applications involving other 
order parameters.

\subsection{Fermionic flows}
\label{sec:fermi-flows}

The effective action $\Gam^{\Lam}$ obtained from the exact flow
equation or from symmetry-conserving truncations thereof
exhibits the same symmetries as the bare action $\cS$.
To analyze spontaneous symmetry breaking, one therefore has 
to add a symmetry-breaking term $\delta\cS$ to the bare action 
and compute the flow of $\Gam^{\Lam}$ in presence of this
term.
In case of spontaneous symmetry breaking an arbitrarily small
symmetry breaking term is promoted to a finite order parameter 
at a scale $\Lam_c$, which survives until the end of the flow.

\begin{figure} 
\centerline{\includegraphics[width = 8cm]{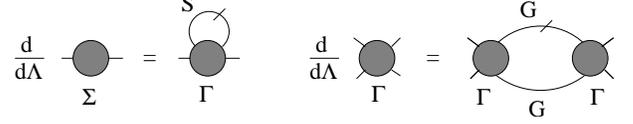}}
\caption{Coupled flow equations for the self-energy and the
 two-particle vertex determining the fermionic flow with
 symmetry breaking.}
\label{fig:katanin}
\end{figure}

A crucial issue is then to find a managable truncation of the 
exact flow equation which captures the essential features of 
the flow into the symmetry broken phase.
This is non-trivial since the effective two-particle interactions
driving the symmetry breaking become large.
Indeed, truncations based on neglecting vertices $\Gam^{(2m)\Lam}$ 
with $m > 2$ in the hierarchy of flow equations fail miserably.
A benchmark for truncations is the requirement that they should
at least provide a decent solution for mean-field models.
This requirement is met by an approximation introduced by
\textcite{Katanin04a} to implement Ward identities in truncated 
flow equations.
Katanin's truncation, which was described already in Sec.~II.C,
consists of two coupled flow equations for the self-energy
$\Sg^{\Lam}$ and the two-particle vertex $\Gam^{(4)\Lam}$,
see Fig.~\ref{fig:katanin}.
They are almost identical to the first two equations in the
hierarchy described in Sec.~II.C, with $\Gam^{(6)\Lam} = 0$, 
but in the flow equation for $\Gam^{(4)\Lam}$ the single-scale 
propagator $S^{\Lam}$ is replaced by 
$\partial_{\Lam} G^{\Lam} = S^{\Lam} + 
 G^{\Lam} \partial_{\Lam} \Sg^{\Lam} G^{\Lam}$.
This modification takes tadpole contributions obtained from
contractions of the three-particle vertex $\Gam^{(6)\Lam}$ into
account.

It is easy to see that the Katanin truncation solves mean-field 
models for symmetry breaking such as the Stoner model for 
ferromagnetism or the reduced BCS model exactly 
\cite{Salmhofer04}.
The exact self-energy in such models is given by the 
Hartree-Fock term $\Sg = V G$ (schematically), where $V$ is the
bare interaction, and the two-particle vertex by a ladder sum
of the form 
$\Gam^{(4)} = V (1 - GG V)^{-1}$.
These equations hold also in presence of a cutoff $\Lam$.
Applying $\Lam$-derivatives one finds immediately that 
$\Sg^{\Lam}$ and $\Gam^{(4)\Lam}$ obey flow equations 
of the (schematic) form
$\partial_{\Lam} \Sg^{\Lam} = \Gam^{(4)\Lam} S^{\Lam}$
and 
$\partial_{\Lam} \Gam^{(4)\Lam} = \Gam^{(4)\Lam} 
 \partial_{\Lam}(G^{\Lam} G^{\Lam}) \Gam^{(4)\Lam}$,
which corresponds exactly to Katanin's truncation.

To be more specific, we now consider the case of singlet 
superconductivity, where the continuous $U(1)$ symmetry 
associated with charge conservation is spontaneously
broken, while spin-rotation invariance remains conserved.
Superconductivity can be induced by adding a term of the
form
\begin{equation}
 \delta\cS = \sum_k \left[
 \Delta_0(k) \psib_{\up}(k) \psib_{\down}(-k) +
 \Delta_0^*(k) \psi_{\down}(-k) \psi_{\up}(k) \right] \; ,
\end{equation}
with a (generally complex) external pairing field 
$\Delta_0(k)$, to the bare action.
It is convenient to use Nambu spinors $\Psi_{\alf}(k)$ and 
$\Psib_{\alf}(k)$ with 
$\Psib_{+}(k) = \psib_{\up}(k)$,
$\Psi_{+}(k)  = \psi_{\up}(k)$,
$\Psib_{-}(k) = \psi_{\down}(-k)$,
$\Psi_{-}(k)  = \psib_{\down}(-k)$.
The effective action as a functional of the Nambu fields,
truncated beyond two-particle terms, has the form
\begin{widetext}
\begin{eqnarray}
 \Gam^{\Lam}[\Psi,\Psib] &=& 
 \Gam^{(0)\Lam} -
 \sum_k \sum_{\alf_1,\alf_2} \Gam_{\alf_1 \alf_2}^{(2)\Lam}(k) \,
 \Psib_{\alf_1}(k) \Psi_{\alf_2}(k) \nonumber \\
 &+& \frac{1}{4} \sum_{k_1,\dots,k_4} \sum_{\alf_1,\dots,\alf_4}
 \Gam_{\alf_1 \alf_2 \alf_3 \alf_4}^{(4)\Lam}(k_1,k_2,k_3,k_4) \,
 \Psib_{\alf_1}(k_1) \Psib_{\alf_2}(k_2) 
 \Psi_{\alf_3}(k_3) \Psi_{\alf_4}(k_4) \; .
\label{Gam_nambu}
\end{eqnarray}
\end{widetext}
Due to spin-rotation invariance only terms with an equal
number of $\Psi$ and $\Psib$ fields contribute.
The Nambu propagator
$\bG^{\Lam} = (\bGam^{(2)\Lam})^{-1}$ 
can be written as a $2 \times 2$ matrix of the form
\begin{equation}
 \bG^{\Lam}(k) =
 \left( \begin{array}{cc}
 G_{++}^{\Lam}(k) & G_{+-}^{\Lam}(k) \\
 G_{-+}^{\Lam}(k) & G_{--}^{\Lam}(k)
 \end{array} \right) = 
 \left( \begin{array}{cc}
 G^{\Lam}(k) & F^{\Lam}(k) \\
 F^{*\Lam}(k) & -G^{\Lam}(-k)
 \end{array} \right) .
\label{G_nambu}
\end{equation}

It is instructive to discuss the flow of the superconducting 
gap and the two-particle vertex for the reduced BCS model
\cite{Salmhofer04}, which is defined by an action of the form
\begin{eqnarray}
 \cS[\psi,\psib] &=& \!
 \sum_{k,\sg} (-ik_0 + \xi_{\bk}) \, 
 \psib_{\sg}(k) \psi_{\sg}(k)
 \nonumber \\ 
 &+& \! \sum_{k,k'} V(k,k') \,
 \psib_{\up}(k) \psib_{\down}(-k) 
 \psi_{\down}(-k') \psi_{\up}(k') . \hskip 8mm
\end{eqnarray}
Note that the interaction is restricted to particles with 
strictly opposite momenta and spins. It is well-known that 
mean-field theory solves this model exactly in the 
thermodynamical limit \cite{Haag62,Muehlschlegel62}.
The restricted momentum dependence of the bare interaction
carries over to similar restrictions for the effective 
two-particle vertex $\Gam^{(4)\Lam}$ in Eq.~(\ref{Gam_nambu}).
Only two independent components appear, namely
\begin{eqnarray}
 V^{\Lam}(k,k') &=& \Gam_{+-+-}^{(4)\Lam}(k,k',k',k) \; , \\
 W^{\Lam}(k,k') &=& \Gam_{++--}^{(4)\Lam}(k,k',k',k) \; .
\label{VW}
\end{eqnarray}
The first component is a normal interaction between two
particles, and its initial value $V^{\Lam_0}(k,k')$ is the 
bare interaction.
The second component is an anomalous term describing the
creation of four particles. It is initially zero, but is
generated by charge symmetry breaking terms in the course
of the flow.
Another anomalous term describing the destruction of four
particles is given by the complex conjugate of $W^{\Lam}(k,k')$.
The diagonal element of the Nambu self-energy vanishes for
the reduced BCS model, while the off-diagonal element is
given by the gap function $\Delta^{\Lam}(k)$.

\begin{figure}[ht]
\centerline{\includegraphics[width = 9cm]{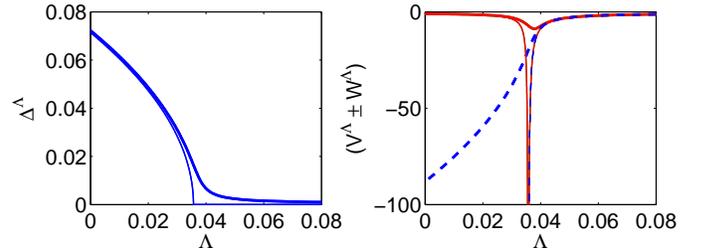}}
\caption{(Color online)
 Flow for a reduced BCS model with a constant density
 of states at zero temperature; 
 the band width is one and the bare interaction $V=-0.3$.
 Left: Flow of the gap $\Delta^{\Lam}$; the thick line is
 for a bare gap $\Delta_0 = 2.4 \cdot 10^{-4}$ and the thin
 line for $\Delta_0 = 6 \cdot 10^{-8}$, in units of the band
 width. 
 Right: Flow of the linear combinations $V^{\Lam} + W^{\Lam}$
 (solid lines) and $V^{\Lam} - W^{\Lam}$ (dashed lines) of 
 normal and anomalous vertices. Thick lines are again for
 $\Delta_0 = 2.4 \cdot 10^{-4}$ and thin lines for 
 $\Delta_0 = 6 \cdot 10^{-8}$.}
\label{fig:gapflow}
\end{figure}

For the special case of a momentum-independent s-wave
interaction $V$, the flow equations obtained from the
procedure described above are particularly simple.
Choosing a momentum-independent and real bare gap 
$\Delta_0 > 0$, the flowing quantities $\Delta^{\Lam}$, 
$V^{\Lam}$ and $W^{\Lam}$ are real and momentum-independent, 
too. Their (exact) flow is given by
\begin{equation}
 \frac{d}{d\Lam} \Delta^{\Lam} = 
 - (V^{\Lam} + W^{\Lam}) \sum_k \frac{d}{d\Lam} \left. 
 F^{\Lam} \right|_{\Delta^{\Lam} \; {\rm fixed}} \; ,
\end{equation}
where
$\frac{d}{d\Lam} F^{\Lam}|_{\Delta^{\Lam} \; {\rm fixed}}$ 
is the anomalous Nambu single-scale propagator, and
\begin{eqnarray}
 \frac{d}{d\Lam} (V^{\Lam} \pm W^{\Lam}) &=&
 - (V^{\Lam} \pm W^{\Lam})^2 \nonumber \\
 &\times& \sum_k \frac{d}{d\Lam}
 \left[ |G^{\Lam}(k)|^2 \mp |F^{\Lam}(k)|^2 \right] 
 \, . \hskip 8mm
\end{eqnarray}
A typical flow for an attractive bare interaction $V<0$
is shown in Fig.~\ref{fig:gapflow},
for two different choices of the bare gap $\Delta_0$.
The gap increases monotonically from the initial value 
$\Delta_0$ upon lowering $\Lam$, and reaches a finite value 
$\Delta \gg \Delta_0$ for $\Lam \to 0$.
A finite $\Delta_0$ regularizes the square-root singularity
in the gap flow at $\Lam = \Lam_c$.
The normal vertex $V^{\Lam}$ reaches a large negative value 
at the critical scale $\Lam_c$, while the anomalous vertex
$W^{\Lam}$ becomes large and positive.
The linear combination $V^{\Lam} + W^{\Lam}$, which drives
the gap flow, is also strongly negative at $\Lam_c$, but it
saturates at a moderately negative value for $\Lam$ below 
$\Lam_c$.
By contrast, $V^{\Lam} - W^{\Lam}$ decreases monotonically 
and reaches a final value of order $1/\Delta_0$ for 
$\Lam \to 0$, which diverges for $\Delta_0 \to 0$.
This divergence is the mean-field remnant of the Goldstone 
mode associated with the broken continuous symmetry.

In the case of a discrete broken symmetry, the effective 
interaction becomes large only at the critical scale, while
no large components remain for $\Lam \to 0$.
This has been exemplified in a study of the RG flow of a 
mean-field model for a commensurate charge density wave
\cite{Gersch05}.

The performance of the Katanin truncation for models with full
(not reduced) interactions has not yet been fully explored,
since an accurate parametrization of the flowing vertex is 
quite demanding.
However, the results obtained so far are encouraging.
Staying with superconductivity as a an example, the (Nambu) 
vertex contains 16 components, most of them corresponding to
anomalous interactions. In addition to the anomalous terms
appearing already in the reduced BCS model, there are 
anomalous interactions corresponding to the creation of three 
particles and destruction of one particle, and vice versa 
\cite{Salmhofer04,Gersch08}. 
Making full use of spin-rotation invariance, all the Nambu 
components can be actually expressed by only three independent
functions of momenta and frequencies \cite{Eberlein10}.
The main challenge is an adequate parametrization of the 
(three-fold) momentum and frequency dependence, since 
singularities associated with symmetry-breaking and the 
Goldstone mode appear in the course of the flow.
Surprisingly, in a test case study for the weakly attractive 
Hubbard model, a rather crude parametrization using the 
$N$-patch discretization described in Sec.~III turned out to 
yield a reasonable flow into the superconducting phase, with 
results for the gap in good agreement with results obtained
earlier by other means \cite{Gersch08}.
This is encouraging, but the low energy fluctuations are 
clearly not well described in such a parametrization.
To deal with the singular momentum and frequency dependence
in the Cooper channel (and possibly also in the forward
scattering channel), the channel decomposition devised by
\textcite{Husemann09a} seems very useful, since it allows one
to isolate singular dependences in functions of only one
momentum and frequency variable, similar to a description of
singular interactions by exchange bosons.
The channel decomposition has been formally extended already 
to the superconducting state \cite{Eberlein10}, but a concrete
calculation beyond mean-field models has not yet been performed.


In systems with a first order phase transition one may miss
the symmetry broken phase if one tests only for local 
stability of the symmetric phase by offering a small symmetry 
breaking field, since the latter may be metastable.
However, one can escape from the metastable state by adding
a scale dependent symmetry breaking counterterm $R^{\Lam}$
to the effective action, which has to be choosen sufficiently 
large at the beginning of the flow and fades out for 
$\Lam \to 0$, such that the system is ultimately not modified.
Formally this is just another choice of regularization within
the general framework described in Sec.~II.B.
The counterterm method has been implemented for the exactly
soluble test case of a charge-density wave mean-field model 
by \textcite{Gersch06}.

Popular approximations also beyond mean-field theory can be
retrieved from the functional RG by keeping a suitable
subset of contributions.
In particular, the Eliashberg theory for frequency-dependent
(usually phonon-induced) pairing interactions can be obtained 
as an approximation to the exact flow equations both in the
symmetric \cite{Tsai05} and in the symmetry-broken state 
\cite{Honerkamp05}.
This is achieved by keeping the Cooper channel for zero total 
momentum and frequency and the crossed particle-hole channel 
for zero transfer in the flow of the interaction, and the Fock 
term for the self-energy.

\subsection{Flows with Hubbard-Stratonovich fields}
\label{sec:hs-flows}

Collective order parameter fluctuations associated with spontaneous
symmetry breaking in interacting many-body systems are often treated
by introducing an auxiliary order parameter field via a 
Hubbard-Stratonovich transformation \cite{Popov87}.
A combination of the functional RG with the Hubbard-Stratonovich route 
to spontaneous symmetry breaking in an interacting Fermi system was
first used by \textcite{Baier04}. They studied the formation of an
antiferromagnetic state in the repulsive two-dimensional Hubbard
model at half-filling and managed to recover the low-energy collective 
behavior (described by a non-linear sigma model) from a truncated 
set of coupled flow equations for the fermions and the order parameter
field.
In the following we describe the method for the case of a superfluid
phase, summarizing the work of several groups.

We consider an interacting continuum or lattice Fermi system with a
local attraction $V < 0$. 
For continuum systems a suitable ultraviolet regularization is
necessary.
A local attraction can act only between particles with opposite 
spin and leads to singlet pairing.
It is thus natural to decouple this interaction by a 
Hubbard-Stratonovich transformation with a complex bosonic field
$\phi(q)$ corresponding to the bilinear composite of fermionic
fields $V \sum_k \psi_{\down}(-k) \psi_{\up}(k+q)$.
This leads to an action of the form
\begin{eqnarray} \label{S_super}
 \cS[\phi,\psi,\psib] &=&
 - \sum_{k,\sg} \psib_{\sg}(k) 
 \left(ik_0 - \xi_{\bk} \right) \psi_{\sg}(k)
 \nonumber \\
 &+& \frac{m_b}{2} \sum_q \phi^*(q) \phi(q) 
 \nonumber \\
 &+& \sum_{k,q} \left[
 \psib_{\up}(k+q) \psib_{\down}(-k) \phi(q) +
 {\rm h.c.} \right] \, , \hskip 5mm
\end{eqnarray}
where $\phi^*$ is the complex conjugate of $\phi$ and $m_b = -1/V > 0$.

Spontaneous symmetry breaking can now be studied by using the flow
equation for the effective action $\Gam^{\Lam}[\phi,\psi,\psib]$
for coupled bosonic and fermionic fields derived in Sec.~II.G.
Relatively simple truncations capture several non-trivial fluctuation 
effects.
Effective interactions beyond quartic order in the fields are
generally neglected. Also boson-fermion vertices beyond the order
appearing already in the bare action are discarded.
The truncations are usually formulated as an ansatz for the 
effective average action 
\begin{eqnarray}
 \Gam_R^{\Lam}[\phi,\psi,\psib] &=& 
 \Gam^{\Lam}[\phi,\psi,\psib] - \mbox{regulator term}
 \nonumber \\
 &=& \Gam_b^{\Lam}[\phi] + \Gam_f^{\Lam}[\psi,\psib] +
 \Gam_{bf}^{\Lam}[\phi,\psi,\psib] \, , \hskip 5mm
\end{eqnarray}
which obeys the initial condition $\Gam_R^{\Lam_0} = \cS$, see 
Sec.~II.G.

The ansatz used for the bosonic part is guided by the usual
strategy of a double expansion in $\phi$ and gradients
(see, for example, \textcite{Tetradis94}):
\begin{equation} \label{4b:Gam_b}
 \Gam_b^{\Lam}[\phi] = 
 \sum_x \, U_{\rm loc}^{\Lam}(\phi(x)) \, + \,
 \mbox{gradient terms} \, ,
\end{equation}
where $x = (x_0,x_1,\dots,x_d)$ collects imaginary time and
real space coordinates. 
Note that we use the same letter $\phi$ for the real space and
momentum space representations of the bosonic field.
The shape of the local potential $U_{\rm loc}^{\Lam}(\phi)$
depends on the scale. For $\Lam$ above a critical scale $\Lam_c$
it has the convex form
\begin{equation}
 U_{\rm loc}^{\Lam}(\phi) = 
 m_b^{\Lam} |\phi|^2 + u^{\Lam} |\phi|^4 \, ,
\end{equation}
with a minimum at $\phi = 0$.
For $\Lam < \Lam_c$ the potential assumes a mexican hat shape
\begin{equation}
 U_{\rm loc}^{\Lam}(\phi) = 
 u^{\Lam} \left[ |\phi|^2 - |\alf^{\Lam}|^2 \right]^2 \, ,
\end{equation}
with a circle of minima at $|\phi| = |\alf^{\Lam}|$, where 
$\alf^{\Lam}$ is the (flowing) bosonic order parameter.
The regime $\Lam > \Lam_c$ is called the {\em symmetric} 
regime. At $\Lam = \Lam_c$ the bosonic mass $m_b$ vanishes.
In the {\em symmetry-broken} regime, for $\Lam < \Lam_c$, the order
parameter $\alf^{\Lam}$ rises continuously from zero to a finite 
value. Its flow can be computed by tracing the minimum of the
flowing potential $U_{\rm loc}^{\Lam}$ or, equivalently, by the 
condition that the bosonic one-point vertex $\Gam_b^{(1)\Lam}$
vanishes.

For the gradient terms in $\Gam_b^{\Lam}[\phi]$ various choices
have been made.
The simplest one \cite{Birse05,Diehl07a,Krippa07} compatible with 
the $U(1)$ symmetry has the form of an inverse bare propagator for 
free bosons, 
\begin{eqnarray} \label{gradterms}
 && \sum_x \left[ 
 Z_b^{\Lam} \phi^*(x) \partial_{x_0} \phi(x) -
 A_b^{\Lam} \phi^*(x) \nabla^2 \phi(x) \right] =
 \nonumber \\
 && \sum_q \phi^*(q) \left[-i Z_b^{\Lam} q_0 + A_b^{\Lam} \bq^2 
   \right] \phi(q) \, ,
\end{eqnarray}
where $\nabla = (\partial_{x_1},\dots,\partial_{x_d})$.
For lattice fermions one may replace $\bq^2$ by a periodic
dispersion $\om_{\bq}$ which is proportional to $\bq^2$ only
at small $\bq$ \cite{Strack08}.
The term linear in $q_0$ is absent in particle-hole symmetric 
systems \cite{Strack08}, such that contributions of order
$q_0^2$ become important.
Additional gradient terms have to be taken into account to
fully capture the effects of the Goldstone mode, as discussed
below.

The normal fermionic part of the effective action is usually 
kept in its bare form, sometimes adjusted by renormalization 
factors for the frequency and momentum dependences. 
In the symmetry broken regime, an anomalous term is generated, 
such that $\Gam_f^{\Lam}$ becomes
\begin{eqnarray} \label{ansatzGam_f}
 \Gam_f^{\Lam}[\psi,\psib] &=& 
 - \sum_{k,\sg} \psib_{\sg}(k) 
 (iZ_f^{\Lam} k_0 - A_f^{\Lam} \xi_{\bk}) \psi_{\sg}(k) 
 \nonumber \\
 &+& \sum_k \left[ 
 \Delta^{\Lam}(k) \psib_{\up}(k) \psib_{\down}(-k) +
 {\rm h.c.} \right] \, .
\end{eqnarray}
For a local interaction the $k$-dependence of the gap function 
$\Delta^{\Lam}(k)$ is very weak and in simple truncations fully 
absent.
Quartic terms corresponding to effective two-fermion interactions
are absent in the bare action by virtue of the Hubbard-Stratonovich
decoupling, but are generated again in the course of the flow.
These generated terms are neglected in lowest order truncations,
and sometimes they are treated by a dynamical decoupling procedure
called dynamical bosonization, see below. 

For the effective boson-fermion interaction one also maintains
the bare form of a local 3-point function,
\begin{eqnarray}
 \Gam_{bf}^{\Lam}[\phi,\psi,\psib] &=& 
 \sum_{k,q} g^{\Lam} \left[ 
 \psib_{\up}(k+q) \psib_{\down}(-k) \phi(q) +
 {\rm h.c.} \right]
 \nonumber \\
 &+& \mbox{anomalous terms} \; ,
\end{eqnarray}
with anomalous terms of the form $\psi\psi\phi$ and 
$\psib\psib\phi^*$ contributing only in the symmetry-broken 
regime. The coupling $g^{\Lam}$ is frequently referred to as 
``Yukawa coupling''.
The anomalous terms in the boson-fermion interaction are usually
neglected. If taken into account, they remain indeed rather small
\cite{Strack08}.

Instead of using a $U(1)$-symmetric ansatz for the effective action,
one may also start from the hierarchy of flow equations for the
vertex functions and implement the $U(1)$-symmetry by Ward identities
\cite{Bartosch09}. 

Even with the simple ansatz (\ref{gradterms}) for the bosonic 
gradient terms, the effective action described above yields sensible 
results not only at weak coupling, but actually in the entire regime 
from BCS superfluidity to Bose Einstein condensation of tightly 
bound pairs \cite{Diehl07a}.
In particular, the transition temperature $T_c$ increases 
exponentially with the interaction in the weak coupling regime, 
reaches a maximum, and finally saturates in the strong coupling
limit, as it should.

The bosonic interaction $u^{\Lam}$ and also the bosonic 
renormalization factors $A_b^{\Lam}$ and $Z_b^{\Lam}$ vanish in
the limit $\Lam \to 0$ \cite{Birse05,Krippa07}.
This fluctuation effect reflects the drastic renormalization of 
longitudinal order parameter correlations, which are well-known
from the interacting Bose gas in dimensions $d \leq 3$ 
(see, for example, \textcite{Pistolesi04}).
Note that for a nodeless gap function the low-energy behavior 
of a fermionic superfluid is equivalent to that of an interacting
Bose gas, since fermionic excitations are fully gapped.
However, with the simple ansatz (\ref{gradterms}) the transverse
order parameter fluctuations corresponding to the Goldstone mode
are also strongly renormalized, which is not correct.
To distinguish between longitudinal and transverse fluctuations,
one may fix the phase of the order parameter $\alf^{\Lam}$ such
that $\alf^{\Lam}$ is real, decompose the complex order parameter 
field in real and imaginary parts $\phi(q) = \sg(q) + i\pi(q)$ 
with $\sg(-q) = \sg^*(q)$ and $\pi(-q) = \pi^*(q)$, and introduce
different renormalization factors for $\sg$ and $\pi$ fields
\cite{Pistolesi04}.
Using this decomposition, the correct infrared behavior was 
obtained by \textcite{Strack08} where, however, the cancellation
of singular contributions to the renormalization factors for
the transverse $\pi$ fields was implemented by hand.
To capture this cancellation intrinsically, one has to include 
an additional $U(1)$ symmetric gradient term of the form 
$[\sg(\partial_{x_0},\nabla)\sg + \pi(\partial_{x_0},\nabla)\pi]^2$ 
\cite{Tetradis94,Strack09}.

The fermionic flow based on the Katanin truncation described in
Sec.~IV.A reproduces the exact solution of the reduced BCS model 
(and other mean-field models). 
Within the truncation described above, the bosonized flow yields 
a reasonable solution without artificial features, but the gap 
comes out a bit too small.
The reason for this is the truncation of $U_{\rm loc}^{\Lam}(\phi)$ 
at quartic order.
To recover the exact solution, one has to keep all orders in $\phi$
\cite{Strack08}.

The ansatz (\ref{ansatzGam_f}) for $\Gam_f^{\Lam}[\psi,\psib]$ 
neglects the generation of fermionic interactions by the flow. 
In particular, quartic (two-fermion) interactions are
generated by box diagrams with four boson-fermion vertices.
These terms contain contributions from particle-hole fluctuations
which, among other effects, lead to a significant reduction of 
the transition temperature.
The (re-)generated two-fermion interaction can be decoupled
at each step in the flow by a procedure called dynamical
bosonization \cite{Gies02,Gies04,Floerchinger09d}.
A general two-fermion interaction cannot be decoupled (exactly)
by a single Hubbard-Stratonovich field, such that several fields
may be needed to obtain accurate results.
Dynamical bosonization was used to include effects from
particle-hole fluctuations in attractively interacting Fermi
systems by \textcite{Floerchinger08a}.

Following the work of \textcite{Baier04} on the repulsive
Hubbard model at half-filling,
functional RG flow equations with Hubbard-Stratonovich fields
were also applied to the Hubbard model away from half-filling.
Commensurate and incommensurate antiferromagnetic fluctuations
were investigated \cite{Krahl09a}.
More importantly, it was clarified how the generation of 
$d$-wave pairing from antiferromagnetic fluctuations can be
captured by a bosonized flow \cite{Krahl09b}, and the flow 
was continued into the symmetry broken phase, with coupled 
order parameter fields describing antiferromagnetism and 
$d$-wave superconductivity \cite{Friederich10, Friederich11}.

Compared to the purely fermionic RG described in Sec.~IV.A,
the treatment of order parameter fluctuations is facilitated
considerably by the Hubbard-Stratonovich field.
On the other hand, fluctuation effects associated with other
channels (the particle-hole channel in case of superfluidity)
look more complicated.
For systems with competing instabilities the choice of an
adequate Hubbard-Stratonovich field becomes problematic,
since the fermionic interaction can be decoupled in different
ways, which, in combination with truncations, may lead to 
ambiguities in the results.
In general, several Hubbard-Stratonovich fields must be used, 
and the analysis done in Section \ref{sec:III} indicates which ones
are the most important. The decomposition of the interaction in 
\cite{Husemann09a} allows to switch to Hubbard-Stratonovich fields
after the fermionic flow has been performed down to a certain scale, 
and may thus be used to combine the two flow representations. 


\section{QUANTUM CRITICALITY}
\label{sec:V}

Instabilities of the normal metallic state lead to a rich variety
of quantum phase transitions \cite{Sachdev99} in the ground state 
of interacting electron systems, which can be tuned by a control
parameter such as pressure, doping, or a magnetic field.
Most interesting are {\em continuous} transitions which lead to
quantum critical fluctuations \cite{Belitz05}.
Near a quantum critical point (QCP) electronic excitations are
strongly scattered by order parameter fluctuations such that
Fermi liquid theory breaks down \cite{Vojta03,Loehneysen07}.
Quantum critical fluctuations are therefore frequently invoked
as a mechanism for non-Fermi liquid behavior observed in strongly 
correlated electron compounds.

Quantum phase transitions in interacting Fermi systems are
traditionally described by an effective order parameter theory
pioneered by \textcite{Hertz76} and \textcite{Millis93}.
An order parameter field $\phi$ is introduced by a 
Hubbard-Stratonovich decoupling of the fermionic interaction, 
and the fermionic fields are subsequently integrated out.
The resulting effective action for the order parameter is 
truncated at quartic order and analyzed by standard scaling
and RG techniques.
However, more recent studies revealed that the Hertz-Millis
approach is not always applicable, especially in low-dimensional 
systems \cite{Belitz05,Loehneysen07}.
Since electronic excitations in a metal are gapless, integrating 
out the electrons can lead to singular interactions in the 
effective order parameter action which cannot be approximated 
by a local quartic term. 
The nature of the problem was identified and essential aspects
of its solution were presented first for disordered ferromagnets
by \textcite{Kirkpatrick96}, and later elaborated on by
\textcite{Belitz01a,Belitz01b}.
For clean ferromagnets, \textcite{Belitz97} showed that 
Hertz-Millis theory breaks down, and no continuous quantum
phase transition can exist, in any dimension $d \leq 3$;
the transition is generically of first order \cite{Belitz99}.
The Hertz-Millis approach was also shown to be invalid for
the quantum antiferromagnetic transition in two dimensions
\cite{Abanov03,Abanov04,Metlitski10b}.
In that case a continuous transition survives, but the QCP
becomes non-Gaussian.

Applications of the functional RG to quantum phase transitions
and quantum criticality in interacting Fermi systems have
appeared only recently.
In Sec.~V.A we explain how the Hertz-Millis theory fits into
the functional RG framework and we review some extensions
relying on an effective order parameter action truncated at
quartic or hexatic order.
An application of a non-perturbative truncation, where all
orders in $\phi$ are (and must be) kept, is discussed in 
Sec.~V.B.
Finally, in Sec.~V.C we briefly discuss the possibility to 
study coupled flow equations for fermions and their 
critical order parameter fluctuations in the functional RG 
framework, and we refer to first steps in this direction.


\subsection{Hertz-Millis theory}
\label{sec:hertz}

In his seminal work on quantum phase transitions in metallic
electron systems, \textcite{Hertz76} proposed to decouple the
electron-electron interaction by introducing an order parameter
field $\phi$ via a Hubbard-Stratonovich transformation.
The resulting action is quadratic in the fermionic variables
$\psi$ and $\psib$, which can therefore be integrated out.
One thus obtains an effective action which depends only on $\phi$.
Truncating at quartic order in $\phi$, and discarding
irrelevant momentum and frequency dependences (in the sense
of standard power counting), leads to the Hertz action.
\begin{eqnarray} \label{S_hertz}
 \cS[\phi] &=& \cS^{(0)} + \sum_q \phi(-q) \left[ 
 A \bq^2 + Z \frac{|q_0|}{|\bq|^{z-2}} \right] \phi(q) 
 \nonumber \\
 &+& \sum_x U_{\rm loc}(\phi(x)) \, ,
\end{eqnarray}
where $\cS^{(0)}$ is a field-independent term, and
\begin{equation}
 U_{\rm loc}(\phi) = r \phi^2 + u \phi^4 \, .
\end{equation}
We write our equations for the case of a real scalar order
parameter for simplicity, using (again) the same letter
$\phi$ for the real and momentum space representations
of the field.
Except for the frequency dependence, the action has the
form of a $\phi^4$ theory for thermal phase transitions.
The frequency dependent term stems from low-energy particle-hole 
excitations.
Here the dynamical exponent $z$ is an integer number $\geq 2$
depending on the type of transition.
Tuning the parameter $r$ one can approach the phase
transition, in particular the quantum phase transition at
$T = 0$.

The action (\ref{S_hertz}) has been analyzed by standard scaling
and RG techniques. 
Due to the frequency dependence, the scaling behavior at the
QCP corresponds to a system with an effective dimensionality 
$d_{\rm eff} = d + z$, where $d$ is the spatial dimension 
\cite{Hertz76}.
As a consequence, the QCP appears to be Gaussian in three- 
and even in two-dimensional systems.
An important insight by \textcite{Millis93} was that the 
$\phi^4$ term in the action is nevertheless crucial to obtain 
the correct temperature dependences near the QCP.
He derived the temperature dependence of the correlation
length $\xi$ and other quantities by using a perturbative
RG with a mixed momentum and frequency cutoff.
From a functional RG perspective, Millis' scaling theory
can be viewed as a simple truncation of the effective average 
action $\Gam_R^{\Lam}[\phi]$ evolving from $\cS[\phi]$,
namely
\begin{eqnarray} \label{Gam_millis}
 \Gam_R^{\Lam}[\phi] &=& \Gam^{(0)\Lam} + 
 \sum_q \phi(-q) \left[ 
 A^{\Lam} \bq^2 + Z^{\Lam} \frac{|q_0|}{|\bq|^{z-2}} \right] 
 \phi(q) \nonumber \\
 &+& \sum_x U_{\rm loc}^{\Lam}(\phi(x)) \, ,
\end{eqnarray}
where
\begin{equation}
 U_{\rm loc}^{\Lam}(\phi) = 
 r^{\Lam} \phi^2 + u^{\Lam} \phi^4 \, ,
\end{equation}
and $\Lam$ parametrizes a mixed momentum and frequency cutoff.
The flow equations for the parameters in Eq.~(\ref{Gam_millis})
are obtained by inserting $\Gam_R^{\Lam}[\phi]$ in the exact
functional flow equation (\ref{floweqGamRb}) and comparing 
coefficients.
Due to the local form of the $\phi^4$ interaction, the 
self-energy is momentum and frequency independent such that 
the parameters $A^{\Lam}$ and $Z^{\Lam}$ remain invariant.
The flow of $u^{\Lam}$, which is driven by a contribution of
order $(u^{\Lam})^2$, is important only in the marginal case 
$d + z = 4$ and near the thermal phase transition at $T_c > 0$.
Hence, most of Millis' results on the region around the QCP 
in the phase diagram are based on an analysis of the flow of
$r^{\Lam}$ and the thermodynamic potential
$\Omega^{\Lam} = T \Gam^{(0)\Lam}$ 
(for a review, see \textcite{Loehneysen07}).

Various extensions of Millis' analysis were derived within
the functional RG framework.
In particular, an extension to the symmetry-broken phase
was formulated, for cases where the broken symmetry is
{\em discrete} and does not gap out the fermionic excitations 
\cite{Jakubczyk08}.
One such case is a nematic transition
driven by a Pomeranchuk instability of interacting electrons 
on a lattice, where the discrete point-group symmetry of the
lattice is spontaneously broken \cite{Fradkin10}.
The symmetry-broken regime was described by the ansatz
(\ref{Gam_millis}) for $\Gam_R^{\Lam}[\phi]$, with a quartic 
local potential which has a minimum away from zero:
\begin{equation}
 U_{\rm loc}^{\Lam}(\phi) = 
 u^{\Lam} \left[ \phi^2 - (\phi_0^{\Lam})^2 \right]^2 \, .
\end{equation}
The resulting flow equations were used to compute $T_c$ and
the Ginzburg temperature $T_G^<$ below $T_c$ as a function
of the control parameter $r$. 
To access the non-Gaussian thermal critical regime near $T_c$,
it is crucial to take the flow of the quartic coupling
$u^{\Lam}$ into account.
The parameters $A^{\Lam}$ and $Z^{\Lam}$ are now scale dependent, too. 
While $Z^{\Lam}$ remains almost invariant, the flow of
$A^{\Lam}$ is important near $T_c$ and gives rise to an
anomalous scaling dimension.
A main result of the calculation was that the leading
$r$-dependence of $T_c$ is the same as that
of the Ginzburg temperatures below and above $T_c$ (the 
latter was calculated by \textcite{Millis93}), but a fairly
large Ginzburg region opens in two dimensions 
\cite{Jakubczyk08,Bauer11}.


In another extension a $\phi^6$-interaction was included
in $U_{\rm loc}^{\Lam}$ to study a possible change of the order
of the transition by fluctuations \cite{Jakubczyk09a}, 
as well as quantum tricritical points in metals
\cite{Jakubczyk10}.

Note that
the extensions mentioned above are based on perturbative 
truncations resulting in flow equations with few running 
couplings, which could have been obtained also by more 
conventional RG methods.


\subsection{Full potential flow}
\label{sec:fullpot}

We now review an application to a problem where the effective
action cannot be truncated at any finite order in $\phi$, 
such that the possibility to make non-perturbative truncations
becomes crucial \cite{Jakubczyk09b}.

The problem arises when asking how a nematic transition 
caused by a $d$-wave Pomeranchuk instability in two dimensions
is affected by fluctuations.
Such a transition can be modelled by tight-binding electrons
on a square lattice with an attractive $d$-wave forward scattering 
interaction \cite{Metzner03}:
\begin{equation} \label{f-model}
 H = \sum_{\bk} \eps_{\bk} n_{\bk} +
 \frac{1}{2L} \sum_{\bk,\bk',\bq} f_{\bk\bk'}(\bq) \,
 n_{\bk}(\bq) \, n_{\bk'}(-\bq) \; ,
\end{equation}
where $n_{\bk}(\bq) = \sum_{\sg} 
 c^{\dag}_{\bk-\bq/2,\sg} c^{\phantom\dag}_{\bk+\bq/2,\sg}$
and $L$ is the number of lattice sites.
The interaction has the form
\begin{equation}
 f_{\bk\bk'}(\bq) = - g(\bq) d_{\bk} d_{\bk'} \; ,
\end{equation}
where $d_{\bk} = \cos k_x - \cos k_y$ is a form factor with
$d_{x^2-y^2}$ symmetry. The coupling function $g(\bq) \geq 0$
has a maximum at $\bq={\bf 0}$ and is restricted to small momentum 
transfers by a cutoff $\Lam_0$. 
For sufficiently large $g = g({\bf 0})$ the interaction drives a 
$d$-wave Pomeranchuk instability leading to a nematic state with 
broken orientation symmetry, which can be described by the 
order parameter
\begin{equation}
 \phi = \frac{g}{L} \sum_{\bk} d_{\bk} \bra n_{\bk} \ket
 \; .
\end{equation}
In the plane spanned by the chemical potential and temperature 
a nematic phase is formed below a dome-shaped transition line 
$T_c(\mu)$ with a maximal transition temperature near Van Hove 
filling. In mean-field theory, the phase transition is usually 
first order near the edges of the transition line, that is, 
where $T_c$ is relatively low, and second order at the roof of 
the dome \cite{Kee03,Khavkine04,Yamase05}.

Introducing an order parameter field via a Hubbard-Stratonovich
transformation, integrating out the fermions, and keeping only
the leading momentum and frequency dependences for small $\bq$
and small $q_0/|\bq|$ leads to a Hertz-type action $\cS[\phi]$
of the form (\ref{S_hertz}), with $z=3$ and a local potential 
given by the mean-field potential
\begin{equation}
 U_{\rm loc}(\phi) = 
 \frac{\phi^2}{2g} - \frac{2T}{L} \sum_{\bk}
 \ln\left( 1 + e^{-(\eps_{\bk} - \phi d_{\bk} - \mu)/T}
 \right) \; .
\end{equation}
At low temperatures,
the coefficients of a Landau expansion of $U_{\rm loc}(\phi)$ in 
powers of the field, 
$U(\phi) = a_0 + a_2 \phi^2 + a_4 \phi^4 + \dots$,  
are typically negative for all exponents $2m \geq 4$.
Hence, $\cS[\phi]$ and consequently also the effective action
$\Gam_R^{\Lam}[\phi]$ cannot be truncated at any finite order
in $\phi$.
Fortunately, for bosonic fields the functional RG allows also 
for non-perturbative approximations, where one expands only in 
gradients, and not in powers of $\phi$ \cite{Berges02}.
In particular, one can use the ansatz (\ref{Gam_millis}) for
$\Gam_R^{\Lam}[\phi]$ without expanding the local potential
$U_{\rm loc}^{\Lam}(\phi)$.
The flow of $U_{\rm loc}^{\Lam}(\phi)$ is then determined by
a {\em partial} differential equation which contains a second
derivative of the potential with respect to $\phi$.

In Fig.~15 an exemplary plot of the evolution of the flowing 
effective potential $U_{\rm loc}^{\Lam}(\phi)$ is shown for 
$\Lam$ ranging from the ultraviolet cutoff 
$\Lam_0 = e^{-1} \approx 0.37$ to the final value $\Lam = 0$.
The flow has been computed for electrons on a square lattice 
with nearest neighbor hopping $t=1$, next-to-nearest neighbor
hopping $t'=-\frac{1}{6}$, and a coupling strength $g=0.8$.
The initial (mean-field) potential has a minimum at
$\phi_0 = 0.112$.
The final potential exhibits spontaneous symmetry breaking 
with an order parameter $\phi_0 = 0.102$.
Fluctations shift $\phi_0$ toward a slightly smaller value
compared to the mean-field solutions.
The flat shape of $U_{\rm loc}^{\Lam}(\phi)$ for 
$\phi \leq \phi_0$ at $\Lam = 0$ is imposed by the 
convexity property of the grand canonical potential.
\begin{figure}[ht]
\begin{center}
\includegraphics[width=7cm]{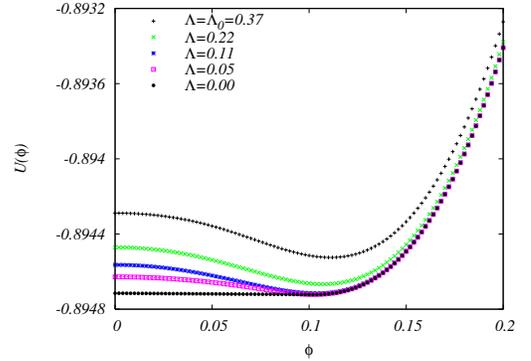}
\caption{(Color online)
 Flowing effective potential $U_{\rm loc}^{\Lam}(\phi)$ 
 for various values of $\Lam$ between $\Lam_0 = e^{-1}$ 
 and $0$, at $\mu = -0.78$ and $T = 0.05$ \cite{Jakubczyk09b}.}
\end{center}
\label{fig:nematicflow}
\end{figure}
The transition line between normal and symmetry-broken phases
is shown in Fig.~16 for two choices of $\Lam_0$. 
Compared to the corresponding mean-field result, the 
transition temperature is suppressed, with a larger reduction
for larger $\Lambda_0$ (corresponding to a larger phase space
for fluctuations).
For $\Lambda_0 = 1$ the transition is continuous down to $T=0$, 
leading to quantum critical points at the edges of the nematic 
dome.
Increasing $\Lambda_0$ further (or reducing $g$), one may even
eliminate the nematic phase completely from the phase diagram
\cite{Yamase11}.
\begin{figure} 
\begin{center}
\includegraphics[width=7cm]{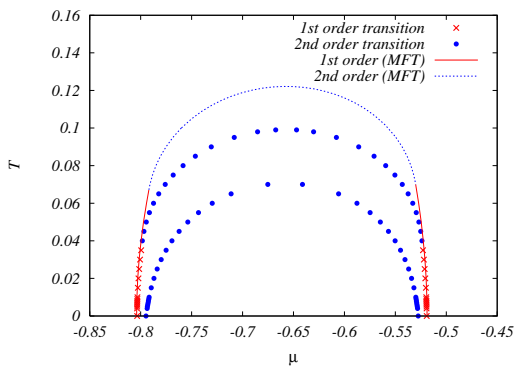}
\caption{(Color online) Critical temperatures versus chemical
 potential for $\Lambda_0 = e^{-1}$ (larger dome with dots and 
 crosses) and $\Lambda_0 = 1$ (smaller dome with dots).
 The mean-field transition line is also shown for comparison
 \cite{Jakubczyk09b}.}
\end{center}
\label{fig:nematicphase}
\end{figure}

\vspace*{5mm}

\subsection{Coupled flow of fermions and order parameter fluctuations}
\label{sec:coupledflow}

There are various systems where integrating out the electrons leads
to an effective order parameter action with singular interactions
which cannot be approximated by a local coupling
\cite{Belitz05,Loehneysen07}.
In such cases it can be advantageous to keep the electrons in the 
action, treating the coupled system consisting of electrons and their
order parameter fluctuations.
Several coupled boson-fermion systems exhibiting quantum criticality 
have already been analyzed by various methods; see, for example, 
\textcite{Vojta00,Belitz01a,Belitz01b,Abanov03,Rech06,Kaul08,Huh08,
Metlitski10a,Metlitski10b}.
The interplay of bosonic and fermionic infrared singularities
at the quantum critical point poses an interesting problem.

The functional RG for coupled bosons and fermions described in 
Sec.~II.G provides a suitable framework to study such problems.
So far, it has not been applied to quantum phase transitions 
in metallic electron systems. 
However, encouraging works on relativistic field-theoretic models 
with gapless bosons and fermions have already appeared.
For example, functional RG flow equations have been used to study
the Gross-Neveu model \cite{Rosa01}, 
quantum electrodynamics \cite{Gies04a},
and supersymmetric Wess-Zumino models \cite{Gies09}.
In the context of condensed matter physics, a toy model for a
semimetal-to-superfluid quantum phase transition has been studied
by coupled flow equations for the electrons and the superfluid
order parameter \cite{Strack10,Obert11}.
In dimensions $d<3$ the fermions and the order parameter
fluctuations acquire anomalous scaling dimensions at the QCP
of that model, leading to non-Fermi liquid behavior and non-Gaussian 
criticality.

It will be interesting to devise suitable truncations of the 
coupled boson-fermion flow equations for magnetic and nematic 
quantum phase transitions in low-dimensional metallic systems, 
where many open questions need to be clarified.


\section{CORRELATION EFFECTS IN QUANTUM WIRES AND QUANTUM DOTS}
\label{sec:VI}

As our last application of the functional RG to correlated fermion 
systems we discuss many-body effects in quantum wires and dots.
The focus is on transport through such systems which are coupled 
to two or more semi-infinite leads.  
While in many of the above applications it was crucial 
to devise an approximation scheme in which the flow of the two-particle 
vertex (the effective two-particle interaction) was properly described, 
in the ones reviewed in this section the physics is dominated 
by the flow of the self-energy.
We start out with a 
brief discussion of quantum transport through a region  containing 
correlations in Sec.~\ref{subsec:QT}. To study transport beyond the 
linear response regime the functional RG was recently extended 
to Fermi systems out of equilibrium. In Sec.~\ref{subsec:nonequfRG} 
we review the main steps of this generalization. 
After discussing the most elementary example of linear transport through an 
inhomogeneous correlated quantum wire -- a chain with a single local 
impurity -- in Secs.~\ref{subsec:LL1}-\ref{subsubnum}, we show how the 
functional RG can be used (i) to describe transport on all energy scales 
for more complex setups (Sec.~\ref{resotunsec}), (ii) to identify 
unconventional low-energy fixed points of such systems 
(Sec.~\ref{Yjunctionsec}), and (iii) to study finite bias 
non-equilibrium transport (Sec.~\ref{subsecnonequiwire}). 
As an example of the application of the functional RG to quantum dots, in  
Sec.~\ref{subsubchargefluc} we consider an interacting chain of only 
three lattice sites corresponding to a dot. 

\subsection{Quantum transport}
\label{subsec:QT}

Experimental progress has made it possible to measure transport 
through mesoscopic regions like one-dimensional (1D) 
quantum wires of up to micrometer length and  ``zero-dimensional'' 
quantum dots. The experiments provide evidence for correlation effects 
\cite{Deshpande10,Hanson07}.\footnote{Correlation effects in effectively 
1D electronic systems are also studied using photoemission. For a recent 
review, see \textcite{Grioni09}.} 
It is a great theoretical challenge to describe transport 
when correlations in the mesoscopic system are important. 
Usually the leads to which this correlated region is connected 
are modeled as non-interacting.  
A general formal expression in terms of Keldysh Green functions 
for the  current $I$ through an {\it interacting} system coupled 
to two leads (indices $L,R$) in the stationary state 
was presented by \textcite{Meir92}. Either for specific 
models or applying certain approximations to the two-particle 
interaction for each channel $\zeta$ it can be brought into a 
Landauer-B\"uttiker type form \cite{Landauer57,Buettiker86}
\begin{equation}
 \label{Landauer}
 I_{\zeta} = \frac{1}{2 \pi} \int {\mathcal T}_{\zeta}(\epsilon,T,V_b)  
 \left[f_L(\epsilon)-f_R(\epsilon)\right] d\epsilon \; ,
\end{equation} 
where we set $e=1=\hbar$ such that the conductance quantum per channel 
is given by $1/(2 \pi)$.
Here ${\mathcal T}_{\zeta}$ is an {\it effective} transmission 
probability, $V_b=\mu_L-\mu_R$ the bias voltage, and  $f_{L/R}$
are Fermi functions with the chemical potentials of 
the left and right leads $\mu_{L/R}$.
For the transport through a non-interacting system ${\mathcal T}_{\zeta}$ 
is the single-particle transmission probability.
The goal is to compute ${\mathcal T}_{\zeta}$ in the presence of correlations.
 Here  truncations of the functional RG flow equations which 
lead to {\it frequency independent} self-energies are considered.  
In this case using Eq.\ (\ref{Landauer}), with ${\mathcal T}_{\zeta}$ being
proportional to the ``contact-to-contact'' matrix element 
of the (retarded) one-particle Green function (see below) does 
not present an additional approximation as current vertex 
corrections vanish \cite{Oguri01,Enss05a}.  
In this approximation the two-particle interaction 
affects the transport only via the renormalized self-energy which acts 
as an additional, $T$ and $V_b$ dependent 
scattering potential
on non-interacting electrons \cite{Oguri01,Langer61}.
For the {\it linear} conductance $g_{\zeta}(T)$ the transmission probability 
enters only at zero bias $V_b = 0$,
\begin{equation}
 \label{linearcond}
 g_{\zeta}(T) = \frac{1}{2 \pi}\int {\mathcal T}_{\zeta}(\epsilon,T,0)  
 \left ( -\frac{\partial f}{\partial \epsilon}\right )d\epsilon \; ,
\end{equation}
i.e.\ $  {\mathcal T}_{\zeta}(\epsilon,T,0) $ is an {\it equilibrium} property.
At zero temperature Eq.\ (\ref{linearcond}) simplifies further to 
$g_{\zeta}(0) = {\mathcal T}_{\zeta}(\mu,0,0)/(2 \pi) $.

\subsection{Functional RG in non-equilibrium}
\label{subsec:nonequfRG}

Recently, the functional RG approach was extended to study steady state 
non-equilibrium transport through quantum wires and dots in the presence 
of a finite bias voltage given by the difference of the chemical potentials 
of the left and right leads $V_b = \mu_L-\mu_R$ 
\cite{Jakobs03,Gezzi07,Jakobs07a,Jakobs10a,Karrasch10a,Jakobs10b,Karrasch10b}. 
The basic idea behind this extension is the use of real time or real frequency
Green functions on the Keldysh contour \cite{Rammer86} instead of Matsubara 
Green functions.
As usual in diagrammatic approaches based on Keldysh Green functions one 
assumes that the initial statistical operator does not contain any correlations 
\cite{Rammer86}.   
One can then either use a functional integral formulation \cite{Kamenev04}
of the non-equilibrium many-body problem \cite{Gezzi07} or a purely 
diagrammatic approach \cite{Jakobs03,Jakobs07a} to derive the (same) flow 
equations for the self-energy and higher order vertex functions. 
Although the method allows to work with two-time Green functions and to study 
transient dynamics, in the current implementation of the functional RG in 
non-equilibrium for fermions the system is {\it assumed} to be in the steady 
state. For interacting bosons functional RG was also used to study dynamics 
\cite{Gasenzer08,Kloss11}.
On a technical level and compared to the equilibrium Matsubara functional RG 
in the steady state the Keldysh structure only leads to an additional index 
(the so-called Keldysh index $\pm$ referring to the upper and lower branch of 
the Keldysh contour) to be added to the set of quantum numbers.    

One of the main challenges of the non-equilibrium functional RG is to devise 
cutoff schemes which do not violate causality and general Kubo-Martin-Schwinger 
(KMS) relations \cite{Jakobs10b,Jakobs10c} after truncation of the infinite 
hierarchy of flow equations. For a general cutoff fulfilling the requirements 
discussed in Sect. \ref{sec:flowparameters} it is only guaranteed that 
causality and KMS relations hold up to the truncation order, e.g.~first order
for the level-1 truncation. In fact, an
infrared (real) frequency cutoff similar to Eq.\ (\ref{cutfun}) violates 
causality in second order \cite{Jakobs03,Gezzi07}. Its breaking constitutes 
a severe problem as relations connecting the Keldysh contour matrix elements 
of the Green function cannot be used. In particular, 
one cannot rotate from the $G^{a,b}$ Green function, with Keldysh indices 
$a,b = \pm$, to retarded, advanced, and Keldysh Green functions as it is usually 
done \cite{Rammer86}. The momentum cutoff scheme used in other sections of this 
review avoids this problem but is not suitable for models with broken translational 
invariance. We will return to this issue and discuss appropriate cutoff schemes 
when reviewing applications of non-equilibrium functional RG to transport through 
wires in Sec.~\ref{subsecnonequiwire} and dots in Sec.~\ref{subsubchargefluc}.

Besides its ability to be applied to non-equilibrium problems the real time 
(or real frequency) Keldysh functional RG has a distinct advantage even in 
equilibrium: the analytic continuation from Matsubara to real frequencies 
can be avoided. In fact, computing the real frequency dependence of observables 
obtained by a numerical solution of the Matsubara RG flow equations 
truncated on a level which includes a flowing two-particle vertex and self-energy
of initially unknown frequency structure presents a formidable task 
\cite{Karrasch08a,Karrasch10d} as it is known from other imaginary time quantum 
many-body methods such as quantum Monte-Carlo.      
This advantage was utilized in the real time functional RG study of the 
single-impurity Anderson model \cite{Jakobs10a,Jakobs10b} in which a frequency 
dependent two-particle vertex and self-energy (complete level-2 truncation) 
were kept (for the imaginary time analog of this study, see Sec. \ref{subsubspinfluc}, 
\textcite{Hedden04}, and \textcite{Karrasch08a}), and can be expected to be useful 
also in functional RG studies of other models.

\subsection{Impurities in Luttinger liquids}

The metallic state of correlated fermions in one dimension is a non-Fermi liquid. 
It falls into the Luttinger liquid (LL) class \cite{Haldane80}. 
This state of matter is characterized by a power-law decay of space-time 
correlation functions with interaction dependent exponents 
\cite{Luttinger63,Mattis65,Luther74,Mattis74,Giamarchi04,Schoenhammer05} and 
spin-charge separation \cite{Dzyaloshinskii74,Meden92,Voit93}.
In particular, after Fourier transforming, 
the $2 k_F$-component of the density-density response function 
shows a power-law divergence \cite{Mattis74,Apel82} for repulsive interactions, 
instead of a logarithmic one for non-interacting electrons (Lindhard 
function in 1D). This indicates that 
any local inhomogeneity 
with a non-vanishing $2 k_F$ component strongly affects the
low-energy physics and thus the transport 
characteristics of a LL at low temperatures.  

Perturbation theory is insufficient to describe the interaction effects 
as it fails to capture the RG flow of the inhomogeneity appearing even to 
first order in the interaction and higher order diagrams diverge.
As will become clear below the functional RG approach captures the 
impurity flow and does {\it not} require a simplified modeling of the 
inhomogeneity. 

\subsubsection{A single local impurity--the local sine-Gordon model}
\label{subsec:LL1}

To understand the effect of a single impurity on the low-energy
physics of a {\it spinless, infinite} LL (absence of non-interacting leads), 
bosonization  was used first (for a review of this method see 
e.g.~\textcite{vonDelft98}, \textcite{Giamarchi04}, and \textcite{Schoenhammer05}). 
Within this approach the Fourier components of fermionic density
operators are split into their chiral parts which
obey Bose commutation relations in the low energy 
subspace \cite{Tomonaga50}. In this limit the 
kinetic energy and the two-particle interaction can be written as 
bilinears in these operators \cite{Tomonaga50}, while a single-particle 
scattering term (impurity) generally
takes a complicated form in the bosonic degrees of freedom. It 
simplifies if only the pure forward scattering $\tilde V(0)$ 
and backward scattering $\tilde V(2k_F)$ contributions 
are kept. Using bosonization to obtain results beyond this 
approximate modelling of the impurity is rather involved. 

The forward 
scattering term is linear in the bosons and can easily be treated
leading to a  phase shift. This is irrelevant for the physics described
 in the following.  
The backward scattering term consists of the cosine of a local 
bosonic field and the resulting Hamiltonian is known 
from field theory  as the local sine-Gordon (LSG) model. One can 
either use the exact Bethe ansatz solution \cite{Fendley95} of 
this model or a bosonic RG which is perturbative in $\tilde V(2k_F)$ 
\cite{Kane92,Furusaki93a} to obtain 
analytical results. Alternatively numerical methods can be 
applied to the LSG model \cite{Moon93,Egger95}. 
This led to a complete picture of the RG scaling of $\tilde V(2k_F)$ 
which has direct consequences for the linear conductance. 
The RG flow connects two fixed points, 
the perfect chain fixed point with conductance $g=K/(2 \pi)$ 
and the open chain fixed point with $g=0$. 
In 1D spinless fermion systems only a single transport channel
exists and we thus suppress the channel index $\zeta$ from now on.
Here $K>0$ denotes the so-called 
LL parameter which depends on the underlying model of the 
quantum wire and its parameters such as the strength and range of 
the two-particle interaction as well as 
the band filling.  Independently of 
the model considered $K<1$ for repulsive interactions and $K>1$ for 
attractive ones, while the 
non-interacting case corresponds to $K=1$. 
The corrections to the fixed point conductances are  
power laws $s^{\gamma_{p/o}}$ with the infrared energy scale $s$ 
(e.g. the temperature $T$ or the energy cutoff $\Lambda$ in a 
RG procedure). 
The exponents are {\it independent} 
of the bare impurity strength and given by $\gamma_p=2(K-1)$ and 
$\gamma_o=2(1/K-1)$, respectively. 
The sign of the scaling exponents implies that the open chain fixed point 
is stable for repulsive interactions and unstable for attractive ones. 
The opposite holds for the perfect chain fixed point. 
These insights confirmed earlier indications that  
impurities with a backscattering 
component strongly alter the low-energy physics of LLs 
with  repulsive interactions \cite{Mattis74,Apel82}. 
The behavior can be summarized by saying 
that even a weak single impurity grows and eventually
cuts the chain into two parts with open boundary conditions at the
end points.   

The exponent $\gamma_o$ characterizing (for repulsive 
two-particle interactions) the suppression of $g$ on small scales is 
twice the scaling exponent of the local single-particle 
spectral function of a LL close to an open boundary. 
This can be understood by viewing transport across the 
impurity as an end-to-end tunneling between two semi-infinite 
LLs (see e.g. \textcite{Kane92}).

Bosonization was not only used for an infinite LL wire 
but also for the experimentally more relevant case in which an 
interacting wire  containing a single 
impurity is contacted to two semi-infinite
non-interacting leads. Contacts generically lead to single-particle 
backscattering and thus have an effect similar to the impurity (see below). 
To disentangle the effect of the contacts and the impurity one often 
models the contacts such that they do not lead to any backscattering.
In this case and in the absence of the impurity the conductance takes the 
maximal value $1/(2 \pi)$ (instead of $K/(2 \pi)$ for an infinite LL). 
Using bosonization this can either be achieved 
within the so-called {\it local Luttinger liquid} picture 
\cite{Safi95,Maslov95,Ponomarenko95,Janzen06}
or by appropriate boundary conditions for the bosonic 
fields \cite{Egger97,Egger00}. The fixed points and scaling
exponents turned out to be the same as in the LSG model. 

For weak two-particle interactions the problem of a single impurity in a 
LL was also studied using a fermionic RG \cite{Matveev93,Yue94}. In this 
approach a flow equation for the transmission coefficient at $k_F$ is 
derived using poor man's RG. An extension of this method to interactions of 
arbitrary strength was presented by \textcite{Aristov08} and \textcite{Aristov09}.   

Remarkably, no intermediate fixed points appear within the LSG model and the 
crossover between the perfect and open chain is characterized by a one-parameter 
scaling function \cite{Kane92,Moon93,Fendley95,Egger00}. In one spatial
dimension a general impurity can be described  by matrix elements $V_{k,k'}$ 
(only $ V_{k_F,-k_F}= \tilde V(2 k_F)$  is kept in the LSG model). 
The RG analysis for such an impurity 
involves the coupling of all matrix elements $V_{k,k'}$ in the flow and 
one might wonder if this leads to an intermediate fixed point absent in 
the LSG model. To lowest order in the impurity strength 
the flow of $V_{k_F,-k_F}$ upon lowering the cutoff $\Lambda$ is driven by 
$V_{k_F,-k_F}$ itself (see the next section). For repulsive two-particle 
interactions $V_{k_F,-k_F}$ increases and the 
perturbative RG breaks down. Now the other couplings, absent in the LSG model,
might become relevant and eventually cut off the flow of $V_{k_F,-k_F}$ to large 
values, that is towards the open chain fixed point. That this does not happen 
can nicely be revealed within a functional RG approach. 
Before reviewing how this question was approached (see Sec.~\ref{subsubnum})
we first present the most elementary functional RG flow equation to tackle 
inhomogeneous LLs and analytically show that it leads to the scaling behavior 
known from bosonization in the limit of weak impurities.

\subsubsection{The functional RG approach to the single  impurity problem}

\label{subsubsimp}

As in the previous applications because of the necessary truncations the 
functional RG can (presently) only be used for small to intermediate
two-particle interactions. For the application to inhomogeneous LLs  
it is crucial that it is non-perturbative in the single-particle 
inhomogeneity. Here the focus is on a description 
in which the RG flow and the interaction dependent exponents 
characterizing the physics close to fixed points are kept at least to 
leading order in the interaction. Following the discussion
in the last paragraph of the preceeding subsection 
the feedback of the impurity into 
the flow of the self-energy dominates the physics to be studied.
It is thus advantageous to use the one-particle irreducible functional RG 
scheme in which the full propagator including the self-energy 
appears on the right hand side of the flow equations. For spinless fermions 
in a {\it homogeneous} wire the electron-electron interaction is renormalized 
only by a finite amount of order interaction squared \cite{Solyom79}.
A single impurity does not alter this. To obtain the fixed point value 
of the effective interaction to leading order one can 
therefore replace the flowing two-particle 
vertex by the antisymmetrized bare interaction corresponding to the 
level-1 truncation introduced in Sec.\ \ref{sec:truncations}. 
An improvement which includes the flow of the static two-particle 
vertex is reviewed in Sec.~\ref{subsubnum}. After presenting the 
most elementary functional RG flow equation for the self-energy we show 
that it leads to the correct scaling properties for a weak impurity.
  
As one deals with systems in which translational 
invariance is broken it is advantageous to introduce the infrared 
cutoff $\Lambda$ in {\it frequency} space.  To set up the 
functional RG flow equations for the self-energy the propagator 
$G_0$ of the non-interacting Hamiltonian $H_0$ 
containing only the kinetic energy
is replaced by  $G_0^{\Lambda}(i \omega_n) = \Theta^{\Lambda}(\omega_n)
G_0(i \omega_n)$
with a function $\Theta^{\Lambda}$ which is 
unity for $|\omega_n| \gg \Lambda$ and vanishes for 
$|\omega_n| \ll \Lambda$.
More specifically 
\begin{eqnarray}
\mbox{} \Theta^\Lambda(\omega_n) & =
 \left\{  \begin{array}{ll}  
    0  \, ,& |\omega_n| \leq \Lambda-  \pi T  \\
    \frac{1}{2} + \frac{|\omega_n|-\Lambda}{2\pi T}
    \, , & \Lambda - \pi T \leq |\omega_n | \leq  \Lambda +\pi T   \\
    1  \, ,& |\omega_n| \geq  \Lambda  + \pi T
  \end{array} \right. \nonumber \\
\label{cutfun}
\end{eqnarray}
was used \cite{Enss05b}, where $\Lambda$ starts at $\infty$ and goes 
down to $0$.\footnote{A significant speed-up of the numerical solution of the  
differential flow equations can be achieved using the alternative cutoff 
scheme discussed in the Appendix of \textcite{Andergassen06b}.} 
For $T \to 0$,  
$\Theta^{\Lambda}$ becomes a sharp $\Theta$-function and Matsubara frequencies 
with $|\omega| < \Lambda$ are suppressed. 

In this subsection a general continuum or lattice model of spinless fermions
is considered
with one-particle states $\left| \alpha \right>$, which in the
following will be either local states $\left| x \right>$, where $x=j$ 
 is the site
index for the lattice model, with lattice constant $a=1$, or
momentum states $\left| k_n \right>$ with $k_n=2\pi n/L$. 
A general two-particle interaction 
Eq.\ (\ref{bareinteraction}) is assumed and an impurity term  
$V_{\rm imp}= \sum_{\alpha,\beta} V_{\alpha,\beta} \, \psi^{\dag}_{\alpha} \psi^{}_{\beta} $.
The flow for the self-energy Eq.\ (\ref{floweqSigma}) in the 
level-1 truncation reads \cite{Meden02b}
\begin{eqnarray}
&& \hspace{-.5cm}  \frac{d}{d \Lambda}\Sigma^{\Lambda}_{\alpha,\beta}  =  
 T \sum_{\omega_l} e^{i\omega_l 0^+} \!\!
 \sum_{\gamma,\delta} \Bigg\{   
 \bigg[  1- G_0^{\Lambda}(i \omega_l) 
 \Sigma^{\Lambda} \bigg]^{-1}  \nonumber \\*  
&&   \times  \frac{d G_0^{\Lambda}(i \omega_l)}{d \Lambda}
 \bigg[  1- \Sigma^{\Lambda}
    G_0^{\Lambda}(i \omega_l) \bigg]^{-1}
 \Bigg\}_{\delta,\gamma}  U_{\alpha,\gamma,\beta,\delta} \; , \hskip 5mm
 \label{volker:sflow}
\end{eqnarray}
where $G_0^{\Lambda}$ and $\Sigma^{\Lambda}$ are matrices. 
The initial condition is given by
$\Sigma^{\Lambda=\infty}_{\alpha,\beta} = V_{\alpha,\beta}$ and
$U_{\alpha,\beta,\gamma,\delta}$ denotes the antisymmetrized bare 
two-particle vertex. 

At temperature $T=0$ and applying Eq.\ (\ref{morrislemma}) to products of $\Theta$- 
and $\delta$-functions, Eq.\ (\ref{volker:sflow}) simplifies to   
\begin{eqnarray}
 \frac{d}{d \Lambda}\Sigma^{\Lambda}_{\alpha,\beta} = 
 - \frac{1}{2 \pi}
 \sum_{\omega = \pm \Lambda} \sum_{\gamma,\delta}
 U_{\alpha,\gamma,\beta,\delta} \, 
 \tilde G_{\delta,\gamma}^{\Lambda}(i \omega) \, e^{i \omega 0^+},
 \label{volker:sigmalater}
\end{eqnarray}
where 
\begin{eqnarray}
\tilde G^{\Lambda}(i\omega) = \left[ Q_0( i \omega) -
  \Sigma^{\Lambda} \right]^{-1}
\label{volker:fullgdef}
\end{eqnarray}
is the full propagator for the cutoff dependent 
self-energy and $Q_0=(G_0)^{-1}$. 
The convergence factor $e^{i \omega 0^+}$ 
in Eq.\ (\ref{volker:sigmalater}) is relevant only for determining 
the flow from $\Lambda=\infty$ down to some arbitrarily large 
$\Lambda_0$. For $\Lambda_0$ much larger  than the band 
width this high  energy part of the flow can be computed 
analytically leading to the initial condition 
$ \Sigma^{\Lambda_0}_{\alpha,\beta} = V_{\alpha,\beta} + 
 \sum_{\gamma} U_{\alpha,\gamma,\beta,\gamma}/2 $.

Within the present scheme $\Sigma^{\Lambda}_{\alpha,\beta}$ is 
frequency independent and can be considered as 
a flowing effective impurity potential.
To obtain an approximation for the Green function of the original 
cutoff free problem for arbitrary impurity parameters
one has to determine the self-energy  $\Sigma^{\Lambda}_{\alpha,\beta}$  
at $\Lambda=0$ by numerically solving the set of differential equations 
(\ref{volker:sflow}) or (\ref{volker:sigmalater}). To compute the 
right hand side of the flow equations one has to invert the matrix 
Eq.\ (\ref{volker:fullgdef}), i.e., to solve the problem of a single 
particle moving in the effective scattering potential
$\Sigma^{\Lambda}_{\alpha,\beta}$.
                
Because of the matrix inversion involved in calculating 
the right hand side of Eq.\ (\ref{volker:sigmalater}) the flow equations 
can be solved analytically only in limiting cases, one being the situation 
of a weak impurity. One then works in momentum space and considers 
$\Sigma^{\Lambda}_{k,k'}$ for $k\ne k'$. The term linear in
$\Sigma^{\Lambda}$ presents the leading approximation in the expansion
of $\tilde G^{\Lambda}$ on the right hand side of Eq.\ (\ref{volker:sigmalater}) 
and one obtains \cite{Meden02b}
\begin{eqnarray}
 \frac{d}{d \Lambda}\Sigma^{\Lambda}_{k,k'}  &=&  - \frac{1}{2 \pi} 
\sum_{k_1,k_2} U_{k,k_1,k',k_2}
 \nonumber \\*
&& \mbox{} \hspace{-2.cm}\times \left[   \frac{1}{i \Lambda - \xi_{k_2} }
  \Sigma^{\Lambda}_{k_2,k_1} 
 \frac{1}{i \Lambda - \xi_{k_1} }  + \left( \Lambda \to - \Lambda \right) \right] \; ,
\label{volker:expansion}
\end{eqnarray}
where $\xi_k = \epsilon_k - \mu$ with the one-particle dispersion
$\epsilon_k$. The antisymmetrized two-body matrix element 
$U_{k,k_1,k',k_2}$  contains a momentum conserving Kronecker
delta, where $k+k_1=k'+k_2$  modulo the reciprocal lattice vector
$ 2\pi n$ for the lattice model  
with $n=0,\pm 1$, when the four momenta are in the first Brillouin 
zone. The umklapp processes $n=\pm 1$ involve low
energy excitations only for special fillings--half filling for the
nearest neighbor hopping model discussed later.

First models are considered for which 
umklapp processes are absent. To determine the
backscattering properties of the self-energy one can put $k=k_F$ 
and $k'=-k_F-q$ with $|q| \ll k_F$. In order to read off the
dominant behavior for small $\Lambda$  the remaining
summation variable is shifted $k_1=-k_F+\tilde k_1$
\begin{eqnarray}
&& \frac{d}{d \Lambda}\Sigma^{\Lambda}_{k_F,-k_F-q}  =  - \frac{1}{2 \pi} 
\sum_{\tilde k_1} U_{k_F,-k_F+\tilde k_1,-k_F-q,k_F+q+\tilde k_1}
 \nonumber \\*
&& \times \Big[   \frac{1}{i \Lambda - \xi_{k_F+q+\tilde k_1} }
  \Sigma^{\Lambda}_{k_F+q+\tilde k_1,-k_F+\tilde k_1} 
 \frac{1}{i \Lambda - \xi_{-k_F+\tilde k_1} }  \nonumber \\* 
&& +\left( \Lambda \to - \Lambda \right) \Big].
\label{volker:expansion2}
\end{eqnarray} 
For $|\tilde k_1|\ll k_F$ one can linearize the dispersion
$\xi_{-k_F+\tilde k_1}\approx -v_F \tilde k_1$ and $\xi_{k_F+q+\tilde
  k_1}\approx v_F (\tilde k_1+q)$, 
where $v_F$ denotes the Fermi velocity.
In the thermodynamic limit the two $G_0$ factors for $q=0$ and
$\Lambda \to 0$ are
proportional to $\delta(\tilde k_1)$. 
 If only the singular contributions
are kept the differential equation for $\Sigma^{\Lambda}_{k_F,-k_F}$
reads 
\begin{eqnarray}
\frac{d}{d \Lambda}\Sigma^{\Lambda}_{k_F,-k_F} =  
  -\frac{\hat U_{k_F,-k_F,k_F,-k_F }}{2 \pi v_F}  
\frac{1}{\Lambda} \, \Sigma^{\Lambda}_{k_F,-k_F} \, 
\label{volker:sigmakfmkf}
\end{eqnarray}
where $\hat  U_{k_F,-k_F,k_F,-k_F } =  L U_{k_F,-k_F,k_F,-k_F }$
is independent of the
system size. For the continuum model $\hat  U_{k_F,-k_F,k_F,-k_F}=
\tilde U(0)-\tilde U(2k_F)$, where $\tilde U(k)$ is the Fourier
transform of the two-particle potential $U(x-x')$.
 This leads to the scaling relation
\begin{eqnarray}
\Sigma^{\Lambda}_{k_F,-k_F} \sim \left( \frac{1}{\Lambda} 
\right)^{ \hat U_{k_F,-k_F,k_F,-k_F } /(2 \pi v_F)} \; .
\label{volker:sigmascal}
\end{eqnarray}
As Eq.\ (\ref{volker:sigmakfmkf}) was derived by
expanding the Green function $G^{\Lambda}$ in powers of the self-energy, 
the scaling behavior  Eq.\ (\ref{volker:sigmascal}) can be trusted
only as long as $\Sigma^{\Lambda}$ stays small. Thus a single-particle 
backscattering term is a relevant perturbation for repulsive interactions 
and an irrelevant one for attractive ones consistent with
the bosonization result. 

Equation (\ref{volker:sigmakfmkf})  holds even in the half-filled band
case for the nearest neighbor 
hopping lattice model. The additional singular contribution due to 
umklapp scattering is proportional to $\hat U_{k_F,k_F,-k_F,-k_F}$ 
which vanishes because of the antisymmetry of the matrix element 
\cite{Meden02b}.

Neglecting all interaction effects beyond the renormalization of the 
impurity potential and using the Born approximation the correction to 
the perfect chain conductance is given by the self-energy squared. 
The corresponding exponent $ \hat U_{k_F,-k_F,k_F,-k_F } /(\pi v_F)$ 
agrees to leading order in the interaction \cite{Schoenhammer05} with 
$\gamma_p=2(K-1)$ obtained within bosonization. 

The opposite limit of a weak link can be treated analytically as well
leading to results consistent to those of the bosonization 
approach with an exponent characterizing the deviation from the open chain 
fixed point conductance $g=0$ which agrees to leading order in the
interaction with $\gamma_o=2(1/K-1)$ \cite{Meden02b}. 

Next the numerical solution of the 
RG flow for a specific lattice model with 
arbitrary impurity strength is discussed. This allows us to address the 
question of additional fixed points.

\subsubsection{Basic wire model}

\label{basicmodel}

In the following the 
tight-binding model of spinless fermions with nearest
neighbor interaction supplemented by an impurity is considered. 
The Hamiltonian is given by $ H = H_{\rm kin} + H_{\rm int} + H_{\rm imp} $ 
with kinetic energy  
\begin{equation}
\label{volker:kindef}
 H_{\rm kin} = -\sum_{j=-\infty}^{\infty} \big( \,
 c^{\dag}_{j+1} c_j^{\phantom\dag} + c^{\dag}_j \, c_{j+1}^{\phantom\dag}
 \, \big) \; , 
\end{equation}
where $c^{\dag}_j$ 
and $c_j^{\phantom\dag}$ are the creation and annihilation operators on 
site $j$, respectively. The
corresponding non-interacting dispersion is 
 $\epsilon_k=-2\cos k$.
The interaction is restricted to electrons on $N$ neighboring sites \cite{Enss05b}
\begin{eqnarray}
\label{volker:intdef}
\mbox{} \hspace{-.5cm} H_{\rm int} & = & \sum_{j=1}^{N-1} U_{j,j+1}  \left[ n_j - \nu(n,U)
\right]  \left[ n_{j+1} - \nu(n,U) \right] , \hskip 5mm
\end{eqnarray}
with the local density operator $n_j = c^{\dag}_j \,
c_j^{\phantom\dag}$.  The two regions of the lattice with $j < 1$ and
$j>N$ 
constitute the semi-infinite non-interacting leads.
To model contacts which do not lead to single-particle 
backscattering (see above) the interaction $U_{j,j+1}$ 
between electrons on sites $j$
and $j+1$ is allowed to depend on the position. 
A conductance $g=1/(2 \pi)$ in the absence of single-particle
impurities is {\it only} achieved if $U_{j,j+1}$ is taken as a smoothly 
increasing function of $j$ starting form zero at the bond $(1,2)$ and 
approaching a constant bulk value $U$ over a sufficiently large number 
of bonds. Equally, the $U_{j,j+1}$ are switched off 
close to the bond $(N-1,N)$.
The results are independent of the shape of the envelope
function as long as it is sufficiently smooth. 
An abrupt two-particle inhomogeneity acts similarly to a
single-particle impurity. 
A detailed disussion on the effect of the spatial variation 
of the two-particle interaction was presented by  
\textcite{Meden03b} and \textcite{Janzen06}. 

In Eq.\ (\ref{volker:intdef}) the density $n_j$ is shifted by 
a parameter $\nu(n,U)$, which depends on the filling factor $n$ 
and the bulk interaction $U$. This is equivalent to introducing 
an additional  one-particle potential which can compensate
the Hartree potential in the bulk of the interacting wire.
In the half-filled band case $\nu(1/2,U)=1/2$ \cite{Enss05b}.

\begin{figure}[tb]
\begin{center}
\includegraphics[width=0.40\textwidth,clip]{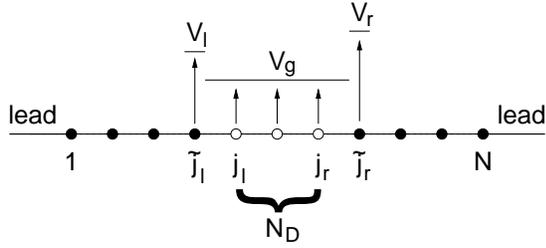}
\end{center}
\vspace{-0.3cm}
\caption[]{Schematic plot of the quantum dot situation, where the
  barriers are modeled by two site impurities \cite{Enss05b}. 
\label{fig:doublebar}}
\end{figure}

The general form of the impurity part of the Hamiltonian is 
\begin{equation}
\label{volker:impdef}
 H_{\rm imp} = \sum_{j,j'} 
 V_{j,j'}^{\phantom\dag} \; c^{\dag}_{j} \, c_{j'}^{\phantom\dag} \; ,
\end{equation}
where $V_{j,j'}$ is a static potential.
Site impurities are given by  
$V_{j,j'} = V_j \, \delta_{j,j'} $.
For a single site impurity  $ V_j = V \, \delta_{j,j_0} $ 
$j_0$ is chosen to be far away from both leads.
Impurities close to the contacts were discussed by \textcite{Furusaki96} and 
\textcite{Enss05b}.
Resonant tunneling can be studied considering two site 
impurities of strengths $V_l$ and $V_r$ on the sites
$\tilde j_l=j_l-1$ and $\tilde j_r=j_r+1$. 
The $N_D$ sites between $j_l$ and $j_r$ define
a quantum dot. The effect of a
gate voltage restricted to the dot region is described by a constant 
$V_g$ on sites $j_l$ to $j_r$. 
This situation is sketched in Fig.~\ref{fig:doublebar}.
Hopping impurities are achieved setting  
$V_{j,j'} = V_{j',j} = - t_{j,j+1} \, \delta_{j',j+1} $.
For the special case of a single hopping impurity,
$t_{j,j+1} = (t'-1) \, \delta_{j,j_0} $,
the unit hopping amplitude is replaced by $t'$ on the bond linking
the sites $j_0$ and $j_0+1$. In the double-barrier problem 
 a hopping $t_l$ across the bond $(\tilde j_l,j_l)$  and $t_r$
across $(j_r,\tilde j_r)$ is considered. 

The {\it homogeneous} model $H=H_{\rm kin} + H_{\rm int}$ with a constant
interaction $U$ across {\it all bonds}--not only the ones within $[1,N]$--can  
be solved exactly by the Bethe ansatz 
and $K$ is determined by a system
 of coupled integral equations \cite{Haldane80}. In 
the half-filled case they can be solved analytically leading to  
\begin{equation}
\label{volker:BetheAnsatz}
 K = \left[\frac{2}{\pi} \, 
 \arccos \left(-\frac{U}{2} \right) \right]^{-1} 
\end{equation}
for $|U| \leq 2$. At other fillings the integral equations
can be solved numerically. 
The model shows LL behavior for all 
particle densities $n$ and any interaction strength except at 
half filling for $|U| > 2$ were either phase separation sets in (for 
$U< -2$) or the system orders into a charge-density-wave state 
(for $U>2$).  

Due to the presence of the semi-infinite 
leads the direct calculation  of  the non-interacting  
propagator related to $H_{\rm kin} + H_{\rm imp}$ 
requires the inversion of an infinite matrix. Using a
standard projection technique \cite{Taylor72} it can be 
reduced to the inversion of an $N \times N$ matrix. 
The leads then provide an additional $\omega_n$-dependent 
diagonal one-particle potential on sites $1$ and $N$ \cite{Enss05b}
\begin{eqnarray}
\label{volker:leadpotdef}
 V_{j,j'}^{\rm lead}(i\omega_n) & = &
 \frac{i\omega_n+\mu}{2} \left( 1 - 
 \sqrt{1 - \frac{4}{(i\omega_n+\mu)^2}} \, \right) \nonumber \\
&& \times  \delta_{j,j'} \left( \delta_{1,j} + \delta_{N,j} \right) \; .
\end{eqnarray}
Since the interaction is only non-vanishing
on the bonds between the sites $1$ to $N$ the problem including the
semi-infinite leads is this way reduced to the problem of an 
$N$-site chain.  In the functional RG it is then advantageous to 
replace the {\it projected} 
non-interacting propagator $G_0$ {\it including} the impurity 
by a cutoff dependent one \cite{Enss05b}.

\subsubsection{Numerical solution of improved flow equations}

\label{subsubnum}

With a minor increase in the numerical effort one can go beyond 
Eq.\ (\ref{volker:sflow}) for the flow of 
the self-energy and include a $\Lambda$-dependent static 
interaction $U^\Lambda$  \cite{Andergassen04}. 
Its flow equation is derived from  
the general one for the two-particle 
vertex Eq.\ (\ref{floweqGamma4}) applying the following approximations:
(i) The three-particle vertex is set to zero. 
(ii) All frequencies are set to zero. (iii) The feedback of the inhomogeneity 
on the flow of the interaction is neglected. (iv) The interaction is assumed 
to remain of nearest-neighbor form. 
Then $U^\Lambda$ obeys a simple differential equation
\cite{Andergassen04},
\begin{eqnarray}
\label{RGU}
\frac{d}{d\Lambda} U^\Lambda = h(\Lambda) \left( U^\Lambda \right)^2 \; ,
\end{eqnarray}
where the function $h(\Lambda)$ depends only on the cutoff $\Lambda$ and 
the Fermi momentum $k_F$.
The solution of the flow equation is lengthy for arbitrary fillings 
\cite{Andergassen04} but has a simple form for half-filling
\begin{eqnarray}
\label{flussU}
 U^{\Lambda} = \frac{U}
 {1 + 
 \left(\Lambda - \frac{2 + \Lambda^2}{\sqrt{4 + \Lambda^2}} \right) \, 
 U/(2\pi)} \; .
\end{eqnarray} 
These approximations are sufficient to correctly describe the RG flow 
of the two-particle vertex on the Fermi surface of the homogeneous system 
\cite{Andergassen04} to second order, as it is usually done in the 
so-called g-ology method \cite{Solyom79}. For the inhomogeneous LLs 
the flow of the effective interaction leads to improved results 
for the scaling exponents (see below).

Within these approximations and using the projection of the leads,
the self-energy at $T \geq 0$ is a frequency-independent tridiagonal 
matrix in real space
determined by  the  flow equation 
($j,j\pm 1 \in [1,N]$)
\begin{eqnarray}
 \frac{\partial}{\partial \Lambda} \Sigma_{j,j}^\Lambda  = 
    -\frac{1}{2\pi}
    \sum_{|\omega_n| \approx \Lambda} \sum_{r=\pm 1}
    U_{j,j+r}^\Lambda \left[ \frac{1}{Q_0(i\omega_n)-\Theta^\Lambda(\omega_n)
       \Sigma^\Lambda} \right. \nonumber \\*  \left. \times    
      Q_0(i\omega_n) \, \frac{1}{Q_0(i\omega_n)-\Theta^\Lambda(\omega_n)
      \Sigma^\Lambda } \right]_{j+r,j+r} \; , \nonumber\\
 \frac{\partial}{\partial \Lambda}     \Sigma_{j,j\pm 1}^\Lambda  = 
    \frac{1}{2\pi}
    \sum_{|\omega_n| \approx \Lambda} U_{j,j\pm 1}^\Lambda 
\left[ \frac{1}{Q_0(i\omega_n)-\Theta^\Lambda(\omega_n) \Sigma^\Lambda }
  \right.  \nonumber \\* 
\left. \times Q_0(i\omega_n) \, \frac{1}{Q_0(i\omega_n)-\Theta^\Lambda(\omega_n)
\Sigma^\Lambda} \right]_{j,j\pm 1} \; , \nonumber
\end{eqnarray}
\vspace{-.5cm}
\begin{equation}
\vspace{-.2cm}
\label{volker:sigmaflowtg0}
\end{equation} 
where the matrix $Q_0=(G_0)^{-1}$ is the inverse of
the projected non-interacting propagator with impurity.
The symbol $|\omega_n| \approx \Lambda$ stands for taking the positive 
as well as negative frequency with absolute value closest to 
$\Lambda$.\footnote{To achieve this result the cutoff scheme discussed in the 
Appendix of \textcite{Andergassen06b} was used.}  
The initial conditions  for $\Sigma$ at $\Lambda = \Lambda_0 \to \infty$ 
are independent of the precise realization of the inhomogeneity and read
$\Sigma^{\Lambda_0}_{j,j} =  
\left[1/2- \nu(n,U)\right] \left( U_{j-1,j} +
  U_{j,j+1} \right) $ and 
$\Sigma^{\Lambda_0}_{j,j\pm 1}  =0 $. 
The frequency
dependence of the self-energy which appears in the exact solution
in order $U^2$ is {\it not} captured by this scheme.
Thus only the leading order is completely kept in the flow of $\Sigma$.

The coupled flow equations can be solved numerically by an algorithm 
which approximately scales as $N$ \cite{Andergassen04}. Typically 
systems of $10^4$ lattice sites 
were considered, roughly corresponding to the length of quantum wires 
accessible to transport experiments.  For the interacting wire of finite 
length the energy scale $ \delta_N = v_F / N$ forms a cutoff for any RG flow. 
The flowing self-energy Eq.\ (\ref{volker:sigmaflowtg0}) 
depends on the three scales $T$, $\delta_N$, and $\Lambda$. 
Saturation of $\Sigma^{\Lambda}$ for $\Lambda \lessapprox T$ 
or $\Lambda \lessapprox \delta_N$ sets in ``automatically'' 
in contrast to more intuitive RG schemes in 
which the flowing couplings depend on $\Lambda$ only and 
the flow is stopped ``by hand'' by replacing $\Lambda \to T$ or 
$\Lambda \to \delta_N$, respectively \cite{Kane92,Yue94}.

\begin{figure}[tb]
\begin{center}
\includegraphics[width=0.40\textwidth,clip]{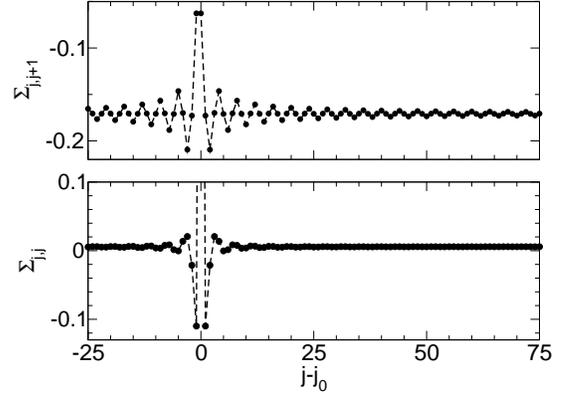}
\end{center}
\vspace{-0.3cm}
\caption[]{Self-energy near a site impurity of strength $V=1.5$
 filling $n=1/4$, and interaction $U=1$; the impurity 
 is located at $j_0=513$ with $N = 1025$ sites. (Data 
taken from \textcite{Andergassen04}.)
\label{fig:self1}}
\end{figure}

\begin{figure}[htb]
\begin{center}
\includegraphics[width=0.49\textwidth,clip]{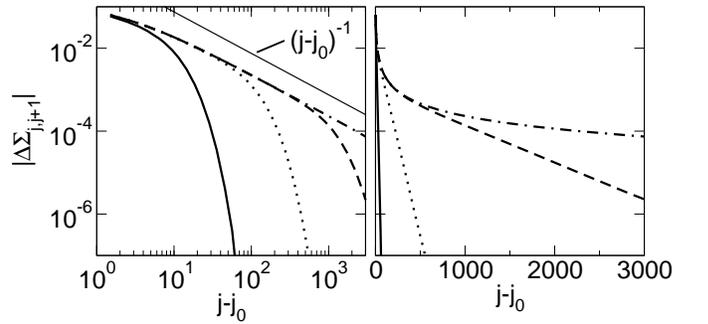}
\end{center}
\vspace{-0.3cm}
\caption[]{ Decay of the oscillatory part of the off-diagonal matrix
  element of the self-energy away from a single hopping impurity at
  bond $j_0,j_0+1$. Results for $t'=0.1$, $j_0=5000$, $N=10^4$, $U=1$,
  $n=1/2$, and different temperatures $T=10^{-1}$ (solid line),
  $T=10^{-2}$ (dotted line), $T=10^{-3}$ (dashed line), and
  $T=10^{-4}$ (dashed-dotted line) are presented.  The left panel
  shows the data on a log-log scale, the right panel on a linear-log scale.
  For comparison the left panel contains a power-law $(j-j_0)^{-1}$
  (thin solid line) \cite{Enss05b}. \label{fig:self2}}
\end{figure}

Figure \ref{fig:self1} shows the self-energy $\Sigma$  at the end of the RG 
flow, for $T=0$ in the vicinity of a site impurity
of intermediate strength. Both 
the onsite energy $\Sigma_{j,j}$ as well as the hopping $\Sigma_{j,j+1}$ become 
oscillatory functions with wave number $2 k_F$ and a decaying amplitude. 
The asymptotic value of $\Sigma_{j,j+1}$ away from
the impurity leads to a broadening of 
the band due to the interaction. A more detailed analysis of the oscillatory 
part $|\Delta \Sigma_{j,j+1}|= \left| \Sigma_{j,j+ 1} - \bar \Sigma_{\rm off} \right|$,
with the spatial average $\bar \Sigma_{\rm off}$,  
for different $T>0$ is presented in Fig.\  \ref{fig:self2}. The left panel 
shows that for $|j-j_0| \gtrapprox 10$ it decays as $1/|j-j_0|$ up to a thermal 
length scale $\sim 1/T$ (provided $T > \delta_N$)
beyond which the decay becomes exponential; see the right panel.
For $U>0$ this is the generic behavior for large bare impurities or on 
asymptotical large length scales.
It is the scattering off such a long-ranged oscillatory 
potential--so-called Wigner-von Neumann potential \cite{Reed75}--which leads 
to the power-law suppression of the conductance and the local spectral weight. 
One can analytically show that the amplitude of the asymptotic $1/|j-j_0|$ 
decay determines the exponent \cite{Barnabe05a}. 
By virtue of the RG flow this amplitude--and thus the 
exponent--becomes independent of the impurity strength. For this reason 
first order perturbation theory fails. It also leads to an 
oscillatory self-energy which decays as $1/|j-j_0|$ but with an amplitude 
which depends on the bare impurity strength \cite{Meden02b} incorrectly 
leading to a power-law with an impurity dependent exponent. 
The idea of an oscillatory decaying potential is 
similarly inherent to a poor man's fermionic RG 
approach \cite{Yue94}. Often these oscillations of the effective 
renormalized potential are referred to as Friedel oscillations. 
This is misleading as this term is reserved to the spatial 
oscillations of the {\it electron density.} In particular, in 
an inhomogeneous LL the effective potential decays as $|j-j_0|^{-1}$, 
while the density oscillations asymptotically decay as $|j-j_0|^{-K}$ \cite{Egger95}. 
The latter can also be shown within the functional RG 
formalism presented here \cite{Andergassen04}.
The application of the {\it self-consistent} Hartree-Fock approximation 
leads to an oscillatory self-energy with a constant amplitude and thus to a 
charge density wave state \cite{Meden02b}. This is an unphysical artifact of 
the approximation. 

Using scattering theory \cite{Enss05b} one can show that the effective 
transmission ${\mathcal T}(\epsilon,T)$ is given by the $(1,N)$ matrix
 element of the 
single-particle Green function $ {\mathcal T}(\epsilon_k,T)=4 \sin^2 k \, 
|\langle N| G(\epsilon_k+i0)|1\rangle|^2$. Via the $T$-dependent self-energy 
(see Fig.~\ref{fig:self2}) 
$G$ and thus $\mathcal T$ carries a temperature dependence. A typical example 
for the $T$-dependence of the linear conductance $g(T)$ for a strong local impurity 
is shown as the solid line in Fig.\ \ref{fig:reso}. It clearly follows the expected 
power-law behavior for $\delta_N \lessapprox T \ll B$ with the band width $B=4$. 
The $T^{-1}$ scaling at larger $T$ is a band effect. 
For $-0.5 \leq U \leq 1.5$ and fillings $n=1/2$ as well as  $1/4$ the exponent extracted 
(see lower panel of Fig.\ \ref{fig:reso}) 
agrees well with the one of the LSG model  $\gamma_o=2(1/K-1)$, with $K$ 
taken from the Bethe ansatz. Even for $U=1.5$ the
relative error is less than 5 percent (see Fig.\ 5 of \textcite{Enss05b}). 
Higher order corrections in $U$ present in the numerical solution of the 
flow equations (\ref{flussU}) and (\ref{volker:sigmaflowtg0})
clearly improve the result over the one of the perturbative (in the impurity strength) 
analytical solution of Sec.\ \ref{subsubsimp} which yields a purely linear exponent. 
We emphasize that this improvement is not systematic  
as second and higher order terms are only partly kept in the RG. 
A similar agreement can be found for $\gamma_p$ \cite{Enss05b}.

\begin{figure}[htb]
\begin{center}
\includegraphics[width=0.4\textwidth,clip]{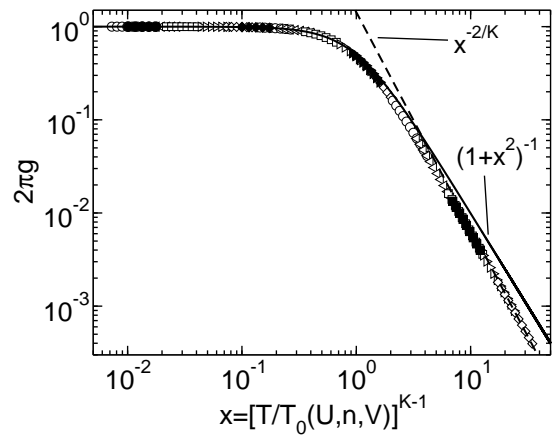}
\end{center}
\vspace{-0.3cm}
\caption[]{One-parameter scaling plot of the conductance. Open
  symbols represent results obtained for $U=0.5$, $n=1/2$, and
  different $T$ and $V$, while filled symbols were calculated for 
  $U=0.851$, $n=1/4$. Both pairs of $U$ and $n$ lead to the same
  $K=0.85$ (within the present approximation). The solid line indicates 
  the non-interacting scaling function $(1+x^2)^{-1}$ and the dashed 
  one the LSG model power-law decay with exponent $-2/K$. 
(Data taken from \textcite{Enss05b}.)   
 \label{fig:scaling}}
\end{figure}

Within the LSG model no intermediate fixed points appear 
which is reflected by one-parameter scaling
$g = \tilde g_K(x)/2 \pi$ with $x=\left[ T/T_0\right]^{K-1} $ and  
a non-universal scale $T_0$ \cite{Kane92,Moon93,Fendley95,Egger00}. 
For appropriately chosen $T_0$ data for different 
$T$ and $\tilde V(2k_F)$ but fixed $K$ can be collapsed onto
 the $K$-dependent 
scaling function $\tilde g_K(x)$. It has the limiting behavior 
$\tilde g_K(x) \propto 1-x^2$ for $x \to 0$ and $\tilde g_K(x) \propto x^{-2/K}$ 
for $x \to \infty$. 
One can perform a similar scaling with data from the numerical solution 
of the flow equations for the microscopic lattice model 
considering different $V$ and $T$ as well as 
two sets of $(U,n)$ leading to the same LL parameter \cite{Enss05b}. 
The perfect collaps of the data of Fig.\ \ref{fig:scaling} 
shows that the improved description of the impurity flow 
beyond the single amplitude approximation 
inherent to the LSG model does not lead to additional 
fixed points. 
The functional RG scaling function shows a sensible $K$ dependence. 
The exponent of the large $x$ power-law decay is smaller than the 
non-interacting one $-2$ (solid line at large $x$) and very close to the 
LSG model exponent $-2/K$ shown as the dashed line in Fig.\ \ref{fig:scaling}. 
This has to be contrasted to the $K$ {\it independent} (non-interacting) 
scaling function $\tilde g = (1+x^2)^{-1}$ resulting from the poor man's 
fermionic RG \cite{Yue94}; solid line in Fig.\ \ref{fig:scaling}. 

The functional RG results for a single local impurity in a LL show 
that the LSG model describes the physics (two fixed points, exponents, 
one-parameter scaling) of a broader class of models.
The same approach was also used to study the persistent current through 
a LL ring with a local impurity prierced by a magnetic 
flux \cite{Meden03a,Meden03b,Gendiar09}   
Aspects resulting from the {\it spin} degree of freedom of electrons 
were discussed by  \textcite{Andergassen06a} and \textcite{Andergassen06b}.

\subsubsection{Resonant tunneling} 

\label{resotunsec}

We next review the results on resonant transport through a double barrier--defining 
an interacting quantum dot embedded in a LL \cite{Meden05,Enss05b}. The setup is 
sketched in Fig.\ \ref{fig:doublebar}. 
The linear conductance $g$ is characterized 
by a hierarchy of energy scales. The functional RG is a unique tool to access this 
problem as it provides reliable results on all scales. 
For a fixed dot size $N_D$ and fixed barriers $V_{l/r}$ 
(or $t_{l/r}$) the dot can be tuned to 
resonance varying $V_g$. Only for symmetric dots with $V_l=V_r$ (or $t_l=t_r$) 
the peak conductance becomes ``perfect'' $g_p=1/(2\pi)$. For asymmetric 
barriers a backscattering component of the single-particle inhomogeneity remains, 
leading to a reduced $g_p$. 
Due to the interaction backscattering grows during a RG procedure
and on asymptotic 
scales the conductance  vanishes with scaling exponent $\gamma_o$. The same 
holds away from resonance regardless of the ratio $V_l/V_r$ (or $t_l/t_r$). 
Thus the non-interacting resonance of finite width  
either disappears (asymmetric barriers) or turns 
into a resonance of {\it zero 
width} (symmetric barriers). A rich $T$-dependence is found on 
resonance and for symmetric barriers on which we now focus. 
Without loss of generality we only consider site impurities as barriers.  

\begin{figure}[htb]
\begin{center}
\includegraphics[width=0.4\textwidth,clip]{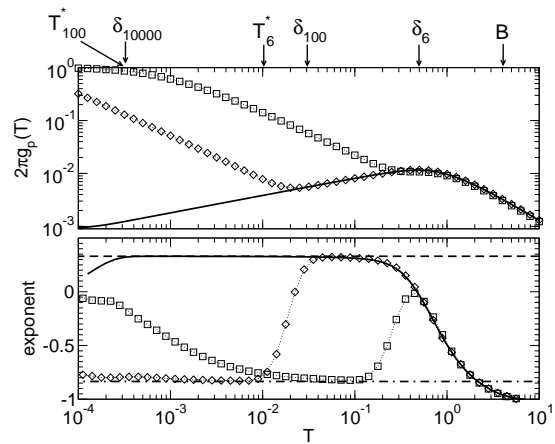}
\end{center}
\vspace{-0.3cm}
\caption[]{Upper panel: $g_p(T)$ for
  $U=0.5$, $N=10^4$, $V_{l/r}=10$, $n=1/2$, and 
  $N=6$ (squares), $100$ (diamonds). The arrows indicate the relevant energy 
  scales $B$, $\delta_{N_D}$, $T^\ast_{N_D}$ and $\delta_N$.  
  The solid curve shows $g(T)/2$ for a single barrier with $V=10$ and $U=0.5$, 
  $N=10^4$. Lower panel: Logarithmic
  derivative of the conductance. Dashed line: $\gamma_o$; dashed-dotted
  line: $\gamma_o/2-1$. (Data taken from \textcite{Enss05b}.) \label{fig:reso}}
\end{figure}

The functional RG procedure can directly be applied to the double 
barrier problem. The dot parameters only enter via the non-interacting 
propagator. Figure \ref{fig:reso} shows the peak conductance $g_p(T)$ for 
a dot with high barriers and two different dot sizes. The relevant 
energy scales $B$, $\delta_{N_D}=v_F/N_D$, 
$T^\ast_{N_D}$ (see below), and $\delta_N$ are indicated 
by the arrows. For $\delta_{N_D} \lessapprox T $ the two barriers behave as 
independent impurities. Using scattering theory one can show that in
this case $g_p$ is 
obtained by adding the resistances of the two barriers
 \cite{Enss05b,Jakobs07b}. This 
explains why for $N_D=100$, for which this temperature regime is clearly developed, 
$g_p(T)$ agrees to the solid line obtained by taking $g(T)/2$ of a single site 
impurity of equal heigth as used for the double barrier. Note that it
is a non-trivial 
fact that in this temperature regime the individual resistances can be added 
to give the total resistance. In the presence of {\it inelastic processes} 
one would of course expect this result (resistors in series) but they
 are absent in the  
mesoscopic setup--the ones resulting from the electron-electron interaction
are suppressed by the approximations. In fact, the case of three barriers
constitutes an example for which adding resistances does 
no longer hold \cite{Jakobs07b}. 
For $T \lessapprox \delta_{N_D}$ the width of 
$-\partial f/\partial \epsilon$ is smaller than $\delta_{N_D}$ 
and only the resonance peak around $\epsilon=0$ of 
${\mathcal T}(\epsilon,T)$ contributes to the integral in 
Eq.\ (\ref{linearcond}). The width $w$ of this peak  vanishes as 
$T^{\gamma_o/2}/N_D$ \cite{Enss05b} leading to $g_p(T) \propto
T^{\gamma_o/2-1}$. The lower bound of this scaling regime,
first discussed using bosonization \cite{Furusaki93b,Furusaki98}, is reached
when $T$  equals  
$w$, i.e.\ at $T_{N_D}^\ast \propto N_D^{-1/(1-\gamma_o/2)}$.
 For $T < T_{N_D}^\ast$, 
$2 \pi g_p$ approaches $1$. For $T$ reaching $\delta_N$ any power-law scaling in 
$T$ with an interaction dependent exponent 
is cut off by the finite size of the interacting part of the quantum wire. 
In addition of identifying the different temperature regimes the functional RG 
approach allows to (i) quantify the size of the crossover regime--typically 
half an order of magnitude--and to (ii) obtain results for ``non-universal'' 
regimes as e.g.\ realized for $N_D=6$ and $\delta_{N_D} < T \ll B$. For dots with 
weak barriers and sufficiently large $N_D$ only the regime with scaling exponent 
$\gamma_o/2-1$ is realized  and for weak barriers and small $N_D$ none of the 
above power-law regimes emerges \cite{Meden05,Enss05b}.  
Resonant transport in LLs was also studied by bosonization 
\cite{Furusaki93b,Furusaki98}, poor man's fermionic RG \cite{Nazarov03,Polyakov03}, 
and numerically \cite{Huegle04}.  

The temperature dependence of the peak conductance of resonant tunneling nicely 
exemplifies that the functional RG approach provides sensible results on 
{\it all energy scales} even for problems with a hierachy of scales. 
Other examples from this class are situations in which the wire-lead contacts 
are not modeled as being ``perfect'' \cite{Jakobs07b} and models in which the 
leads and contacts are described in a more realistic way  \cite{Waechter09}.

\subsubsection{Y-junctions}

\label{Yjunctionsec}

\begin{figure}[tbh]
\begin{center}
\includegraphics[width=0.16\textwidth,clip]{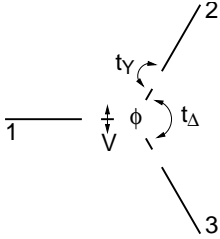}
\end{center}
\vspace{-0.6cm}
\caption[]{Sketch of the symmetric Y-junction of three quantum wires \cite{Barnabe05}. 
\label{fig:skizze}}
\end{figure}

\begin{figure}[htb]
\begin{center}
\includegraphics[width=0.45\textwidth,clip]{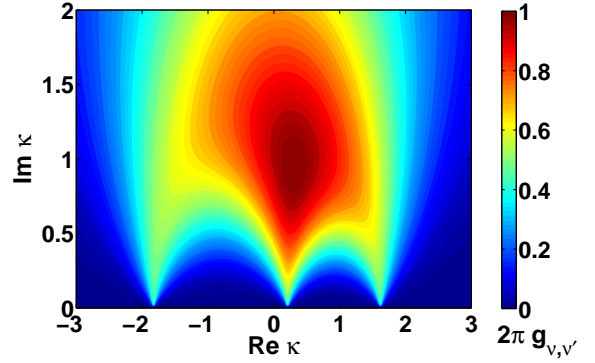}
\end{center}
\vspace{-0.3cm}
\caption[]{(Color online) The non-interacting conductance $2 \pi g_{\nu,\nu'}$ 
(cyclic indices) as a function of the complex parameter $\kappa$ which in 
turn is a function of the junction parameters $t_{\rm Y}$, $V$, and 
$t_{\triangle}$. The flux is $\phi=0.4 \pi$. 
\label{fig:U0cond}}
\end{figure}

The power of the functional RG approach to uncover unconventional fixed points 
and the related interesting physics was exemplified by discussing a specific 
junction of three 1D wires, a so-called Y-junction. The three 
LL wires (index $\nu=1,2,3$)  each of length $N$ and coupled to a non-interacting 
semi-infinite lead via a ``perfect'' contact are described by the basic model 
discussed in Sec.\ \ref{basicmodel}. 
The {\it symmetric} junction {\it pierced by a magnetic flux} $\phi$, is sketched in 
Fig.\ \ref{fig:skizze} and given by 
\begin{eqnarray}
\label{hamjunct}
H_{\rm Y} & = & - t_{\rm Y} \sum_{\nu=1}^{3}
 \left( c_{1,\nu}^\dag c_{0,\nu} + \mbox{h.c.} \right) + 
V \sum_{\nu=1}^{3} n_{0,\nu} \nonumber \\* 
&& - t_{\triangle} \sum_{\nu=1}^{3}  \left( e^{i \phi/3} 
c_{0,\nu}^\dag c_{0,\nu+1} +
  \mbox{h.c.} \right)     \; ,
\end{eqnarray}
where the wire indices 4 and 1 are identified. The junction is  
characterized by the three parameters $t_{\rm Y}$, $V$, and $ t_{\triangle}$.  
Using scattering  theory \cite{Barnabe05,Barnabe05a,Enss05b}, 
the $U=0$ conductance  from wire $\nu$ to wire $\nu'$ can be written as
\begin{eqnarray} 
\label{trans}
2 \pi g_{\nu,\nu'} = \frac{4 \left( \mbox{Im} \, \kappa \right)^2 
\left| e^{-i \phi} - \kappa \right|^2}{\left|
  \kappa^3-3 \kappa+ 2 \cos{\phi} \right|^2} \; ,
\end{eqnarray}
with a {\it single complex parameter} 
$\kappa=(-V-t_{\rm Y}^2 \, \hat G^0_{1,1})/|t_{\triangle}|$. The Green 
function $\hat G^0$ is obtained for one of the equivalent  
disconnected ($t_{\rm Y}=0$) wires and $\hat G_{1,1}^0 \in 
{\mathbb C}$ denotes its diagonal matrix element taken at the first 
site $j=1$. It is evaluated at energy $\epsilon +i0$ with $\epsilon \to 0$.
Equation (\ref{trans}) holds if $\nu$ and $\nu'$ are in cyclic order and is 
independent of the pair considered;  $g_{\nu',\nu}$ follows by replacing 
$\phi \to - \phi$. 
If $\phi$ does not correspond to an integer multiple of $\pi$ and for 
generic junction parameters the conductance from $\nu$ to $\nu'$ 
{\it differs} from the one with reversed indices indicating the breaking of 
time-reversal symmetry. This constitues the most interesting 
situation and we focus on such fluxes. In Fig.\ \ref{fig:U0cond} the 
conductance from wire $\nu$ to $\nu'$ (cyclic) for $\phi = 0.4 \pi$ is
shown for the 
upper half of the complex $\kappa$-plane. For restored time-reversal symmetry 
the largest conductance allowed by the unitarity of the scattering matrix is 
$2 \pi g_{\nu,\nu'}=4/9$ (denoted the ``perfect junction value'' in the following); 
even for optimized parameters a reflection of $1/9$ is unavoidable. 

\begin{figure}[htb]
\begin{center}
\includegraphics[width=0.4\textwidth,clip]{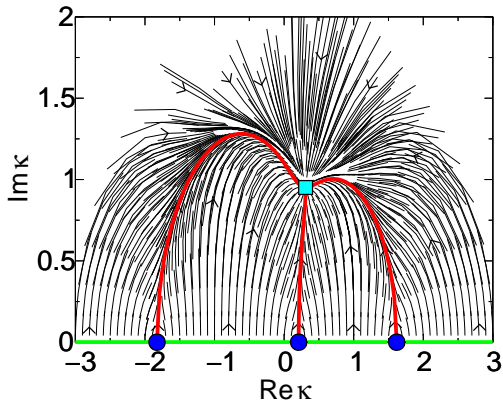}
\end{center}
\vspace{-0.3cm}
\caption[]{(Color online) Flow of $\kappa$ for $U=-1$, $n=1/2$, and $\phi=0.4 \pi$. 
Arrows indicate the direction for $U<0$. For $U>0$ it is reversed. 
For details see the text \cite{Barnabe05}. 
\label{fig:Yscal}}
\end{figure}

For $U \neq 0$ the Y-junction can straightforwardly be treated within  
the functional RG based approximation scheme \cite{Barnabe05}. We 
here focus on $T=0$. To compute the conductance from  
Eq.\ (\ref{trans}), $\hat G^0$ must be replaced by the 
{\it auxiliary} Green function  $\hat G$ obtained by 
considering $\Sigma$ (at the end of the RG flow for the full system) 
as an effective potential for a {\it single disconnected} wire 
setting $t_{\rm Y}=0$ \cite{Barnabe05}. 
Via the RG flow of $\Sigma$, $\hat G$ 
develops a dependence on ($t_{\rm Y},t_{\triangle},V$), $U$, and 
$\delta_N=v_F/N$. The latter energy scale is a natural infrared 
cutoff--in contrast to the flow parameter $\Lambda$ which is 
artificial and sent to 0. 
A comprehencive picture of the low-energy physics 
is obtained from the dependence of $\kappa$ on  
$\delta_N$. In Fig.\ \ref{fig:Yscal} each line is 
for a fixed set of junction  parameters and $\delta_N$  as a 
variable. The flux is chosen as $\phi= 0.4 \pi$ and the arrows 
indicate the direction of decreasing $\delta_N$. 
As $\mbox{Im} \, \kappa$ has the opposite sign of $\mbox{Im} \,
\hat G_{1,1} <0$ it is restricted to positive values.

\begin{figure}[tbh]
\begin{center}
\includegraphics[width=0.35\textwidth,clip]{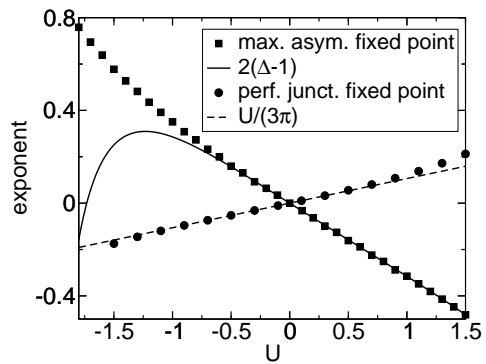}
\end{center}
\vspace{-.3cm}
\caption[]{Scaling exponents of the Y-junction close to the fixed 
points \cite{Barnabe05}.\label{fig:Yexp}}
\end{figure}

Equation (\ref{trans}) allows for {\it four} distinguished conductance situations 
(see Fig.\ \ref{fig:U0cond}): (i) on the line $\mbox{Im} \, \kappa = 0$, 
$g_{\nu,\nu'}=g_{\nu',\nu}=0$ for almost all  $\mbox{Re} \, \kappa$; 
(ii) it is interrupted by three points having flux-dependent positions with 
the conductance $2 \pi g_{\nu,\nu'}=2 \pi g_{\nu',\nu}=4/9$; 
(iii) for a specific flux-dependent $\kappa$ one finds $2 \pi g_{\nu,\nu'}= 1$ and
$2 \pi g_{\nu',\nu}= 0$; (iv) $g_{\nu,\nu'}=g_{\nu',\nu}=0$ is also reached for 
$|\kappa| \to \infty$. 
These are the fixed points of the RG flow as is evident from Fig.\ \ref{fig:Yscal}.  
(i) is an interupted line of decoupled chain fixed points with vanishing conductances 
which is stable for $U>0$ and unstable in the opposite case. Analyzing the 
dependence of $g_{\nu,\nu'}$ on $\delta_N$ in its vicinity 
for different $U$ one finds that the scaling exponent is independent of $\phi$ given 
by $\gamma_o$ as obtained for the single impurity. (ii) constitute three perfect 
junction fixed points (circles in Fig.\ \ref{fig:Yscal}).  
For $U>0$ each of the  three  fixed points has a {\it basin of attraction} given 
by one of the three parts of the curve ${\mathcal C}(\phi)$ (curved thick line in 
Fig.\ \ref{fig:Yscal} interrupted by the square) 
on which the reflection $1-2 \pi  g_{\nu,\nu'} - 2 \pi g_{\nu',\nu}$ takes a local minimum. 
The $U$ dependence of the scaling exponent when approaching one of the fixed points 
along its corresponding line is shown in Fig.\ \ref{fig:Yexp} (circles). 
It is independent of $\phi$ and for small $|U|$ it can be fitted by 
$U/(3 \pi)$. These fixed points have not been found by any method which is 
based on bosonization and the exact dependence of their scaling exponent on $K$ 
is presently unknown. Because of the factor $1/3$ in the leading order it must be 
different from the $K$-dependence of any of the exponents discussed so far. 
(iii) The basins of attraction are separated by the maximal asymmetry fixed point 
(maximal breaking of time-reversal symmetry; square in Fig.~\ref{fig:Yscal}). 
For $\phi=\pi/2$ this fixed point was identified by a bosonization based approach 
\cite{Chamon03}, and it was conjectured that the behavior found holds for all 
fluxes different from integer multiples of $\pi$. The functional RG 
results indeed confirm this--at least for small to intermediate $|U|$--as one 
finds this fixed point for {\it all} such $\phi$ and obtains a {\it flux-independent} 
scaling exponent which to leading order agrees to the bosonization result 
$\gamma_{\rm Y}=2(\Delta-1)$ with $\Delta=4K/(3+K^2)$ (see Fig.\ \ref{fig:Yexp}). 
The bosonization exponent shows a {\it non-monotonic} dependence on $K$ 
and thus $U$, which the approximate functional RG  approach does not capture. 
This implies that the maximal asymmetry fixed point is unstable for repulsive 
interactions, and stable for sufficiently small attractive ones but turns 
unstable again for larger attractive interactions. 
(iv) In the mapping of the complex plane onto the Riemann sphere the 
$g=\infty$ fixed point (north pole) is part of the projected line of
decoupled chain fixed points and shows the same stability properties 
and scaling dimension. 
    
The most interesting physics is associated with the perfect junction
fixed points which for $U>0$ each have one stable direction. 
If the junction parameters  of a non-interacting system at fixed $\phi \neq m \pi$, 
$m \in {\mathbb N}_0$ are chosen such that the resulting $\kappa$ lies on 
${\mathcal C}(\phi)$, but not on one of the three special points (ii),  
$g_{\nu,\nu'} \neq  g_{\nu',\nu}$ and the conductance 
indicates the breaking of time-reversal symmetry as expected. 
Turning on an interaction $U>0$ the ``fine-tuned'' system flows to one 
of the perfect chain fixed points with equal ``perfect'' conductances 
$2 \pi g_{\nu,\nu'}=4/9$ and $2\pi g_{\nu',\nu}=4/9$. Therefore, at small energy 
scales the junction conductance does 
{\it no longer indicate the explicit breaking of time-reversal symmetry.} 
For generic junction parameters away from  ${\mathcal C}(\phi)$
one finds related behavior. 
Close to the line of decoupled chain fixed points the relative 
difference $|g_{\nu,\nu'}-g_{\nu',\nu}|/(g_{\nu,\nu'}+g_{\nu',\nu})$ scales as a 
power law in $\delta_N$ with an exponent given by 
$\gamma_o/2$ and thus vanishes if $U>0$. This implies 
that $g_{\nu,\nu'}$ and $g_{\nu',\nu}$ {\it become equal} faster than they go 
to zero. In that sense for $U>0$ and up to the unstable maximal asymmetry 
fixed point, on small scales the conductance does not show the breaking of 
time-reversal symmetry--time reversal symmetry is ``restored'' by the 
interaction.
   
Other types of junctions of an arbitrary number of LL wires were studied 
using functional RG  \cite{Barnabe05a} as well as by the poor man's 
fermionic RG \cite{Lal02,Aristov10} and bosonization based approaches 
\cite{Nayak99,Chen02}.

\subsubsection{Non-equilibrium transport through a contacted wire}

\label{subsecnonequiwire}

Non-equilibrium functional RG was used to study a finite bias transport 
geometry with an impurity-free $N$ site interacting wire contacted to two 
non-interacting semi-infinite leads by tunnel barriers modeled by reduced 
hopping matrix elements as introduced in subsection \ref{basicmodel}: 
$t_{0,1}=(t_L-1)$ and $t_{N,N+1} = (t_R-1)$ \cite{Jakobs07a}.
In equilibrium this model features a local single-particle spectral function 
$\rho_j(\omega)$ which close to the chemical potential, in the vicinity of the 
contacts, and for repulsive interactions is suppressed: 
$\rho_j(\omega) \sim \omega^{\gamma_o}$ \cite{Enss05b}. The
linear conductance behaves as $g(T) \sim T^{\gamma_o}$ which can be understood 
from viewing transport as an end-to-end tunneling between a LL and a Fermi 
liquid lead and using the sum of two resistances as discussed in subsection 
\ref{resotunsec}.

A cutoff scheme which conserves causality to any truncation order 
\cite{Jakobs10b} is given by an imaginary frequency cutoff. 
The Fermi function of the two leads which can be written as a Matsubara sum 
\begin{eqnarray} 
f_{L/R}(\omega) & = & \left[  e^{\beta(\omega-\mu_{L/R})} +1 \right]^{-1} \nonumber \\
\label{imagfermi}
& = & \beta^{-1} \sum_{\omega_n} \frac{e^{i \omega_n 0^+}}{i \omega_n - \omega +\mu_{L/R}}
\end{eqnarray}
is replaced by 
\begin{eqnarray} 
\label{imagcutoff}
f_{L/R}^\Lambda(\omega) = \beta^{-1} \sum_{\omega_n} \frac{\Theta(|\omega_n|-\Lambda)  
e^{i \omega_n 0^+}}{i \omega_n - \omega +\mu_{L/R}} \; .
\end{eqnarray}
Details of this procedure including a discussion of the initial conditions 
and its relation to the temperature flow scheme \cite{Honerkamp01c} are 
presented by \textcite{Jakobs10b}.
Within the lowest-order truncation and after taking the equilibrium limit this 
cutoff implemented for Keldysh Green functions yields the same flow equations as 
the Matsubara functional RG with the frequency cutoff Eq.\ (\ref{cutfun}) 
\cite{Jakobs07a,Jakobs10b,Jakobs10c}.

\begin{figure}[tbh]
\begin{center}
\includegraphics[width=0.4\textwidth,clip]{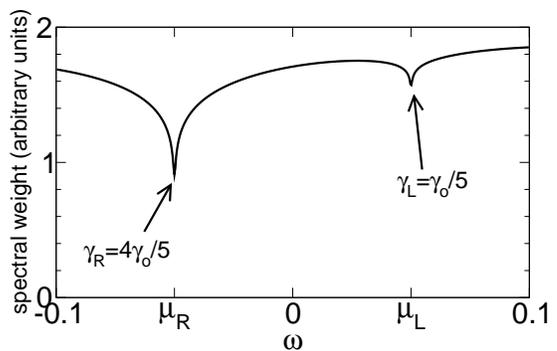}
\end{center}
\vspace{-.3cm}
\caption[]{Suppression of the local one-particle spectral weight as a function
    of energy near the left contact (at site $5$) of an interacting wire 
    driven out of equilibrium by a finite bias current. The parameters      
   are $T = 10^{-4}$, 
    $N=2 \time 10^4$, $U=0.5$, $t_L=0.075$, $t_R=0.15$, and 
    $\mu_{L/R}=\pm 0.05$. (Data taken from \textcite{Jakobs07a}.) 
\label{fig:nonequiwire}}
\end{figure}

In the presence of a finite bias voltage the level-1 truncation scheme (bare 
two-particle vertex) with the cutoff procedure (\ref{imagcutoff}) was applied.
As discussed in Sec.~\ref{subsubsimp}, in equilibrium this is sufficient 
to obtain scaling exponents correctly to leading order in $U$.
For weak tunneling
$\Gamma_{L/R} =  \pi t_{L/R}^2 \rho_{0} \ll 1$,
with   $\rho_{0}$  the 
density of states of the disconnected, non-interacting leads taken at the last 
lattice site, the flow of the retarded non-equilibrium self-energy matrix 
$\Sigma^{{\rm ret},\Lambda}$ is given by a weighted sum of two equilibrium flows 
\begin{eqnarray} 
\label{nonequiflow}
\frac{d}{d \Lambda} \Sigma^{{\rm ret},\Lambda} = 
\sum_{\lambda=L,R} \frac{\Gamma_\lambda}{\Gamma_L + \Gamma_R} 
\left[ \frac{d}{d \Lambda}  \Sigma^{{\rm eq},\Lambda} \right]_{\mu=\mu_\lambda}  \; , 
\end{eqnarray}
where the terms inside the brackets on the right hand side are given by 
Eq.\ (\ref{volker:sigmaflowtg0}) with the chemical potential set to $\mu_L$ or $\mu_R$
respectively (and $U^\Lambda \to U$). As discussed in subsection \ref{subsubnum} each 
such term leads to an oscillatory slowly decaying self-energy originating at the 
inhomogeneity--the tunnel barriers in the present case--and extending into the 
interacting part of the wire.
The two chemical potentials $\mu_{L/R}$ imply two different wave numbers $2 k_F^{(L/R)}$.
Because of the weighting factor $\Gamma_\lambda/(\Gamma_L+\Gamma_R)$ the amplitudes 
of the two superimposed decaying oscillations are generically different and depend
on the strength of the bare inhomogeneity.
The resulting non-equilibrium effect of two different and
$\Gamma_\lambda$-dependent exponents characterizing the scaling 
of the spectral function close to $\mu_L$ and $\mu_R$
goes beyond the naive expectation that the bias voltage 
plays the role of an infrared cutoff scale only (see e.g. \textcite{Schoeller09a}).
In Fig.~\ref{fig:nonequiwire} the local spectral function near the left contact 
and for a restricted energy range around $\mu_{L/R}$ is shown. Due to the finite 
temperature ($T=10^{-4}$) and the finite size of the interacting wire 
($N=2 \times 10^4$) the suppression at $\mu_{L/R}$ 
is incomplete (cut off by $\mbox{max} \{ T,\delta_N \}$), but the difference in 
the exponent is still apparent. 
A detailed analysis shows that the exponents at $\mu_{L/R}$ are given by 
$ \gamma_{L/R} = \Gamma_{L/R}  \gamma_o(\mu_{L/R}) / (\Gamma_L + \Gamma_R)$
where the argument in the open boundary exponent $\gamma_o$ indicates that 
it depends on the band filling and thus the chemical potential. 
After adding a third probe lead these ``non-universal'' exponents can be 
measured in a transport experiment \cite{Jakobs07a,Jakobs10b}.

\subsection{Quantum dots}

A spatially confined system featuring a few energy levels is called a quantum dot.
In a transport geometry the dot is coupled to at least two leads.
Quantum dots  show interesting physics if all relevant energy scales 
(e.g. level-lead couplings and $T$) 
are smaller than the level spacing of the isolated system. Due to the 
strong confinement the two-particle interaction on the dot cannot be neglected 
and leads to phenomena such as Coulomb blockade and the Kondo effect. 

\subsubsection{Spin fluctuations}  

\label{subsubspinfluc}

In the Kondo regime the physics is dominated by {\it spin 
fluctuations.} The virtues and limitations of the functional RG approach to
describe aspects of Kondo physics in and out off equilibrium were extensively 
studied within the single-impurity Anderson model and more complex
variants of the latter
\cite{Hedden04,Andergassen06a,Meden06,Karrasch06,Gezzi07,Karrasch07a,Karrasch07b,
Karrasch08a,Karrasch08b,Weyrauch08,Xu08,Karrasch09a,Kashcheyevs09,Eichler09,
Goldstein09,Bartosch09a,Karrasch10b,Jakobs10a,Jakobs10b,Xu10,Schmidt10,Isidori10}.
As the number of correlated degrees of freedom in quantum dots is
small the static truncation used for LLs was extended to contain
 all second order 
processes including a frequency dependent two-particle vertex and self-energy, 
capturing the full real-space as well as spin structure 
\cite{Hedden04,Karrasch08a,Jakobs10a,Jakobs10b,Karrasch10a,Karrasch10b}.
In fact, the studies of the single-impurity Anderson model constitute 
one of the rare examples of the functional RG approach to correlated Fermi 
systems using a complete level-2 truncation (even supplemented by parts of 
the six-point vertex through the replacement discussed in the second part 
of Sec.\ \ref{sec:truncations}).
Including the frequency dependence clearly improves the results beyond bare 
perturbation theory  of the same (that is second) order but it is presently not 
possible to reach the strong coupling regime in a controlled way. 
A discussion of the problems yet to be solved was presented by \textcite{Karrasch08a}, 
\textcite{Karrasch10a}, \textcite{Karrasch10b}, \textcite{Jakobs10a}, and 
\textcite{Jakobs10b}.
An alternative way to include the full frequency dependence was recently
introduced for another impurity model by \textcite{Schmidt11}. 
In the following a simple quantum dot model dominated by {\it charge fluctuations} 
is discussed. 

\subsubsection{Charge fluctuations in non-equilibrium}

\label{subsubchargefluc}

The quantum dot model belongs to the class of spinless models introduced
in subsection \ref{basicmodel}.
For a three site interacting chain ($N=3$) with $U_{1,2}=U_L$ and $U_{2,3}=U_R$
two hopping impurities are located at bonds $(1,2)$ and $(2,3)$: $t_{1,2}=t_L-1$, 
$t_{2,3}=t_R-1$. Lattice site 2 constitutes a single-level dot which can be adjusted 
in energy by a gate voltage $V_g$ (see Fig.\ \ref{fig:doublebar}). 
An electron on this site interacts with lead electrons via a nearest-neighbor 
coupling $U_{L/R}$ which are otherwise non-interacting. 
Choosing $\nu=1/2$ in Eq.\ (\ref{volker:intdef}), $V_g$=0 corresponds to the 
particle-hole symmetric point with dot occupation $\left< n_2 \right> =1/2$.
This model is a lattice realization of the 
{\it interacting resonant level model} (IRLM). 
The use of a variety of analytical as well as numerical methods 
led to a rather complete understanding of the physics of this model in equilibrium 
(see e.g. \textcite{Borda07} and references therein). 
In addition, the current under a finite bias voltage $\mu_L=V_b/2$ and 
$\mu_R = - V_b/2$ was investigated \cite{Borda07,Doyon07,Boulat08}.  
Field theoretical methods were applied in the {\it scaling limit} in which all 
energy scales are much smaller than the band width $B$. 
In the following the focus is on this limit. 
Functional RG results for the equilibrium and non-equilibrium properties beyond
the scaling limit including a favorable comparison with recent numerical 
time-dependent density-matrix renormalization group data \cite{Boulat08} are 
presented by \textcite{Karrasch10a}, \textcite{Karrasch10b}, and 
\textcite{Karrasch10c}.

First order perturbation theory in $U_{L/R}$ leads to logarithmic terms in the 
self-energy of the form $U_{L/R} \ln{ (t_{L/R}/B)}$
which in the scaling limit become large. 
They indicate the appearance of power laws in $t_{L/R}$ with $U_{L/R}$ 
dependent exponents. 
To uncover them requires a treatment which goes beyond perturbation theory.
In the limit of weak to intermediate two-particle interactions a Keldysh  
functional RG approach to the IRLM in the level-1 truncation leads to a 
comprehensive picture of the physics in and out of equilibrium. 
In particular, it allows to identify the relevant energy scales.

For the present model instead of Eq.~(\ref{imagcutoff}) another cutoff 
scheme suitable for non-equilibrium \cite{Jakobs10a,Jakobs10b} was implemented  
and tested \cite{Karrasch10a,Karrasch10b}. 
In this approach each of the three interacting sites is coupled to its own 
auxiliary lead--in addition to the coupling of sites 1 and 3 to the physical leads. 
The local density of states at the contact points of the auxiliary leads 
is assumed to be energy independent (wide band limit) such that the 
hybridization is energy independent and forms an additional onsite 
``energy'' $i \Lambda$  on each of the three sites. 
The auxiliary couplings are then considered as the cutoff and flow from 
$\Lambda=\infty$, at which regularization is achieved, down to $\Lambda=0$, 
at which the auxiliary leads are decoupled and the original problem is restored. 
One can show that in the lowest order truncation and in the equilibrium limit 
the Keldysh contour flow equations become equal to the equilibrium ones obtained 
using the Matsubara formalism with the (at $T=0$) sharp energy cutoff 
Eq.\ (\ref{cutfun}).  
Similarly to the imaginary frequency cutoff of subsection \ref{subsecnonequiwire} 
it conserves causality even after truncation of the functional RG flow equation 
hierarchy. 
In addition, in the equilibrium limit this so-called {\it reservoir cutoff}
scheme obeys the KMS relation in any truncation order 
\cite{Jakobs10a,Jakobs10b,Jakobs10c}. 

In the scaling limit and to lowest order in $U$ only flow equations
for the hydridizations $\Gamma^\Lambda_\lambda$ with initial values 
$\Gamma_\lambda^{\rm ini} = \pi \rho_{0} t_\lambda^2$ appear ($\lambda=\mbox{L/R}$); 
the flow of the level energies of the sites 1 to 3 is of order $U^2$. 
The renormalized hybridizations set the width of the resonance at $V_g=0$. 
For $\Lambda$ being smaller than the band width the flow equations for the 
rates read ($\Gamma^\Lambda = \Gamma_L^\Lambda +\Gamma_R^\Lambda $) 
\begin{eqnarray}
\label{flowIRLMgeneral} 
\frac{d \Gamma^\Lambda_\lambda }{d \Lambda} = 
-2 \rho_{0} U_\lambda \Gamma^\Lambda_\lambda 
\frac{\Lambda + \Gamma^\Lambda}{(\mu_\lambda - V_g)^2 + 
(\Lambda + \Gamma^\Lambda)^2} \; .
\end{eqnarray} 
They have the approximate solutions
\begin{eqnarray}
\label{renorwidthLR} 
\Gamma_\lambda \approx \Gamma^{\rm ini}_\lambda 
\left( \frac{\Lambda_0}{\mbox{max} \{|\mu_\lambda-V_g|,\Gamma/2 \} } 
\right)^{2 \rho_{0} U_\lambda}  \; . 
\end{eqnarray} 
The scale $\Lambda_0$ is of the order of the band width.
Within the static approximation the current takes the form of the non-interacting 
expression with the bare hybridizations $\Gamma_\lambda^{\rm ini}$ replaced by the 
renormalized ones
\begin{equation}
\label{eq:curr}
I=\frac{1}{\pi}\frac{\Gamma_L\Gamma_R}{\Gamma}
        \left[\text{arctan}\frac{V_b/2-V_g}{\Gamma}+
        \text{arctan}\frac{V_b/2+ V_g}{\Gamma}\right] \; .
\end{equation}
It turns out to be useful \cite{Karrasch10c,Andergassen11} to introduce the two 
scales $T_u^\lambda = \Gamma_\lambda^{\rm ini} \Lambda_0/T_u$, with $T_u = T_u^L + T_u^R$, 
and the asymmetry parameter $c^2 = T_u^L/T_u^R$.  The same flow equation 
can be derived using the so-called real-time RG in frequency space 
\cite{Schoeller09a,Karrasch10c,Andergassen11}. 
Within this approach also the relaxation into the steady state was analyzed 
in detail \cite{Karrasch10c,Andergassen11}. 

We first review the results obtained for  left-right symmetric model with 
$t_L = t_R = t'$ and $U_L = U_R =U$ as well as particle-hole symmetry 
$V_g=0$ \cite{Karrasch10b,Karrasch10c,Andergassen11}.
From Eq.\ (\ref{renorwidthLR}) it follows that in this case the maximum 
of either $|\mu_\lambda|=|V_b|/2$ or $\Gamma$ itself cuts off the RG flow. 
The charge susceptibility $\chi=-\left. d \left<n_2 \right>/d V_g\right|_{V_g=0}$ is 
directly given by the renormalized width $\chi^{-1} = \pi \Gamma$, which at $V_b=0$ and 
to leading order in $U$ ($\Gamma_\lambda \to \Gamma_\lambda^{\rm ini}$ on the right 
hand side of Eq.\ (\ref{renorwidthLR})) gives the scaling relation 
\begin{eqnarray}
\label{scalchi}
\chi \sim (\Gamma^{\rm ini})^{\alpha_\chi-1} \; , \;\;\;  
\alpha_\chi=2\rho_{0} U + {\mathcal O}(U^2) \; .
\end{eqnarray}
In the non-interacting case $\chi \sim  (\Gamma^{\rm ini})^{-1}$ as expected. 
From Eq.\ (\ref{eq:curr}) it follows that the current for 
$T_u^\lambda = \Gamma_\lambda \ll V_b \ll B$ is given by 
\begin{eqnarray}
\label{scalI}
I \sim \Gamma \sim V_b^{- \alpha_I} \; , \;\;\; 
\alpha_I=2\rho_{0} U + {\mathcal O}(U^2) \; .
\end{eqnarray}
Equations (\ref{scalchi}) and (\ref{scalI}) were also obtained using other 
approaches \cite{Doyon07,Borda07,Boulat08} and suggest that the bias voltage 
merely plays the role of an additional infrared cutoff, besides e.g. $\Gamma$ 
or temperature \cite{Borda10}. 
That this is in general not the case is nicely shown by a functional RG treatment 
{\it away} from particle-hole and/or left-right symmetry  
\cite{Karrasch10b,Karrasch10c,Andergassen11}.

The differential conductance $g= dI / dV_b$ has a maximum when $V_g$ crosses 
the chemical potential at $V_g = \pm  V_b/2$ \cite{Karrasch10c,Andergassen11}. 
In the on-resonance case the current for $V \gg \Gamma$  reads
\begin{equation}
\label{eq:IVon}
  I(V_b) \approx \frac{\Gamma_L \Gamma_R}{2 \Gamma} 
= T_u\frac{\left(\frac{T_u}{\Gamma}\right)^{2 \rho_{0} U_L}
    \left(\frac{T_u}{V_b}\right)^{2 \rho_{0} U_R}}
  {c\left(\frac{T_u}{\Gamma}\right)^{2 \rho_{0} U_L}
    +\frac{1}{c}\left(\frac{T_u}{V_b}\right)^{2 \rho_{0} U_R}}\frac{c}{1+c^2} \; .
\end{equation}
The bias voltage dependence is clearly more involved than
 in Eq.\ (\ref{scalI}). 
In particular, simple power-law scaling with exponent $-2 \rho_{0}
U_R$ is only recovered in the 
extreme limits  of either $V_b \ggg T_u$ or $c\gg1$ \cite{Andergassen11} 
as the exponent of the second term in the denominator $2 \rho_{0} U_R$ is small.  
Off resonance (e.g.\ at $V_g=0$) and for $V_b \gg \Gamma$ the current is given by 
\begin{equation}
   \label{eq:IVoff}
  I(V_b) \approx 
  T_u\frac{\left(\frac{T_u}{|V_b/2- V_g|}\right)^{2  \rho_{0}  U_L}
    \left(\frac{T_b}{|V_b/2 +V_g|}\right)^{2  \rho_{0}   U_R}}
        {c\left(\frac{T_u}{|V_b/2 - V_g|}\right)^{2  \rho_{0}  U_L}
        + \frac{1}{c}\left(\frac{T_u}{|V_b/2 + V_g|}\right)^{2 \rho_{0} U_R}}
        \frac{2 c}{1+c^2} \; .
   \end{equation}
The more involved role of $V_b$ is again apparent.
A power law is obtained in the above studied left-right symmetric case or for very 
strong left-right asymmtery ($c \ll 1$ or $c\gg1$) \cite{Andergassen11}. 

This concludes the analysis of the IRLM which shows that the functional RG can be a 
tool to obtain a comprehensive picture of the equilibrium and steady-state 
non-equilibrium physics of a dot model dominated by charge fluctuations. 
The approach allows for an unbiased analysis of the non-equilibrium rates and 
cutoff scales. 


\section{CONCLUSION}
\label{sec:VII}

\subsection{Summary}

The functional RG has proven to be a valuable source of new
approximation schemes for interacting fermion systems.
The heart of the method is an exact flow equation, which 
describes the flow of the effective action as a function of
a suitable flow parameter. 
The flow provides a smooth evolution from the bare action
to the final effective action from which all properties of
the systems can be obtained.
Approximations are obtained by truncating the effective action.
In many cases, rather simple truncations turned out to capture
rather complex many-body phenomena.
Compared to the traditional resummations of perturbation 
theory these approximations have the advantage that infrared
singularities are treated properly due to the built-in RG
structure. 
Approximations derived in the functional RG framework can be 
applied directly to microscopic models, not only to 
renormalizable effective field theories.
Remarkably, the functional RG reviewed here as a computational
tool is very similar to RG approaches used by mathematicians to 
derive general rigorous results for interacting fermion systems.

We have dedicated a large portion of this review to general
features of the functional RG method for interacting Fermi
systems (Sec.~II). 
After defining the relevant generating functionals, we have 
presented a self-contained derivation of the exact functional
flow equation and its expansion leading to an exact hierarchy
of flow equations for vertex functions.
We have reviewed the different choices of flow parameters 
used so far, along with their advantages and disadvantages.
Truncations and their justification by power-counting have
been discussed in detail for translation invariant bulk 
systems, with links to the closely related mathematical 
literature.

In Secs.~III-VI we have reviewed applications of the functional
RG to specific systems.
Most of the approximations used in these sections are based on
relatively simple truncations involving only the flow of the 
two-particle vertex and/or the self-energy.
Nevertheless a rich variety of phenomena associated with low 
energy singularities and instabilities is captured.
Instead of summarizing the content of each section, let us merely
highlight some distinctive features.
Sec.~III reviews functional RG work on the stability analysis of 
two-dimensional electron systems with competing instabilities.
The main advantage of the functional RG based one-loop computation
of the two-particle vertex, compared to other weak-coupling
approximations, is that particle-particle and particle-hole channels 
are treated on equal footing, such that there is no artificial bias 
due to a selection or a different treatment of channels.
In the conventional many-body framework a summation of all parquet
diagrams would be required to achieve this, but a solution of the
parquet equations is extremely difficult.
Spontaneous symmetry breaking, the topic of Sec.~IV, can be treated
either by a purely fermionic flow, or by coupled flow equations
for the fermions and a Hubbard-Stratonovich field for the order
parameter. It seems that a comprehensive treatment of all relevant
fluctuation effects related to symmetry breaking can be achieved.
Applications of the fRG to quantum criticality, reviewed in Sec.~V,
have begun only recently. Approximations beyond Hertz-Millis 
theory can be obtained from non-perturbative truncations of the
effective order parameter action, or by treating fermions and
order parameter fluctuations in a coupled flow instead of 
integrating the fermionic degrees of freedom at once.
While the applications reviewed in Secs.~III-V address translation 
invariant bulk systems, the purpose of Sec.~VI is to show how the 
functional RG can be fruitfully applied to inhomogeneous systems 
such as quantum wires and quantum dots -- in thermal equilibrium 
and also in a non-equilibrium steady state.
A strikingly simple truncation of the flow equation hierarchy
turned out to describe a wealth of non-trivial quantum transport 
properties characterized by low-energy power-laws and complex
crossover phenomena.

\subsection{Future directions}

The number of functional RG based works on interacting Fermi
systems has increased steadily over the last decade, but the
possibilities opened by this approach are far from being 
exhausted.
There are many opportunities and challenges concerning both 
fundamental developments of the method and the extension to 
a broader range of systems.

On the methodological side there are a number of open issues.
In systems with an instability of the normal metallic state,
the flow of the effective interactions is not yet fully 
understood, even on the level of truncations involving only 
the two-particle vertex and the self-energy, since a 
faithful parametrization of singular momentum and frequency
dependences of the vertex is not easy.

The most outstanding challenge is probably to identify
accurate and computable truncations of the exact flow equation 
for strongly interacting systems such as systems close to a 
Mott metal-insulator transition.
It is clear that three-particle and higher order vertices
cannot be discarded in a strongly interacting system.
However, they will usually not lead to qualitative changes
such as new singularities. 
Hence, there is hope that the contribution from many-body 
vertices can be absorbed in the structure appearing already 
on the two-particle level.
After all, many strong coupling phenomena, including the Mott
transition, consist essentially in the formation of two-particle 
bound states.
To capture effects related to strong local correlations, such
as the Mott transition, one may also try to treat higher order
vertices in a local approximation. This would make a link to 
the dynamical mean-field theory (DMFT), where all vertices,
including the self-energy, are approximated by local 
functions \cite{Georges96}.

For systems with strongly interacting order parameter 
fluctuations there are already a number of non-perturbative
approximations for bosonic actions on the market.
The local potential approximation presented in Sec.~V is 
only the simplest one. It can be extended by taking non-local
contributions into account, either in a derivative expansion
\cite{Berges02}, or by including the full momentum or 
frequency dependence up to a certain level in the hierarchy
\cite{Blaizot05}.
Such approximations may be very useful for studying 
incommensurate density wave instabilities in cases where the 
modulation vector of the density wave can be determined only 
after taking fluctuations into account.

Recently, the functional RG was extended to a real time (or 
real frequency) Keldysh functional RG which can be used for
studying correlated Fermi systems in non-equilibrium
\cite{Jakobs03,Gezzi07,Jakobs10b,Karrasch10b}.
First applications, partly reviewed in Sec.~VI, indicate that 
also for these type of problems the functional RG constitutes a 
useful tool of outstanding flexibility. 
So far only non-equilibrium steady states were studied. 
To investigate a time evolution is technically straightforward 
but requires a significantly increased computational effort or
additional approximations.

With few exceptions, applications of the functional RG to 
interacting Fermi systems have so far been limited to purely 
fermionic one-band systems.
There are many extensions of this restricted class of 
systems, where the flexibility of the functional RG can be
fruitfully used in the future.
Multi-band models have been studied already for the pnictide 
superconductors, but there are many more and qualitatively
different models for transition metal oxides with orbital
degrees of freedom.
One may include phonons and analyze the electron-phonon 
interaction effects beyond the Eliashberg theory.
Allowing for disorder one may study the complex interplay 
of interaction and disorder effects.
It is not hard to generalize the exact flow equations for
the extensions listed above. The interesting task is then
to devise suitable truncations.

Last but not least, the functional RG is an ideal many-body
tool to be combined with {\em ab initio}\/ band structure 
calculations. 
A lot of work in the last 15 years has been dedicated to
the ab initio calculation of correlated electron materials 
with the DMFT \cite{Kotliar06,Anisimov10}.
As in DMFT, an arbitrary band structure can be used as
input for a functional RG calculation. 
Furthermore, one can easily implement non-local potentials 
and non-local two-particle interactions.


\begin{acknowledgments}
We are very grateful for fruitful collaborations and/or discussions 
with S. Andergassen, J. Bauer, D. Baeriswyl, C. Castellani,
A. Chubukov, C. Di Castro, J. von Delft, A. Eberlein, T. Enss, 
J. Feldman, R. Gersch, H. Gies, W. Hanke,
C. Husemann, S. Jakobs, P. Jakubczyk,
C. Karrasch, A. Katanin, H. Kn\"orrer, P. Kopietz, D.-H. Lee,
B. Obert, J. Pawlowski, C. Platt, M. Pletyukhov, M. Rice, 
D. Rohe, A. Rosch, H. Schoeller, P. Strack, S. Takei, R. Thomale,
E. Trubowitz, C. Wetterich, P. W\"olfle, and H. Yamase.
All of us greatly benefitted from the DFG research group 
{\em Functional renormalization group in correlated fermion systems} 
(FOR 723).
\end{acknowledgments}


\begin{appendix}

\section{Wick-ordered flow equations}
\label{sec:wick}

In this appendix we present a derivation of Wick ordered
flow equations for fermions, which have been used for 
calculations of instabilities and symmetry-breaking in
the two-dimensional Hubbard model.

{\em Wick ordered} $m$-particle functions $W_m^{\Lam}$ are
generated from the Wick-ordered effective interaction 
\cite{Salmhofer99}
\begin{equation}
 \cW^{\Lam}[\chi,\chib] = 
 e^{\Delta_{\bar G_0^{\Lam}}} \, \cV^{\Lam}[\chi,\chib] \; .
\end{equation}
The exponent in the Wick-ordering factor is the functional
Laplacian
$\Delta_{\bar G_0^{\Lam}} = 
 (\partial_{\chi}, \bar G_0^{\Lam} \partial_{\chib})$ with
$\bar G_0^{\Lam} = G_0 - G_0^{\Lam}$. 
The Wick-ordered interaction converges to $\cV$ for
$\Lam \to 0$, since $\bar G_0^{\Lam}$ vanishes in that limit.
However, the flow equations for $\cW^{\Lam}$ and the corresponding 
$m$-particle functions differ from those for $\cV^{\Lam}$.
The flow equation for the generating functional $\cW^{\Lam}$
reads \cite{Salmhofer99}
\begin{equation} \label{rgewick}
 \frac{d}{d\Lam} \cW^{\Lam} = 
 \frac{1}{2} \, e^{\Delta^{\rm diff}_{\bar G_0^{\Lam}}} \,
 \Delta_{\dot{\bar G}_0^{\Lam}}^{\rm diff} \,
 \cW^{\Lam} \, \cW^{\Lam} \; ,
\end{equation}
where the superscript ''diff'' indicates that the Laplacian
takes one derivate on the first, and the other on the second
factor $\cW^{\Lam}$ on the right-hand side.
This equation is obtained as follows. Using the definition of
$\cW^{\Lam}$ and the flow equation for $\cV^{\Lam}$, one can
write
\begin{eqnarray}
 \frac{d}{d\Lam} {\cW^{\Lam}} &=& 
 \frac{d}{d\Lam} 
     \big( e^{\Delta_{\bar G_0^{\Lam}}} \cV^{\Lam} \big)
 \nonumber \\
 &=& \Delta_{\dot{\bar G}_0^{\Lam}} 
     e^{\Delta_{\bar G_0^{\Lam}}} \cV^{\Lam} 
 \nonumber \\ 
 &+& e^{\Delta_{\bar G_0^{\Lam}}} 
     \big( - \Delta_{\dot{\bar G}_0^{\Lam}} \cV^{\Lam} +
     \frac{1}{2} \, {\Delta_{\dot{\bar G}_0^{\Lam}}^{\rm diff}} \,
     \cV^{\Lam} \cV^{\Lam} \big)
 \nonumber \\ 
 &=& \frac{1}{2} \, e^{\Delta_{\bar G_0^{\Lam}}} \,
     {\Delta_{\dot{\bar G}_0^{\Lam}}^{\rm diff}} \, 
     \cV^{\Lam} \cV^{\Lam} \; .
 \nonumber
\end{eqnarray}
Using the decomposition
$\Delta_{\bar G_0^{\Lam}} = 
 \Delta_{\bar G_0^{\Lam}}^{\rm factor 1} + 
 \Delta_{\bar G_0^{\Lam}}^{\rm factor 2} +
 {\Delta_{\bar G_0^{\Lam}}^{\rm diff}} \;$ 
(when acting on a product), this yields
$\partial_{\Lam} {\cW^{\Lam}} = 
 e^{\Delta_{\bar G_0^{\Lam}}^{\rm diff}} \,
 \frac{1}{2} \, {\Delta_{\dot{\bar G}_0^{\Lam}}^{\rm diff}} 
 \big( e^{\Delta_{\bar G_0^{\Lam}}} \cV^{\Lam} \big) \,
 \big( e^{\Delta_{\bar G_0^{\Lam}}} \cV^{\Lam} \big)$
and thus Eq.~(\ref{rgewick}).

\vskip 1mm

\begin{figure}[ht]
\centerline{\includegraphics[width = 7cm]{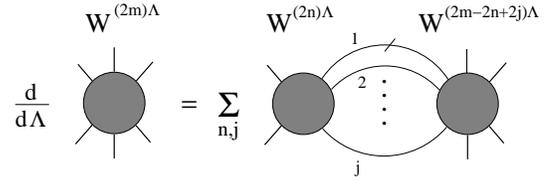}}
\caption{Diagrammatic representation of the flow equations for
 the effective $m$-particle interactions $W^{(2m)\Lam}$ in the 
 Wick-ordered version of the functional RG; the internal line with 
 a dash corresponds to $\partial_{\Lam} \bar G_0^{\Lam}$, the 
 others to $\bar G_0^{\Lam}$; all possible pairings leaving 
 $m$ ingoing and $m$ outgoing external legs have to be summed.}
\label{fig:floweqwick}
\end{figure}

Expanding in powers of Grassmann fields and comparing 
coefficients, one obtains a hierarchy of flow equations for
the $m$-particle functions $W^{(2m)\Lam}$, which is illustrated
diagrammatically in Fig.~\ref{fig:floweqwick}.
The line with the dash is due to contractions generated by 
$\Delta_{\dot{\bar G}_0^{\Lam}}^{\rm diff}$ in Eq.~(\ref{rgewick}), 
the other lines are generated by the exponential of
$\Delta^{\rm diff}_{\bar G_0^{\Lam}}$.
Note that the right-hand side of the Wick-ordered flow equations
is bilinear in the effective interactions, and no tadpole
terms appear. Note also that the propagators connecting the
vertices have support for energies at and \emph{below} the 
cutoff scale $\Lam$, such that the integration region shrinks 
as $\Lam$ decreases. 
One might worry that the low-energy propagators lead to 
infrared divergences even for $\Lam > 0$. This is not the
case, as can be seen from the general infrared power-counting
analysis presented by \textcite{Salmhofer99}.


\section{Details of power counting}
\label{app:powercounting}

\subsection{Propagator bounds}\label{appssec:propbounds}

\newcommand{\xxi}{\eta}

Here we show, using properties of the dispersion function and the Fermi
surface, that
\begin{equation}\label{propbounds}
\snorm{\Lambda} \le 
\Sa + \Sb \log \sfrac{\Lambda_0}{\Lambda} 
\quad \mbox{ and } \quad 
\Vert G^\Lambda\Vert \le \Sc \; \Lambda^{-1}\; 
\end{equation}
where $\Sa, \Sb$ and $\Sc$ are constants that do not depend on $\Lambda$.
In absence of Van Hove singularities, $\Sb = 0$. 
We first consider the case where the self-energy effects
are not taken into account (they are discussed in Subsection \ref{sssec:selfenergy}). 
Then one can simply take 
$
\chi^\Lambda (k) = \chi_> \left(
\sfrac{k_0^2+\xi_{\bk}^2}{\Lambda^2}
\right)
$
where $\chi_> (\xxi)$ is a fixed (i.e.\ $\Lambda$-independent)
increasing function that vanishes at least linearly at $\xxi=0$, 
tends to $1$ as $\xxi \to \infty$ and satisfies
$\chi'_>(\xxi) \le \xxi^{-2}$ for large $\xxi$. 
We can then verify (\ref{propbounds})
by scaling, as follows. The full propagator is $G^\Lambda (k) =(i k_0 -
\xi_{\bk})^{-1} \chi^\Lambda (k)$, so 
$
|G^\Lambda (k)|
\le 
\frac{\Sc}{\Lambda} 
$ 
where $\Sc = \max \{ \xxi^{-1} \chi_> (\xxi^2): \xxi > 0\}$ is finite.
The single-scale propagator is 
\begin{equation}
S^\Lambda (k)
=
- \frac{2}{\Lambda^3} (ik_0 + \xi_{\bk}) \; \chi'_> \left(
\sfrac{k_0^2+\xi_{\bk}^2}{\Lambda^2}
\right) .
\end{equation}
Using that the Matsubara sum is a Riemann sum for the convergent integral 
of $S^\Lambda$ over $k_0$
and introducing the density of states 
$N(E) = \int d^dk \, \delta (\xi_{\bk} - E)$,
we get
\begin{eqnarray}
\snorm{\Lambda} 
\le
\frac{4}{\Lambda^3}
\int dk_0 \int dE \, N(E) \sqrt{k_0^2 + E^2} \, 
\chi'_> \left(
\sfrac{k_0^2+E^2}{\Lambda^2}
\right) , \hskip 3mm
\end{eqnarray}
where the $4$ instead of $2$ gives a (crude) bound for the change from the 
Matsubara sum to the integral  for large enough $\beta$. 
Changing variables to $\rho =(k_0^2 + E^2)^{\frac12} $ and a polar angle
$\vphi$, we obtain
\begin{equation}
\snorm{\Lambda} 
\le
\frac{4}{\Lambda^3}
\int_0^\infty \rho^2 d \rho\; \chi'_> \left(
\sfrac{\rho^2}{\Lambda^2}
\right)
\int_0^{2\pi} d\vphi\; N(\rho \cos \vphi).
\end{equation}
If the density of states $N$ is bounded, using $N(E) \le N_0$ and scaling out
$\Lambda$ implies $\snorm{\Lambda} \le \Sa$, with 
$\Sa = 8 \pi N_0 \int \rho^2 \chi'_> (\rho^2) d\rho < \infty$. 
In presence of a Van Hove point on the Fermi surface, 
$N$ stays bounded in dimensions $d \ge 3$, but diverges logarithmically
for $d=2$. In this case, the $\vphi$-integral contributes an additional factor $\log
\Lambda$. 

Thus (\ref{propbounds}) holds. The hypotheses on $\chi_>$ are satisfied
in particular for the standard strict cutoff functions that vanish identically
near $\xxi=0$, and which are identically $1$ for $\xxi \ge 1$. 
For such cutoffs, the single-scale propagator is nonvanishing only in a
``momentum shell'' of thickness $\Lambda$ around the Fermi surface, 
and the above bounds can also
be obtained by estimating the $\bk$-space volume of this shell
(see also Section \ref{ssec:imppoco}).

\subsection{Power counting}

Here we prove (\ref{poco}) to all orders in the running coupling expansion.
All terms on the right hand side of the flow equation for $\Gamma_r^{(2m)\Lambda}$
are of the form 
\begin{equation}
\frac12 \, {\rm tr} \left(
\bS^\Lambda \, 
\bP^\Lambda
\right)(\km,\bsg) 
=
\frac12 
\int \db l 
\sum_{\alpha,\alpha'}
S^\Lambda_{\alpha,\alpha'} (l) \, 
\hat P^\Lambda_{\alpha,\alpha'} (\km,\bsg;l,-l) ,
\end{equation}
where $\hat P^\Lambda = \hat\Gamma_r^{(2m+2)\Lambda}$ in the first term of (\ref{mqexp}) 
and given by the other summands in (\ref{mqexp}) for the other terms, 
and $\int \db l  = \frac1\beta \sum\limits_{l_0} \int \frac{d^d l}{(2\pi)^d}$ contains
both frequency and momentum summations. 
Taking the absolute values inside the sum and estimating the factor 
$\hat P^\Lambda$ by its maximum $\Vert P^\Lambda \Vert$,
we obtain
\begin{equation}
\Vert 
\sfrac12 \; {\rm tr}\; (
\bS^\Lambda\; 
\bP^\Lambda) 
\Vert
\le
\snorm{\Lambda} 
\;
\Vert P^\Lambda \Vert
\end{equation}
with $\snorm{\Lambda}  =  \max\limits_\alpha \sum\limits_{\alpha'}\;
\int \db k \;
|\hat S^\Lambda_{\alpha,\alpha'} (k) |$.
The second simple inequality that we shall use is that
$\Vert P_1 \ldots P_n \Vert \le \Vert P_1 \Vert \ldots \Vert P_n \Vert$.
It implies bounds for all $\Lower^\Lambda_p$-contributions in terms of
$\Vert G^\Lambda\Vert$ and $\Vert \Gamma_{r_q}^{(2m_q)\Lambda} \Vert$,
so that 
$\Vert  
\sfrac{d}{d\Lambda} \Gamma_r^{(2m)\Lambda}
\Vert
$
is bounded by
\begin{eqnarray}\label{inductivebounds}
&&
\snorm{\Lambda} \;
\Big[
\Vert \Gamma_r^{(2m+2)\Lambda}\Vert
+
\Vert \Four{\Lambda} \Vert\;
\Vert G^\Lambda \Vert\;
\Vert \Gamma_{r-1}^{(2m)\Lambda}\Vert
\nonumber \\
&+&
\sum_{p \ge 2} 
\Vert G^\Lambda \Vert^{p-1}
\sum {}'\;
\Vert \Gamma_{r_1}^{(2m_1)\Lambda}\Vert\;
\ldots
\Vert \Gamma_{r_p}^{(2m_p)\Lambda}\Vert
\Big] . \hskip 5mm
\end{eqnarray}
The power counting is now determined by $\snorm{\Lambda}$ and $\Vert G^\Lambda\Vert$.
Given (\ref{inductivebounds}) and (\ref{propbounds}), the proof of (\ref{poco})
is an effortless induction argument. The inductive scheme proceeds in the
standard way of \cite{Polchinski84}, namely upwards in $r \ge 1$ and at
fixed $r$, downwards in $m$, starting at $m=r$, where $\Gamma_r^{(2m+2)\Lambda}=0$. 
The induction start $r=1$ is trivial.
Let $r \ge 2$ and assume (\ref{poco}) to hold for all $(r',m')$ with 
$r' < r$ and for $r'=r$, $m' > m$. 
The right hand side of (\ref{inductivebounds}) contains only terms to which
the inductive hypothesis (\ref{poco}) applies. Inserting it, using (\ref{propbounds}),
and collecting powers in the form $1 - p + \sum_q (2-m_q) = 1-m$
and
$\sum_{q=1}^p (r_q-m_q+1) = r-m$, 
we obtain
\begin{equation}\label{Gadotbound}
\Vert  
\sfrac{d}{d\Lambda} \Gamma_r^{(2m)\Lambda}
\Vert
\le 
\tilde\gamma_r^{(2m)}
{\snorm{\Lambda}}^{r-m+1}
{f_\Lambda}^r \Lambda^{1-m} 
\end{equation}
where the constant $\tilde\gamma_r^{(2m)}$ is a weighted 
sum of products of the $\gamma_{r_q}^{(2m_q)}$.
We now use the initial condition $\Gamma_r^{(2m)\Lambda} = 0$ to write
$\Gamma_r^{(2m)\Lambda} = - \int_\Lambda^{\Lambda_0} d \ell \; 
\frac{d}{d \ell}\Gamma_r^{(2m)\ell}\;$,
take the norm of this equation, and use (\ref{Gadotbound}).
This gives
\begin{equation}
\Vert
\Gamma_r^{(2m)\Lambda}
\Vert
\le
\tilde\gamma_r^{(2m)}
\int_\Lambda^{\Lambda_0} d \ell \; 
{\snorm{\ell}}^{r-m+1}
{f_\ell}^r \ell^{1-m} 
\end{equation}
By definition, $f_{\Lambda} \ge f_{\Lambda'}$
if $\Lambda \le \Lambda'$, so 
$f_\ell \le f_\Lambda$ for all 
$\ell $ in the integration interval. 
Thus the last integral is bounded by 
$\tilde\gamma_r^{(2m)}
{f_\Lambda}^r
\int_\Lambda^{\Lambda_0} d \ell \; 
{\snorm{\ell}}^{r-m+1}
\ell^{1-m}$.
Because $\snorm{\ell}$ is at most logarithmic in $\ell$, and $m \ge 3$,
$\int_\Lambda^{\Lambda_0} d \ell \; 
{\snorm{\ell}}^\alpha
\ell^{1-m}
 \le 
K \Lambda^{2-m} \snorm{\Lambda}^\alpha $
with a constant $K$ that depends on $\alpha$ and $m$. 
This, together with an appropriate choice of $\gamma_r^{(2m)}$, 
completes the induction step. 

For $m=2$, doing the last integral increases the power of the logarithm
by one. This case is discussed in more detail in Subsection \ref{ssec:imppoco}.

For $m=1$, the self-energy term, the same simple bound gives
$\Vert \frac{d}{d\Lambda} \Sigma^\Lambda \Vert
\le
\snorm{\Lambda}\; f_\Lambda$, so the integral gives a contribution 
of order $f_\Lambda$. When a counterterm is used to keep the Fermi surface fixed,
the initial condition for $\Sigma^\Lambda$ at $\Lambda=\Lambda_0$
is given by the counterterm, which needs to be adjusted such that 
at low scales $\Lambda$, $\Sigma^\Lambda (0, \bk) = O(\Lambda)$ 
whenever $\xi_{\bk} = 0$. This leads to the self-consistency relation mentioned
in section \ref{sssec:selfenergy}.

A similar proof can be given in the Wick ordered scheme \cite{Salmhofer98b};
it is even simpler because the double induction used here is replaced by 
single induction on $r$.

A crucial point in obtaining (\ref{Gadotbound}) is that all the 
dependence on $p$ and on the $m_q$ drops out when the power
of $\Lambda$ is collected. It is this property that makes 
many-fermion models with short-range interactions
{\em renormalizable} in the strict quantum-field-theoretical sense.  
The classification in relevant, marginal and irrelevant
terms now also becomes apparent
because for $m \ge 3$, the $\Gamma^{(2m)\Lambda}_r$
grow as $\Lambda$ decreases:
%
%
suppose we add an additional $(2m \ge 6)$-point interaction vertex 
$\tilde V^{(2m)\Lambda_0}$ of order $1$ to the initial interaction
at $\Lambda_0$. Its insertion at a lower scale $\Lambda$ is
a factor $(\frac{\Lambda}{\Lambda_0})^{m-2}$ smaller than that
of the effective $2m$-point-vertex created by the two-particle
interaction. Thus the influence of $\tilde V^{(2m)\Lambda_0}$
wanes at low scales -- it is irrelevant. 
A simple adaptation of the above inductive argument indeed shows
that the inclusion of such additional terms 
with $m \ge 3$ in the interaction at $\Lambda_0$
changes only prefactors in the power counting bounds. 
For $m=2$, this suppressing factor is absent, so that these terms 
are marginal (the more detailed analysis of Section 
\ref{ssec:imppoco} shows how to separate the marginally relevant
from the marginally irrelevant terms). 
Moreover, it is clear that this power counting breaks down when
$\Gamma^{(4)\Lambda}$ develops singularities as a function 
of $\bk$ and $\omega$, because then $f_\Lambda = \infty$. 
Finally, for $m=1$, the scale derivative of the self-energy obtained
by the above argument is of order $\Lambda^{1-m}$,
as in (\ref{Gadotbound}), 
but since $m=1$, this integrates to $O(1)$ instead of $O(\Lambda^{2-m})$
-- this term is relevant. 
To get its size back to $O(\Lambda^{2-m})$ in the momentum shell
where $|\xi(\bk)| \sim \Lambda$, one needs a cancellation by 
a counterterm, as described briefly in Section \ref{sssec:selfenergy}.
In the Taylor expansion 
required to do the cancellation, the derivative of the self-energy 
appears. By the above power counting, this is a marginal term. 
In the Luttinger model, it is really marginal and 
causes the anomalous exponents.
For curved Fermi surfaces in $d \ge 2$ dimensions, it is seen to be
irrelevant by the arguments discussed in 
Section \ref{sssec:OLloops}.

There is a hard problem hidden in the recursion of the constants 
$\gamma_r^{(2m)}$. In the recursion described above, the number of terms
that gets added corresponds to the number of Feynman graphs with $r$ 
vertices, which grows factorially in $r$, so that the bound obtained in
this way 
for $\gamma_r^{(2m)}$ and $c_r^{(2m)}$ is of order $r!$. If saturated,
it would lead to a convergence problem. 
However, due to the fermionic antisymmetry, sign cancellations in the sum
over Feynman diagrams prevent this factorial from arising. For proofs,
we refer interested readers to the literature (see 
\textcite{FMRT92,FKT98,Disertori00,Salmhofer00,FKT02,Pedra08} 
and references therein). In their application to propagators with Fermi
surfaces, these proofs also provide a rigorous basis 
for the use of Fermi surface patches (first used in \cite{FMRT92}
and there called ``sectors''). Patching the Fermi surface has become 
an essential tool also in applications, see Section \ref{sec:III}.

In a typical lattice model, the kinetic energy per particle is bounded, so that
the flow is usually started at the highest value (the bandwidth) of the
kinetic energy, $\Lambda_0$. 
The $\chi^\Lambda$ we used here also cuts off large frequencies.
Thus the starting interaction is
in this case one where the degrees of freedom with frequencies $|k_0|$
above $\Lambda_0$ have already been integrated over. 
This starting action can be obtained by convergent perturbation theory,
see \textcite{Pedra08}. 

\subsection{Improved power counting}\label{ssec:imppoco}

This is a refinement of power counting, valid in a large class of bulk
fermion systems in $d \ge 2$ \cite{FT1,Shankar94,FST1,FST2,FST3,FST4}. 
It is the deeper reason behind the emergence of Fermi liquid behaviour 
and of dominant Cooper pairing tencencies in weakly coupled standard 
fermion systems, and it provides a precise link between Fermi surface 
geometry and scaling properties of the effective $m$-particle vertices 
in general.

We discuss this in the absence of self-energy effects,
to bring out the main effects as clearly as possible. 
(The self-energy changes the Fermi surface; if the interacting 
Fermi surface is regular, the following analysis remains essentially 
unchanged.) We also assume a strict cutoff function, 
i.e.\ $\chi_> (\xxi ) = 0 $ for $\xxi \le (1-\delta)^2$, where 
$0 < \delta < 1/2$ is fixed, and $\chi_>  (\xxi)=1$ for $\xxi \ge 1$.
Again, this choice is not essential; it just simplifies the discussion. 

\subsubsection{Effects of curvature on power counting}

The integral 
$I_\Lambda (k) = \int d p_0 d^dp |S^\Lambda (p)| \, |G^\Lambda (\pm p+k)|$
arises from the trace on the right hand side of the RG equation
when all effective vertices and all but one of the propagators $G^\Lambda$
have been estimated by their maximal values. It
thus determines the maximal possible value of a term on the right hand
side of the RG equation, where the dependence on one external momentum 
is kept. 
In particular, $I_\Lambda$ is directly relevant for the one-loop contributions 
to the flowing four-point vertex. The power counting done above corresponds
to the estimate $I_\Lambda (k) \le \Vert G^\Lambda\Vert \, \snorm{\Lambda}
\le \Lambda^{-1} \snorm{\Lambda}$, so that $\int_\Lambda I_\lambda d\lambda$ 
grows logarithmically in $\Lambda$ for small $\Lambda$.

Since $\Sigma^\lambda =0$,
$S^\Lambda (k) = (ik_0 - \xi_{\bk})^{-1} \, \partial_\Lambda \chi^\Lambda (k)$ 
and 
\begin{equation}\label{Gint}
G^\Lambda (k)
=
\frac{\chi^\Lambda (k)}%
{i k_0 - \xi_{\bk}}
=
G^{\Lambda_0} (k) 
- 
\int_\Lambda^{\Lambda_0} d \lambda\;
S^\lambda (k) .
\end{equation}
The term $G^{\Lambda_0}$ is nonvanishing at large frequencies, but 
not important 
($\Vert G^{\Lambda_0}\Vert \le \Lambda_0^{-1}$, hence a factor 
$\Lambda/\Lambda_0$ smaller than $\Vert G^{\Lambda}\Vert $
when $\Lambda$ gets small. Thus 
$\int d p_0 d^dp |S^\Lambda (p)| \, |G^{\Lambda_0} (p+k)|
\le \snorm{\Lambda} \Lambda_0^{-1}$, hence its integral over $\Lambda$
is bounded by a constant). 
The derivative of the strict cutoff function vanishes unless
$\Lambda (1-\delta) \le |ik_0 - \xi_{\bk}| \le \Lambda$, so
$S^\Lambda (p)$ vanishes unless $|p_0| \le \Lambda$ and 
$\bp $ is in the momentum space shell 
$\mshell{\Lambda} = \left\{ \bk: |\xi_{\bk}| \le \Lambda \right\}$,
and there, $|S^\Lambda (p) |\le \frac{1}{\Lambda^2}$.
The $p_0$-sum in $I_\Lambda$ gives at most $2 \Lambda$, and the 
$\bp$-integral gives the $d$-dimensional volume of the intersection 
$\mshell{\Lambda} \cap (\bk \pm \mshell{\lambda})$
of two momentum space shells, where one is shifted by $\bk$. 
It follows that 
\begin{equation}\label{Iint}
I_\Lambda (k) 
\le 
O(\Lambda_0^{-1})
+
\frac{2}{\Lambda}
\int_\Lambda^{\Lambda_0} \frac{d\lambda}{\lambda^2} \;
\mbox{vol}_d \left(\mshell{\Lambda} \cap (\bk \pm \mshell{\lambda})\right)
.
\end{equation}
This links the scaling behaviour of terms in the RG equation to 
the geometric properties of the Fermi surface. 

Obviously, the volume of the intersection is at most 
as large as the volume of $\mshell{\Lambda}$ itself:
$\mbox{vol}_d (\mshell{\Lambda} \cap (\bk \pm \mshell{\lambda}))
\le \mbox{vol}_d \mshell{\Lambda} \le \mbox{const.} \Lambda$.
Using this in (\ref{Iint}) gives 
the general power counting bound mentioned at the beginning of 
this section, $I_\Lambda (k) \le $ const.$\Lambda^{-1}$. 
Assuming that $\xi_{-\bk} = \xi_{\bk}$, 
this bound is always saturated for $\bk = 0$, 
and also for those $\bk$ for which 
the shift by $\bk$ makes the two shells overlap over a significant
region of the Fermi surface, that is, when $\bk$ is a nesting vector of 
the Fermi surface.

\begin{figure}[ht]
\centerline{\includegraphics[width = 7cm]{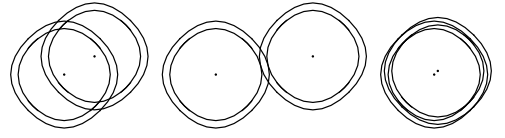}}
\caption{Intersections of a momentum shell around the Fermi surface with
its translate, as arising in loop integrals on the right hand side of the
RG equation. When the Fermi surface is curved, the intersection volume
decreases strongly unless the translating momentum is small.}
\label{fig:schnitte}
\end{figure}

For other values of $\bk$, 
the intersection volume can be much smaller than that of $\mshell{\Lambda}$.
A general definition of non-nesting was given, 
and power counting bounds were derived when it is satisfied, by \textcite{FST1}, 
and extended to the case with Van Hove singularities in \cite{FS1,FS2}.
Here we only cite the result for the case of a strictly convex and positively
curved Fermi surface without Van Hove singularities, discussed also in
the Appendix of \cite{Salmhofer99}. 
In that case, and for $\Lambda \le \lambda \le \vFmin |\bk| $, 
one can show that the volume ratio 
$\frac{{\rm vol}_d\, (\mshell{\Lambda} \cap (\bk \pm \mshell{\lambda}))}%
{{\rm vol}_d\,\mshell{\Lambda}}$
is proportional to 
\begin{eqnarray}
\sfrac{\lambda}{|\bk|\vFmin \; \kappa} 
& \mbox{ if } & 
\bk \not\in \twokF_\lambda 
\label{transversal}\\
\left(\sfrac{\lambda}{\kappa}\right)^{\frac{d-1}{2}}
& \mbox{ if } &
\bk \in\twokF_\lambda.
\label{2kF}
\end{eqnarray}
Here $\vFmin$ is the smallest value of $|\nabla e|$ on the Fermi
surface, $\twokF_\lambda$ is a $O(\lambda)$-neighbourhood
of the set $\left\{2 \bk: \xi_{\bk} = 0 \right\}$ (note that $2\bk$ is 
taken modulo reciprocal lattice vectors), and $\kappa$ denotes the 
minimal curvature on the Fermi surface. 
This is illustrated for $\lambda = \Lambda$ in Fig.~\ref{fig:schnitte}.
In the first case, the intersection is transversal, 
which decreases the intersection volume by the factor in (\ref{transversal}).
The second case corresponds to a $2k_F$-intersection, 
where the curvature in a region of size $\sqrt{\lambda}$ determines 
the intersection volume, corresponding to (\ref{2kF}).
In the third case, $|\bk|$ is so small that the volume of the intersection
is essentially equal to that of $\mshell{\Lambda}$. 

The scale in the flow where the improvements set in is 
determined by the curvature of the Fermi surface, 
because there is really only an improvement if the 
additional factors are smaller than one. 
In cases where the curvature is small on large parts of the Fermi surface,
as in the Hubbard model near to half-filling and at small next-to-nearest hopping,
one thus has an effective nesting at those scales and at those $\bk$ where the quotients
in  (\ref{transversal}) and (\ref{2kF}) are so big that they give a bound
that is bigger than the trivial bound $1$ for the volume ratio.  

Eqs.\  (\ref{transversal}) and (\ref{2kF}) imply that 
for small $|\bk|$, 
\begin{equation}
\int_{0}^{\Lambda_0} I_\Lambda (k) d\Lambda
\le 
\mbox{const. } \log \frac{\Lambda_0}{|\bk| \vFmin}
\end{equation}
(where the constant depends on the curvature of the Fermi surface)
and that the function remains bounded for $|\bk|$ not close to zero
(for details, see \textcite{FST1,Salmhofer98a,Salmhofer99}).
Thus, for convex curved Fermi surfaces, the four-point function can diverge only 
at $\bk = 0$ and there, only logarithmically
(by a similar argument, one can see that it can diverge only at $k_0 =0$).
The particle-particle correction
to the vertex function has exactly this behaviour. In the particle-hole term, 
there is an additional sign cancellation that removes the logarithm. 
The same argument shows that in general,  divergences can occur
only at nesting vectors of the Fermi surface.

\subsubsection{Uniform improvement from overlapping loops}
\label{sssec:OLloops}
An extension of these geometric estimates to two-loop integrals 
of the type $\int d p\; \int d q\; S^\Lambda (p) \, S^{\Lambda'} (q)
S^{\Lambda''} (p\pm q\pm k)$ is very useful for $d \ge 2$:
it is shown in \cite{FST1} that 
in absence of nesting and Van Hove singularities,
such integrals contain a scaling improvement independently of $\bk$. 
Such two-loop integrals associated to graphs with overlapping loops
arise when the RG equation gets iterated; 
the graph classification of \cite{FST1,FST3} shows that in a precise sense,
the overwhelming majority of graphs in the Feynman graph expansion 
contains one or even two such subintegrals, hence becomes subleading
at low scales. 
As is explained in detail in \cite{FST1,Salmhofer98a},
in absence of nesting and Van Hove singularities, this justifies
the particle-particle-ladder aproximation, 
it singles out the 
Hartree-Fock type contributions to the self-energy by scaling arguments, 
and it allows to show that the derivative of the self-energy is 
RG-irrelevant instead of marginal. 

\end{appendix}


\bibliographystyle{apsrmp}


\begin{thebibliography}{324}
\expandafter\ifx\csname natexlab\endcsname\relax\def\natexlab#1{#1}\fi
\expandafter\ifx\csname bibnamefont\endcsname\relax
  \def\bibnamefont#1{#1}\fi
\expandafter\ifx\csname bibfnamefont\endcsname\relax
  \def\bibfnamefont#1{#1}\fi
\expandafter\ifx\csname citenamefont\endcsname\relax
  \def\citenamefont#1{#1}\fi
\expandafter\ifx\csname url\endcsname\relax
  \def\url#1{\texttt{#1}}\fi
\expandafter\ifx\csname urlprefix\endcsname\relax\def\urlprefix{URL }\fi
\providecommand{\bibinfo}[2]{#2}
\providecommand{\eprint}[2][]{\url{#2}}

\bibitem[{\citenamefont{Abanov and Chubukov}(2004)}]{Abanov04}
\bibinfo{author}{\bibnamefont{Abanov}, \bibfnamefont{A.}}, and
  \bibinfo{author}{\bibfnamefont{A.~V.} \bibnamefont{Chubukov}},
  \bibinfo{year}{2004}, \bibinfo{journal}{Phys. Rev. Lett.}
  \textbf{\bibinfo{volume}{93}}, \bibinfo{pages}{255702}.

\bibitem[{\citenamefont{Abanov} \emph{et~al.}(2003)\citenamefont{Abanov,
  Chubukov, and Schmalian}}]{Abanov03}
\bibinfo{author}{\bibnamefont{Abanov}, \bibfnamefont{A.}},
  \bibinfo{author}{\bibfnamefont{A.~V.} \bibnamefont{Chubukov}}, and
  \bibinfo{author}{\bibfnamefont{J.}~\bibnamefont{Schmalian}},
  \bibinfo{year}{2003}, \bibinfo{journal}{Adv. Phys.}
  \textbf{\bibinfo{volume}{52}}, \bibinfo{pages}{119}.

\bibitem[{\citenamefont{Abdesselam and Rivasseau}(1998)}]{Abdesselam98}
\bibinfo{author}{\bibnamefont{Abdesselam}, \bibfnamefont{A.}}, and
  \bibinfo{author}{\bibfnamefont{V.}~\bibnamefont{Rivasseau}},
  \bibinfo{year}{1998}, \bibinfo{journal}{Lett. Math. Phys.}
  \textbf{\bibinfo{volume}{44}}, \bibinfo{pages}{77}.

\bibitem[{\citenamefont{Abrikosov} \emph{et~al.}(1963)\citenamefont{Abrikosov,
  Gorkov, and Dzyaloshinski}}]{Abrikosov63}
\bibinfo{author}{\bibnamefont{Abrikosov}, \bibfnamefont{A.}},
  \bibinfo{author}{\bibfnamefont{L.}~\bibnamefont{Gorkov}}, and
  \bibinfo{author}{\bibfnamefont{I.}~\bibnamefont{Dzyaloshinski}},
  \bibinfo{year}{1963}, \emph{\bibinfo{title}{Methods of Quantum Field Theory
  in Statistical Physics}} (\bibinfo{publisher}{Prentice-Hall, Englewood
  Cliffs, New Jersey}).

\bibitem[{\citenamefont{Andergassen}
  \emph{et~al.}(2006{\natexlab{a}})\citenamefont{Andergassen, Enss, and
  Meden}}]{Andergassen06a}
\bibinfo{author}{\bibnamefont{Andergassen}, \bibfnamefont{S.}},
  \bibinfo{author}{\bibfnamefont{T.}~\bibnamefont{Enss}}, and
  \bibinfo{author}{\bibfnamefont{V.}~\bibnamefont{Meden}},
  \bibinfo{year}{2006}{\natexlab{a}}, \bibinfo{journal}{Phys. Rev. B}
  \textbf{\bibinfo{volume}{73}}, \bibinfo{pages}{153308}.

\bibitem[{\citenamefont{Andergassen}
  \emph{et~al.}(2004)\citenamefont{Andergassen, Enss, Meden, Metzner,
  Schollw{\"o}ck, and Sch{\"o}nhammer}}]{Andergassen04}
\bibinfo{author}{\bibnamefont{Andergassen}, \bibfnamefont{S.}},
  \bibinfo{author}{\bibfnamefont{T.}~\bibnamefont{Enss}},
  \bibinfo{author}{\bibfnamefont{V.}~\bibnamefont{Meden}},
  \bibinfo{author}{\bibfnamefont{W.}~\bibnamefont{Metzner}},
  \bibinfo{author}{\bibfnamefont{U.}~\bibnamefont{Schollw{\"o}ck}}, and
  \bibinfo{author}{\bibfnamefont{K.}~\bibnamefont{Sch{\"o}nhammer}},
  \bibinfo{year}{2004}, \bibinfo{journal}{Phys. Rev. B}
  \textbf{\bibinfo{volume}{70}}, \bibinfo{pages}{075102}.

\bibitem[{\citenamefont{Andergassen}
  \emph{et~al.}(2006{\natexlab{b}})\citenamefont{Andergassen, Enss, Meden,
  Metzner, Schollw{\"o}ck, and Sch{\"o}nhammer}}]{Andergassen06b}
\bibinfo{author}{\bibnamefont{Andergassen}, \bibfnamefont{S.}},
  \bibinfo{author}{\bibfnamefont{T.}~\bibnamefont{Enss}},
  \bibinfo{author}{\bibfnamefont{V.}~\bibnamefont{Meden}},
  \bibinfo{author}{\bibfnamefont{W.}~\bibnamefont{Metzner}},
  \bibinfo{author}{\bibfnamefont{U.}~\bibnamefont{Schollw{\"o}ck}}, and
  \bibinfo{author}{\bibfnamefont{K.}~\bibnamefont{Sch{\"o}nhammer}},
  \bibinfo{year}{2006}{\natexlab{b}}, \bibinfo{journal}{Phys. Rev. B}
  \textbf{\bibinfo{volume}{73}}, \bibinfo{pages}{045125}.

\bibitem[{\citenamefont{Andergassen}
  \emph{et~al.}(2011)\citenamefont{Andergassen, Pletyukhov, Schuricht,
  Schoeller, and Borda}}]{Andergassen11}
\bibinfo{author}{\bibnamefont{Andergassen}, \bibfnamefont{S.}},
  \bibinfo{author}{\bibfnamefont{M.}~\bibnamefont{Pletyukhov}},
  \bibinfo{author}{\bibfnamefont{D.}~\bibnamefont{Schuricht}},
  \bibinfo{author}{\bibfnamefont{H.}~\bibnamefont{Schoeller}}, and
  \bibinfo{author}{\bibfnamefont{L.}~\bibnamefont{Borda}},
  \bibinfo{year}{2011}, \bibinfo{journal}{Phys. Rev. B}
  \textbf{\bibinfo{volume}{83}}, \bibinfo{pages}{205103}.

\bibitem[{\citenamefont{Anders and Schiller}(2005)}]{Anders05}
\bibinfo{author}{\bibnamefont{Anders}, \bibfnamefont{F.}}, and
  \bibinfo{author}{\bibfnamefont{A.}~\bibnamefont{Schiller}},
  \bibinfo{year}{2005}, \bibinfo{journal}{Phys. Rev. Lett.}
  \textbf{\bibinfo{volume}{95}}, \bibinfo{pages}{196801}.

\bibitem[{\citenamefont{Anderson}(1997)}]{Anderson97}
\bibinfo{author}{\bibnamefont{Anderson}, \bibfnamefont{P.~W.}},
  \bibinfo{year}{1997}, \emph{\bibinfo{title}{The Theory of High-Temperature
  Superconductivity}} (\bibinfo{publisher}{Princeton University Press}).

\bibitem[{\citenamefont{Anisimov and Izyumov}(2010)}]{Anisimov10}
\bibinfo{author}{\bibnamefont{Anisimov}, \bibfnamefont{V.}}, and
  \bibinfo{author}{\bibfnamefont{Y.}~\bibnamefont{Izyumov}},
  \bibinfo{year}{2010}, \emph{\bibinfo{title}{Electronic Structure of Strongly
  Correlated Materials}} (\bibinfo{publisher}{Springer, Berlin}).

\bibitem[{\citenamefont{Anisimov} \emph{et~al.}(2009)\citenamefont{Anisimov,
  Korotin, Korotin, Kozhevnikov, Kunes, Shorikov, Skornyakov, and
  Streltsov}}]{Anisimov09}
\bibinfo{author}{\bibnamefont{Anisimov}, \bibfnamefont{V.~I.}},
  \bibinfo{author}{\bibfnamefont{D.~M.} \bibnamefont{Korotin}},
  \bibinfo{author}{\bibfnamefont{M.~A.} \bibnamefont{Korotin}},
  \bibinfo{author}{\bibfnamefont{A.~V.} \bibnamefont{Kozhevnikov}},
  \bibinfo{author}{\bibfnamefont{J.}~\bibnamefont{Kunes}},
  \bibinfo{author}{\bibfnamefont{A.~O.} \bibnamefont{Shorikov}},
  \bibinfo{author}{\bibfnamefont{S.~L.} \bibnamefont{Skornyakov}}, and
  \bibinfo{author}{\bibfnamefont{S.~V.} \bibnamefont{Streltsov}},
  \bibinfo{year}{2009}, \bibinfo{journal}{J. Phys. Cond. Mat.}
  \textbf{\bibinfo{volume}{21}}, \bibinfo{pages}{075602}.

\bibitem[{\citenamefont{Apel and Rice}(1982)}]{Apel82}
\bibinfo{author}{\bibnamefont{Apel}, \bibfnamefont{W.}}, and
  \bibinfo{author}{\bibfnamefont{T.~M.} \bibnamefont{Rice}},
  \bibinfo{year}{1982}, \bibinfo{journal}{Phys. Rev. B}
  \textbf{\bibinfo{volume}{26}}, \bibinfo{pages}{7063}.

\bibitem[{\citenamefont{Aristov} \emph{et~al.}(2010)\citenamefont{Aristov,
  Dmitriev, Gornyi, Kachorovskii, Polyakov, and W{\"o}lfle}}]{Aristov10}
\bibinfo{author}{\bibnamefont{Aristov}, \bibfnamefont{D.~N.}},
  \bibinfo{author}{\bibfnamefont{A.~P.} \bibnamefont{Dmitriev}},
  \bibinfo{author}{\bibfnamefont{I.~V.} \bibnamefont{Gornyi}},
  \bibinfo{author}{\bibfnamefont{V.~Y.} \bibnamefont{Kachorovskii}},
  \bibinfo{author}{\bibfnamefont{D.~G.} \bibnamefont{Polyakov}}, and
  \bibinfo{author}{\bibfnamefont{P.}~\bibnamefont{W{\"o}lfle}},
  \bibinfo{year}{2010}, \bibinfo{journal}{Phys. Rev. Lett.}
  \textbf{\bibinfo{volume}{105}}, \bibinfo{pages}{266404}.

\bibitem[{\citenamefont{Aristov and W{\"o}lfle}(2008)}]{Aristov08}
\bibinfo{author}{\bibnamefont{Aristov}, \bibfnamefont{D.~N.}}, and
  \bibinfo{author}{\bibfnamefont{P.}~\bibnamefont{W{\"o}lfle}},
  \bibinfo{year}{2008}, \bibinfo{journal}{Europhys. Lett.}
  \textbf{\bibinfo{volume}{82}}, \bibinfo{pages}{27001}.

\bibitem[{\citenamefont{Aristov and W{\"o}lfle}(2009)}]{Aristov09}
\bibinfo{author}{\bibnamefont{Aristov}, \bibfnamefont{D.~N.}}, and
  \bibinfo{author}{\bibfnamefont{P.}~\bibnamefont{W{\"o}lfle}},
  \bibinfo{year}{2009}, \bibinfo{journal}{Phys. Rev. B}
  \textbf{\bibinfo{volume}{80}}, \bibinfo{pages}{045109}.

\bibitem[{\citenamefont{Baier} \emph{et~al.}(2004)\citenamefont{Baier, Bick,
  and Wetterich}}]{Baier04}
\bibinfo{author}{\bibnamefont{Baier}, \bibfnamefont{T.}},
  \bibinfo{author}{\bibfnamefont{E.}~\bibnamefont{Bick}}, and
  \bibinfo{author}{\bibfnamefont{C.}~\bibnamefont{Wetterich}},
  \bibinfo{year}{2004}, \bibinfo{journal}{Phys. Rev. B}
  \textbf{\bibinfo{volume}{70}}, \bibinfo{pages}{125111}.

\bibitem[{\citenamefont{Bakrim and Bourbonnais}(2010)}]{Bakrim10}
\bibinfo{author}{\bibnamefont{Bakrim}, \bibfnamefont{H.}}, and
  \bibinfo{author}{\bibfnamefont{C.}~\bibnamefont{Bourbonnais}},
  \bibinfo{year}{2010}, \bibinfo{journal}{Europhys. Lett.}
  \textbf{\bibinfo{volume}{90}}, \bibinfo{pages}{27001}.

\bibitem[{\citenamefont{Balaban} \emph{et~al.}(2010)\citenamefont{Balaban,
  Feldman, Kn{\"{o}}rrer, and Trubowitz}}]{BKFT10}
\bibinfo{author}{\bibnamefont{Balaban}, \bibfnamefont{T.}},
  \bibinfo{author}{\bibfnamefont{J.}~\bibnamefont{Feldman}},
  \bibinfo{author}{\bibfnamefont{H.}~\bibnamefont{Kn{\"{o}}rrer}}, and
  \bibinfo{author}{\bibfnamefont{E.}~\bibnamefont{Trubowitz}},
  \bibinfo{year}{2010}, \bibinfo{journal}{Ann. Henri Poincar\'{e}}
  \textbf{\bibinfo{volume}{11}}, \bibinfo{pages}{151}.

\bibitem[{\citenamefont{Barnab{\'e}-Th{\'e}riault}
  \emph{et~al.}(2005)\citenamefont{Barnab{\'e}-Th{\'e}riault, Sedeki, Meden,
  and Sch{\"o}nhammer}}]{Barnabe05a}
\bibinfo{author}{\bibnamefont{Barnab{\'e}-Th{\'e}riault}, \bibfnamefont{X.}},
  \bibinfo{author}{\bibfnamefont{A.}~\bibnamefont{Sedeki}},
  \bibinfo{author}{\bibfnamefont{V.}~\bibnamefont{Meden}}, and
  \bibinfo{author}{\bibfnamefont{K.}~\bibnamefont{Sch{\"o}nhammer}},
  \bibinfo{year}{2005}, \bibinfo{journal}{Phys. Rev. B}
  \textbf{\bibinfo{volume}{71}}, \bibinfo{pages}{205327}.

\bibitem[{\citenamefont{Barnab\'e-Th\'eriault}
  \emph{et~al.}(2005)\citenamefont{Barnab\'e-Th\'eriault, Sedeki, Meden, and
  Sch{\"o}nhammer}}]{Barnabe05}
\bibinfo{author}{\bibnamefont{Barnab\'e-Th\'eriault}, \bibfnamefont{X.}},
  \bibinfo{author}{\bibfnamefont{A.}~\bibnamefont{Sedeki}},
  \bibinfo{author}{\bibfnamefont{V.}~\bibnamefont{Meden}}, and
  \bibinfo{author}{\bibfnamefont{K.}~\bibnamefont{Sch{\"o}nhammer}},
  \bibinfo{year}{2005}, \bibinfo{journal}{Phys. Rev. Lett.}
  \textbf{\bibinfo{volume}{94}}, \bibinfo{pages}{136405}.

\bibitem[{\citenamefont{Bartosch}
  \emph{et~al.}(2009{\natexlab{a}})\citenamefont{Bartosch, Freire, Cardenas,
  and Kopietz}}]{Bartosch09a}
\bibinfo{author}{\bibnamefont{Bartosch}, \bibfnamefont{L.}},
  \bibinfo{author}{\bibfnamefont{H.}~\bibnamefont{Freire}},
  \bibinfo{author}{\bibfnamefont{J.~J.~R.} \bibnamefont{Cardenas}}, and
  \bibinfo{author}{\bibfnamefont{P.}~\bibnamefont{Kopietz}},
  \bibinfo{year}{2009}{\natexlab{a}}, \bibinfo{journal}{J. Phys.: Condens.
  Matter} \textbf{\bibinfo{volume}{21}}, \bibinfo{pages}{305602}.

\bibitem[{\citenamefont{Bartosch}
  \emph{et~al.}(2009{\natexlab{b}})\citenamefont{Bartosch, Kopietz, and
  Ferraz}}]{Bartosch09}
\bibinfo{author}{\bibnamefont{Bartosch}, \bibfnamefont{L.}},
  \bibinfo{author}{\bibfnamefont{P.}~\bibnamefont{Kopietz}}, and
  \bibinfo{author}{\bibfnamefont{A.}~\bibnamefont{Ferraz}},
  \bibinfo{year}{2009}{\natexlab{b}}, \bibinfo{journal}{Phys. Rev. B}
  \textbf{\bibinfo{volume}{80}}, \bibinfo{pages}{104514}.

\bibitem[{\citenamefont{Bauer} \emph{et~al.}(2011)\citenamefont{Bauer,
  Jakubczyk, and Metzner}}]{Bauer11}
\bibinfo{author}{\bibnamefont{Bauer}, \bibfnamefont{J.}},
  \bibinfo{author}{\bibfnamefont{P.}~\bibnamefont{Jakubczyk}}, and
  \bibinfo{author}{\bibfnamefont{W.}~\bibnamefont{Metzner}},
  \bibinfo{year}{2011}, \bibinfo{journal}{Phys. Rev. B}
  \textbf{\bibinfo{volume}{84}}, \bibinfo{pages}{075122}.

\bibitem[{\citenamefont{Belitz}
  \emph{et~al.}(2001{\natexlab{a}})\citenamefont{Belitz, Kirkpatrick, Mercaldo,
  and Sessions}}]{Belitz01a}
\bibinfo{author}{\bibnamefont{Belitz}, \bibfnamefont{D.}},
  \bibinfo{author}{\bibfnamefont{T.~R.} \bibnamefont{Kirkpatrick}},
  \bibinfo{author}{\bibfnamefont{M.~T.} \bibnamefont{Mercaldo}}, and
  \bibinfo{author}{\bibfnamefont{S.~L.} \bibnamefont{Sessions}},
  \bibinfo{year}{2001}{\natexlab{a}}, \bibinfo{journal}{Phys. Rev. B}
  \textbf{\bibinfo{volume}{63}}, \bibinfo{pages}{174427}.

\bibitem[{\citenamefont{Belitz}
  \emph{et~al.}(2001{\natexlab{b}})\citenamefont{Belitz, Kirkpatrick, Mercaldo,
  and Sessions}}]{Belitz01b}
\bibinfo{author}{\bibnamefont{Belitz}, \bibfnamefont{D.}},
  \bibinfo{author}{\bibfnamefont{T.~R.} \bibnamefont{Kirkpatrick}},
  \bibinfo{author}{\bibfnamefont{M.~T.} \bibnamefont{Mercaldo}}, and
  \bibinfo{author}{\bibfnamefont{S.~L.} \bibnamefont{Sessions}},
  \bibinfo{year}{2001}{\natexlab{b}}, \bibinfo{journal}{Phys. Rev. B}
  \textbf{\bibinfo{volume}{63}}, \bibinfo{pages}{174428}.

\bibitem[{\citenamefont{Belitz} \emph{et~al.}(1997)\citenamefont{Belitz,
  Kirkpatrick, and Vojta}}]{Belitz97}
\bibinfo{author}{\bibnamefont{Belitz}, \bibfnamefont{D.}},
  \bibinfo{author}{\bibfnamefont{T.~R.} \bibnamefont{Kirkpatrick}}, and
  \bibinfo{author}{\bibfnamefont{T.}~\bibnamefont{Vojta}},
  \bibinfo{year}{1997}, \bibinfo{journal}{Phys. Rev. B}
  \textbf{\bibinfo{volume}{55}}, \bibinfo{pages}{9452}.

\bibitem[{\citenamefont{Belitz} \emph{et~al.}(1999)\citenamefont{Belitz,
  Kirkpatrick, and Vojta}}]{Belitz99}
\bibinfo{author}{\bibnamefont{Belitz}, \bibfnamefont{D.}},
  \bibinfo{author}{\bibfnamefont{T.~R.} \bibnamefont{Kirkpatrick}}, and
  \bibinfo{author}{\bibfnamefont{T.}~\bibnamefont{Vojta}},
  \bibinfo{year}{1999}, \bibinfo{journal}{Phys. Rev. Lett.}
  \textbf{\bibinfo{volume}{82}}, \bibinfo{pages}{4707}.

\bibitem[{\citenamefont{Belitz} \emph{et~al.}(2005)\citenamefont{Belitz,
  Kirkpatrick, and Vojta}}]{Belitz05}
\bibinfo{author}{\bibnamefont{Belitz}, \bibfnamefont{D.}},
  \bibinfo{author}{\bibfnamefont{T.~R.} \bibnamefont{Kirkpatrick}}, and
  \bibinfo{author}{\bibfnamefont{T.}~\bibnamefont{Vojta}},
  \bibinfo{year}{2005}, \bibinfo{journal}{Rev. Mod. Phys.}
  \textbf{\bibinfo{volume}{77}}, \bibinfo{pages}{579}.

\bibitem[{\citenamefont{Benfatto and
  Gallavotti}(1990{\natexlab{a}})}]{Benfatto90a}
\bibinfo{author}{\bibnamefont{Benfatto}, \bibfnamefont{G.}}, and
  \bibinfo{author}{\bibfnamefont{G.}~\bibnamefont{Gallavotti}},
  \bibinfo{year}{1990}{\natexlab{a}}, \bibinfo{journal}{Phys. Rev. B}
  \textbf{\bibinfo{volume}{42}}, \bibinfo{pages}{9967}.

\bibitem[{\citenamefont{Benfatto and
  Gallavotti}(1990{\natexlab{b}})}]{Benfatto90b}
\bibinfo{author}{\bibnamefont{Benfatto}, \bibfnamefont{G.}}, and
  \bibinfo{author}{\bibfnamefont{G.}~\bibnamefont{Gallavotti}},
  \bibinfo{year}{1990}{\natexlab{b}}, \bibinfo{journal}{J. Stat. Phys.}
  \textbf{\bibinfo{volume}{59}}, \bibinfo{pages}{541}.

\bibitem[{\citenamefont{Benfatto} \emph{et~al.}(1994)\citenamefont{Benfatto,
  Gallavotti, Procacci, and Scoppola}}]{Benfatto94}
\bibinfo{author}{\bibnamefont{Benfatto}, \bibfnamefont{G.}},
  \bibinfo{author}{\bibfnamefont{G.}~\bibnamefont{Gallavotti}},
  \bibinfo{author}{\bibfnamefont{A.}~\bibnamefont{Procacci}}, and
  \bibinfo{author}{\bibfnamefont{B.}~\bibnamefont{Scoppola}},
  \bibinfo{year}{1994}, \bibinfo{journal}{Comm. Math. Phys.}
  \textbf{\bibinfo{volume}{160}}, \bibinfo{pages}{93}.

\bibitem[{\citenamefont{Benfatto} \emph{et~al.}(2006)\citenamefont{Benfatto,
  Giuliani, and Mastropietro}}]{BGM06}
\bibinfo{author}{\bibnamefont{Benfatto}, \bibfnamefont{G.}},
  \bibinfo{author}{\bibfnamefont{A.}~\bibnamefont{Giuliani}}, and
  \bibinfo{author}{\bibfnamefont{V.}~\bibnamefont{Mastropietro}},
  \bibinfo{year}{2006}, \bibinfo{journal}{Ann. Henri Poincar\' e}
  \textbf{\bibinfo{volume}{7}}, \bibinfo{pages}{809}.

\bibitem[{\citenamefont{Berges} \emph{et~al.}(2002)\citenamefont{Berges,
  Tetradis, and Wetterich}}]{Berges02}
\bibinfo{author}{\bibnamefont{Berges}, \bibfnamefont{J.}},
  \bibinfo{author}{\bibfnamefont{N.}~\bibnamefont{Tetradis}}, and
  \bibinfo{author}{\bibfnamefont{C.}~\bibnamefont{Wetterich}},
  \bibinfo{year}{2002}, \bibinfo{journal}{Phys. Rep.}
  \textbf{\bibinfo{volume}{363}}, \bibinfo{pages}{223}.

\bibitem[{\citenamefont{Binz} \emph{et~al.}(2002)\citenamefont{Binz, Baeriswyl,
  and Dou\c{c}ot}}]{Binz02}
\bibinfo{author}{\bibnamefont{Binz}, \bibfnamefont{B.}},
  \bibinfo{author}{\bibfnamefont{D.}~\bibnamefont{Baeriswyl}}, and
  \bibinfo{author}{\bibfnamefont{B.}~\bibnamefont{Dou\c{c}ot}},
  \bibinfo{year}{2002}, \bibinfo{journal}{Eur. Phys. J. B}
  \textbf{\bibinfo{volume}{25}}, \bibinfo{pages}{69}.

\bibitem[{\citenamefont{Binz} \emph{et~al.}(2003)\citenamefont{Binz, Baeriswyl,
  and Doucot}}]{Binz03}
\bibinfo{author}{\bibnamefont{Binz}, \bibfnamefont{B.}},
  \bibinfo{author}{\bibfnamefont{D.}~\bibnamefont{Baeriswyl}}, and
  \bibinfo{author}{\bibfnamefont{B.}~\bibnamefont{Doucot}},
  \bibinfo{year}{2003}, \bibinfo{journal}{Ann. Phys. (Leipzig)}
  \textbf{\bibinfo{volume}{12}}, \bibinfo{pages}{704}.

\bibitem[{\citenamefont{Birse} \emph{et~al.}(2005)\citenamefont{Birse, Krippa,
  McGovern, and Walet}}]{Birse05}
\bibinfo{author}{\bibnamefont{Birse}, \bibfnamefont{M.~C.}},
  \bibinfo{author}{\bibfnamefont{B.}~\bibnamefont{Krippa}},
  \bibinfo{author}{\bibfnamefont{J.~A.} \bibnamefont{McGovern}}, and
  \bibinfo{author}{\bibfnamefont{N.~R.} \bibnamefont{Walet}},
  \bibinfo{year}{2005}, \bibinfo{journal}{Phys. Lett. B}
  \textbf{\bibinfo{volume}{605}}, \bibinfo{pages}{287}.

\bibitem[{\citenamefont{Blaizot} \emph{et~al.}(2005)\citenamefont{Blaizot,
  Mendez-Galain, and Wschebor}}]{Blaizot05}
\bibinfo{author}{\bibnamefont{Blaizot}, \bibfnamefont{J.-P.}},
  \bibinfo{author}{\bibfnamefont{R.}~\bibnamefont{Mendez-Galain}}, and
  \bibinfo{author}{\bibfnamefont{N.}~\bibnamefont{Wschebor}},
  \bibinfo{year}{2005}, \bibinfo{journal}{Phys. Lett. B}
  \textbf{\bibinfo{volume}{632}}, \bibinfo{pages}{571}.

\bibitem[{\citenamefont{Borda} \emph{et~al.}(2007)\citenamefont{Borda,
  Vlad{\'a}r, and Zawadowski}}]{Borda07}
\bibinfo{author}{\bibnamefont{Borda}, \bibfnamefont{L.}},
  \bibinfo{author}{\bibfnamefont{K.}~\bibnamefont{Vlad{\'a}r}}, and
  \bibinfo{author}{\bibfnamefont{A.}~\bibnamefont{Zawadowski}},
  \bibinfo{year}{2007}, \bibinfo{journal}{Phys. Rev. B}
  \textbf{\bibinfo{volume}{75}}, \bibinfo{pages}{125107}.

\bibitem[{\citenamefont{Borda and Zawadowski}(2010)}]{Borda10}
\bibinfo{author}{\bibnamefont{Borda}, \bibfnamefont{L.}}, and
  \bibinfo{author}{\bibfnamefont{A.}~\bibnamefont{Zawadowski}},
  \bibinfo{year}{2010}, \bibinfo{journal}{Phys. Rev. B}
  \textbf{\bibinfo{volume}{81}}, \bibinfo{pages}{153303}.

\bibitem[{\citenamefont{Boulat} \emph{et~al.}(2008)\citenamefont{Boulat,
  Saleur, and Schmitteckert}}]{Boulat08}
\bibinfo{author}{\bibnamefont{Boulat}, \bibfnamefont{E.}},
  \bibinfo{author}{\bibfnamefont{H.}~\bibnamefont{Saleur}}, and
  \bibinfo{author}{\bibfnamefont{P.}~\bibnamefont{Schmitteckert}},
  \bibinfo{year}{2008}, \bibinfo{journal}{Phys. Rev. Lett.}
  \textbf{\bibinfo{volume}{101}}, \bibinfo{pages}{140601}.

\bibitem[{\citenamefont{Brydges and Wright}(1988)}]{Brydges88}
\bibinfo{author}{\bibnamefont{Brydges}, \bibfnamefont{D.~C.}}, and
  \bibinfo{author}{\bibfnamefont{J.}~\bibnamefont{Wright}},
  \bibinfo{year}{1988}, \bibinfo{journal}{J. Stat. Phys.}
  \textbf{\bibinfo{volume}{51}}, \bibinfo{pages}{435}.

\bibitem[{\citenamefont{B{\"u}ttiker}(1986)}]{Buettiker86}
\bibinfo{author}{\bibnamefont{B{\"u}ttiker}, \bibfnamefont{M.}},
  \bibinfo{year}{1986}, \bibinfo{journal}{Phys. Rev. Lett.}
  \textbf{\bibinfo{volume}{57}}, \bibinfo{pages}{1761}.

\bibitem[{\citenamefont{C.~Karrasch and Meden}(2008)}]{Karrasch08b}
\bibinfo{author}{\bibnamefont{C.~Karrasch}, \bibfnamefont{A.~O.}}, and
  \bibinfo{author}{\bibfnamefont{V.}~\bibnamefont{Meden}},
  \bibinfo{year}{2008}, \bibinfo{journal}{Phys. Rev. B}
  \textbf{\bibinfo{volume}{77}}, \bibinfo{pages}{024571}.

\bibitem[{\citenamefont{Chakravarty}
  \emph{et~al.}(2001)\citenamefont{Chakravarty, Laughlin, Morr, and
  Nayak}}]{Nayak01}
\bibinfo{author}{\bibnamefont{Chakravarty}, \bibfnamefont{S.}},
  \bibinfo{author}{\bibfnamefont{R.~B.} \bibnamefont{Laughlin}},
  \bibinfo{author}{\bibfnamefont{D.~K.} \bibnamefont{Morr}}, and
  \bibinfo{author}{\bibfnamefont{C.}~\bibnamefont{Nayak}},
  \bibinfo{year}{2001}, \bibinfo{journal}{Phys. Rev. B}
  \textbf{\bibinfo{volume}{63}}, \bibinfo{pages}{94503}.

\bibitem[{\citenamefont{Chamon} \emph{et~al.}(2003)\citenamefont{Chamon,
  Oshikawa, and Affleck}}]{Chamon03}
\bibinfo{author}{\bibnamefont{Chamon}, \bibfnamefont{C.}},
  \bibinfo{author}{\bibfnamefont{M.}~\bibnamefont{Oshikawa}}, and
  \bibinfo{author}{\bibfnamefont{I.}~\bibnamefont{Affleck}},
  \bibinfo{year}{2003}, \bibinfo{journal}{Phys. Rev. Lett.}
  \textbf{\bibinfo{volume}{91}}, \bibinfo{pages}{206403}.

\bibitem[{\citenamefont{Chen} \emph{et~al.}(2002)\citenamefont{Chen,
  Trauzettel, and Egger}}]{Chen02}
\bibinfo{author}{\bibnamefont{Chen}, \bibfnamefont{S.}},
  \bibinfo{author}{\bibfnamefont{B.}~\bibnamefont{Trauzettel}}, and
  \bibinfo{author}{\bibfnamefont{R.}~\bibnamefont{Egger}},
  \bibinfo{year}{2002}, \bibinfo{journal}{Phys. Rev. Lett.}
  \textbf{\bibinfo{volume}{89}}, \bibinfo{pages}{226404}.

\bibitem[{\citenamefont{Chitov and S{\'e}n{\'e}chal}(1995)}]{Chitov95}
\bibinfo{author}{\bibnamefont{Chitov}, \bibfnamefont{G.~Y.}}, and
  \bibinfo{author}{\bibfnamefont{D.}~\bibnamefont{S{\'e}n{\'e}chal}},
  \bibinfo{year}{1995}, \bibinfo{journal}{Phys. Rev. B}
  \textbf{\bibinfo{volume}{52}}, \bibinfo{pages}{13487}.

\bibitem[{\citenamefont{Chubukov} \emph{et~al.}(2008)\citenamefont{Chubukov,
  Efremov, and Eremin}}]{Chubukov08}
\bibinfo{author}{\bibnamefont{Chubukov}, \bibfnamefont{A.~V.}},
  \bibinfo{author}{\bibfnamefont{D.}~\bibnamefont{Efremov}}, and
  \bibinfo{author}{\bibfnamefont{I.}~\bibnamefont{Eremin}},
  \bibinfo{year}{2008}, \bibinfo{journal}{Phys. Rev. B}
  \textbf{\bibinfo{volume}{78}}, \bibinfo{pages}{134512}.

\bibitem[{\citenamefont{Daghofer} \emph{et~al.}(2010)\citenamefont{Daghofer,
  Nicholson, Moreo, and Dagotto}}]{Daghofer10}
\bibinfo{author}{\bibnamefont{Daghofer}, \bibfnamefont{M.}},
  \bibinfo{author}{\bibfnamefont{A.}~\bibnamefont{Nicholson}},
  \bibinfo{author}{\bibfnamefont{A.}~\bibnamefont{Moreo}}, and
  \bibinfo{author}{\bibfnamefont{E.}~\bibnamefont{Dagotto}},
  \bibinfo{year}{2010}, \bibinfo{journal}{Phys. Rev. B}
  \textbf{\bibinfo{volume}{81}}, \bibinfo{pages}{014511}.

\bibitem[{\citenamefont{von Delft and Schoeller}(1998)}]{vonDelft98}
\bibinfo{author}{\bibnamefont{von Delft}, \bibfnamefont{J.}}, and
  \bibinfo{author}{\bibfnamefont{H.}~\bibnamefont{Schoeller}},
  \bibinfo{year}{1998}, \bibinfo{journal}{Annalen der Physik}
  \textbf{\bibinfo{volume}{7}}, \bibinfo{pages}{225}.

\bibitem[{\citenamefont{Deshpande} \emph{et~al.}(2010)\citenamefont{Deshpande,
  Bockrath, Glazman, and Yacoby}}]{Deshpande10}
\bibinfo{author}{\bibnamefont{Deshpande}, \bibfnamefont{V.~V.}},
  \bibinfo{author}{\bibfnamefont{M.}~\bibnamefont{Bockrath}},
  \bibinfo{author}{\bibfnamefont{L.~I.} \bibnamefont{Glazman}}, and
  \bibinfo{author}{\bibfnamefont{A.}~\bibnamefont{Yacoby}},
  \bibinfo{year}{2010}, \bibinfo{journal}{Nature}
  \textbf{\bibinfo{volume}{464}}, \bibinfo{pages}{209}.

\bibitem[{\citenamefont{Diehl} \emph{et~al.}(2007)\citenamefont{Diehl, Gies,
  Pawlowski, and Wetterich}}]{Diehl07a}
\bibinfo{author}{\bibnamefont{Diehl}, \bibfnamefont{S.}},
  \bibinfo{author}{\bibfnamefont{H.}~\bibnamefont{Gies}},
  \bibinfo{author}{\bibfnamefont{J.}~\bibnamefont{Pawlowski}}, and
  \bibinfo{author}{\bibfnamefont{C.}~\bibnamefont{Wetterich}},
  \bibinfo{year}{2007}, \bibinfo{journal}{Phys. Rev. A}
  \textbf{\bibinfo{volume}{76}}, \bibinfo{pages}{021602(R)}.

\bibitem[{\citenamefont{Disertori and Rivasseau}(2000)}]{Disertori00}
\bibinfo{author}{\bibnamefont{Disertori}, \bibfnamefont{M.}}, and
  \bibinfo{author}{\bibfnamefont{V.}~\bibnamefont{Rivasseau}},
  \bibinfo{year}{2000}, \bibinfo{journal}{Comm. Math. Phys.}
  \textbf{\bibinfo{volume}{215}}, \bibinfo{pages}{291}.

\bibitem[{\citenamefont{Doyon}(2007)}]{Doyon07}
\bibinfo{author}{\bibnamefont{Doyon}, \bibfnamefont{B.}}, \bibinfo{year}{2007},
  \bibinfo{journal}{Phys. Rev. Lett.} \textbf{\bibinfo{volume}{99}},
  \bibinfo{pages}{076806}.

\bibitem[{\citenamefont{Dzyaloshinskii}(1987)}]{Dzyaloshinskii87}
\bibinfo{author}{\bibnamefont{Dzyaloshinskii}, \bibfnamefont{I.~E.}},
  \bibinfo{year}{1987}, \bibinfo{journal}{Sov. Phys. JETP}
  \textbf{\bibinfo{volume}{66}}, \bibinfo{pages}{848}.

\bibitem[{\citenamefont{Dzyaloshinskii and Larkin}(1974)}]{Dzyaloshinskii74}
\bibinfo{author}{\bibnamefont{Dzyaloshinskii}, \bibfnamefont{I.~E.}}, and
  \bibinfo{author}{\bibfnamefont{A.~I.} \bibnamefont{Larkin}},
  \bibinfo{year}{1974}, \bibinfo{journal}{Sov. Phys.-JETP}
  \textbf{\bibinfo{volume}{38}}, \bibinfo{pages}{202}.

\bibitem[{\citenamefont{Eberlein and Metzner}(2010)}]{Eberlein10}
\bibinfo{author}{\bibnamefont{Eberlein}, \bibfnamefont{A.}}, and
  \bibinfo{author}{\bibfnamefont{W.}~\bibnamefont{Metzner}},
  \bibinfo{year}{2010}, \bibinfo{journal}{Prog. Theor. Phys.}
  \textbf{\bibinfo{volume}{124}}, \bibinfo{pages}{471}.

\bibitem[{\citenamefont{Egger and Grabert}(1995)}]{Egger95}
\bibinfo{author}{\bibnamefont{Egger}, \bibfnamefont{R.}}, and
  \bibinfo{author}{\bibfnamefont{H.}~\bibnamefont{Grabert}},
  \bibinfo{year}{1995}, \bibinfo{journal}{Phys. Rev. Lett.}
  \textbf{\bibinfo{volume}{75}}, \bibinfo{pages}{3505}.

\bibitem[{\citenamefont{Egger and Grabert}(1997)}]{Egger97}
\bibinfo{author}{\bibnamefont{Egger}, \bibfnamefont{R.}}, and
  \bibinfo{author}{\bibfnamefont{H.}~\bibnamefont{Grabert}},
  \bibinfo{year}{1997}, \bibinfo{journal}{Phys. Rev. Lett.}
  \textbf{\bibinfo{volume}{79}}, \bibinfo{pages}{3463}.

\bibitem[{\citenamefont{Egger} \emph{et~al.}(2000)\citenamefont{Egger, Grabert,
  Koutouza, Saleur, and Siano}}]{Egger00}
\bibinfo{author}{\bibnamefont{Egger}, \bibfnamefont{R.}},
  \bibinfo{author}{\bibfnamefont{H.}~\bibnamefont{Grabert}},
  \bibinfo{author}{\bibfnamefont{A.}~\bibnamefont{Koutouza}},
  \bibinfo{author}{\bibfnamefont{H.}~\bibnamefont{Saleur}}, and
  \bibinfo{author}{\bibfnamefont{F.}~\bibnamefont{Siano}},
  \bibinfo{year}{2000}, \bibinfo{journal}{Phys. Rev. Lett.}
  \textbf{\bibinfo{volume}{84}}, \bibinfo{pages}{3682}.

\bibitem[{\citenamefont{Eichler} \emph{et~al.}(2009)\citenamefont{Eichler,
  Deblock, Weiss, Karrasch, Meden, Sch{\"o}nenberger, and
  Bouchiat}}]{Eichler09}
\bibinfo{author}{\bibnamefont{Eichler}, \bibfnamefont{A.}},
  \bibinfo{author}{\bibfnamefont{R.}~\bibnamefont{Deblock}},
  \bibinfo{author}{\bibfnamefont{M.}~\bibnamefont{Weiss}},
  \bibinfo{author}{\bibfnamefont{C.}~\bibnamefont{Karrasch}},
  \bibinfo{author}{\bibfnamefont{V.}~\bibnamefont{Meden}},
  \bibinfo{author}{\bibfnamefont{C.}~\bibnamefont{Sch{\"o}nenberger}}, and
  \bibinfo{author}{\bibfnamefont{H.}~\bibnamefont{Bouchiat}},
  \bibinfo{year}{2009}, \bibinfo{journal}{Phys. Rev. B}
  \textbf{\bibinfo{volume}{79}}, \bibinfo{pages}{161407(R)}.

\bibitem[{\citenamefont{Ellwanger and Wetterich}(1994)}]{Ellwanger94}
\bibinfo{author}{\bibnamefont{Ellwanger}, \bibfnamefont{U.}}, and
  \bibinfo{author}{\bibfnamefont{C.}~\bibnamefont{Wetterich}},
  \bibinfo{year}{1994}, \bibinfo{journal}{Nucl. Phys. B}
  \textbf{\bibinfo{volume}{423}}, \bibinfo{pages}{137}.

\bibitem[{\citenamefont{Enss}(2005)}]{Enss05a}
\bibinfo{author}{\bibnamefont{Enss}, \bibfnamefont{T.}}, \bibinfo{year}{2005},
  \emph{\bibinfo{title}{Ph.D. thesis}} (\bibinfo{publisher}{University
  Stuttgart}).

\bibitem[{\citenamefont{Enss} \emph{et~al.}(2005)\citenamefont{Enss, Meden,
  Andergassen, Barnab\'e-Th\'eriault, Metzner, and Sch{\"o}nhammer}}]{Enss05b}
\bibinfo{author}{\bibnamefont{Enss}, \bibfnamefont{T.}},
  \bibinfo{author}{\bibfnamefont{V.}~\bibnamefont{Meden}},
  \bibinfo{author}{\bibfnamefont{S.}~\bibnamefont{Andergassen}},
  \bibinfo{author}{\bibfnamefont{X.}~\bibnamefont{Barnab\'e-Th\'eriault}},
  \bibinfo{author}{\bibfnamefont{W.}~\bibnamefont{Metzner}}, and
  \bibinfo{author}{\bibfnamefont{K.}~\bibnamefont{Sch{\"o}nhammer}},
  \bibinfo{year}{2005}, \bibinfo{journal}{Phys. Rev. B}
  \textbf{\bibinfo{volume}{71}}, \bibinfo{pages}{155401}.

\bibitem[{\citenamefont{Feldman}
  \emph{et~al.}(1998{\natexlab{a}})\citenamefont{Feldman, Kn{\"{o}}rrer, and
  Trubowitz}}]{FKT98}
\bibinfo{author}{\bibnamefont{Feldman}, \bibfnamefont{J.}},
  \bibinfo{author}{\bibfnamefont{H.}~\bibnamefont{Kn{\"{o}}rrer}}, and
  \bibinfo{author}{\bibfnamefont{E.}~\bibnamefont{Trubowitz}},
  \bibinfo{year}{1998}{\natexlab{a}}, \bibinfo{journal}{Commun. Math. Phys.}
  \textbf{\bibinfo{volume}{195}}, \bibinfo{pages}{465}.

\bibitem[{\citenamefont{Feldman} \emph{et~al.}(2002)\citenamefont{Feldman,
  Kn{\"{o}}rrer, and Trubowitz}}]{FKT02}
\bibinfo{author}{\bibnamefont{Feldman}, \bibfnamefont{J.}},
  \bibinfo{author}{\bibfnamefont{H.}~\bibnamefont{Kn{\"{o}}rrer}}, and
  \bibinfo{author}{\bibfnamefont{E.}~\bibnamefont{Trubowitz}},
  \bibinfo{year}{2002}, \emph{\bibinfo{title}{Fermionic Functional Integrals
  and the Renormalization Group}} (\bibinfo{publisher}{American Mathematical
  Society, Providence, RI}).

\bibitem[{\citenamefont{Feldman} \emph{et~al.}(2003)\citenamefont{Feldman,
  Kn{\"{o}}rrer, and Trubowitz}}]{FKT03}
\bibinfo{author}{\bibnamefont{Feldman}, \bibfnamefont{J.}},
  \bibinfo{author}{\bibfnamefont{H.}~\bibnamefont{Kn{\"{o}}rrer}}, and
  \bibinfo{author}{\bibfnamefont{E.}~\bibnamefont{Trubowitz}},
  \bibinfo{year}{2003}, \bibinfo{journal}{Rev. Math. Phys.}
  \textbf{\bibinfo{volume}{15}}, \bibinfo{pages}{949,995,1039,1121}.

\bibitem[{\citenamefont{Feldman} \emph{et~al.}(2004)\citenamefont{Feldman,
  Kn{\"{o}}rrer, and Trubowitz}}]{FKT04}
\bibinfo{author}{\bibnamefont{Feldman}, \bibfnamefont{J.}},
  \bibinfo{author}{\bibfnamefont{H.}~\bibnamefont{Kn{\"{o}}rrer}}, and
  \bibinfo{author}{\bibfnamefont{E.}~\bibnamefont{Trubowitz}},
  \bibinfo{year}{2004}, \bibinfo{journal}{Commun. Math. Phys.}
  \textbf{\bibinfo{volume}{247}}, \bibinfo{pages}{1,49,113,179,195,243}.

\bibitem[{\citenamefont{Feldman} \emph{et~al.}(1986)\citenamefont{Feldman,
  Magnen, Rivasseau, and S\'{e}n\'{e}or}}]{FMRS}
\bibinfo{author}{\bibnamefont{Feldman}, \bibfnamefont{J.}},
  \bibinfo{author}{\bibfnamefont{J.}~\bibnamefont{Magnen}},
  \bibinfo{author}{\bibfnamefont{V.}~\bibnamefont{Rivasseau}}, and
  \bibinfo{author}{\bibfnamefont{R.}~\bibnamefont{S\'{e}n\'{e}or}},
  \bibinfo{year}{1986}, \bibinfo{journal}{Commun. Math. Phys.}
  \textbf{\bibinfo{volume}{103}}, \bibinfo{pages}{67}.

\bibitem[{\citenamefont{Feldman} \emph{et~al.}(1992)\citenamefont{Feldman,
  Magnen, Rivasseau, and Trubowitz}}]{FMRT92}
\bibinfo{author}{\bibnamefont{Feldman}, \bibfnamefont{J.}},
  \bibinfo{author}{\bibfnamefont{J.}~\bibnamefont{Magnen}},
  \bibinfo{author}{\bibfnamefont{V.}~\bibnamefont{Rivasseau}}, and
  \bibinfo{author}{\bibfnamefont{E.}~\bibnamefont{Trubowitz}},
  \bibinfo{year}{1992}, \bibinfo{journal}{Helv. Physica Acta}
  \textbf{\bibinfo{volume}{65}}, \bibinfo{pages}{679}.

\bibitem[{\citenamefont{Feldman and Salmhofer}(2008{\natexlab{a}})}]{FS1}
\bibinfo{author}{\bibnamefont{Feldman}, \bibfnamefont{J.}}, and
  \bibinfo{author}{\bibfnamefont{M.}~\bibnamefont{Salmhofer}},
  \bibinfo{year}{2008}{\natexlab{a}}, \bibinfo{journal}{Rev. Math. Phys.}
  \textbf{\bibinfo{volume}{20}}, \bibinfo{pages}{233}.

\bibitem[{\citenamefont{Feldman and Salmhofer}(2008{\natexlab{b}})}]{FS2}
\bibinfo{author}{\bibnamefont{Feldman}, \bibfnamefont{J.}}, and
  \bibinfo{author}{\bibfnamefont{M.}~\bibnamefont{Salmhofer}},
  \bibinfo{year}{2008}{\natexlab{b}}, \bibinfo{journal}{Rev. Math. Phys.}
  \textbf{\bibinfo{volume}{20}}, \bibinfo{pages}{275}.

\bibitem[{\citenamefont{Feldman} \emph{et~al.}(1996)\citenamefont{Feldman,
  Salmhofer, and Trubowitz}}]{FST1}
\bibinfo{author}{\bibnamefont{Feldman}, \bibfnamefont{J.}},
  \bibinfo{author}{\bibfnamefont{M.}~\bibnamefont{Salmhofer}}, and
  \bibinfo{author}{\bibfnamefont{E.}~\bibnamefont{Trubowitz}},
  \bibinfo{year}{1996}, \bibinfo{journal}{J. Stat. Phys.}
  \textbf{\bibinfo{volume}{84}}, \bibinfo{pages}{1209}.

\bibitem[{\citenamefont{Feldman}
  \emph{et~al.}(1998{\natexlab{b}})\citenamefont{Feldman, Salmhofer, and
  Trubowitz}}]{FST2}
\bibinfo{author}{\bibnamefont{Feldman}, \bibfnamefont{J.}},
  \bibinfo{author}{\bibfnamefont{M.}~\bibnamefont{Salmhofer}}, and
  \bibinfo{author}{\bibfnamefont{E.}~\bibnamefont{Trubowitz}},
  \bibinfo{year}{1998}{\natexlab{b}}, \bibinfo{journal}{Comm. Pure Appl. Math.}
  \textbf{\bibinfo{volume}{LI}}, \bibinfo{pages}{1133}.

\bibitem[{\citenamefont{Feldman} \emph{et~al.}(1999)\citenamefont{Feldman,
  Salmhofer, and Trubowitz}}]{FST3}
\bibinfo{author}{\bibnamefont{Feldman}, \bibfnamefont{J.}},
  \bibinfo{author}{\bibfnamefont{M.}~\bibnamefont{Salmhofer}}, and
  \bibinfo{author}{\bibfnamefont{E.}~\bibnamefont{Trubowitz}},
  \bibinfo{year}{1999}, \bibinfo{journal}{Comm. Pure Appl. Math.}
  \textbf{\bibinfo{volume}{LII}}, \bibinfo{pages}{273}.

\bibitem[{\citenamefont{Feldman} \emph{et~al.}(2000)\citenamefont{Feldman,
  Salmhofer, and Trubowitz}}]{FST4}
\bibinfo{author}{\bibnamefont{Feldman}, \bibfnamefont{J.}},
  \bibinfo{author}{\bibfnamefont{M.}~\bibnamefont{Salmhofer}}, and
  \bibinfo{author}{\bibfnamefont{E.}~\bibnamefont{Trubowitz}},
  \bibinfo{year}{2000}, \bibinfo{journal}{Comm. Pure Appl. Math.}
  \textbf{\bibinfo{volume}{LIII}}, \bibinfo{pages}{1350}.

\bibitem[{\citenamefont{Feldman and Trubowitz}(1990)}]{FT1}
\bibinfo{author}{\bibnamefont{Feldman}, \bibfnamefont{J.}}, and
  \bibinfo{author}{\bibfnamefont{E.}~\bibnamefont{Trubowitz}},
  \bibinfo{year}{1990}, \bibinfo{journal}{Helv. Physica Acta}
  \textbf{\bibinfo{volume}{63}}, \bibinfo{pages}{157}.

\bibitem[{\citenamefont{Feldman and Trubowitz}(1991)}]{FT2}
\bibinfo{author}{\bibnamefont{Feldman}, \bibfnamefont{J.}}, and
  \bibinfo{author}{\bibfnamefont{E.}~\bibnamefont{Trubowitz}},
  \bibinfo{year}{1991}, \bibinfo{journal}{Helv. Physica Acta}
  \textbf{\bibinfo{volume}{64}}, \bibinfo{pages}{213}.

\bibitem[{\citenamefont{Fendley} \emph{et~al.}(1995)\citenamefont{Fendley,
  Ludwig, and Saleur}}]{Fendley95}
\bibinfo{author}{\bibnamefont{Fendley}, \bibfnamefont{P.}},
  \bibinfo{author}{\bibfnamefont{A.~W.~W.} \bibnamefont{Ludwig}}, and
  \bibinfo{author}{\bibfnamefont{H.}~\bibnamefont{Saleur}},
  \bibinfo{year}{1995}, \bibinfo{journal}{Phys. Rev. Lett.}
  \textbf{\bibinfo{volume}{74}}, \bibinfo{pages}{3005}.

\bibitem[{\citenamefont{Floerchinger}
  \emph{et~al.}(2008)\citenamefont{Floerchinger, Scherer, Diehl, and
  Wetterich}}]{Floerchinger08a}
\bibinfo{author}{\bibnamefont{Floerchinger}, \bibfnamefont{S.}},
  \bibinfo{author}{\bibfnamefont{M.}~\bibnamefont{Scherer}},
  \bibinfo{author}{\bibfnamefont{S.}~\bibnamefont{Diehl}}, and
  \bibinfo{author}{\bibfnamefont{C.}~\bibnamefont{Wetterich}},
  \bibinfo{year}{2008}, \bibinfo{journal}{Phys. Rev. B}
  \textbf{\bibinfo{volume}{78}}, \bibinfo{pages}{174528}.

\bibitem[{\citenamefont{Floerchinger and Wetterich}(2009)}]{Floerchinger09d}
\bibinfo{author}{\bibnamefont{Floerchinger}, \bibfnamefont{S.}}, and
  \bibinfo{author}{\bibfnamefont{C.}~\bibnamefont{Wetterich}},
  \bibinfo{year}{2009}, \bibinfo{journal}{Phys. Lett. B}
  \textbf{\bibinfo{volume}{680}}, \bibinfo{pages}{371}.

\bibitem[{\citenamefont{Fradkin}(1991)}]{Fradkin91}
\bibinfo{author}{\bibnamefont{Fradkin}, \bibfnamefont{E.}},
  \bibinfo{year}{1991}, \emph{\bibinfo{title}{Field Theories of Condensed
  Matter Systems}} (\bibinfo{publisher}{Addison-Wesley, Redwood City,
  California}).

\bibitem[{\citenamefont{Fradkin} \emph{et~al.}(2010)\citenamefont{Fradkin,
  Kivelson, Lawler, Eisenstein, and Mackenzie}}]{Fradkin10}
\bibinfo{author}{\bibnamefont{Fradkin}, \bibfnamefont{E.}},
  \bibinfo{author}{\bibfnamefont{S.~A.} \bibnamefont{Kivelson}},
  \bibinfo{author}{\bibfnamefont{M.~J.} \bibnamefont{Lawler}},
  \bibinfo{author}{\bibfnamefont{J.~P.} \bibnamefont{Eisenstein}}, and
  \bibinfo{author}{\bibfnamefont{A.~P.} \bibnamefont{Mackenzie}},
  \bibinfo{year}{2010}, \bibinfo{journal}{Annu. Rev. Condens. Matter Phys.}
  \textbf{\bibinfo{volume}{1}}, \bibinfo{pages}{153}.

\bibitem[{\citenamefont{Friederich}
  \emph{et~al.}(2010)\citenamefont{Friederich, Krahl, and
  Wetterich}}]{Friederich10}
\bibinfo{author}{\bibnamefont{Friederich}, \bibfnamefont{S.}},
  \bibinfo{author}{\bibfnamefont{H.~C.} \bibnamefont{Krahl}}, and
  \bibinfo{author}{\bibfnamefont{C.}~\bibnamefont{Wetterich}},
  \bibinfo{year}{2010}, \bibinfo{journal}{Phys. Rev. B}
  \textbf{\bibinfo{volume}{81}}, \bibinfo{pages}{235108}.

\bibitem[{\citenamefont{Friederich}
  \emph{et~al.}(2011)\citenamefont{Friederich, Krahl, and
  Wetterich}}]{Friederich11}
\bibinfo{author}{\bibnamefont{Friederich}, \bibfnamefont{S.}},
  \bibinfo{author}{\bibfnamefont{H.~C.} \bibnamefont{Krahl}}, and
  \bibinfo{author}{\bibfnamefont{C.}~\bibnamefont{Wetterich}},
  \bibinfo{year}{2011}, \bibinfo{journal}{Phys. Rev. B}
  \textbf{\bibinfo{volume}{83}}, \bibinfo{pages}{155125}.

\bibitem[{\citenamefont{Fulde}(1991)}]{Fulde91}
\bibinfo{author}{\bibnamefont{Fulde}, \bibfnamefont{P.}}, \bibinfo{year}{1991},
  \emph{\bibinfo{title}{Electron Correlations in Molecules and Solids}}
  (\bibinfo{publisher}{Springer, Heidelberg}).

\bibitem[{\citenamefont{Furukawa} \emph{et~al.}(1998)\citenamefont{Furukawa,
  Rice, and Salmhofer}}]{Furukawa98}
\bibinfo{author}{\bibnamefont{Furukawa}, \bibfnamefont{N.}},
  \bibinfo{author}{\bibfnamefont{T.~M.} \bibnamefont{Rice}}, and
  \bibinfo{author}{\bibfnamefont{M.}~\bibnamefont{Salmhofer}},
  \bibinfo{year}{1998}, \bibinfo{journal}{Phys. Rev. Lett.}
  \textbf{\bibinfo{volume}{81}}, \bibinfo{pages}{3195}.

\bibitem[{\citenamefont{Furusaki}(1998)}]{Furusaki98}
\bibinfo{author}{\bibnamefont{Furusaki}, \bibfnamefont{A.}},
  \bibinfo{year}{1998}, \bibinfo{journal}{Phys. Rev. B}
  \textbf{\bibinfo{volume}{57}}, \bibinfo{pages}{7141}.

\bibitem[{\citenamefont{Furusaki and
  Nagaosa}(1993{\natexlab{a}})}]{Furusaki93a}
\bibinfo{author}{\bibnamefont{Furusaki}, \bibfnamefont{A.}}, and
  \bibinfo{author}{\bibfnamefont{N.}~\bibnamefont{Nagaosa}},
  \bibinfo{year}{1993}{\natexlab{a}}, \bibinfo{journal}{Phys. Rev. B}
  \textbf{\bibinfo{volume}{47}}, \bibinfo{pages}{4631}.

\bibitem[{\citenamefont{Furusaki and
  Nagaosa}(1993{\natexlab{b}})}]{Furusaki93b}
\bibinfo{author}{\bibnamefont{Furusaki}, \bibfnamefont{A.}}, and
  \bibinfo{author}{\bibfnamefont{N.}~\bibnamefont{Nagaosa}},
  \bibinfo{year}{1993}{\natexlab{b}}, \bibinfo{journal}{Phys. Rev. B}
  \textbf{\bibinfo{volume}{47}}, \bibinfo{pages}{3827}.

\bibitem[{\citenamefont{Furusaki and Nagaosa}(1996)}]{Furusaki96}
\bibinfo{author}{\bibnamefont{Furusaki}, \bibfnamefont{A.}}, and
  \bibinfo{author}{\bibfnamefont{N.}~\bibnamefont{Nagaosa}},
  \bibinfo{year}{1996}, \bibinfo{journal}{Phys. Rev. B}
  \textbf{\bibinfo{volume}{54}}, \bibinfo{pages}{R5239}.

\bibitem[{\citenamefont{Gasenzer and Pawlowski}(2008)}]{Gasenzer08}
\bibinfo{author}{\bibnamefont{Gasenzer}, \bibfnamefont{T.}}, and
  \bibinfo{author}{\bibfnamefont{J.~M.} \bibnamefont{Pawlowski}},
  \bibinfo{year}{2008}, \bibinfo{journal}{Phys. Lett. B}
  \textbf{\bibinfo{volume}{670}}, \bibinfo{pages}{135}.

\bibitem[{\citenamefont{Gawedzki and Kupiainen}(1985)}]{GKGN}
\bibinfo{author}{\bibnamefont{Gawedzki}, \bibfnamefont{K.}}, and
  \bibinfo{author}{\bibfnamefont{A.}~\bibnamefont{Kupiainen}},
  \bibinfo{year}{1985}, \bibinfo{journal}{Commun. Math. Phys.}
  \textbf{\bibinfo{volume}{102}}, \bibinfo{pages}{1}.

\bibitem[{\citenamefont{Gendiar} \emph{et~al.}(2009)\citenamefont{Gendiar,
  Krcmar, and Weyrauch}}]{Gendiar09}
\bibinfo{author}{\bibnamefont{Gendiar}, \bibfnamefont{A.}},
  \bibinfo{author}{\bibfnamefont{R.}~\bibnamefont{Krcmar}}, and
  \bibinfo{author}{\bibfnamefont{M.}~\bibnamefont{Weyrauch}},
  \bibinfo{year}{2009}, \bibinfo{journal}{Phys. Rev. B}
  \textbf{\bibinfo{volume}{79}}, \bibinfo{pages}{205118}.

\bibitem[{\citenamefont{Georges} \emph{et~al.}(1996)\citenamefont{Georges,
  Kotliar, Krauth, and Rozenberg}}]{Georges96}
\bibinfo{author}{\bibnamefont{Georges}, \bibfnamefont{A.}},
  \bibinfo{author}{\bibfnamefont{G.}~\bibnamefont{Kotliar}},
  \bibinfo{author}{\bibfnamefont{W.}~\bibnamefont{Krauth}}, and
  \bibinfo{author}{\bibfnamefont{M.~J.} \bibnamefont{Rozenberg}},
  \bibinfo{year}{1996}, \bibinfo{journal}{Rev. Mod. Phys.}
  \textbf{\bibinfo{volume}{68}}, \bibinfo{pages}{13}.

\bibitem[{\citenamefont{Gersch} \emph{et~al.}(2008)\citenamefont{Gersch,
  Honerkamp, and Metzner}}]{Gersch08}
\bibinfo{author}{\bibnamefont{Gersch}, \bibfnamefont{R.}},
  \bibinfo{author}{\bibfnamefont{C.}~\bibnamefont{Honerkamp}}, and
  \bibinfo{author}{\bibfnamefont{W.}~\bibnamefont{Metzner}},
  \bibinfo{year}{2008}, \bibinfo{journal}{New J. Phys.}
  \textbf{\bibinfo{volume}{10}}, \bibinfo{pages}{045003}.

\bibitem[{\citenamefont{Gersch} \emph{et~al.}(2005)\citenamefont{Gersch,
  Honerkamp, Rohe, and Metzner}}]{Gersch05}
\bibinfo{author}{\bibnamefont{Gersch}, \bibfnamefont{R.}},
  \bibinfo{author}{\bibfnamefont{C.}~\bibnamefont{Honerkamp}},
  \bibinfo{author}{\bibfnamefont{D.}~\bibnamefont{Rohe}}, and
  \bibinfo{author}{\bibfnamefont{W.}~\bibnamefont{Metzner}},
  \bibinfo{year}{2005}, \bibinfo{journal}{Eur. Phys. J. B}
  \textbf{\bibinfo{volume}{48}}, \bibinfo{pages}{349}.

\bibitem[{\citenamefont{Gersch} \emph{et~al.}(2006)\citenamefont{Gersch, Reiss,
  and Honerkamp}}]{Gersch06}
\bibinfo{author}{\bibnamefont{Gersch}, \bibfnamefont{R.}},
  \bibinfo{author}{\bibfnamefont{J.}~\bibnamefont{Reiss}}, and
  \bibinfo{author}{\bibfnamefont{C.}~\bibnamefont{Honerkamp}},
  \bibinfo{year}{2006}, \bibinfo{journal}{New J. Phys.}
  \textbf{\bibinfo{volume}{8}}, \bibinfo{pages}{320}.

\bibitem[{\citenamefont{Gezzi} \emph{et~al.}(2007)\citenamefont{Gezzi,
  Pruschke, and Meden}}]{Gezzi07}
\bibinfo{author}{\bibnamefont{Gezzi}, \bibfnamefont{R.}},
  \bibinfo{author}{\bibfnamefont{T.}~\bibnamefont{Pruschke}}, and
  \bibinfo{author}{\bibfnamefont{V.}~\bibnamefont{Meden}},
  \bibinfo{year}{2007}, \bibinfo{journal}{Phys. Rev. B}
  \textbf{\bibinfo{volume}{75}}, \bibinfo{pages}{045324}.

\bibitem[{\citenamefont{Giamarchi}(2004)}]{Giamarchi04}
\bibinfo{author}{\bibnamefont{Giamarchi}, \bibfnamefont{T.}},
  \bibinfo{year}{2004}, \emph{\bibinfo{title}{Quantum Physics in One
  Dimension}} (\bibinfo{publisher}{Oxford University Press}).

\bibitem[{\citenamefont{Gies and Jaeckel}(2004)}]{Gies04a}
\bibinfo{author}{\bibnamefont{Gies}, \bibfnamefont{H.}}, and
  \bibinfo{author}{\bibfnamefont{J.}~\bibnamefont{Jaeckel}},
  \bibinfo{year}{2004}, \bibinfo{journal}{Phys. Rev. Lett.}
  \textbf{\bibinfo{volume}{93}}, \bibinfo{pages}{110405}.

\bibitem[{\citenamefont{Gies} \emph{et~al.}(2009)\citenamefont{Gies,
  Synatschke, and Wipf}}]{Gies09}
\bibinfo{author}{\bibnamefont{Gies}, \bibfnamefont{H.}},
  \bibinfo{author}{\bibfnamefont{F.}~\bibnamefont{Synatschke}}, and
  \bibinfo{author}{\bibfnamefont{A.}~\bibnamefont{Wipf}}, \bibinfo{year}{2009},
  \bibinfo{journal}{Phys. Rev. D} \textbf{\bibinfo{volume}{80}},
  \bibinfo{pages}{101701}.

\bibitem[{\citenamefont{Gies and Wetterich}(2002)}]{Gies02}
\bibinfo{author}{\bibnamefont{Gies}, \bibfnamefont{H.}}, and
  \bibinfo{author}{\bibfnamefont{C.}~\bibnamefont{Wetterich}},
  \bibinfo{year}{2002}, \bibinfo{journal}{Phys. Rev. D}
  \textbf{\bibinfo{volume}{65}}, \bibinfo{pages}{065001}.

\bibitem[{\citenamefont{Gies and Wetterich}(2004)}]{Gies04}
\bibinfo{author}{\bibnamefont{Gies}, \bibfnamefont{H.}}, and
  \bibinfo{author}{\bibfnamefont{C.}~\bibnamefont{Wetterich}},
  \bibinfo{year}{2004}, \bibinfo{journal}{Phys. Rev. D}
  \textbf{\bibinfo{volume}{69}}, \bibinfo{pages}{025001}.

\bibitem[{\citenamefont{Goldstein} \emph{et~al.}(2009)\citenamefont{Goldstein,
  Berkovits, Gefen, and Weidenm{\"u}ller}}]{Goldstein09}
\bibinfo{author}{\bibnamefont{Goldstein}, \bibfnamefont{M.}},
  \bibinfo{author}{\bibfnamefont{R.}~\bibnamefont{Berkovits}},
  \bibinfo{author}{\bibfnamefont{Y.}~\bibnamefont{Gefen}}, and
  \bibinfo{author}{\bibfnamefont{H.~A.} \bibnamefont{Weidenm{\"u}ller}},
  \bibinfo{year}{2009}, \bibinfo{journal}{Phys. Rev. B}
  \textbf{\bibinfo{volume}{79}}, \bibinfo{pages}{125307}.

\bibitem[{\citenamefont{Gonzalez} \emph{et~al.}(1996)\citenamefont{Gonzalez,
  Guinea, and Vozmediano}}]{Guinea}
\bibinfo{author}{\bibnamefont{Gonzalez}, \bibfnamefont{J.}},
  \bibinfo{author}{\bibfnamefont{F.}~\bibnamefont{Guinea}}, and
  \bibinfo{author}{\bibfnamefont{M.~A.~H.} \bibnamefont{Vozmediano}},
  \bibinfo{year}{1996}, \bibinfo{journal}{Europhys. Lett.}
  \textbf{\bibinfo{volume}{34}}, \bibinfo{pages}{711}.

\bibitem[{\citenamefont{Graser} \emph{et~al.}(2009)\citenamefont{Graser, Maier,
  Hirschfeld, and Scalapino}}]{Graser09}
\bibinfo{author}{\bibnamefont{Graser}, \bibfnamefont{S.}},
  \bibinfo{author}{\bibfnamefont{T.~A.} \bibnamefont{Maier}},
  \bibinfo{author}{\bibfnamefont{P.~J.} \bibnamefont{Hirschfeld}}, and
  \bibinfo{author}{\bibfnamefont{D.~J.} \bibnamefont{Scalapino}},
  \bibinfo{year}{2009}, \bibinfo{journal}{New J. Phys.}
  \textbf{\bibinfo{volume}{11}}, \bibinfo{pages}{025016}.

\bibitem[{\citenamefont{Grioni} \emph{et~al.}(2009)\citenamefont{Grioni, Pons,
  and Frantzeskakis}}]{Grioni09}
\bibinfo{author}{\bibnamefont{Grioni}, \bibfnamefont{M.}},
  \bibinfo{author}{\bibfnamefont{S.}~\bibnamefont{Pons}}, and
  \bibinfo{author}{\bibfnamefont{E.}~\bibnamefont{Frantzeskakis}},
  \bibinfo{year}{2009}, \bibinfo{journal}{J. Phys.: Condens. Matter}
  \textbf{\bibinfo{volume}{21}}, \bibinfo{pages}{023201}.

\bibitem[{\citenamefont{Grote} \emph{et~al.}(2002)\citenamefont{Grote,
  K{\"o}rding, and Wegner}}]{Grote02}
\bibinfo{author}{\bibnamefont{Grote}, \bibfnamefont{I.}},
  \bibinfo{author}{\bibfnamefont{E.}~\bibnamefont{K{\"o}rding}}, and
  \bibinfo{author}{\bibfnamefont{F.}~\bibnamefont{Wegner}},
  \bibinfo{year}{2002}, \bibinfo{journal}{J. Low Temp. Phys.}
  \textbf{\bibinfo{volume}{126}}, \bibinfo{pages}{1385}.

\bibitem[{\citenamefont{Haag}(1962)}]{Haag62}
\bibinfo{author}{\bibnamefont{Haag}, \bibfnamefont{R.}}, \bibinfo{year}{1962},
  \bibinfo{journal}{Nuovo Cimento} \textbf{\bibinfo{volume}{25}},
  \bibinfo{pages}{287}.

\bibitem[{\citenamefont{Halboth and Metzner}(2000{\natexlab{a}})}]{Halboth00a}
\bibinfo{author}{\bibnamefont{Halboth}, \bibfnamefont{C.~J.}}, and
  \bibinfo{author}{\bibfnamefont{W.}~\bibnamefont{Metzner}},
  \bibinfo{year}{2000}{\natexlab{a}}, \bibinfo{journal}{Phys. Rev. B}
  \textbf{\bibinfo{volume}{61}}, \bibinfo{pages}{7364}.

\bibitem[{\citenamefont{Halboth and Metzner}(2000{\natexlab{b}})}]{Halboth00b}
\bibinfo{author}{\bibnamefont{Halboth}, \bibfnamefont{C.~J.}}, and
  \bibinfo{author}{\bibfnamefont{W.}~\bibnamefont{Metzner}},
  \bibinfo{year}{2000}{\natexlab{b}}, \bibinfo{journal}{Phys. Rev. Lett.}
  \textbf{\bibinfo{volume}{85}}, \bibinfo{pages}{5162}.

\bibitem[{\citenamefont{Haldane}(1980)}]{Haldane80}
\bibinfo{author}{\bibnamefont{Haldane}, \bibfnamefont{F.~D.~M.}},
  \bibinfo{year}{1980}, \bibinfo{journal}{Phys. Rev. Lett.}
  \textbf{\bibinfo{volume}{45}}, \bibinfo{pages}{1358}.

\bibitem[{\citenamefont{Hankyevych}
  \emph{et~al.}(2003)\citenamefont{Hankyevych, Kyung, and
  Termblay}}]{Hankyevich2003}
\bibinfo{author}{\bibnamefont{Hankyevych}, \bibfnamefont{V.}},
  \bibinfo{author}{\bibfnamefont{B.}~\bibnamefont{Kyung}}, and
  \bibinfo{author}{\bibfnamefont{A.-M.~S.} \bibnamefont{Termblay}},
  \bibinfo{year}{2003}, \bibinfo{journal}{Phys. Rev. B}
  \textbf{\bibinfo{volume}{68}}, \bibinfo{pages}{214405}.

\bibitem[{\citenamefont{Hanson} \emph{et~al.}(2007)\citenamefont{Hanson,
  Kouwenhoven, Petta, Tarucha, and Vandersypen}}]{Hanson07}
\bibinfo{author}{\bibnamefont{Hanson}, \bibfnamefont{R.}},
  \bibinfo{author}{\bibfnamefont{L.~P.} \bibnamefont{Kouwenhoven}},
  \bibinfo{author}{\bibfnamefont{J.~R.} \bibnamefont{Petta}},
  \bibinfo{author}{\bibfnamefont{S.}~\bibnamefont{Tarucha}}, and
  \bibinfo{author}{\bibfnamefont{L.~M.~K.} \bibnamefont{Vandersypen}},
  \bibinfo{year}{2007}, \bibinfo{journal}{Rev. Mod. Phys.}
  \textbf{\bibinfo{volume}{79}}, \bibinfo{pages}{1217}.

\bibitem[{\citenamefont{Hedden} \emph{et~al.}(2004)\citenamefont{Hedden, Meden,
  Pruschke, and Sch{\"o}nhammer}}]{Hedden04}
\bibinfo{author}{\bibnamefont{Hedden}, \bibfnamefont{R.}},
  \bibinfo{author}{\bibfnamefont{V.}~\bibnamefont{Meden}},
  \bibinfo{author}{\bibfnamefont{T.}~\bibnamefont{Pruschke}}, and
  \bibinfo{author}{\bibfnamefont{K.}~\bibnamefont{Sch{\"o}nhammer}},
  \bibinfo{year}{2004}, \bibinfo{journal}{J. Phys.: Condensed Matter}
  \textbf{\bibinfo{volume}{16}}, \bibinfo{pages}{5279}.

\bibitem[{\citenamefont{Herbut}(2006)}]{herbut}
\bibinfo{author}{\bibnamefont{Herbut}, \bibfnamefont{I.~F.}},
  \bibinfo{year}{2006}, \bibinfo{journal}{Phys. Rev. Lett.}
  \textbf{\bibinfo{volume}{97}}, \bibinfo{pages}{146401}.

\bibitem[{\citenamefont{Hertz}(1976)}]{Hertz76}
\bibinfo{author}{\bibnamefont{Hertz}, \bibfnamefont{J.~A.}},
  \bibinfo{year}{1976}, \bibinfo{journal}{Phys. Rev. B}
  \textbf{\bibinfo{volume}{14}}, \bibinfo{pages}{1165}.

\bibitem[{\citenamefont{Hewson}(1994)}]{Hewson94}
\bibinfo{author}{\bibnamefont{Hewson}, \bibfnamefont{A.~C.}},
  \bibinfo{year}{1994}, \bibinfo{journal}{Adv. Phys.}
  \textbf{\bibinfo{volume}{43}}, \bibinfo{pages}{543}.

\bibitem[{\citenamefont{Hirschfeld and Scalapino}(2010)}]{Hirschfeld10}
\bibinfo{author}{\bibnamefont{Hirschfeld}, \bibfnamefont{P.~J.}}, and
  \bibinfo{author}{\bibfnamefont{D.~J.} \bibnamefont{Scalapino}},
  \bibinfo{year}{2010}, \bibinfo{journal}{Physics}
  \textbf{\bibinfo{volume}{3}}, \bibinfo{pages}{64}.

\bibitem[{\citenamefont{Honerkamp}(2001)}]{Honerkamp01a}
\bibinfo{author}{\bibnamefont{Honerkamp}, \bibfnamefont{C.}},
  \bibinfo{year}{2001}, \bibinfo{journal}{Eur. Phys. J. B}
  \textbf{\bibinfo{volume}{21}}, \bibinfo{pages}{81}.

\bibitem[{\citenamefont{Honerkamp}(2003)}]{Honerkamp03tria}
\bibinfo{author}{\bibnamefont{Honerkamp}, \bibfnamefont{C.}},
  \bibinfo{year}{2003}, \bibinfo{journal}{Phys. Rev. B}
  \textbf{\bibinfo{volume}{68}}, \bibinfo{pages}{104510}.

\bibitem[{\citenamefont{Honerkamp}(2005)}]{Honerkamp05a}
\bibinfo{author}{\bibnamefont{Honerkamp}, \bibfnamefont{C.}},
  \bibinfo{year}{2005}, \bibinfo{journal}{Phys. Rev. B}
  \textbf{\bibinfo{volume}{72}}, \bibinfo{pages}{115103}.

\bibitem[{\citenamefont{Honerkamp}(2008)}]{Honerkamp08}
\bibinfo{author}{\bibnamefont{Honerkamp}, \bibfnamefont{C.}},
  \bibinfo{year}{2008}, \bibinfo{journal}{Phys. Rev. Lett.}
  \textbf{\bibinfo{volume}{100}}, \bibinfo{pages}{146404}.

\bibitem[{\citenamefont{Honerkamp} \emph{et~al.}(2007)\citenamefont{Honerkamp,
  Fu, and Lee}}]{Honerkamp07}
\bibinfo{author}{\bibnamefont{Honerkamp}, \bibfnamefont{C.}},
  \bibinfo{author}{\bibfnamefont{H.~C.} \bibnamefont{Fu}}, and
  \bibinfo{author}{\bibfnamefont{D.-H.} \bibnamefont{Lee}},
  \bibinfo{year}{2007}, \bibinfo{journal}{Phys. Rev. B}
  \textbf{\bibinfo{volume}{75}}, \bibinfo{pages}{014503}.

\bibitem[{\citenamefont{Honerkamp and
  Hofstetter}(2004)}]{HonerkampHofstetter04}
\bibinfo{author}{\bibnamefont{Honerkamp}, \bibfnamefont{C.}}, and
  \bibinfo{author}{\bibfnamefont{W.}~\bibnamefont{Hofstetter}},
  \bibinfo{year}{2004}, \bibinfo{journal}{Phys. Rev. Lett.}
  \textbf{\bibinfo{volume}{92}}, \bibinfo{pages}{170403}.

\bibitem[{\citenamefont{Honerkamp} \emph{et~al.}(2004)\citenamefont{Honerkamp,
  Rohe, Andergassen, and Enss}}]{Honerkamp04}
\bibinfo{author}{\bibnamefont{Honerkamp}, \bibfnamefont{C.}},
  \bibinfo{author}{\bibfnamefont{D.}~\bibnamefont{Rohe}},
  \bibinfo{author}{\bibfnamefont{S.}~\bibnamefont{Andergassen}}, and
  \bibinfo{author}{\bibfnamefont{T.}~\bibnamefont{Enss}}, \bibinfo{year}{2004},
  \bibinfo{journal}{Phys. Rev. B} \textbf{\bibinfo{volume}{70}},
  \bibinfo{pages}{235115}.

\bibitem[{\citenamefont{Honerkamp and
  Salmhofer}(2001{\natexlab{a}})}]{Honerkamp01c}
\bibinfo{author}{\bibnamefont{Honerkamp}, \bibfnamefont{C.}}, and
  \bibinfo{author}{\bibfnamefont{M.}~\bibnamefont{Salmhofer}},
  \bibinfo{year}{2001}{\natexlab{a}}, \bibinfo{journal}{Phys. Rev. B}
  \textbf{\bibinfo{volume}{64}}, \bibinfo{pages}{184516}.

\bibitem[{\citenamefont{Honerkamp and
  Salmhofer}(2001{\natexlab{b}})}]{Honerkamp01b}
\bibinfo{author}{\bibnamefont{Honerkamp}, \bibfnamefont{C.}}, and
  \bibinfo{author}{\bibfnamefont{M.}~\bibnamefont{Salmhofer}},
  \bibinfo{year}{2001}{\natexlab{b}}, \bibinfo{journal}{Phys. Rev. Lett.}
  \textbf{\bibinfo{volume}{87}}, \bibinfo{pages}{187004}.

\bibitem[{\citenamefont{Honerkamp and Salmhofer}(2003)}]{Honerkamp03}
\bibinfo{author}{\bibnamefont{Honerkamp}, \bibfnamefont{C.}}, and
  \bibinfo{author}{\bibfnamefont{M.}~\bibnamefont{Salmhofer}},
  \bibinfo{year}{2003}, \bibinfo{journal}{Phys. Rev. B}
  \textbf{\bibinfo{volume}{67}}, \bibinfo{pages}{174504}.

\bibitem[{\citenamefont{Honerkamp and Salmhofer}(2005)}]{Honerkamp05}
\bibinfo{author}{\bibnamefont{Honerkamp}, \bibfnamefont{C.}}, and
  \bibinfo{author}{\bibfnamefont{M.}~\bibnamefont{Salmhofer}},
  \bibinfo{year}{2005}, \bibinfo{journal}{Prog. Theor. Phys.}
  \textbf{\bibinfo{volume}{113}}, \bibinfo{pages}{1145}.

\bibitem[{\citenamefont{Honerkamp} \emph{et~al.}(2001)\citenamefont{Honerkamp,
  Salmhofer, Furukawa, and Rice}}]{Honerkamp01d}
\bibinfo{author}{\bibnamefont{Honerkamp}, \bibfnamefont{C.}},
  \bibinfo{author}{\bibfnamefont{M.}~\bibnamefont{Salmhofer}},
  \bibinfo{author}{\bibfnamefont{N.}~\bibnamefont{Furukawa}}, and
  \bibinfo{author}{\bibfnamefont{T.~M.} \bibnamefont{Rice}},
  \bibinfo{year}{2001}, \bibinfo{journal}{Phys. Rev. B}
  \textbf{\bibinfo{volume}{63}}, \bibinfo{pages}{035109}.

\bibitem[{\citenamefont{Honerkamp} \emph{et~al.}(2002)\citenamefont{Honerkamp,
  Salmhofer, and Rice}}]{Honerkamp02}
\bibinfo{author}{\bibnamefont{Honerkamp}, \bibfnamefont{C.}},
  \bibinfo{author}{\bibfnamefont{M.}~\bibnamefont{Salmhofer}}, and
  \bibinfo{author}{\bibfnamefont{T.~M.} \bibnamefont{Rice}},
  \bibinfo{year}{2002}, \bibinfo{journal}{Eur. Phys. J. B}
  \textbf{\bibinfo{volume}{27}}, \bibinfo{pages}{127}.

\bibitem[{\citenamefont{H{\"u}gle and Egger}(2004)}]{Huegle04}
\bibinfo{author}{\bibnamefont{H{\"u}gle}, \bibfnamefont{S.}}, and
  \bibinfo{author}{\bibfnamefont{R.}~\bibnamefont{Egger}},
  \bibinfo{year}{2004}, \bibinfo{journal}{Europhys. Lett.}
  \textbf{\bibinfo{volume}{66}}, \bibinfo{pages}{565}.

\bibitem[{\citenamefont{Huh and Sachdev}(2008)}]{Huh08}
\bibinfo{author}{\bibnamefont{Huh}, \bibfnamefont{Y.}}, and
  \bibinfo{author}{\bibfnamefont{S.}~\bibnamefont{Sachdev}},
  \bibinfo{year}{2008}, \bibinfo{journal}{Phys. Rev. B}
  \textbf{\bibinfo{volume}{78}}, \bibinfo{pages}{064512}.

\bibitem[{\citenamefont{Husemann and Salmhofer}(2009)}]{Husemann09a}
\bibinfo{author}{\bibnamefont{Husemann}, \bibfnamefont{C.}}, and
  \bibinfo{author}{\bibfnamefont{M.}~\bibnamefont{Salmhofer}},
  \bibinfo{year}{2009}, \bibinfo{journal}{Phys. Rev. B}
  \textbf{\bibinfo{volume}{79}}, \bibinfo{pages}{195125}.

\bibitem[{\citenamefont{Ikeda} \emph{et~al.}(2010)\citenamefont{Ikeda, Arita,
  and Kunes}}]{Ikeda10}
\bibinfo{author}{\bibnamefont{Ikeda}, \bibfnamefont{H.}},
  \bibinfo{author}{\bibfnamefont{R.}~\bibnamefont{Arita}}, and
  \bibinfo{author}{\bibfnamefont{J.}~\bibnamefont{Kunes}},
  \bibinfo{year}{2010}, \bibinfo{journal}{Phys. Rev. B}
  \textbf{\bibinfo{volume}{81}}, \bibinfo{pages}{054502}.

\bibitem[{\citenamefont{Ishida} \emph{et~al.}(2009)\citenamefont{Ishida, Nakai,
  and Hosono}}]{Ishida09}
\bibinfo{author}{\bibnamefont{Ishida}, \bibfnamefont{K.}},
  \bibinfo{author}{\bibfnamefont{Y.}~\bibnamefont{Nakai}}, and
  \bibinfo{author}{\bibfnamefont{H.}~\bibnamefont{Hosono}},
  \bibinfo{year}{2009}, \bibinfo{journal}{J. Phys. Soc. Jpn.}
  \textbf{\bibinfo{volume}{78}}, \bibinfo{pages}{062001}.

\bibitem[{\citenamefont{Isidori} \emph{et~al.}(2010)\citenamefont{Isidori,
  Rosen, Bartosch, Hofstetter, and Kopietz}}]{Isidori10}
\bibinfo{author}{\bibnamefont{Isidori}, \bibfnamefont{A.}},
  \bibinfo{author}{\bibfnamefont{D.}~\bibnamefont{Rosen}},
  \bibinfo{author}{\bibfnamefont{L.}~\bibnamefont{Bartosch}},
  \bibinfo{author}{\bibfnamefont{W.}~\bibnamefont{Hofstetter}}, and
  \bibinfo{author}{\bibfnamefont{P.}~\bibnamefont{Kopietz}},
  \bibinfo{year}{2010}, \bibinfo{journal}{Phys. Rev. B}
  \textbf{\bibinfo{volume}{81}}, \bibinfo{pages}{235120}.

\bibitem[{\citenamefont{Jakobs}(2003)}]{Jakobs03}
\bibinfo{author}{\bibnamefont{Jakobs}, \bibfnamefont{S.~G.}},
  \bibinfo{year}{2003}, \emph{\bibinfo{title}{Diploma Thesis}}
  (\bibinfo{publisher}{RWTH Aachen University}).

\bibitem[{\citenamefont{Jakobs}(2010)}]{Jakobs10b}
\bibinfo{author}{\bibnamefont{Jakobs}, \bibfnamefont{S.~G.}},
  \bibinfo{year}{2010}, \emph{\bibinfo{title}{PhD Thesis}}
  (\bibinfo{publisher}{RWTH Aachen University}).

\bibitem[{\citenamefont{Jakobs}
  \emph{et~al.}(2007{\natexlab{a}})\citenamefont{Jakobs, Meden, and
  Schoeller}}]{Jakobs07a}
\bibinfo{author}{\bibnamefont{Jakobs}, \bibfnamefont{S.~G.}},
  \bibinfo{author}{\bibfnamefont{V.}~\bibnamefont{Meden}}, and
  \bibinfo{author}{\bibfnamefont{H.}~\bibnamefont{Schoeller}},
  \bibinfo{year}{2007}{\natexlab{a}}, \bibinfo{journal}{Phys. Rev. Lett.}
  \textbf{\bibinfo{volume}{99}}, \bibinfo{pages}{150603}.

\bibitem[{\citenamefont{Jakobs}
  \emph{et~al.}(2007{\natexlab{b}})\citenamefont{Jakobs, Meden, Schoeller, and
  Enss}}]{Jakobs07b}
\bibinfo{author}{\bibnamefont{Jakobs}, \bibfnamefont{S.~G.}},
  \bibinfo{author}{\bibfnamefont{V.}~\bibnamefont{Meden}},
  \bibinfo{author}{\bibfnamefont{H.}~\bibnamefont{Schoeller}}, and
  \bibinfo{author}{\bibfnamefont{T.}~\bibnamefont{Enss}},
  \bibinfo{year}{2007}{\natexlab{b}}, \bibinfo{journal}{Phys. Rev. B}
  \textbf{\bibinfo{volume}{75}}, \bibinfo{pages}{035126}.

\bibitem[{\citenamefont{Jakobs}
  \emph{et~al.}(2010{\natexlab{a}})\citenamefont{Jakobs, Pletyukhov, and
  Schoeller}}]{Jakobs10a}
\bibinfo{author}{\bibnamefont{Jakobs}, \bibfnamefont{S.~G.}},
  \bibinfo{author}{\bibfnamefont{M.}~\bibnamefont{Pletyukhov}}, and
  \bibinfo{author}{\bibfnamefont{H.}~\bibnamefont{Schoeller}},
  \bibinfo{year}{2010}{\natexlab{a}}, \bibinfo{journal}{Phys. Rev. B}
  \textbf{\bibinfo{volume}{81}}, \bibinfo{pages}{195109}.

\bibitem[{\citenamefont{Jakobs}
  \emph{et~al.}(2010{\natexlab{b}})\citenamefont{Jakobs, Pletyukhov, and
  Schoeller}}]{Jakobs10c}
\bibinfo{author}{\bibnamefont{Jakobs}, \bibfnamefont{S.~G.}},
  \bibinfo{author}{\bibfnamefont{M.}~\bibnamefont{Pletyukhov}}, and
  \bibinfo{author}{\bibfnamefont{H.}~\bibnamefont{Schoeller}},
  \bibinfo{year}{2010}{\natexlab{b}}, \bibinfo{journal}{J. Phys. A: Math.
  Theor.} \textbf{\bibinfo{volume}{43}}, \bibinfo{pages}{103001}.

\bibitem[{\citenamefont{Jakubczyk}(2009)}]{Jakubczyk09a}
\bibinfo{author}{\bibnamefont{Jakubczyk}, \bibfnamefont{P.}},
  \bibinfo{year}{2009}, \bibinfo{journal}{Phys. Rev. B}
  \textbf{\bibinfo{volume}{79}}, \bibinfo{pages}{125115}.

\bibitem[{\citenamefont{Jakubczyk} \emph{et~al.}(2010)\citenamefont{Jakubczyk,
  Bauer, and Metzner}}]{Jakubczyk10}
\bibinfo{author}{\bibnamefont{Jakubczyk}, \bibfnamefont{P.}},
  \bibinfo{author}{\bibfnamefont{J.}~\bibnamefont{Bauer}}, and
  \bibinfo{author}{\bibfnamefont{W.}~\bibnamefont{Metzner}},
  \bibinfo{year}{2010}, \bibinfo{journal}{Phys. Rev. B}
  \textbf{\bibinfo{volume}{82}}, \bibinfo{pages}{045103}.

\bibitem[{\citenamefont{Jakubczyk} \emph{et~al.}(2009)\citenamefont{Jakubczyk,
  Metzner, and Yamase}}]{Jakubczyk09b}
\bibinfo{author}{\bibnamefont{Jakubczyk}, \bibfnamefont{P.}},
  \bibinfo{author}{\bibfnamefont{W.}~\bibnamefont{Metzner}}, and
  \bibinfo{author}{\bibfnamefont{H.}~\bibnamefont{Yamase}},
  \bibinfo{year}{2009}, \bibinfo{journal}{Phys. Rev. Lett.}
  \textbf{\bibinfo{volume}{103}}, \bibinfo{pages}{220602}.

\bibitem[{\citenamefont{Jakubczyk} \emph{et~al.}(2008)\citenamefont{Jakubczyk,
  Strack, Katanin, and Metzner}}]{Jakubczyk08}
\bibinfo{author}{\bibnamefont{Jakubczyk}, \bibfnamefont{P.}},
  \bibinfo{author}{\bibfnamefont{P.}~\bibnamefont{Strack}},
  \bibinfo{author}{\bibfnamefont{A.~A.} \bibnamefont{Katanin}}, and
  \bibinfo{author}{\bibfnamefont{W.}~\bibnamefont{Metzner}},
  \bibinfo{year}{2008}, \bibinfo{journal}{Phys. Rev. B}
  \textbf{\bibinfo{volume}{77}}, \bibinfo{pages}{195120}.

\bibitem[{\citenamefont{Janzen} \emph{et~al.}(2006)\citenamefont{Janzen, Meden,
  and Sch{\"o}nhammer}}]{Janzen06}
\bibinfo{author}{\bibnamefont{Janzen}, \bibfnamefont{K.}},
  \bibinfo{author}{\bibfnamefont{V.}~\bibnamefont{Meden}}, and
  \bibinfo{author}{\bibfnamefont{K.}~\bibnamefont{Sch{\"o}nhammer}},
  \bibinfo{year}{2006}, \bibinfo{journal}{Phys. Rev. B}
  \textbf{\bibinfo{volume}{74}}, \bibinfo{pages}{085301}.

\bibitem[{\citenamefont{Kamenev}(2004)}]{Kamenev04}
\bibinfo{author}{\bibnamefont{Kamenev}, \bibfnamefont{A.}},
  \bibinfo{year}{2004}, \emph{\bibinfo{title}{Les Houches, Volume Session LX}}
  (\bibinfo{publisher}{edited by H. Bouchiat, Y. Gefen, S. Gu{\'e}ron, G.
  Montambaux, and J. Dalibard (Elsevier, North-Holland,Amsterdam}).

\bibitem[{\citenamefont{Kane and Fisher}(1992)}]{Kane92}
\bibinfo{author}{\bibnamefont{Kane}, \bibfnamefont{C.~L.}}, and
  \bibinfo{author}{\bibfnamefont{M.~P.~A.} \bibnamefont{Fisher}},
  \bibinfo{year}{1992}, \bibinfo{journal}{Phys. Rev. B}
  \textbf{\bibinfo{volume}{46}}, \bibinfo{pages}{15233}.

\bibitem[{\citenamefont{Karrasch}(2010)}]{Karrasch10b}
\bibinfo{author}{\bibnamefont{Karrasch}, \bibfnamefont{C.}},
  \bibinfo{year}{2010}, \emph{\bibinfo{title}{PhD Thesis}}
  (\bibinfo{publisher}{RWTH Aachen University}).

\bibitem[{\citenamefont{Karrasch}
  \emph{et~al.}(2010{\natexlab{a}})\citenamefont{Karrasch, Andergassen,
  Pletyukhov, Schuricht, Borda, Meden, and Schoeller}}]{Karrasch10c}
\bibinfo{author}{\bibnamefont{Karrasch}, \bibfnamefont{C.}},
  \bibinfo{author}{\bibfnamefont{S.}~\bibnamefont{Andergassen}},
  \bibinfo{author}{\bibfnamefont{M.}~\bibnamefont{Pletyukhov}},
  \bibinfo{author}{\bibfnamefont{D.}~\bibnamefont{Schuricht}},
  \bibinfo{author}{\bibfnamefont{L.}~\bibnamefont{Borda}},
  \bibinfo{author}{\bibfnamefont{V.}~\bibnamefont{Meden}}, and
  \bibinfo{author}{\bibfnamefont{H.}~\bibnamefont{Schoeller}},
  \bibinfo{year}{2010}{\natexlab{a}}, \bibinfo{journal}{Europhys. Lett.}
  \textbf{\bibinfo{volume}{90}}, \bibinfo{pages}{30003}.

\bibitem[{\citenamefont{Karrasch} \emph{et~al.}(2006)\citenamefont{Karrasch,
  Enss, and Meden}}]{Karrasch06}
\bibinfo{author}{\bibnamefont{Karrasch}, \bibfnamefont{C.}},
  \bibinfo{author}{\bibfnamefont{T.}~\bibnamefont{Enss}}, and
  \bibinfo{author}{\bibfnamefont{V.}~\bibnamefont{Meden}},
  \bibinfo{year}{2006}, \bibinfo{journal}{Phys. Rev. B}
  \textbf{\bibinfo{volume}{73}}, \bibinfo{pages}{235337}.

\bibitem[{\citenamefont{Karrasch}
  \emph{et~al.}(2007{\natexlab{a}})\citenamefont{Karrasch, Hecht, Weichselbaum,
  von Delft, Oreg, and Meden}}]{Karrasch07a}
\bibinfo{author}{\bibnamefont{Karrasch}, \bibfnamefont{C.}},
  \bibinfo{author}{\bibfnamefont{T.}~\bibnamefont{Hecht}},
  \bibinfo{author}{\bibfnamefont{A.}~\bibnamefont{Weichselbaum}},
  \bibinfo{author}{\bibfnamefont{J.}~\bibnamefont{von Delft}},
  \bibinfo{author}{\bibfnamefont{Y.}~\bibnamefont{Oreg}}, and
  \bibinfo{author}{\bibfnamefont{V.}~\bibnamefont{Meden}},
  \bibinfo{year}{2007}{\natexlab{a}}, \bibinfo{journal}{New J. Phys.}
  \textbf{\bibinfo{volume}{9}}, \bibinfo{pages}{123}.

\bibitem[{\citenamefont{Karrasch}
  \emph{et~al.}(2007{\natexlab{b}})\citenamefont{Karrasch, Hecht, Weichselbaum,
  Oreg, von Delft, and Meden}}]{Karrasch07b}
\bibinfo{author}{\bibnamefont{Karrasch}, \bibfnamefont{C.}},
  \bibinfo{author}{\bibfnamefont{T.}~\bibnamefont{Hecht}},
  \bibinfo{author}{\bibfnamefont{A.}~\bibnamefont{Weichselbaum}},
  \bibinfo{author}{\bibfnamefont{Y.}~\bibnamefont{Oreg}},
  \bibinfo{author}{\bibfnamefont{J.}~\bibnamefont{von Delft}}, and
  \bibinfo{author}{\bibfnamefont{V.}~\bibnamefont{Meden}},
  \bibinfo{year}{2007}{\natexlab{b}}, \bibinfo{journal}{Phys. Rev. Lett.}
  \textbf{\bibinfo{volume}{98}}, \bibinfo{pages}{186802}.

\bibitem[{\citenamefont{Karrasch} \emph{et~al.}(2008)\citenamefont{Karrasch,
  Hedden, Peters, Pruschke, Sch{\"o}nhammer, and Meden}}]{Karrasch08a}
\bibinfo{author}{\bibnamefont{Karrasch}, \bibfnamefont{C.}},
  \bibinfo{author}{\bibfnamefont{R.}~\bibnamefont{Hedden}},
  \bibinfo{author}{\bibfnamefont{R.}~\bibnamefont{Peters}},
  \bibinfo{author}{\bibfnamefont{T.}~\bibnamefont{Pruschke}},
  \bibinfo{author}{\bibfnamefont{K.}~\bibnamefont{Sch{\"o}nhammer}}, and
  \bibinfo{author}{\bibfnamefont{V.}~\bibnamefont{Meden}},
  \bibinfo{year}{2008}, \bibinfo{journal}{J. Phys.: Condensed Matter}
  \textbf{\bibinfo{volume}{20}}, \bibinfo{pages}{345205}.

\bibitem[{\citenamefont{Karrasch and Meden}(2009)}]{Karrasch09a}
\bibinfo{author}{\bibnamefont{Karrasch}, \bibfnamefont{C.}}, and
  \bibinfo{author}{\bibfnamefont{V.}~\bibnamefont{Meden}},
  \bibinfo{year}{2009}, \bibinfo{journal}{Phys. Rev. B}
  \textbf{\bibinfo{volume}{79}}, \bibinfo{pages}{045110}.

\bibitem[{\citenamefont{Karrasch}
  \emph{et~al.}(2010{\natexlab{b}})\citenamefont{Karrasch, Meden, and
  Sch{\"o}nhammer}}]{Karrasch10d}
\bibinfo{author}{\bibnamefont{Karrasch}, \bibfnamefont{C.}},
  \bibinfo{author}{\bibfnamefont{V.}~\bibnamefont{Meden}}, and
  \bibinfo{author}{\bibfnamefont{K.}~\bibnamefont{Sch{\"o}nhammer}},
  \bibinfo{year}{2010}{\natexlab{b}}, \bibinfo{journal}{Phys. Rev. B}
  \textbf{\bibinfo{volume}{82}}, \bibinfo{pages}{125114}.

\bibitem[{\citenamefont{Karrasch}
  \emph{et~al.}(2010{\natexlab{c}})\citenamefont{Karrasch, Pletyukhov, Borda,
  and Meden}}]{Karrasch10a}
\bibinfo{author}{\bibnamefont{Karrasch}, \bibfnamefont{C.}},
  \bibinfo{author}{\bibfnamefont{M.}~\bibnamefont{Pletyukhov}},
  \bibinfo{author}{\bibfnamefont{L.}~\bibnamefont{Borda}}, and
  \bibinfo{author}{\bibfnamefont{V.}~\bibnamefont{Meden}},
  \bibinfo{year}{2010}{\natexlab{c}}, \bibinfo{journal}{Phys. Rev. B}
  \textbf{\bibinfo{volume}{81}}, \bibinfo{pages}{125122}.

\bibitem[{\citenamefont{Kashcheyevs}
  \emph{et~al.}(2009)\citenamefont{Kashcheyevs, Karrasch, Hecht, Weichselbaum,
  Meden, and Schiller}}]{Kashcheyevs09}
\bibinfo{author}{\bibnamefont{Kashcheyevs}, \bibfnamefont{V.}},
  \bibinfo{author}{\bibfnamefont{C.}~\bibnamefont{Karrasch}},
  \bibinfo{author}{\bibfnamefont{T.}~\bibnamefont{Hecht}},
  \bibinfo{author}{\bibfnamefont{A.}~\bibnamefont{Weichselbaum}},
  \bibinfo{author}{\bibfnamefont{V.}~\bibnamefont{Meden}}, and
  \bibinfo{author}{\bibfnamefont{A.}~\bibnamefont{Schiller}},
  \bibinfo{year}{2009}, \bibinfo{journal}{Phys. Rev. Lett.}
  \textbf{\bibinfo{volume}{102}}, \bibinfo{pages}{136805}.

\bibitem[{\citenamefont{Katanin}(2004)}]{Katanin04a}
\bibinfo{author}{\bibnamefont{Katanin}, \bibfnamefont{A.~A.}},
  \bibinfo{year}{2004}, \bibinfo{journal}{Phys. Rev. B}
  \textbf{\bibinfo{volume}{70}}, \bibinfo{pages}{115109}.

\bibitem[{\citenamefont{Katanin}(2009)}]{Katanin09}
\bibinfo{author}{\bibnamefont{Katanin}, \bibfnamefont{A.~A.}},
  \bibinfo{year}{2009}, \bibinfo{journal}{Phys. Rev. B}
  \textbf{\bibinfo{volume}{79}}, \bibinfo{pages}{235119}.

\bibitem[{\citenamefont{Katanin and Kampf}(2003)}]{Katanin03}
\bibinfo{author}{\bibnamefont{Katanin}, \bibfnamefont{A.~A.}}, and
  \bibinfo{author}{\bibfnamefont{A.~P.} \bibnamefont{Kampf}},
  \bibinfo{year}{2003}, \bibinfo{journal}{Phys. Rev. B}
  \textbf{\bibinfo{volume}{68}}, \bibinfo{pages}{195101}.

\bibitem[{\citenamefont{Katanin and Kampf}(2004)}]{Katanin04b}
\bibinfo{author}{\bibnamefont{Katanin}, \bibfnamefont{A.~A.}}, and
  \bibinfo{author}{\bibfnamefont{A.~P.} \bibnamefont{Kampf}},
  \bibinfo{year}{2004}, \bibinfo{journal}{Phys. Rev. Lett.}
  \textbf{\bibinfo{volume}{93}}, \bibinfo{pages}{106406}.

\bibitem[{\citenamefont{Kaul and Sachdev}(2008)}]{Kaul08}
\bibinfo{author}{\bibnamefont{Kaul}, \bibfnamefont{R.~K.}}, and
  \bibinfo{author}{\bibfnamefont{S.}~\bibnamefont{Sachdev}},
  \bibinfo{year}{2008}, \bibinfo{journal}{Phys. Rev. B}
  \textbf{\bibinfo{volume}{77}}, \bibinfo{pages}{155105}.

\bibitem[{\citenamefont{Kee} \emph{et~al.}(2003)\citenamefont{Kee, Kim, and
  Chung}}]{Kee03}
\bibinfo{author}{\bibnamefont{Kee}, \bibfnamefont{H.-Y.}},
  \bibinfo{author}{\bibfnamefont{E.~H.} \bibnamefont{Kim}}, and
  \bibinfo{author}{\bibfnamefont{C.-H.} \bibnamefont{Chung}},
  \bibinfo{year}{2003}, \bibinfo{journal}{Phys. Rev. B}
  \textbf{\bibinfo{volume}{68}}, \bibinfo{pages}{245109}.

\bibitem[{\citenamefont{Kehrein}(2006)}]{Kehrein06}
\bibinfo{author}{\bibnamefont{Kehrein}, \bibfnamefont{S.}},
  \bibinfo{year}{2006}, \emph{\bibinfo{title}{The flow equation approach to
  many-particle systems}} (\bibinfo{publisher}{Springer, Berlin}).

\bibitem[{\citenamefont{Khavkine} \emph{et~al.}(2004)\citenamefont{Khavkine,
  Chung, Oganesyan, and Kee}}]{Khavkine04}
\bibinfo{author}{\bibnamefont{Khavkine}, \bibfnamefont{I.}},
  \bibinfo{author}{\bibfnamefont{C.-H.} \bibnamefont{Chung}},
  \bibinfo{author}{\bibfnamefont{V.}~\bibnamefont{Oganesyan}}, and
  \bibinfo{author}{\bibfnamefont{H.-Y.} \bibnamefont{Kee}},
  \bibinfo{year}{2004}, \bibinfo{journal}{Phys. Rev. B}
  \textbf{\bibinfo{volume}{70}}, \bibinfo{pages}{155110}.

\bibitem[{\citenamefont{Kino and Fukuyama}(1996)}]{kino}
\bibinfo{author}{\bibnamefont{Kino}, \bibfnamefont{H.}}, and
  \bibinfo{author}{\bibfnamefont{H.}~\bibnamefont{Fukuyama}},
  \bibinfo{year}{1996}, \bibinfo{journal}{J. Phys. Soc. Jpn.}
  \textbf{\bibinfo{volume}{65}}, \bibinfo{pages}{2158}.

\bibitem[{\citenamefont{Kirkpatrick and Belitz}(1996)}]{Kirkpatrick96}
\bibinfo{author}{\bibnamefont{Kirkpatrick}, \bibfnamefont{T.~R.}}, and
  \bibinfo{author}{\bibfnamefont{D.}~\bibnamefont{Belitz}},
  \bibinfo{year}{1996}, \bibinfo{journal}{Phys. Rev. B}
  \textbf{\bibinfo{volume}{53}}, \bibinfo{pages}{14364}.

\bibitem[{\citenamefont{Klironomos and Tsai}(2006)}]{Klironomos06}
\bibinfo{author}{\bibnamefont{Klironomos}, \bibfnamefont{F.~D.}}, and
  \bibinfo{author}{\bibfnamefont{S.~W.} \bibnamefont{Tsai}},
  \bibinfo{year}{2006}, \bibinfo{journal}{Phys. Rev. B}
  \textbf{\bibinfo{volume}{74}}, \bibinfo{pages}{205109}.

\bibitem[{\citenamefont{Klironomos and Tsai}(2007)}]{Klironomos07}
\bibinfo{author}{\bibnamefont{Klironomos}, \bibfnamefont{F.~D.}}, and
  \bibinfo{author}{\bibfnamefont{S.~W.} \bibnamefont{Tsai}},
  \bibinfo{year}{2007}, \bibinfo{journal}{Phys. Rev. Lett}
  \textbf{\bibinfo{volume}{99}}, \bibinfo{pages}{100401}.

\bibitem[{\citenamefont{Kloss and Kopietz}(2011)}]{Kloss11}
\bibinfo{author}{\bibnamefont{Kloss}, \bibfnamefont{T.}}, and
  \bibinfo{author}{\bibfnamefont{P.}~\bibnamefont{Kopietz}},
  \bibinfo{year}{2011}, \bibinfo{journal}{Phys. Rev. B}
  \textbf{\bibinfo{volume}{83}}, \bibinfo{pages}{205118}.

\bibitem[{\citenamefont{Kopietz} \emph{et~al.}(2010)\citenamefont{Kopietz,
  Bartosch, and Sch{\"u}tz}}]{Kopietz10}
\bibinfo{author}{\bibnamefont{Kopietz}, \bibfnamefont{P.}},
  \bibinfo{author}{\bibfnamefont{L.}~\bibnamefont{Bartosch}}, and
  \bibinfo{author}{\bibfnamefont{F.}~\bibnamefont{Sch{\"u}tz}},
  \bibinfo{year}{2010}, \emph{\bibinfo{title}{Introduction to the Functional
  Renormalization Group}} (\bibinfo{publisher}{Springer, Berlin}).

\bibitem[{\citenamefont{Kopietz and Busche}(2001)}]{Kopietz01}
\bibinfo{author}{\bibnamefont{Kopietz}, \bibfnamefont{P.}}, and
  \bibinfo{author}{\bibfnamefont{T.}~\bibnamefont{Busche}},
  \bibinfo{year}{2001}, \bibinfo{journal}{Phys. Rev. B}
  \textbf{\bibinfo{volume}{64}}, \bibinfo{pages}{155101}.

\bibitem[{\citenamefont{Kotliar} \emph{et~al.}(2006)\citenamefont{Kotliar,
  Savrasov, Haule, Oudovenko, Parcollet, and Marianetti}}]{Kotliar06}
\bibinfo{author}{\bibnamefont{Kotliar}, \bibfnamefont{G.}},
  \bibinfo{author}{\bibfnamefont{S.~Y.} \bibnamefont{Savrasov}},
  \bibinfo{author}{\bibfnamefont{K.}~\bibnamefont{Haule}},
  \bibinfo{author}{\bibfnamefont{V.~S.} \bibnamefont{Oudovenko}},
  \bibinfo{author}{\bibfnamefont{O.}~\bibnamefont{Parcollet}}, and
  \bibinfo{author}{\bibfnamefont{C.~A.} \bibnamefont{Marianetti}},
  \bibinfo{year}{2006}, \bibinfo{journal}{Rev. Mod. Phys.}
  \textbf{\bibinfo{volume}{78}}, \bibinfo{pages}{865}.

\bibitem[{\citenamefont{Krahl}
  \emph{et~al.}(2009{\natexlab{a}})\citenamefont{Krahl, Friederich, and
  Wetterich}}]{Krahl09a}
\bibinfo{author}{\bibnamefont{Krahl}, \bibfnamefont{H.~C.}},
  \bibinfo{author}{\bibfnamefont{S.}~\bibnamefont{Friederich}}, and
  \bibinfo{author}{\bibfnamefont{C.}~\bibnamefont{Wetterich}},
  \bibinfo{year}{2009}{\natexlab{a}}, \bibinfo{journal}{Phys. Rev. B}
  \textbf{\bibinfo{volume}{80}}, \bibinfo{pages}{014436}.

\bibitem[{\citenamefont{Krahl}
  \emph{et~al.}(2009{\natexlab{b}})\citenamefont{Krahl, M{\"u}ller, and
  Wetterich}}]{Krahl09b}
\bibinfo{author}{\bibnamefont{Krahl}, \bibfnamefont{H.~C.}},
  \bibinfo{author}{\bibfnamefont{J.~A.} \bibnamefont{M{\"u}ller}}, and
  \bibinfo{author}{\bibfnamefont{C.}~\bibnamefont{Wetterich}},
  \bibinfo{year}{2009}{\natexlab{b}}, \bibinfo{journal}{Phys. Rev. B}
  \textbf{\bibinfo{volume}{79}}, \bibinfo{pages}{094526}.

\bibitem[{\citenamefont{Krippa}(2007)}]{Krippa07}
\bibinfo{author}{\bibnamefont{Krippa}, \bibfnamefont{B.}},
  \bibinfo{year}{2007}, \bibinfo{journal}{Eur. Phys. J. A}
  \textbf{\bibinfo{volume}{31}}, \bibinfo{pages}{734}.

\bibitem[{\citenamefont{Kuroki} \emph{et~al.}(2008)\citenamefont{Kuroki, Onari,
  Arita, Usui, Tanaka, Kontani, and Aoki}}]{Kuroki08}
\bibinfo{author}{\bibnamefont{Kuroki}, \bibfnamefont{K.}},
  \bibinfo{author}{\bibfnamefont{S.}~\bibnamefont{Onari}},
  \bibinfo{author}{\bibfnamefont{R.}~\bibnamefont{Arita}},
  \bibinfo{author}{\bibfnamefont{H.}~\bibnamefont{Usui}},
  \bibinfo{author}{\bibfnamefont{Y.}~\bibnamefont{Tanaka}},
  \bibinfo{author}{\bibfnamefont{H.}~\bibnamefont{Kontani}}, and
  \bibinfo{author}{\bibfnamefont{H.}~\bibnamefont{Aoki}}, \bibinfo{year}{2008},
  \bibinfo{journal}{Phys. Rev. Lett.} \textbf{\bibinfo{volume}{101}},
  \bibinfo{pages}{087004}.

\bibitem[{\citenamefont{Lal} \emph{et~al.}(2002)\citenamefont{Lal, Rao, and
  Sen}}]{Lal02}
\bibinfo{author}{\bibnamefont{Lal}, \bibfnamefont{S.}},
  \bibinfo{author}{\bibfnamefont{S.}~\bibnamefont{Rao}}, and
  \bibinfo{author}{\bibfnamefont{D.}~\bibnamefont{Sen}}, \bibinfo{year}{2002},
  \bibinfo{journal}{Phys. Rev. B} \textbf{\bibinfo{volume}{66}},
  \bibinfo{pages}{165327}.

\bibitem[{\citenamefont{Landauer}(1957)}]{Landauer57}
\bibinfo{author}{\bibnamefont{Landauer}, \bibfnamefont{R.}},
  \bibinfo{year}{1957}, \bibinfo{journal}{IBM J. Res. Dev.}
  \textbf{\bibinfo{volume}{1}}, \bibinfo{pages}{223}.

\bibitem[{\citenamefont{Langer and Ambegaokar}(1961)}]{Langer61}
\bibinfo{author}{\bibnamefont{Langer}, \bibfnamefont{J.~S.}}, and
  \bibinfo{author}{\bibfnamefont{V.}~\bibnamefont{Ambegaokar}},
  \bibinfo{year}{1961}, \bibinfo{journal}{Phys. Rev.}
  \textbf{\bibinfo{volume}{121}}, \bibinfo{pages}{1090}.

\bibitem[{\citenamefont{L{\"a}uchli}
  \emph{et~al.}(2004)\citenamefont{L{\"a}uchli, Honerkamp, and
  Rice}}]{Laeuchli04}
\bibinfo{author}{\bibnamefont{L{\"a}uchli}, \bibfnamefont{A.}},
  \bibinfo{author}{\bibfnamefont{C.}~\bibnamefont{Honerkamp}}, and
  \bibinfo{author}{\bibfnamefont{T.}~\bibnamefont{Rice}}, \bibinfo{year}{2004},
  \bibinfo{journal}{Phys. Rev. Lett.} \textbf{\bibinfo{volume}{92}},
  \bibinfo{pages}{037006}.

\bibitem[{\citenamefont{Lederer} \emph{et~al.}(1987)\citenamefont{Lederer,
  Montambaux, and Poilblanc}}]{Lederer}
\bibinfo{author}{\bibnamefont{Lederer}, \bibfnamefont{P.}},
  \bibinfo{author}{\bibfnamefont{G.}~\bibnamefont{Montambaux}}, and
  \bibinfo{author}{\bibfnamefont{D.}~\bibnamefont{Poilblanc}},
  \bibinfo{year}{1987}, \bibinfo{journal}{J. Phys. (Paris)}
  \textbf{\bibinfo{volume}{48}}, \bibinfo{pages}{1613}.

\bibitem[{\citenamefont{Ledowski and Kopietz}(2003)}]{Ledowski03}
\bibinfo{author}{\bibnamefont{Ledowski}, \bibfnamefont{S.}}, and
  \bibinfo{author}{\bibfnamefont{P.}~\bibnamefont{Kopietz}},
  \bibinfo{year}{2003}, \bibinfo{journal}{J. Phys. Cond. Matt.}
  \textbf{\bibinfo{volume}{15}}, \bibinfo{pages}{4779}.

\bibitem[{\citenamefont{Ledowski and
  Kopietz}(2007{\natexlab{a}})}]{Ledowski07a}
\bibinfo{author}{\bibnamefont{Ledowski}, \bibfnamefont{S.}}, and
  \bibinfo{author}{\bibfnamefont{P.}~\bibnamefont{Kopietz}},
  \bibinfo{year}{2007}{\natexlab{a}}, \bibinfo{journal}{Phys. Rev. B}
  \textbf{\bibinfo{volume}{75}}, \bibinfo{pages}{045134}.

\bibitem[{\citenamefont{Ledowski and
  Kopietz}(2007{\natexlab{b}})}]{Ledowski07b}
\bibinfo{author}{\bibnamefont{Ledowski}, \bibfnamefont{S.}}, and
  \bibinfo{author}{\bibfnamefont{P.}~\bibnamefont{Kopietz}},
  \bibinfo{year}{2007}{\natexlab{b}}, \bibinfo{journal}{Phys. Rev. B}
  \textbf{\bibinfo{volume}{76}}, \bibinfo{pages}{121403(R)}.

\bibitem[{\citenamefont{Ledowski} \emph{et~al.}(2005)\citenamefont{Ledowski,
  Kopietz, and Ferraz}}]{Ledowski05}
\bibinfo{author}{\bibnamefont{Ledowski}, \bibfnamefont{S.}},
  \bibinfo{author}{\bibfnamefont{P.}~\bibnamefont{Kopietz}}, and
  \bibinfo{author}{\bibfnamefont{A.}~\bibnamefont{Ferraz}},
  \bibinfo{year}{2005}, \bibinfo{journal}{Phys. Rev. B}
  \textbf{\bibinfo{volume}{71}}, \bibinfo{pages}{235106}.

\bibitem[{\citenamefont{Lee} \emph{et~al.}(2006)\citenamefont{Lee, Nagaosa, and
  Wen}}]{Lee06}
\bibinfo{author}{\bibnamefont{Lee}, \bibfnamefont{P.~A.}},
  \bibinfo{author}{\bibfnamefont{N.}~\bibnamefont{Nagaosa}}, and
  \bibinfo{author}{\bibfnamefont{X.~G.} \bibnamefont{Wen}},
  \bibinfo{year}{2006}, \bibinfo{journal}{Rev. Mod. Phys.}
  \textbf{\bibinfo{volume}{78}}, \bibinfo{pages}{17}.

\bibitem[{\citenamefont{Lesniewski}(1987)}]{Lesniewski87}
\bibinfo{author}{\bibnamefont{Lesniewski}, \bibfnamefont{A.}},
  \bibinfo{year}{1987}, \bibinfo{journal}{Commun. Math. Phys.}
  \textbf{\bibinfo{volume}{108}}, \bibinfo{pages}{437}.

\bibitem[{\citenamefont{Litim}(2001)}]{Litim01}
\bibinfo{author}{\bibnamefont{Litim}, \bibfnamefont{D.}}, \bibinfo{year}{2001},
  \bibinfo{journal}{Phys. Rev. D} \textbf{\bibinfo{volume}{64}},
  \bibinfo{pages}{105007}.

\bibitem[{\citenamefont{von Loehneysen} \emph{et~al.}(2007)\citenamefont{von
  Loehneysen, Rosch, Vojta, and W{\"o}lfle}}]{Loehneysen07}
\bibinfo{author}{\bibnamefont{von Loehneysen}, \bibfnamefont{H.}},
  \bibinfo{author}{\bibfnamefont{A.}~\bibnamefont{Rosch}},
  \bibinfo{author}{\bibfnamefont{M.}~\bibnamefont{Vojta}}, and
  \bibinfo{author}{\bibfnamefont{P.}~\bibnamefont{W{\"o}lfle}},
  \bibinfo{year}{2007}, \bibinfo{journal}{Rev. Mod. Phys.}
  \textbf{\bibinfo{volume}{79}}, \bibinfo{pages}{1015}.

\bibitem[{\citenamefont{Lopez-Sancho}
  \emph{et~al.}(2009)\citenamefont{Lopez-Sancho, de~Juan, and
  Vozmediano}}]{lopez-sancho}
\bibinfo{author}{\bibnamefont{Lopez-Sancho}, \bibfnamefont{M.~P.}},
  \bibinfo{author}{\bibfnamefont{F.}~\bibnamefont{de~Juan}}, and
  \bibinfo{author}{\bibfnamefont{M.~A.~H.} \bibnamefont{Vozmediano}},
  \bibinfo{year}{2009}, \bibinfo{journal}{Phys. Rev. B}
  \textbf{\bibinfo{volume}{79}}, \bibinfo{pages}{075413}.

\bibitem[{\citenamefont{Luther and Peschel}(1974)}]{Luther74}
\bibinfo{author}{\bibnamefont{Luther}, \bibfnamefont{A.}}, and
  \bibinfo{author}{\bibfnamefont{I.}~\bibnamefont{Peschel}},
  \bibinfo{year}{1974}, \bibinfo{journal}{Phys. Rev. B}
  \textbf{\bibinfo{volume}{9}}, \bibinfo{pages}{2911}.

\bibitem[{\citenamefont{Luttinger}(1963)}]{Luttinger63}
\bibinfo{author}{\bibnamefont{Luttinger}, \bibfnamefont{J.~M.}},
  \bibinfo{year}{1963}, \bibinfo{journal}{J. Math. Phys.}
  \textbf{\bibinfo{volume}{4}}, \bibinfo{pages}{1154}.

\bibitem[{\citenamefont{Maier} \emph{et~al.}(2009)\citenamefont{Maier, Graser,
  Scalapino, and Hirschfeld}}]{Maier09}
\bibinfo{author}{\bibnamefont{Maier}, \bibfnamefont{T.~A.}},
  \bibinfo{author}{\bibfnamefont{S.}~\bibnamefont{Graser}},
  \bibinfo{author}{\bibfnamefont{D.~J.} \bibnamefont{Scalapino}}, and
  \bibinfo{author}{\bibfnamefont{P.~J.} \bibnamefont{Hirschfeld}},
  \bibinfo{year}{2009}, \bibinfo{journal}{Phys. Rev. B}
  \textbf{\bibinfo{volume}{79}}, \bibinfo{pages}{224510}.

\bibitem[{\citenamefont{Maslov and Stone}(1995)}]{Maslov95}
\bibinfo{author}{\bibnamefont{Maslov}, \bibfnamefont{D.~L.}}, and
  \bibinfo{author}{\bibfnamefont{M.}~\bibnamefont{Stone}},
  \bibinfo{year}{1995}, \bibinfo{journal}{Phys. Rev. B}
  \textbf{\bibinfo{volume}{52}}, \bibinfo{pages}{R5539}.

\bibitem[{\citenamefont{Mathey} \emph{et~al.}(2006)\citenamefont{Mathey, Tsai,
  and {Castro Neto}}}]{Mathey06}
\bibinfo{author}{\bibnamefont{Mathey}, \bibfnamefont{L.}},
  \bibinfo{author}{\bibfnamefont{S.~W.} \bibnamefont{Tsai}}, and
  \bibinfo{author}{\bibfnamefont{A.~H.} \bibnamefont{{Castro Neto}}},
  \bibinfo{year}{2006}, \bibinfo{journal}{Phys. Rev. Lett.}
  \textbf{\bibinfo{volume}{97}}, \bibinfo{pages}{030601}.

\bibitem[{\citenamefont{Mathey} \emph{et~al.}(2007)\citenamefont{Mathey, Tsai,
  and {Castro Neto}}}]{Mathey07}
\bibinfo{author}{\bibnamefont{Mathey}, \bibfnamefont{L.}},
  \bibinfo{author}{\bibfnamefont{S.~W.} \bibnamefont{Tsai}}, and
  \bibinfo{author}{\bibfnamefont{A.~H.} \bibnamefont{{Castro Neto}}},
  \bibinfo{year}{2007}, \bibinfo{journal}{Phys. Rev. B}
  \textbf{\bibinfo{volume}{75}}, \bibinfo{pages}{174516}.

\bibitem[{\citenamefont{Mattis}(1974)}]{Mattis74}
\bibinfo{author}{\bibnamefont{Mattis}, \bibfnamefont{D.~C.}},
  \bibinfo{year}{1974}, \bibinfo{journal}{J. Math. Phys.}
  \textbf{\bibinfo{volume}{15}}, \bibinfo{pages}{609}.

\bibitem[{\citenamefont{Mattis and Lieb}(1965)}]{Mattis65}
\bibinfo{author}{\bibnamefont{Mattis}, \bibfnamefont{D.~C.}}, and
  \bibinfo{author}{\bibfnamefont{E.~H.} \bibnamefont{Lieb}},
  \bibinfo{year}{1965}, \bibinfo{journal}{J. Math. Phys.}
  \textbf{\bibinfo{volume}{6}}, \bibinfo{pages}{304}.

\bibitem[{\citenamefont{Matveev} \emph{et~al.}(1993)\citenamefont{Matveev, Yue,
  and Glazman}}]{Matveev93}
\bibinfo{author}{\bibnamefont{Matveev}, \bibfnamefont{K.~A.}},
  \bibinfo{author}{\bibfnamefont{D.}~\bibnamefont{Yue}}, and
  \bibinfo{author}{\bibfnamefont{L.~I.} \bibnamefont{Glazman}},
  \bibinfo{year}{1993}, \bibinfo{journal}{Phys. Rev. Lett.}
  \textbf{\bibinfo{volume}{71}}, \bibinfo{pages}{3351}.

\bibitem[{\citenamefont{Mazin} \emph{et~al.}(2008)\citenamefont{Mazin, Singh,
  Johannes, and Du}}]{Mazin08}
\bibinfo{author}{\bibnamefont{Mazin}, \bibfnamefont{I.~I.}},
  \bibinfo{author}{\bibfnamefont{D.~J.} \bibnamefont{Singh}},
  \bibinfo{author}{\bibfnamefont{M.~D.} \bibnamefont{Johannes}}, and
  \bibinfo{author}{\bibfnamefont{M.~H.} \bibnamefont{Du}},
  \bibinfo{year}{2008}, \bibinfo{journal}{Phys. Rev. Lett.}
  \textbf{\bibinfo{volume}{101}}, \bibinfo{pages}{057003}.

\bibitem[{\citenamefont{McKenzie}(1997)}]{mckenzie}
\bibinfo{author}{\bibnamefont{McKenzie}, \bibfnamefont{R.~H.}},
  \bibinfo{year}{1997}, \bibinfo{journal}{Science}
  \textbf{\bibinfo{volume}{31}}, \bibinfo{pages}{820}.

\bibitem[{\citenamefont{Meden} \emph{et~al.}(2005)\citenamefont{Meden, Enss,
  Andergassen, Metzner, and Sch{\"o}nhammer}}]{Meden05}
\bibinfo{author}{\bibnamefont{Meden}, \bibfnamefont{V.}},
  \bibinfo{author}{\bibfnamefont{T.}~\bibnamefont{Enss}},
  \bibinfo{author}{\bibfnamefont{S.}~\bibnamefont{Andergassen}},
  \bibinfo{author}{\bibfnamefont{W.}~\bibnamefont{Metzner}}, and
  \bibinfo{author}{\bibfnamefont{K.}~\bibnamefont{Sch{\"o}nhammer}},
  \bibinfo{year}{2005}, \bibinfo{journal}{Phys. Rev. B}
  \textbf{\bibinfo{volume}{71}}, \bibinfo{pages}{041302}.

\bibitem[{\citenamefont{Meden and Marquardt}(2006)}]{Meden06}
\bibinfo{author}{\bibnamefont{Meden}, \bibfnamefont{V.}}, and
  \bibinfo{author}{\bibfnamefont{F.}~\bibnamefont{Marquardt}},
  \bibinfo{year}{2006}, \bibinfo{journal}{Phys. Rev. Lett.}
  \textbf{\bibinfo{volume}{96}}, \bibinfo{pages}{146801}.

\bibitem[{\citenamefont{Meden} \emph{et~al.}(2002)\citenamefont{Meden, Metzner,
  Schollw{\"o}ck, and Sch{\"o}nhammer}}]{Meden02b}
\bibinfo{author}{\bibnamefont{Meden}, \bibfnamefont{V.}},
  \bibinfo{author}{\bibfnamefont{W.}~\bibnamefont{Metzner}},
  \bibinfo{author}{\bibfnamefont{U.}~\bibnamefont{Schollw{\"o}ck}}, and
  \bibinfo{author}{\bibfnamefont{K.}~\bibnamefont{Sch{\"o}nhammer}},
  \bibinfo{year}{2002}, \bibinfo{journal}{J. Low Temp. Phys.}
  \textbf{\bibinfo{volume}{126}}, \bibinfo{pages}{1147}.

\bibitem[{\citenamefont{Meden and
  Schollw{\"o}ck}(2003{\natexlab{a}})}]{Meden03b}
\bibinfo{author}{\bibnamefont{Meden}, \bibfnamefont{V.}}, and
  \bibinfo{author}{\bibfnamefont{U.}~\bibnamefont{Schollw{\"o}ck}},
  \bibinfo{year}{2003}{\natexlab{a}}, \bibinfo{journal}{Phys. Rev. B}
  \textbf{\bibinfo{volume}{67}}, \bibinfo{pages}{193303}.

\bibitem[{\citenamefont{Meden and
  Schollw{\"o}ck}(2003{\natexlab{b}})}]{Meden03a}
\bibinfo{author}{\bibnamefont{Meden}, \bibfnamefont{V.}}, and
  \bibinfo{author}{\bibfnamefont{U.}~\bibnamefont{Schollw{\"o}ck}},
  \bibinfo{year}{2003}{\natexlab{b}}, \bibinfo{journal}{Phys. Rev. B}
  \textbf{\bibinfo{volume}{67}}, \bibinfo{pages}{035106}.

\bibitem[{\citenamefont{Meden and Sch{\"o}nhammer}(1992)}]{Meden92}
\bibinfo{author}{\bibnamefont{Meden}, \bibfnamefont{V.}}, and
  \bibinfo{author}{\bibfnamefont{K.}~\bibnamefont{Sch{\"o}nhammer}},
  \bibinfo{year}{1992}, \bibinfo{journal}{Phys. Rev. B}
  \textbf{\bibinfo{volume}{46}}, \bibinfo{pages}{15753}.

\bibitem[{\citenamefont{Meir and Wingreen}(1992)}]{Meir92}
\bibinfo{author}{\bibnamefont{Meir}, \bibfnamefont{Y.}}, and
  \bibinfo{author}{\bibfnamefont{N.~S.} \bibnamefont{Wingreen}},
  \bibinfo{year}{1992}, \bibinfo{journal}{Phys. Rev. Lett.}
  \textbf{\bibinfo{volume}{68}}, \bibinfo{pages}{2512}.

\bibitem[{\citenamefont{Meng} \emph{et~al.}(2010)\citenamefont{Meng, Lang,
  Wessel, Assaad, and Muramatsu}}]{Meng2010}
\bibinfo{author}{\bibnamefont{Meng}, \bibfnamefont{Z.~Y.}},
  \bibinfo{author}{\bibfnamefont{T.~C.} \bibnamefont{Lang}},
  \bibinfo{author}{\bibfnamefont{S.}~\bibnamefont{Wessel}},
  \bibinfo{author}{\bibfnamefont{F.~F.} \bibnamefont{Assaad}}, and
  \bibinfo{author}{\bibfnamefont{A.}~\bibnamefont{Muramatsu}},
  \bibinfo{year}{2010}, \bibinfo{journal}{Nature}
  \textbf{\bibinfo{volume}{464}}, \bibinfo{pages}{847}.

\bibitem[{\citenamefont{Metlitski and
  Sachdev}(2010{\natexlab{a}})}]{Metlitski10b}
\bibinfo{author}{\bibnamefont{Metlitski}, \bibfnamefont{M.}}, and
  \bibinfo{author}{\bibfnamefont{S.}~\bibnamefont{Sachdev}},
  \bibinfo{year}{2010}{\natexlab{a}}, \bibinfo{journal}{Phys. Rev. B}
  \textbf{\bibinfo{volume}{82}}, \bibinfo{pages}{075128}.

\bibitem[{\citenamefont{Metlitski and
  Sachdev}(2010{\natexlab{b}})}]{Metlitski10a}
\bibinfo{author}{\bibnamefont{Metlitski}, \bibfnamefont{M.}}, and
  \bibinfo{author}{\bibfnamefont{S.}~\bibnamefont{Sachdev}},
  \bibinfo{year}{2010}{\natexlab{b}}, \bibinfo{journal}{Phys. Rev. B}
  \textbf{\bibinfo{volume}{82}}, \bibinfo{pages}{075127}.

\bibitem[{\citenamefont{Metlitski and
  Sachdev}(2010{\natexlab{c}})}]{Metlitski10}
\bibinfo{author}{\bibnamefont{Metlitski}, \bibfnamefont{M.~A.}}, and
  \bibinfo{author}{\bibfnamefont{S.}~\bibnamefont{Sachdev}},
  \bibinfo{year}{2010}{\natexlab{c}}, \bibinfo{journal}{New J. Phys.}
  \textbf{\bibinfo{volume}{12}}, \bibinfo{pages}{105007}.

\bibitem[{\citenamefont{Metzner} \emph{et~al.}(1998)\citenamefont{Metzner,
  Castellani, and {Di Castro}}}]{Metzner98}
\bibinfo{author}{\bibnamefont{Metzner}, \bibfnamefont{W.}},
  \bibinfo{author}{\bibfnamefont{C.}~\bibnamefont{Castellani}}, and
  \bibinfo{author}{\bibfnamefont{C.}~\bibnamefont{{Di Castro}}},
  \bibinfo{year}{1998}, \bibinfo{journal}{Adv. Phys.}
  \textbf{\bibinfo{volume}{47}}, \bibinfo{pages}{317}.

\bibitem[{\citenamefont{Metzner} \emph{et~al.}(2003)\citenamefont{Metzner,
  Rohe, and Andergassen}}]{Metzner03}
\bibinfo{author}{\bibnamefont{Metzner}, \bibfnamefont{W.}},
  \bibinfo{author}{\bibfnamefont{D.}~\bibnamefont{Rohe}}, and
  \bibinfo{author}{\bibfnamefont{S.}~\bibnamefont{Andergassen}},
  \bibinfo{year}{2003}, \bibinfo{journal}{Phys. Rev. Lett.}
  \textbf{\bibinfo{volume}{91}}, \bibinfo{pages}{066402}.

\bibitem[{\citenamefont{Millis}(1993)}]{Millis93}
\bibinfo{author}{\bibnamefont{Millis}, \bibfnamefont{A.~J.}},
  \bibinfo{year}{1993}, \bibinfo{journal}{Phys. Rev. B}
  \textbf{\bibinfo{volume}{48}}, \bibinfo{pages}{7183}.

\bibitem[{\citenamefont{Mitra} \emph{et~al.}(2006)\citenamefont{Mitra, Takei,
  Kim, and Millis}}]{Mitra06}
\bibinfo{author}{\bibnamefont{Mitra}, \bibfnamefont{A.}},
  \bibinfo{author}{\bibfnamefont{S.}~\bibnamefont{Takei}},
  \bibinfo{author}{\bibfnamefont{Y.}~\bibnamefont{Kim}}, and
  \bibinfo{author}{\bibfnamefont{A.}~\bibnamefont{Millis}},
  \bibinfo{year}{2006}, \bibinfo{journal}{Phys. Rev. Lett.}
  \textbf{\bibinfo{volume}{97}}, \bibinfo{pages}{236808}.

\bibitem[{\citenamefont{Miyake} \emph{et~al.}(2010)\citenamefont{Miyake,
  Nakamura, Arita, and Imada}}]{Miyake10}
\bibinfo{author}{\bibnamefont{Miyake}, \bibfnamefont{T.}},
  \bibinfo{author}{\bibfnamefont{K.}~\bibnamefont{Nakamura}},
  \bibinfo{author}{\bibfnamefont{R.}~\bibnamefont{Arita}}, and
  \bibinfo{author}{\bibfnamefont{M.}~\bibnamefont{Imada}},
  \bibinfo{year}{2010}, \bibinfo{journal}{J. Phys. Soc. Jpn.}
  \textbf{\bibinfo{volume}{79}}, \bibinfo{pages}{044705}.

\bibitem[{\citenamefont{Moon} \emph{et~al.}(1993)\citenamefont{Moon, Yi, Kane,
  Girvin, and Fisher}}]{Moon93}
\bibinfo{author}{\bibnamefont{Moon}, \bibfnamefont{K.}},
  \bibinfo{author}{\bibfnamefont{H.}~\bibnamefont{Yi}},
  \bibinfo{author}{\bibfnamefont{C.~L.} \bibnamefont{Kane}},
  \bibinfo{author}{\bibfnamefont{S.~M.} \bibnamefont{Girvin}}, and
  \bibinfo{author}{\bibfnamefont{M.~P.~A.} \bibnamefont{Fisher}},
  \bibinfo{year}{1993}, \bibinfo{journal}{Phys. Rev. Lett.}
  \textbf{\bibinfo{volume}{71}}, \bibinfo{pages}{4381}.

\bibitem[{\citenamefont{Morita} \emph{et~al.}(2002)\citenamefont{Morita,
  Watanabe, and Imada}}]{Morita02}
\bibinfo{author}{\bibnamefont{Morita}, \bibfnamefont{H.}},
  \bibinfo{author}{\bibfnamefont{S.}~\bibnamefont{Watanabe}}, and
  \bibinfo{author}{\bibfnamefont{M.}~\bibnamefont{Imada}},
  \bibinfo{year}{2002}, \bibinfo{journal}{J. Phys. Soc. Jpn.}
  \textbf{\bibinfo{volume}{71}}, \bibinfo{pages}{2109}.

\bibitem[{\citenamefont{Morris}(1994)}]{Morris94}
\bibinfo{author}{\bibnamefont{Morris}, \bibfnamefont{T.~R.}},
  \bibinfo{year}{1994}, \bibinfo{journal}{Int. J. Mod. Phys. A}
  \textbf{\bibinfo{volume}{9}}, \bibinfo{pages}{2411}.

\bibitem[{\citenamefont{M{\"u}hlschlegel}(1962)}]{Muehlschlegel62}
\bibinfo{author}{\bibnamefont{M{\"u}hlschlegel}, \bibfnamefont{B.}},
  \bibinfo{year}{1962}, \bibinfo{journal}{J. Math. Phys.}
  \textbf{\bibinfo{volume}{3}}, \bibinfo{pages}{522}.

\bibitem[{\citenamefont{Nayak} \emph{et~al.}(1999)\citenamefont{Nayak, Fisher,
  Ludwig, and Lin}}]{Nayak99}
\bibinfo{author}{\bibnamefont{Nayak}, \bibfnamefont{C.}},
  \bibinfo{author}{\bibfnamefont{M.~P.~A.} \bibnamefont{Fisher}},
  \bibinfo{author}{\bibfnamefont{A.~W.~W.} \bibnamefont{Ludwig}}, and
  \bibinfo{author}{\bibfnamefont{H.~H.} \bibnamefont{Lin}},
  \bibinfo{year}{1999}, \bibinfo{journal}{Phys. Rev. B}
  \textbf{\bibinfo{volume}{59}}, \bibinfo{pages}{15694}.

\bibitem[{\citenamefont{Nazarov and Glazman}(2003)}]{Nazarov03}
\bibinfo{author}{\bibnamefont{Nazarov}, \bibfnamefont{Y.~V.}}, and
  \bibinfo{author}{\bibfnamefont{L.~I.} \bibnamefont{Glazman}},
  \bibinfo{year}{2003}, \bibinfo{journal}{Phys. Rev. Lett.}
  \textbf{\bibinfo{volume}{91}}, \bibinfo{pages}{126804}.

\bibitem[{\citenamefont{Negele and Orland}(1987)}]{Negele87}
\bibinfo{author}{\bibnamefont{Negele}, \bibfnamefont{J.~W.}}, and
  \bibinfo{author}{\bibfnamefont{H.}~\bibnamefont{Orland}},
  \bibinfo{year}{1987}, \emph{\bibinfo{title}{Quantum Many-Particle Systems}}
  (\bibinfo{publisher}{Addison-Wesley, Reading}).

\bibitem[{\citenamefont{Neumayr and Metzner}(2003)}]{Neumayr2003}
\bibinfo{author}{\bibnamefont{Neumayr}, \bibfnamefont{A.}}, and
  \bibinfo{author}{\bibfnamefont{W.}~\bibnamefont{Metzner}},
  \bibinfo{year}{2003}, \bibinfo{journal}{Phys. Rev. B}
  \textbf{\bibinfo{volume}{67}}, \bibinfo{pages}{035112}.

\bibitem[{\citenamefont{Nickel} \emph{et~al.}(2005)\citenamefont{Nickel,
  Duprat, Bourbonnais, and Dupuis}}]{Nickel05}
\bibinfo{author}{\bibnamefont{Nickel}, \bibfnamefont{J.~C.}},
  \bibinfo{author}{\bibfnamefont{R.}~\bibnamefont{Duprat}},
  \bibinfo{author}{\bibfnamefont{C.}~\bibnamefont{Bourbonnais}}, and
  \bibinfo{author}{\bibfnamefont{N.}~\bibnamefont{Dupuis}},
  \bibinfo{year}{2005}, \bibinfo{journal}{Phys. Rev. Lett.}
  \textbf{\bibinfo{volume}{95}}, \bibinfo{pages}{247001}.

\bibitem[{\citenamefont{Nickel} \emph{et~al.}(2006)\citenamefont{Nickel,
  Duprat, Bourbonnais, and Dupuis}}]{Nickel06}
\bibinfo{author}{\bibnamefont{Nickel}, \bibfnamefont{J.~C.}},
  \bibinfo{author}{\bibfnamefont{R.}~\bibnamefont{Duprat}},
  \bibinfo{author}{\bibfnamefont{C.}~\bibnamefont{Bourbonnais}}, and
  \bibinfo{author}{\bibfnamefont{N.}~\bibnamefont{Dupuis}},
  \bibinfo{year}{2006}, \bibinfo{journal}{Phys. Rev. B}
  \textbf{\bibinfo{volume}{73}}, \bibinfo{pages}{165126}.

\bibitem[{\citenamefont{Norman}(2008)}]{Norman08}
\bibinfo{author}{\bibnamefont{Norman}, \bibfnamefont{M.~R.}},
  \bibinfo{year}{2008}, \bibinfo{journal}{Physics}
  \textbf{\bibinfo{volume}{1}}, \bibinfo{pages}{21}.

\bibitem[{\citenamefont{Nozi{\`e}res}(1964)}]{Nozieres64}
\bibinfo{author}{\bibnamefont{Nozi{\`e}res}, \bibfnamefont{P.}},
  \bibinfo{year}{1964}, \emph{\bibinfo{title}{Theory of Interacting Fermi
  Systems}} (\bibinfo{publisher}{Benjamin, Amsterdam}).

\bibitem[{\citenamefont{Obert} \emph{et~al.}(2011)\citenamefont{Obert, Takei,
  and Metzner}}]{Obert11}
\bibinfo{author}{\bibnamefont{Obert}, \bibfnamefont{B.}},
  \bibinfo{author}{\bibfnamefont{S.}~\bibnamefont{Takei}}, and
  \bibinfo{author}{\bibfnamefont{W.}~\bibnamefont{Metzner}},
  \bibinfo{year}{2011}, \bibinfo{journal}{Ann. Phys. (Berlin)}
  \textbf{\bibinfo{volume}{523}}, \bibinfo{pages}{621}.

\bibitem[{\citenamefont{Oguri}(2001)}]{Oguri01}
\bibinfo{author}{\bibnamefont{Oguri}, \bibfnamefont{A.}}, \bibinfo{year}{2001},
  \bibinfo{journal}{J. Phys. Soc. Jpn.} \textbf{\bibinfo{volume}{70}},
  \bibinfo{pages}{2666}.

\bibitem[{\citenamefont{Okamoto} \emph{et~al.}(2010)\citenamefont{Okamoto, S{\'
  e}n{\'e}chal, Civelli, and Tremblay}}]{Okamoto10}
\bibinfo{author}{\bibnamefont{Okamoto}, \bibfnamefont{S.}},
  \bibinfo{author}{\bibfnamefont{D.}~\bibnamefont{S{\' e}n{\'e}chal}},
  \bibinfo{author}{\bibfnamefont{M.}~\bibnamefont{Civelli}}, and
  \bibinfo{author}{\bibfnamefont{A.}~\bibnamefont{Tremblay}},
  \bibinfo{year}{2010}, \bibinfo{journal}{Phys. Rev. B}
  \textbf{\bibinfo{volume}{82}}, \bibinfo{pages}{180511}.

\bibitem[{\citenamefont{Ossadnik} \emph{et~al.}(2008)\citenamefont{Ossadnik,
  Honerkamp, Rice, and Sigrist}}]{Ossadnik08}
\bibinfo{author}{\bibnamefont{Ossadnik}, \bibfnamefont{M.}},
  \bibinfo{author}{\bibfnamefont{C.}~\bibnamefont{Honerkamp}},
  \bibinfo{author}{\bibfnamefont{T.~M.} \bibnamefont{Rice}}, and
  \bibinfo{author}{\bibfnamefont{M.}~\bibnamefont{Sigrist}},
  \bibinfo{year}{2008}, \bibinfo{journal}{Phys. Rev. Lett.}
  \textbf{\bibinfo{volume}{101}}, \bibinfo{pages}{256405}.

\bibitem[{\citenamefont{Pedra and Salmhofer}(2008)}]{Pedra08}
\bibinfo{author}{\bibnamefont{Pedra}, \bibfnamefont{W.}}, and
  \bibinfo{author}{\bibfnamefont{M.}~\bibnamefont{Salmhofer}},
  \bibinfo{year}{2008}, \bibinfo{journal}{Commun. Math. Phys.}
  \textbf{\bibinfo{volume}{282}}, \bibinfo{pages}{797}.

\bibitem[{\citenamefont{Pistolesi} \emph{et~al.}(2004)\citenamefont{Pistolesi,
  Castellani, {Di Castro}, and Strinati}}]{Pistolesi04}
\bibinfo{author}{\bibnamefont{Pistolesi}, \bibfnamefont{F.}},
  \bibinfo{author}{\bibfnamefont{C.}~\bibnamefont{Castellani}},
  \bibinfo{author}{\bibfnamefont{C.}~\bibnamefont{{Di Castro}}}, and
  \bibinfo{author}{\bibfnamefont{G.~C.} \bibnamefont{Strinati}},
  \bibinfo{year}{2004}, \bibinfo{journal}{Phys. Rev. B}
  \textbf{\bibinfo{volume}{69}}, \bibinfo{pages}{024513}.

\bibitem[{\citenamefont{Platt} \emph{et~al.}(2009)\citenamefont{Platt,
  Honerkamp, and Hanke}}]{Platt09}
\bibinfo{author}{\bibnamefont{Platt}, \bibfnamefont{C.}},
  \bibinfo{author}{\bibfnamefont{C.}~\bibnamefont{Honerkamp}}, and
  \bibinfo{author}{\bibfnamefont{W.}~\bibnamefont{Hanke}},
  \bibinfo{year}{2009}, \bibinfo{journal}{New. J. Phys.}
  \textbf{\bibinfo{volume}{11}}, \bibinfo{pages}{055058}.

\bibitem[{\citenamefont{Platt}
  \emph{et~al.}(2011{\natexlab{a}})\citenamefont{Platt, Thomale, and
  Hanke}}]{Platt10}
\bibinfo{author}{\bibnamefont{Platt}, \bibfnamefont{C.}},
  \bibinfo{author}{\bibfnamefont{R.}~\bibnamefont{Thomale}}, and
  \bibinfo{author}{\bibfnamefont{W.}~\bibnamefont{Hanke}},
  \bibinfo{year}{2011}{\natexlab{a}}, \bibinfo{journal}{Ann. Phys. (Berlin)}
  \textbf{\bibinfo{volume}{523}}, \bibinfo{pages}{638}.

\bibitem[{\citenamefont{Platt}
  \emph{et~al.}(2011{\natexlab{b}})\citenamefont{Platt, Thomale, and
  Hanke}}]{Platt11}
\bibinfo{author}{\bibnamefont{Platt}, \bibfnamefont{C.}},
  \bibinfo{author}{\bibfnamefont{R.}~\bibnamefont{Thomale}}, and
  \bibinfo{author}{\bibfnamefont{W.}~\bibnamefont{Hanke}},
  \bibinfo{year}{2011}{\natexlab{b}}, \bibinfo{journal}{Phys. Rev. B}
  \textbf{\bibinfo{volume}{84}}, \bibinfo{pages}{235121}.

\bibitem[{\citenamefont{Polchinski}(1984)}]{Polchinski84}
\bibinfo{author}{\bibnamefont{Polchinski}, \bibfnamefont{J.}},
  \bibinfo{year}{1984}, \bibinfo{journal}{Nucl. Phys. B}
  \textbf{\bibinfo{volume}{231}}, \bibinfo{pages}{269}.

\bibitem[{\citenamefont{Polchinski}(1993)}]{Polchinski93}
\bibinfo{author}{\bibnamefont{Polchinski}, \bibfnamefont{J.}},
  \bibinfo{year}{1993}, \emph{\bibinfo{title}{Proceedings of 1993 Theoretical
  Advanced Studies Institute in Elementary Particle Physics}}
  (\bibinfo{publisher}{edited by J. Harvey and J. Polchinski, World Scientific,
  Singapore}).

\bibitem[{\citenamefont{Polyakov and Gornyi}(2003)}]{Polyakov03}
\bibinfo{author}{\bibnamefont{Polyakov}, \bibfnamefont{D.~G.}}, and
  \bibinfo{author}{\bibfnamefont{I.~V.} \bibnamefont{Gornyi}},
  \bibinfo{year}{2003}, \bibinfo{journal}{Phys. Rev. B}
  \textbf{\bibinfo{volume}{68}}, \bibinfo{pages}{035421}.

\bibitem[{\citenamefont{Pomeranchuk}(1959)}]{Pomeranchuk58}
\bibinfo{author}{\bibnamefont{Pomeranchuk}, \bibfnamefont{I.~J.}},
  \bibinfo{year}{1959}, \bibinfo{journal}{Sov. Phys. JETP}
  \textbf{\bibinfo{volume}{8}}, \bibinfo{pages}{361}.

\bibitem[{\citenamefont{Ponomarenko}(1995)}]{Ponomarenko95}
\bibinfo{author}{\bibnamefont{Ponomarenko}, \bibfnamefont{V.~V.}},
  \bibinfo{year}{1995}, \bibinfo{journal}{Phys. Rev. B}
  \textbf{\bibinfo{volume}{52}}, \bibinfo{pages}{R8666}.

\bibitem[{\citenamefont{Popov}(1987)}]{Popov87}
\bibinfo{author}{\bibnamefont{Popov}, \bibfnamefont{V.~N.}},
  \bibinfo{year}{1987}, \emph{\bibinfo{title}{Functional integrals and
  collective excitations}} (\bibinfo{publisher}{Cambridge University Press,
  Cambridge}).

\bibitem[{\citenamefont{Raghu} \emph{et~al.}(2008)\citenamefont{Raghu, Qi,
  Honerkamp, and Zhang}}]{Raghu08}
\bibinfo{author}{\bibnamefont{Raghu}, \bibfnamefont{S.}},
  \bibinfo{author}{\bibfnamefont{X.-L.} \bibnamefont{Qi}},
  \bibinfo{author}{\bibfnamefont{C.}~\bibnamefont{Honerkamp}}, and
  \bibinfo{author}{\bibfnamefont{S.-C.} \bibnamefont{Zhang}},
  \bibinfo{year}{2008}, \bibinfo{journal}{Phys. Rev. Lett.}
  \textbf{\bibinfo{volume}{100}}, \bibinfo{pages}{156401}.

\bibitem[{\citenamefont{Rammer and Smith}(1986)}]{Rammer86}
\bibinfo{author}{\bibnamefont{Rammer}, \bibfnamefont{J.}}, and
  \bibinfo{author}{\bibfnamefont{H.}~\bibnamefont{Smith}},
  \bibinfo{year}{1986}, \bibinfo{journal}{Rev. Mod. Phys.}
  \textbf{\bibinfo{volume}{58}}, \bibinfo{pages}{323}.

\bibitem[{\citenamefont{Rech} \emph{et~al.}(2006)\citenamefont{Rech, Pepin, and
  Chubukov}}]{Rech06}
\bibinfo{author}{\bibnamefont{Rech}, \bibfnamefont{J.}},
  \bibinfo{author}{\bibfnamefont{C.}~\bibnamefont{Pepin}}, and
  \bibinfo{author}{\bibfnamefont{A.~V.} \bibnamefont{Chubukov}},
  \bibinfo{year}{2006}, \bibinfo{journal}{Phys. Rev. B}
  \textbf{\bibinfo{volume}{74}}, \bibinfo{pages}{195126}.

\bibitem[{\citenamefont{Reed and Simon}(1975)}]{Reed75}
\bibinfo{author}{\bibnamefont{Reed}, \bibfnamefont{M.}}, and
  \bibinfo{author}{\bibfnamefont{B.}~\bibnamefont{Simon}},
  \bibinfo{year}{1975}, \emph{\bibinfo{title}{Methods of Modern Mathematical
  Physics vols 3-4}} (\bibinfo{publisher}{Academic Press}).

\bibitem[{\citenamefont{Reiss} \emph{et~al.}(2007)\citenamefont{Reiss, Rohe,
  and Metzner}}]{Reiss07}
\bibinfo{author}{\bibnamefont{Reiss}, \bibfnamefont{J.}},
  \bibinfo{author}{\bibfnamefont{D.}~\bibnamefont{Rohe}}, and
  \bibinfo{author}{\bibfnamefont{W.}~\bibnamefont{Metzner}},
  \bibinfo{year}{2007}, \bibinfo{journal}{Phys. Rev. B}
  \textbf{\bibinfo{volume}{75}}, \bibinfo{pages}{075110}.

\bibitem[{\citenamefont{Reuther and Thomale}(2011)}]{reuther2}
\bibinfo{author}{\bibnamefont{Reuther}, \bibfnamefont{J.}}, and
  \bibinfo{author}{\bibfnamefont{R.}~\bibnamefont{Thomale}},
  \bibinfo{year}{2011}, \bibinfo{journal}{Phys. Rev. B}
  \textbf{\bibinfo{volume}{83}}, \bibinfo{pages}{024402}.

\bibitem[{\citenamefont{Reuther and W{\"o}lfle}(2010)}]{reuther1}
\bibinfo{author}{\bibnamefont{Reuther}, \bibfnamefont{J.}}, and
  \bibinfo{author}{\bibfnamefont{P.}~\bibnamefont{W{\"o}lfle}},
  \bibinfo{year}{2010}, \bibinfo{journal}{Phys. Rev. B}
  \textbf{\bibinfo{volume}{81}}, \bibinfo{pages}{144410}.

\bibitem[{\citenamefont{Reuther} \emph{et~al.}(2011)\citenamefont{Reuther,
  W{\"o}lfle, Darradi, Brenig, Arlego, and Richter}}]{reuther3}
\bibinfo{author}{\bibnamefont{Reuther}, \bibfnamefont{J.}},
  \bibinfo{author}{\bibfnamefont{P.}~\bibnamefont{W{\"o}lfle}},
  \bibinfo{author}{\bibfnamefont{R.}~\bibnamefont{Darradi}},
  \bibinfo{author}{\bibfnamefont{W.}~\bibnamefont{Brenig}},
  \bibinfo{author}{\bibfnamefont{M.}~\bibnamefont{Arlego}}, and
  \bibinfo{author}{\bibfnamefont{J.}~\bibnamefont{Richter}},
  \bibinfo{year}{2011}, \bibinfo{journal}{Phys. Rev. B}
  \textbf{\bibinfo{volume}{83}}, \bibinfo{pages}{064416}.

\bibitem[{\citenamefont{Rohe and Metzner}(2005)}]{Rohe05}
\bibinfo{author}{\bibnamefont{Rohe}, \bibfnamefont{D.}}, and
  \bibinfo{author}{\bibfnamefont{W.}~\bibnamefont{Metzner}},
  \bibinfo{year}{2005}, \bibinfo{journal}{Phys. Rev. B}
  \textbf{\bibinfo{volume}{71}}, \bibinfo{pages}{115116}.

\bibitem[{\citenamefont{Rosa} \emph{et~al.}(2001)\citenamefont{Rosa, Vitale,
  and Wetterich}}]{Rosa01}
\bibinfo{author}{\bibnamefont{Rosa}, \bibfnamefont{L.}},
  \bibinfo{author}{\bibfnamefont{P.}~\bibnamefont{Vitale}}, and
  \bibinfo{author}{\bibfnamefont{C.}~\bibnamefont{Wetterich}},
  \bibinfo{year}{2001}, \bibinfo{journal}{Phys. Rev. Lett.}
  \textbf{\bibinfo{volume}{86}}, \bibinfo{pages}{958}.

\bibitem[{\citenamefont{Rosch} \emph{et~al.}(2001)\citenamefont{Rosch, Kroha,
  and W{\"o}lfle}}]{Rosch01}
\bibinfo{author}{\bibnamefont{Rosch}, \bibfnamefont{A.}},
  \bibinfo{author}{\bibfnamefont{J.}~\bibnamefont{Kroha}}, and
  \bibinfo{author}{\bibfnamefont{P.}~\bibnamefont{W{\"o}lfle}},
  \bibinfo{year}{2001}, \bibinfo{journal}{Phys. Rev. Lett.}
  \textbf{\bibinfo{volume}{87}}, \bibinfo{pages}{156802}.

\bibitem[{\citenamefont{Sachdev}(1999)}]{Sachdev99}
\bibinfo{author}{\bibnamefont{Sachdev}, \bibfnamefont{S.}},
  \bibinfo{year}{1999}, \emph{\bibinfo{title}{Quantum Phase Transitions}}
  (\bibinfo{publisher}{Cambridge University Press, Cambridge}).

\bibitem[{\citenamefont{Safi and Schulz}(1995)}]{Safi95}
\bibinfo{author}{\bibnamefont{Safi}, \bibfnamefont{I.}}, and
  \bibinfo{author}{\bibfnamefont{H.~J.} \bibnamefont{Schulz}},
  \bibinfo{year}{1995}, \bibinfo{journal}{Phys. Rev. B}
  \textbf{\bibinfo{volume}{52}}, \bibinfo{pages}{R17040}.

\bibitem[{\citenamefont{Sahebsara and S{\'e}n{\'e}chal}(2008)}]{Sahebsara08}
\bibinfo{author}{\bibnamefont{Sahebsara}, \bibfnamefont{P.}}, and
  \bibinfo{author}{\bibfnamefont{D.}~\bibnamefont{S{\'e}n{\'e}chal}},
  \bibinfo{year}{2008}, \bibinfo{journal}{Phys. Rev. Lett.}
  \textbf{\bibinfo{volume}{103}}, \bibinfo{pages}{136402}.

\bibitem[{\citenamefont{Salmhofer}(1998{\natexlab{a}})}]{Salmhofer98a}
\bibinfo{author}{\bibnamefont{Salmhofer}, \bibfnamefont{M.}},
  \bibinfo{year}{1998}{\natexlab{a}}, \bibinfo{journal}{Rev. Math. Phys.}
  \textbf{\bibinfo{volume}{10}}, \bibinfo{pages}{553}.

\bibitem[{\citenamefont{Salmhofer}(1998{\natexlab{b}})}]{Salmhofer98b}
\bibinfo{author}{\bibnamefont{Salmhofer}, \bibfnamefont{M.}},
  \bibinfo{year}{1998}{\natexlab{b}}, \bibinfo{journal}{Comm. Math. Phys.}
  \textbf{\bibinfo{volume}{194}}, \bibinfo{pages}{249}.

\bibitem[{\citenamefont{Salmhofer}(1999)}]{Salmhofer99}
\bibinfo{author}{\bibnamefont{Salmhofer}, \bibfnamefont{M.}},
  \bibinfo{year}{1999}, \emph{\bibinfo{title}{Renormalization: An
  Introduction}} (\bibinfo{publisher}{Springer, Heidelberg}).

\bibitem[{\citenamefont{Salmhofer}(2007)}]{Salmhofer07}
\bibinfo{author}{\bibnamefont{Salmhofer}, \bibfnamefont{M.}},
  \bibinfo{year}{2007}, \bibinfo{journal}{Annalen der Physik}
  \textbf{\bibinfo{volume}{16}}, \bibinfo{pages}{171}.

\bibitem[{\citenamefont{Salmhofer and Honerkamp}(2001)}]{Salmhofer01}
\bibinfo{author}{\bibnamefont{Salmhofer}, \bibfnamefont{M.}}, and
  \bibinfo{author}{\bibfnamefont{C.}~\bibnamefont{Honerkamp}},
  \bibinfo{year}{2001}, \bibinfo{journal}{Prog. Theor. Phys.}
  \textbf{\bibinfo{volume}{105}}, \bibinfo{pages}{1}.

\bibitem[{\citenamefont{Salmhofer} \emph{et~al.}(2004)\citenamefont{Salmhofer,
  Honerkamp, Metzner, and Lauscher}}]{Salmhofer04}
\bibinfo{author}{\bibnamefont{Salmhofer}, \bibfnamefont{M.}},
  \bibinfo{author}{\bibfnamefont{C.}~\bibnamefont{Honerkamp}},
  \bibinfo{author}{\bibfnamefont{W.}~\bibnamefont{Metzner}}, and
  \bibinfo{author}{\bibfnamefont{O.}~\bibnamefont{Lauscher}},
  \bibinfo{year}{2004}, \bibinfo{journal}{Prog. Theor. Phys.}
  \textbf{\bibinfo{volume}{112}}, \bibinfo{pages}{943}.

\bibitem[{\citenamefont{Salmhofer and Wieczerkowski}(2000)}]{Salmhofer00}
\bibinfo{author}{\bibnamefont{Salmhofer}, \bibfnamefont{M.}}, and
  \bibinfo{author}{\bibfnamefont{C.}~\bibnamefont{Wieczerkowski}},
  \bibinfo{year}{2000}, \bibinfo{journal}{J. Stat. Phys.}
  \textbf{\bibinfo{volume}{99}}, \bibinfo{pages}{557}.

\bibitem[{\citenamefont{Schmidt and W{\"o}lfle}(2010)}]{Schmidt10}
\bibinfo{author}{\bibnamefont{Schmidt}, \bibfnamefont{H.}}, and
  \bibinfo{author}{\bibfnamefont{P.}~\bibnamefont{W{\"o}lfle}},
  \bibinfo{year}{2010}, \bibinfo{journal}{Annalen der Physik}
  \textbf{\bibinfo{volume}{19}}, \bibinfo{pages}{60}.

\bibitem[{\citenamefont{Schmidt and Enss}(2011)}]{Schmidt11}
\bibinfo{author}{\bibnamefont{Schmidt}, \bibfnamefont{R.}}, and
  \bibinfo{author}{\bibfnamefont{T.}~\bibnamefont{Enss}}, \bibinfo{year}{2011},
  \bibinfo{journal}{Phys. Rev. A} \textbf{\bibinfo{volume}{83}},
  \bibinfo{pages}{063620}.

\bibitem[{\citenamefont{Schoeller}(2000)}]{Schoeller00b}
\bibinfo{author}{\bibnamefont{Schoeller}, \bibfnamefont{H.}},
  \bibinfo{year}{2000}, \emph{\bibinfo{title}{Low-Dimensional Systems:
  Interaction and Transport Properties}} (\bibinfo{publisher}{edited by T.
  Brandes, Springer, Berlin}).

\bibitem[{\citenamefont{Schoeller}(2009)}]{Schoeller09a}
\bibinfo{author}{\bibnamefont{Schoeller}, \bibfnamefont{H.}},
  \bibinfo{year}{2009}, \bibinfo{journal}{Eur. Phys. J. Special Topics}
  \textbf{\bibinfo{volume}{168}}, \bibinfo{pages}{179}.

\bibitem[{\citenamefont{Schoeller and K{\"o}nig}(2000)}]{Schoeller00a}
\bibinfo{author}{\bibnamefont{Schoeller}, \bibfnamefont{H.}}, and
  \bibinfo{author}{\bibfnamefont{J.}~\bibnamefont{K{\"o}nig}},
  \bibinfo{year}{2000}, \bibinfo{journal}{Phys. Rev. Lett.}
  \textbf{\bibinfo{volume}{84}}, \bibinfo{pages}{3686}.

\bibitem[{\citenamefont{Sch{\"o}nhammer}(2005)}]{Schoenhammer05}
\bibinfo{author}{\bibnamefont{Sch{\"o}nhammer}, \bibfnamefont{K.}},
  \bibinfo{year}{2005}, \emph{\bibinfo{title}{Interacting Electrons in Low
  Dimensions}} (\bibinfo{publisher}{edited by D. Baeriswyl, Kluwer Academic
  Publishers, Dordrecht}).

\bibitem[{\citenamefont{Schulz}(1987)}]{Schulz87}
\bibinfo{author}{\bibnamefont{Schulz}, \bibfnamefont{H.~J.}},
  \bibinfo{year}{1987}, \bibinfo{journal}{Europhys. Lett.}
  \textbf{\bibinfo{volume}{4}}, \bibinfo{pages}{609}.

\bibitem[{\citenamefont{Sch{\"u}tz}
  \emph{et~al.}(2005)\citenamefont{Sch{\"u}tz, Bartosch, and
  Kopietz}}]{Schuetz05}
\bibinfo{author}{\bibnamefont{Sch{\"u}tz}, \bibfnamefont{F.}},
  \bibinfo{author}{\bibfnamefont{L.}~\bibnamefont{Bartosch}}, and
  \bibinfo{author}{\bibfnamefont{P.}~\bibnamefont{Kopietz}},
  \bibinfo{year}{2005}, \bibinfo{journal}{Phys. Rev. B}
  \textbf{\bibinfo{volume}{72}}, \bibinfo{pages}{035107}.

\bibitem[{\citenamefont{Sch{\"u}tz and Kopietz}(2006)}]{Schuetz06}
\bibinfo{author}{\bibnamefont{Sch{\"u}tz}, \bibfnamefont{F.}}, and
  \bibinfo{author}{\bibfnamefont{P.}~\bibnamefont{Kopietz}},
  \bibinfo{year}{2006}, \bibinfo{journal}{J. Phys. A: Math. Gen.}
  \textbf{\bibinfo{volume}{39}}, \bibinfo{pages}{8205}.

\bibitem[{\citenamefont{Shankar}(1991)}]{Shankar91}
\bibinfo{author}{\bibnamefont{Shankar}, \bibfnamefont{R.}},
  \bibinfo{year}{1991}, \bibinfo{journal}{Physica A}
  \textbf{\bibinfo{volume}{177}}, \bibinfo{pages}{530}.

\bibitem[{\citenamefont{Shankar}(1994)}]{Shankar94}
\bibinfo{author}{\bibnamefont{Shankar}, \bibfnamefont{R.}},
  \bibinfo{year}{1994}, \bibinfo{journal}{Rev. Mod. Phys.}
  \textbf{\bibinfo{volume}{66}}, \bibinfo{pages}{129}.

\bibitem[{\citenamefont{S{\'o}lyom}(1979)}]{Solyom79}
\bibinfo{author}{\bibnamefont{S{\'o}lyom}, \bibfnamefont{J.}},
  \bibinfo{year}{1979}, \bibinfo{journal}{Adv. Phys.}
  \textbf{\bibinfo{volume}{28}}, \bibinfo{pages}{201}.

\bibitem[{\citenamefont{Strack}(2009)}]{Strack09}
\bibinfo{author}{\bibnamefont{Strack}, \bibfnamefont{P.}},
  \bibinfo{year}{2009}, \emph{\bibinfo{title}{Ph.D. thesis}}
  (\bibinfo{publisher}{University Stuttgart}).

\bibitem[{\citenamefont{Strack} \emph{et~al.}(2008)\citenamefont{Strack,
  Gersch, and Metzner}}]{Strack08}
\bibinfo{author}{\bibnamefont{Strack}, \bibfnamefont{P.}},
  \bibinfo{author}{\bibfnamefont{R.}~\bibnamefont{Gersch}}, and
  \bibinfo{author}{\bibfnamefont{W.}~\bibnamefont{Metzner}},
  \bibinfo{year}{2008}, \bibinfo{journal}{Phys. Rev. B}
  \textbf{\bibinfo{volume}{78}}, \bibinfo{pages}{014522}.

\bibitem[{\citenamefont{Strack} \emph{et~al.}(2010)\citenamefont{Strack, Takei,
  and Metzner}}]{Strack10}
\bibinfo{author}{\bibnamefont{Strack}, \bibfnamefont{P.}},
  \bibinfo{author}{\bibfnamefont{S.}~\bibnamefont{Takei}}, and
  \bibinfo{author}{\bibfnamefont{W.}~\bibnamefont{Metzner}},
  \bibinfo{year}{2010}, \bibinfo{journal}{Phys. Rev. B}
  \textbf{\bibinfo{volume}{81}}, \bibinfo{pages}{125103}.

\bibitem[{\citenamefont{Tam} \emph{et~al.}(2006)\citenamefont{Tam, Tsai, and
  Campbell}}]{Tam06}
\bibinfo{author}{\bibnamefont{Tam}, \bibfnamefont{K.~M.}},
  \bibinfo{author}{\bibfnamefont{S.~W.} \bibnamefont{Tsai}}, and
  \bibinfo{author}{\bibfnamefont{D.~K.} \bibnamefont{Campbell}},
  \bibinfo{year}{2006}, \bibinfo{journal}{Phys. Rev. Lett.}
  \textbf{\bibinfo{volume}{96}}, \bibinfo{pages}{036408}.

\bibitem[{\citenamefont{Tam}
  \emph{et~al.}(2007{\natexlab{a}})\citenamefont{Tam, Tsai, Campbell, and
  {Castro Neto}}}]{Tam07C}
\bibinfo{author}{\bibnamefont{Tam}, \bibfnamefont{K.-M.}},
  \bibinfo{author}{\bibfnamefont{S.-W.} \bibnamefont{Tsai}},
  \bibinfo{author}{\bibfnamefont{D.~K.} \bibnamefont{Campbell}}, and
  \bibinfo{author}{\bibfnamefont{A.~H.} \bibnamefont{{Castro Neto}}},
  \bibinfo{year}{2007}{\natexlab{a}}, \bibinfo{journal}{Phys. Rev. B}
  \textbf{\bibinfo{volume}{75}}, \bibinfo{pages}{161103 (R)}.

\bibitem[{\citenamefont{Tam}
  \emph{et~al.}(2007{\natexlab{b}})\citenamefont{Tam, Tsai, Campbell, and
  {Castro Neto}}}]{Tam07L}
\bibinfo{author}{\bibnamefont{Tam}, \bibfnamefont{K.-M.}},
  \bibinfo{author}{\bibfnamefont{S.-W.} \bibnamefont{Tsai}},
  \bibinfo{author}{\bibfnamefont{D.~K.} \bibnamefont{Campbell}}, and
  \bibinfo{author}{\bibfnamefont{A.~H.} \bibnamefont{{Castro Neto}}},
  \bibinfo{year}{2007}{\natexlab{b}}, \bibinfo{journal}{Phys. Rev. B}
  \textbf{\bibinfo{volume}{75}}, \bibinfo{pages}{195119}.

\bibitem[{\citenamefont{Taylor}(2000)}]{Taylor72}
\bibinfo{author}{\bibnamefont{Taylor}, \bibfnamefont{J.~R.}},
  \bibinfo{year}{2000}, \emph{\bibinfo{title}{Scattering Theory}}
  (\bibinfo{publisher}{Dover Publications}).

\bibitem[{\citenamefont{Tetradis and Wetterich}(1994)}]{Tetradis94}
\bibinfo{author}{\bibnamefont{Tetradis}, \bibfnamefont{N.}}, and
  \bibinfo{author}{\bibfnamefont{C.}~\bibnamefont{Wetterich}},
  \bibinfo{year}{1994}, \bibinfo{journal}{Nucl. Phys. B}
  \textbf{\bibinfo{volume}{422}}, \bibinfo{pages}{541}.

\bibitem[{\citenamefont{Thomale}
  \emph{et~al.}(2011{\natexlab{a}})\citenamefont{Thomale, Platt, Hanke, and
  Bernevig}}]{Thomale10}
\bibinfo{author}{\bibnamefont{Thomale}, \bibfnamefont{R.}},
  \bibinfo{author}{\bibfnamefont{C.}~\bibnamefont{Platt}},
  \bibinfo{author}{\bibfnamefont{W.}~\bibnamefont{Hanke}}, and
  \bibinfo{author}{\bibfnamefont{B.~A.} \bibnamefont{Bernevig}},
  \bibinfo{year}{2011}{\natexlab{a}}, \bibinfo{journal}{Phys. Rev. Lett.}
  \textbf{\bibinfo{volume}{106}}, \bibinfo{pages}{187003}.

\bibitem[{\citenamefont{Thomale}
  \emph{et~al.}(2011{\natexlab{b}})\citenamefont{Thomale, Platt, Hanke, Hu, and
  Bernevig}}]{Thomale11}
\bibinfo{author}{\bibnamefont{Thomale}, \bibfnamefont{R.}},
  \bibinfo{author}{\bibfnamefont{C.}~\bibnamefont{Platt}},
  \bibinfo{author}{\bibfnamefont{W.}~\bibnamefont{Hanke}},
  \bibinfo{author}{\bibfnamefont{J.}~\bibnamefont{Hu}}, and
  \bibinfo{author}{\bibfnamefont{B.~A.} \bibnamefont{Bernevig}},
  \bibinfo{year}{2011}{\natexlab{b}}, \bibinfo{journal}{Phys. Rev. Lett.}
  \textbf{\bibinfo{volume}{107}}, \bibinfo{pages}{117001}.

\bibitem[{\citenamefont{Thomale} \emph{et~al.}(2009)\citenamefont{Thomale,
  Platt, Hu, Honerkamp, and Bernevig}}]{Thomale09}
\bibinfo{author}{\bibnamefont{Thomale}, \bibfnamefont{R.}},
  \bibinfo{author}{\bibfnamefont{C.}~\bibnamefont{Platt}},
  \bibinfo{author}{\bibfnamefont{J.}~\bibnamefont{Hu}},
  \bibinfo{author}{\bibfnamefont{C.}~\bibnamefont{Honerkamp}}, and
  \bibinfo{author}{\bibfnamefont{B.~A.} \bibnamefont{Bernevig}},
  \bibinfo{year}{2009}, \bibinfo{journal}{Phys. Rev. B}
  \textbf{\bibinfo{volume}{80}}, \bibinfo{pages}{180505}.

\bibitem[{\citenamefont{Tomonaga}(1950)}]{Tomonaga50}
\bibinfo{author}{\bibnamefont{Tomonaga}, \bibfnamefont{S.}},
  \bibinfo{year}{1950}, \bibinfo{journal}{Prog. Theor. Phys.}
  \textbf{\bibinfo{volume}{5}}, \bibinfo{pages}{544}.

\bibitem[{\citenamefont{Tsai} \emph{et~al.}(2005)\citenamefont{Tsai, {Castro
  Neto}, Shankar, and Campbell}}]{Tsai05}
\bibinfo{author}{\bibnamefont{Tsai}, \bibfnamefont{S.~W.}},
  \bibinfo{author}{\bibfnamefont{A.~H.} \bibnamefont{{Castro Neto}}},
  \bibinfo{author}{\bibfnamefont{R.}~\bibnamefont{Shankar}}, and
  \bibinfo{author}{\bibfnamefont{D.~K.} \bibnamefont{Campbell}},
  \bibinfo{year}{2005}, \bibinfo{journal}{Phys. Rev. B}
  \textbf{\bibinfo{volume}{72}}, \bibinfo{pages}{054531}.

\bibitem[{\citenamefont{Tsai and Marston}(2001)}]{Tsai01}
\bibinfo{author}{\bibnamefont{Tsai}, \bibfnamefont{S.~W.}}, and
  \bibinfo{author}{\bibfnamefont{J.~B.} \bibnamefont{Marston}},
  \bibinfo{year}{2001}, \bibinfo{journal}{Can. J. Phys.}
  \textbf{\bibinfo{volume}{79}}, \bibinfo{pages}{1463}.

\bibitem[{\citenamefont{Voit}(1993)}]{Voit93}
\bibinfo{author}{\bibnamefont{Voit}, \bibfnamefont{J.}}, \bibinfo{year}{1993},
  \bibinfo{journal}{Phys. Rev. B} \textbf{\bibinfo{volume}{47}},
  \bibinfo{pages}{6740}.

\bibitem[{\citenamefont{Vojta}(2003)}]{Vojta03}
\bibinfo{author}{\bibnamefont{Vojta}, \bibfnamefont{M.}}, \bibinfo{year}{2003},
  \bibinfo{journal}{Rep. Prog. Phys.} \textbf{\bibinfo{volume}{66}},
  \bibinfo{pages}{2069}.

\bibitem[{\citenamefont{Vojta} \emph{et~al.}(2000)\citenamefont{Vojta, Zhang,
  and Sachdev}}]{Vojta00}
\bibinfo{author}{\bibnamefont{Vojta}, \bibfnamefont{M.}},
  \bibinfo{author}{\bibfnamefont{Y.}~\bibnamefont{Zhang}}, and
  \bibinfo{author}{\bibfnamefont{S.}~\bibnamefont{Sachdev}},
  \bibinfo{year}{2000}, \bibinfo{journal}{Phys. Rev. Lett.}
  \textbf{\bibinfo{volume}{85}}, \bibinfo{pages}{4940}.

\bibitem[{\citenamefont{W{\"a}chter}
  \emph{et~al.}(2009)\citenamefont{W{\"a}chter, Meden, and
  Sch{\"o}nhammer}}]{Waechter09}
\bibinfo{author}{\bibnamefont{W{\"a}chter}, \bibfnamefont{P.}},
  \bibinfo{author}{\bibfnamefont{V.}~\bibnamefont{Meden}}, and
  \bibinfo{author}{\bibfnamefont{K.}~\bibnamefont{Sch{\"o}nhammer}},
  \bibinfo{year}{2009}, \bibinfo{journal}{J. Phys.: Condensed Matter}
  \textbf{\bibinfo{volume}{21}}, \bibinfo{pages}{215608}.

\bibitem[{\citenamefont{Wang}
  \emph{et~al.}(2009{\natexlab{a}})\citenamefont{Wang, Zhai, and
  Lee}}]{Wang09a}
\bibinfo{author}{\bibnamefont{Wang}, \bibfnamefont{F.}},
  \bibinfo{author}{\bibfnamefont{H.}~\bibnamefont{Zhai}}, and
  \bibinfo{author}{\bibfnamefont{D.-H.} \bibnamefont{Lee}},
  \bibinfo{year}{2009}{\natexlab{a}}, \bibinfo{journal}{Europhys. Lett.}
  \textbf{\bibinfo{volume}{85}}, \bibinfo{pages}{37005}.

\bibitem[{\citenamefont{Wang} \emph{et~al.}(2010)\citenamefont{Wang, Zhai, and
  Lee}}]{Wang10}
\bibinfo{author}{\bibnamefont{Wang}, \bibfnamefont{F.}},
  \bibinfo{author}{\bibfnamefont{H.}~\bibnamefont{Zhai}}, and
  \bibinfo{author}{\bibfnamefont{D.-H.} \bibnamefont{Lee}},
  \bibinfo{year}{2010}, \bibinfo{journal}{Phys. Rev. B}
  \textbf{\bibinfo{volume}{81}}, \bibinfo{pages}{184512}.

\bibitem[{\citenamefont{Wang}
  \emph{et~al.}(2009{\natexlab{b}})\citenamefont{Wang, Zhai, Ran, Vishwanath,
  and Lee}}]{Wang09b}
\bibinfo{author}{\bibnamefont{Wang}, \bibfnamefont{F.}},
  \bibinfo{author}{\bibfnamefont{H.}~\bibnamefont{Zhai}},
  \bibinfo{author}{\bibfnamefont{Y.}~\bibnamefont{Ran}},
  \bibinfo{author}{\bibfnamefont{A.}~\bibnamefont{Vishwanath}}, and
  \bibinfo{author}{\bibfnamefont{D.-H.} \bibnamefont{Lee}},
  \bibinfo{year}{2009}{\natexlab{b}}, \bibinfo{journal}{Phys. Rev. Lett.}
  \textbf{\bibinfo{volume}{102}}, \bibinfo{pages}{047005}.

\bibitem[{\citenamefont{Wegner}(1994)}]{Wegner94}
\bibinfo{author}{\bibnamefont{Wegner}, \bibfnamefont{F.}},
  \bibinfo{year}{1994}, \bibinfo{journal}{Annalen der Physik (Leipzig)}
  \textbf{\bibinfo{volume}{3}}, \bibinfo{pages}{77}.

\bibitem[{\citenamefont{Wegner and Houghton}(1973)}]{Wegner73}
\bibinfo{author}{\bibnamefont{Wegner}, \bibfnamefont{F.~J.}}, and
  \bibinfo{author}{\bibfnamefont{A.}~\bibnamefont{Houghton}},
  \bibinfo{year}{1973}, \bibinfo{journal}{Phys. Rev. A}
  \textbf{\bibinfo{volume}{8}}, \bibinfo{pages}{401}.

\bibitem[{\citenamefont{Wetterich}(1993)}]{Wetterich93}
\bibinfo{author}{\bibnamefont{Wetterich}, \bibfnamefont{C.}},
  \bibinfo{year}{1993}, \bibinfo{journal}{Phys. Lett. B}
  \textbf{\bibinfo{volume}{301}}, \bibinfo{pages}{90}.

\bibitem[{\citenamefont{Weyrauch and Sibold}(2008)}]{Weyrauch08}
\bibinfo{author}{\bibnamefont{Weyrauch}, \bibfnamefont{M.}}, and
  \bibinfo{author}{\bibfnamefont{D.}~\bibnamefont{Sibold}},
  \bibinfo{year}{2008}, \bibinfo{journal}{Phys. Rev. B}
  \textbf{\bibinfo{volume}{77}}, \bibinfo{pages}{125309}.

\bibitem[{\citenamefont{Wieczerkowski}(1988)}]{Wieczerkowski88}
\bibinfo{author}{\bibnamefont{Wieczerkowski}, \bibfnamefont{C.}},
  \bibinfo{year}{1988}, \bibinfo{journal}{Comm. Math. Phys.}
  \textbf{\bibinfo{volume}{120}}, \bibinfo{pages}{149}.

\bibitem[{\citenamefont{Wilson and Kogut}(1974)}]{Wilson74}
\bibinfo{author}{\bibnamefont{Wilson}, \bibfnamefont{K.~G.}}, and
  \bibinfo{author}{\bibfnamefont{J.~B.} \bibnamefont{Kogut}},
  \bibinfo{year}{1974}, \bibinfo{journal}{Phys. Rep.}
  \textbf{\bibinfo{volume}{12}}, \bibinfo{pages}{7}.

\bibitem[{\citenamefont{Xu} \emph{et~al.}(2008)\citenamefont{Xu, Gao, and
  Xiong}}]{Xu08}
\bibinfo{author}{\bibnamefont{Xu}, \bibfnamefont{Q.~Q.}},
  \bibinfo{author}{\bibfnamefont{B.~L.} \bibnamefont{Gao}}, and
  \bibinfo{author}{\bibfnamefont{S.~J.} \bibnamefont{Xiong}},
  \bibinfo{year}{2008}, \bibinfo{journal}{Physica B}
  \textbf{\bibinfo{volume}{403}}, \bibinfo{pages}{2468}.

\bibitem[{\citenamefont{Xu} \emph{et~al.}(2010)\citenamefont{Xu, Gao, and
  Xiong}}]{Xu10}
\bibinfo{author}{\bibnamefont{Xu}, \bibfnamefont{Q.~Q.}},
  \bibinfo{author}{\bibfnamefont{B.~L.} \bibnamefont{Gao}}, and
  \bibinfo{author}{\bibfnamefont{S.~J.} \bibnamefont{Xiong}},
  \bibinfo{year}{2010}, \bibinfo{journal}{Int. J. Mod. Phys. B}
  \textbf{\bibinfo{volume}{24}}, \bibinfo{pages}{575}.

\bibitem[{\citenamefont{Yamase}(2009)}]{Yamase09}
\bibinfo{author}{\bibnamefont{Yamase}, \bibfnamefont{H.}},
  \bibinfo{year}{2009}, \bibinfo{journal}{Phys. Rev. Lett.}
  \textbf{\bibinfo{volume}{102}}, \bibinfo{pages}{116404}.

\bibitem[{\citenamefont{Yamase} \emph{et~al.}(2011)\citenamefont{Yamase,
  Jakubczyk, and Metzner}}]{Yamase11}
\bibinfo{author}{\bibnamefont{Yamase}, \bibfnamefont{H.}},
  \bibinfo{author}{\bibfnamefont{P.}~\bibnamefont{Jakubczyk}}, and
  \bibinfo{author}{\bibfnamefont{W.}~\bibnamefont{Metzner}},
  \bibinfo{year}{2011}, \bibinfo{journal}{Phys. Rev. B}
  \textbf{\bibinfo{volume}{83}}, \bibinfo{pages}{125121}.

\bibitem[{\citenamefont{Yamase and Metzner}(2006)}]{Yamase06}
\bibinfo{author}{\bibnamefont{Yamase}, \bibfnamefont{H.}}, and
  \bibinfo{author}{\bibfnamefont{W.}~\bibnamefont{Metzner}},
  \bibinfo{year}{2006}, \bibinfo{journal}{Phys. Rev. B}
  \textbf{\bibinfo{volume}{73}}, \bibinfo{pages}{214517}.

\bibitem[{\citenamefont{Yamase and Metzner}(2007)}]{Yamase07}
\bibinfo{author}{\bibnamefont{Yamase}, \bibfnamefont{H.}}, and
  \bibinfo{author}{\bibfnamefont{W.}~\bibnamefont{Metzner}},
  \bibinfo{year}{2007}, \bibinfo{journal}{Phys. Rev. B}
  \textbf{\bibinfo{volume}{75}}, \bibinfo{pages}{155117}.

\bibitem[{\citenamefont{Yamase} \emph{et~al.}(2005)\citenamefont{Yamase,
  Oganesyan, and Metzner}}]{Yamase05}
\bibinfo{author}{\bibnamefont{Yamase}, \bibfnamefont{H.}},
  \bibinfo{author}{\bibfnamefont{V.}~\bibnamefont{Oganesyan}}, and
  \bibinfo{author}{\bibfnamefont{W.}~\bibnamefont{Metzner}},
  \bibinfo{year}{2005}, \bibinfo{journal}{Phys. Rev. B}
  \textbf{\bibinfo{volume}{72}}, \bibinfo{pages}{035114}.

\bibitem[{\citenamefont{Yoshioka} \emph{et~al.}(2009)\citenamefont{Yoshioka,
  Koga, and Kawakami}}]{Yoshioka09}
\bibinfo{author}{\bibnamefont{Yoshioka}, \bibfnamefont{T.}},
  \bibinfo{author}{\bibfnamefont{A.}~\bibnamefont{Koga}}, and
  \bibinfo{author}{\bibfnamefont{N.}~\bibnamefont{Kawakami}},
  \bibinfo{year}{2009}, \bibinfo{journal}{Phys. Rev. Lett.}
  \textbf{\bibinfo{volume}{103}}, \bibinfo{pages}{036401}.

\bibitem[{\citenamefont{Yue} \emph{et~al.}(1994)\citenamefont{Yue, Glazman, and
  Matveev}}]{Yue94}
\bibinfo{author}{\bibnamefont{Yue}, \bibfnamefont{D.}},
  \bibinfo{author}{\bibfnamefont{L.~I.} \bibnamefont{Glazman}}, and
  \bibinfo{author}{\bibfnamefont{K.~A.} \bibnamefont{Matveev}},
  \bibinfo{year}{1994}, \bibinfo{journal}{Phys. Rev. B}
  \textbf{\bibinfo{volume}{49}}, \bibinfo{pages}{1966}.

\bibitem[{\citenamefont{Zanchi}(2001)}]{Zanchi01}
\bibinfo{author}{\bibnamefont{Zanchi}, \bibfnamefont{D.}},
  \bibinfo{year}{2001}, \bibinfo{journal}{Europhys. Lett.}
  \textbf{\bibinfo{volume}{55}}, \bibinfo{pages}{376}.

\bibitem[{\citenamefont{Zanchi and Schulz}(1997)}]{Zanchi97}
\bibinfo{author}{\bibnamefont{Zanchi}, \bibfnamefont{D.}}, and
  \bibinfo{author}{\bibfnamefont{H.~J.} \bibnamefont{Schulz}},
  \bibinfo{year}{1997}, \bibinfo{journal}{Z. Phys. B}
  \textbf{\bibinfo{volume}{103}}, \bibinfo{pages}{339}.

\bibitem[{\citenamefont{Zanchi and Schulz}(1998)}]{Zanchi98}
\bibinfo{author}{\bibnamefont{Zanchi}, \bibfnamefont{D.}}, and
  \bibinfo{author}{\bibfnamefont{H.~J.} \bibnamefont{Schulz}},
  \bibinfo{year}{1998}, \bibinfo{journal}{Europhys. Lett.}
  \textbf{\bibinfo{volume}{44}}, \bibinfo{pages}{235}.

\bibitem[{\citenamefont{Zanchi and Schulz}(2000)}]{Zanchi00}
\bibinfo{author}{\bibnamefont{Zanchi}, \bibfnamefont{D.}}, and
  \bibinfo{author}{\bibfnamefont{H.~J.} \bibnamefont{Schulz}},
  \bibinfo{year}{2000}, \bibinfo{journal}{Phys. Rev. B}
  \textbf{\bibinfo{volume}{61}}, \bibinfo{pages}{13609}.

\end{thebibliography}

\end{document}